\documentclass[apjl,iop]{emulateapj}
\usepackage{comment}
\usepackage{ifthen}
\usepackage[switch]{lineno}

\newcommand{\forloop}[5][1]%
{%
\setcounter{#2}{#3}%
\ifthenelse{#4}%
	{%
	#5%
	\addtocounter{#2}{#1}%
	\forloop[#1]{#2}{\value{#2}}{#4}{#5}%
	}%
	{%
	}%
}%

\newcommand{\ctbd}[1]{}

\newcommand{\lc}{light curve}
\newcommand{\lcs}{light curves}
\newcommand{\Lc}{Light curve}

\newcommand{\masy}{\ensuremath{\rm mas\,yr^{-1}}}
\newcommand{\kms}{\ensuremath{\rm km\,s^{-1}}}
\newcommand{\ms}{\ensuremath{\rm m\,s^{-1}}}

\newcommand{\gcmc}{\ensuremath{\rm g\,cm^{-3}}}

\newcommand{\teff}{\ensuremath{T_{\rm eff}}}
\newcommand{\logg}{\ensuremath{\log{g}}}
\newcommand{\vsini}{\ensuremath{v \sin{i}}}
\newcommand{\feh}{\ensuremath{\rm [Fe/H]}}

\newcommand{\vmac}{\ensuremath{v_{\rm mac}}}
\newcommand{\vmic}{\ensuremath{v_{\rm mic}}}

\newcommand{\rsun}{\ensuremath{R_\sun}}
\newcommand{\msun}{\ensuremath{M_\sun}}
\newcommand{\lsun}{\ensuremath{L_\sun}}

\newcommand{\rstar}{\ensuremath{R_\star}}
\newcommand{\mstar}{\ensuremath{M_\star}}
\newcommand{\lstar}{\ensuremath{L_\star}}

\newcommand{\teffstar}{\ensuremath{T_{\rm eff\star}}}
\newcommand{\rhostar}{\ensuremath{\rho_\star}}
\newcommand{\loggstar}{\ensuremath{\log{g_{\star}}}}

\newcommand{\rpl}{\ensuremath{R_{p}}}
\newcommand{\mpl}{\ensuremath{M_{p}}}

\newcommand{\rhopl}{\ensuremath{\rho_{p}}}

\newcommand{\arstar}{\ensuremath{a/\rstar}}
\newcommand{\zrstar}{\ensuremath{\zeta/\rstar}}

\newcommand{\rjup}{\ensuremath{R_{\rm J}}}
\newcommand{\mjup}{\ensuremath{M_{\rm J}}}

\newcommand{\refsecl}[1]{\mbox{Section \ref{sec:#1}}}

\newcommand{\reftabl}[1]{Table~\ref{tab:#1}}

\newcommand{\loopand}{\ifnum\value{planetcounter}=2 and \else\fi}
\newcommand{\loopcomma}{\ifnum\value{planetcounter}<2 ,\else. \fi}
\newcommand{\loopcommanoperiod}{\ifnum\value{planetcounter}<2 ,\else \space\fi}
\newcommand{\loopcommanospace}{\ifnum\value{planetcounter}<2 ,\else \fi}

\newcommand{\hatcurhtrxxxxA}{HATS598-015}                      
\newcommand{\hatcurfieldxxxxA}{\ensuremath{string}}            
\newcommand{\hatcurCCraxxxxA}{\ensuremath{05^{\mathrm h}22^{\mathrm m}09.16{\mathrm s}}}                     
\newcommand{\hatcurCCdecxxxxA}{\ensuremath{-30{\arcdeg}58{\arcmin}15.0{\arcsec}}}                    
\newcommand{\hatcurCCmagxxxxA}{13.593}                         
\newcommand{\hatcurCCtwomassxxxxA}{2MASS~05220915-3058150}     
\newcommand{\hatcurCCgscxxxxA}{GSC~7048-01851}                 
\newcommand{\hatcurCCtassmvxxxxA}{\ensuremath{13.593\pm0.030}} 
\newcommand{\hatcurCCtassmvshortxxxxA}{\ensuremath{13.6}}      
\newcommand{\hatcurCCtassmBxxxxA}{\ensuremath{14.471\pm0.050}} 
\newcommand{\hatcurCCtassmBshortxxxxA}{\ensuremath{14.5}}      
\newcommand{\hatcurCCtassmIxxxxA}{\ensuremath{nff\pmnff}}      
\newcommand{\hatcurCCtassmIshortxxxxA}{\ensuremath{0.0}}       
\newcommand{\hatcurCCtassmgxxxxA}{\ensuremath{13.973\pm0.030}} 
\newcommand{\hatcurCCtassmgshortxxxxA}{\ensuremath{14.0}}      
\newcommand{\hatcurCCtassmrxxxxA}{\ensuremath{13.301\pm0.020}} 
\newcommand{\hatcurCCtassmrshortxxxxA}{\ensuremath{13.3}}      
\newcommand{\hatcurCCtassmixxxxA}{\ensuremath{13.094\pm0.070}} 
\newcommand{\hatcurCCtassmishortxxxxA}{\ensuremath{13.1}}      
\newcommand{\hatcurCCtwomassJmagxxxxA}{\ensuremath{12.064\pm0.026}} 
\newcommand{\hatcurCCtwomassHmagxxxxA}{\ensuremath{11.646\pm0.022}} 
\newcommand{\hatcurCCtwomassKmagxxxxA}{\ensuremath{11.556\pm0.023}} 
\newcommand{\hatcurCCcitJmagxxxxA}{\ensuremath{12.073\pm0.026}} 
\newcommand{\hatcurCCcitHmagxxxxA}{\ensuremath{11.640\pm0.023}} 
\newcommand{\hatcurCCcitKmagxxxxA}{\ensuremath{11.580\pm0.023}} 
\newcommand{\hatcurCCbbJmagxxxxA}{\ensuremath{12.134\pm0.028}} 
\newcommand{\hatcurCCbbHmagxxxxA}{\ensuremath{11.662\pm0.023}} 
\newcommand{\hatcurCCbbKmagxxxxA}{\ensuremath{11.600\pm0.023}} 
\newcommand{\hatcurCCesoJmagxxxxA}{\ensuremath{12.138\pm0.030}} 
\newcommand{\hatcurCCesoHmagxxxxA}{\ensuremath{11.658\pm0.027}} 
\newcommand{\hatcurCCesoKmagxxxxA}{\ensuremath{11.598\pm0.024}} 
\newcommand{\hatcurCCesoJHmagxxxxA}{\ensuremath{0.480\pm0.038}} 
\newcommand{\hatcurCCesoJKmagxxxxA}{\ensuremath{0.540\pm0.038}} 
\newcommand{\hatcurCCesoHKmagxxxxA}{\ensuremath{0.060\pm0.036}} 
\newcommand{\hatcurLCdipxxxxA}{\ensuremath{2.8}}               
\newcommand{\hatcurLCrprstarxxxxA}{\ensuremath{0.1487\pm0.0015}} 
\newcommand{\hatcurLCbsqxxxxA}{\ensuremath{0.023_{-0.016}^{+0.034}}} 
\newcommand{\hatcurLCimpxxxxA}{\ensuremath{0.152_{-0.070}^{+0.087}}} 
\newcommand{\hatcurLCzetaxxxxA}{\ensuremath{18.536\pm0.081}}   
\newcommand{\hatcurLCdurxxxxA}{\ensuremath{0.12441\pm0.00076}} 
\newcommand{\hatcurLCdurshortxxxxA}{\ensuremath{0.1244}}       
\newcommand{\hatcurLCdurhrxxxxA}{\ensuremath{2.986\pm0.018}}   
\newcommand{\hatcurLCdurhrshortxxxxA}{\ensuremath{2.986}}      
\newcommand{\hatcurLCqxxxxA}{\ensuremath{0.02830\pm0.00018}}   
\newcommand{\hatcurLCqshortxxxxA}{\ensuremath{0.028}}          
\newcommand{\hatcurLCingdurxxxxA}{\ensuremath{0.01644\pm0.00057}} 
\newcommand{\hatcurLCPxxxxA}{\ensuremath{4.3888481\pm0.0000049}} 
\newcommand{\hatcurLCPprecxxxxA}{\ensuremath{4.3888481}}       
\newcommand{\hatcurLCPshortxxxxA}{\ensuremath{4.3888}}         
\newcommand{\hatcurLCTxxxxA}{\ensuremath{2457627.31177\pm0.00028}} 
\newcommand{\hatcurLCTAxxxxA}{\ensuremath{2456543.2663\pm0.0012}} 
\newcommand{\hatcurLCTBxxxxA}{\ensuremath{2457688.75565\pm0.00030}} 
\newcommand{\hatcurLCrhoxxxxA}{\ensuremath{2.029_{-0.130}^{+0.070}}} 
\newcommand{\hatcurSMEiteffxxxxA}{\ensuremath{5268\pm86}}      
\newcommand{\hatcurSMEizfehxxxxA}{\ensuremath{0.160\pm0.052}}  
\newcommand{\hatcurSMEizfehshortxxxxA}{\ensuremath{0.16}}      
\newcommand{\hatcurSMEiloggxxxxA}{\ensuremath{4.75\pm0.17}}    
\newcommand{\hatcurSMEivsinxxxxA}{\ensuremath{0.5\pm1.2}}      
\newcommand{\hatcurSMEivmacxxxxA}{\ensuremath{3.21\pm0.13}}            
\newcommand{\hatcurSMEivmicxxxxA}{\ensuremath{0.827\pm0.039}}            
\newcommand{\hatcurSMEiiteffxxxxA}{\ensuremath{5099\pm61}}     
\newcommand{\hatcurSMEiizfehxxxxA}{\ensuremath{0.050\pm0.041}} 
\newcommand{\hatcurSMEiizfehshortxxxxA}{\ensuremath{0.05}}     
\newcommand{\hatcurSMEiiloggxxxxA}{\ensuremath{4.527\pm0.018}} 
\newcommand{\hatcurSMEiivsinxxxxA}{\ensuremath{1.11\pm0.82}}   
\newcommand{\hatcurSMEiivmacxxxxA}{\ensuremath{2.948\pm0.093}}            
\newcommand{\hatcurSMEiivmicxxxxA}{\ensuremath{0.747\pm0.030}}            
\newcommand{\hatcurLBizxxxxA}{\ensuremath{0.3050}}             
\newcommand{\hatcurLBiizxxxxA}{\ensuremath{0.2769}}            
\newcommand{\hatcurLBiixxxxA}{\ensuremath{0.3873}}             
\newcommand{\hatcurLBiiixxxxA}{\ensuremath{0.2588}}            
\newcommand{\hatcurLBiIxxxxA}{\ensuremath{0.3601}}             
\newcommand{\hatcurLBiiIxxxxA}{\ensuremath{0.2648}}            
\newcommand{\hatcurLBigxxxxA}{\ensuremath{0.7513}}             
\newcommand{\hatcurLBiigxxxxA}{\ensuremath{0.0782}}            
\newcommand{\hatcurLBirxxxxA}{\ensuremath{0.5115}}             
\newcommand{\hatcurLBiirxxxxA}{\ensuremath{0.2239}}            
\newcommand{\hatcurLBiRxxxxA}{\ensuremath{0.4772}}             
\newcommand{\hatcurLBiiRxxxxA}{\ensuremath{0.2344}}            
\newcommand{\hatcurLBikepxxxxA}{\ensuremath{0.1000}}           
\newcommand{\hatcurLBiikepxxxxA}{\ensuremath{0.1000}}          
\newcommand{\hatcurISOmxxxxA}{\ensuremath{0.830\pm0.020}}      
\newcommand{\hatcurISOmshortxxxxA}{\ensuremath{0.83}}          
\newcommand{\hatcurISOmlongxxxxA}{\ensuremath{0.830\pm0.020}}  
\newcommand{\hatcurISOrxxxxA}{\ensuremath{0.831_{-0.011}^{+0.015}}} 
\newcommand{\hatcurISOrshortxxxxA}{\ensuremath{0.83}}          
\newcommand{\hatcurISOrlongxxxxA}{\ensuremath{0.831_{-0.011}^{+0.015}}} 
\newcommand{\hatcurISOrhoxxxxA}{\ensuremath{2.049_{-0.101}^{+0.055}}} 
\newcommand{\hatcurISOrholongxxxxA}{\ensuremath{2.049_{-0.101}^{+0.055}}} 
\newcommand{\hatcurISOloggxxxxA}{\ensuremath{4.519\pm0.012}}   
\newcommand{\hatcurISOlumxxxxA}{\ensuremath{0.426\pm0.028}}    
\newcommand{\hatcurISOlumshortxxxxA}{\ensuremath{0.43}}        
\newcommand{\hatcurISOmvxxxxA}{\ensuremath{5.897\pm0.081}}     
\newcommand{\hatcurISOvixxxxA}{\ensuremath{0.875\pm0.013}}     
\newcommand{\hatcurISOagexxxxA}{\ensuremath{10.6\pm1.7}}       
\newcommand{\hatcurISOsigmaxxxxA}{\ensuremath{0.00050\pm0.00013}} 
\newcommand{\hatcurISOMJxxxxA}{\ensuremath{4.398\pm0.056}}     
\newcommand{\hatcurISOMHxxxxA}{\ensuremath{3.928\pm0.050}}     
\newcommand{\hatcurISOMKxxxxA}{\ensuremath{3.851\pm0.049}}     
\newcommand{\hatcurISOJKxxxxA}{\ensuremath{0.550\pm0.010}}     
\newcommand{\hatcurISOspecxxxxA}{G}                            
\newcommand{\hatcurRVKxxxxA}{\ensuremath{32.9\pm8.4}}          
\newcommand{\hatcurRVrkxxxxA}{\ensuremath{0\pm0}}              
\newcommand{\hatcurRVrhxxxxA}{\ensuremath{0\pm0}}              
\newcommand{\hatcurRVkxxxxA}{\ensuremath{0\pm0}}               
\newcommand{\hatcurRVhxxxxA}{\ensuremath{0\pm0}}               
\newcommand{\hatcurRVtronexxxxA}{\ensuremath{0\pm0}}           
\newcommand{\hatcurRVtrtwoxxxxA}{\ensuremath{0\pm0}}           
\newcommand{\hatcurRVgammaAxxxxA}{\ensuremath{22077.5\pm8.3}}  
\newcommand{\hatcurRVjitterAxxxxA}{\ensuremath{25.0\pm7.2}}    
\newcommand{\hatcurRVjittertwosiglimAxxxxA}{\ensuremath{<38.3}} 
\newcommand{\hatcurRVfitrmsAxxxxA}{\ensuremath{0.0}}           
\newcommand{\hatcurRVgammaBxxxxA}{\ensuremath{22052.7\pm7.7}}  
\newcommand{\hatcurRVjitterBxxxxA}{\ensuremath{0\pm11}}        
\newcommand{\hatcurRVjittertwosiglimBxxxxA}{\ensuremath{<28.1}} 
\newcommand{\hatcurRVfitrmsBxxxxA}{\ensuremath{0.0}}           
\newcommand{\hatcurRVeccenxxxxA}{\ensuremath{0\pm0}}           
\newcommand{\hatcurRVeccentwosiglimxxxxA}{\ensuremath{<0.000}} 
\newcommand{\hatcurRVomegaxxxxA}{\ensuremath{0\pm0}}           
\newcommand{\hatcurPPixxxxA}{\ensuremath{89.32\pm0.34}}        
\newcommand{\hatcurPPgxxxxA}{\ensuremath{4.0\pm1.1}}           
\newcommand{\hatcurPPloggxxxxA}{\ensuremath{2.60\pm0.13}}      
\newcommand{\hatcurPParxxxxA}{\ensuremath{12.77_{-0.21}^{+0.11}}} 
\newcommand{\hatcurPParelxxxxA}{\ensuremath{0.04932\pm0.00039}} 
\newcommand{\hatcurPPrhoxxxxA}{\ensuremath{0.164\pm0.046}}     
\newcommand{\hatcurPPmxxxxA}{\ensuremath{0.235\pm0.060}}       
\newcommand{\hatcurPPmshortxxxxA}{\ensuremath{0.24}}           
\newcommand{\hatcurPPmlongxxxxA}{\ensuremath{0.235\pm0.060}}   
\newcommand{\hatcurPPmexxxxA}{\ensuremath{75\pm19}}            
\newcommand{\hatcurPPmeshortxxxxA}{\ensuremath{74.7}}          
\newcommand{\hatcurPPmelongxxxxA}{\ensuremath{75\pm19}}        
\newcommand{\hatcurPPrxxxxA}{\ensuremath{1.202\pm0.028}}       
\newcommand{\hatcurPPrshortxxxxA}{\ensuremath{1.20}}           
\newcommand{\hatcurPPrlongxxxxA}{\ensuremath{1.202\pm0.028}}   
\newcommand{\hatcurPPrexxxxA}{\ensuremath{13.48\pm0.31}}       
\newcommand{\hatcurPPreshortxxxxA}{\ensuremath{13.5}}          
\newcommand{\hatcurPPrelongxxxxA}{\ensuremath{13.48\pm0.31}}   
\newcommand{\hatcurPPmrcorrxxxxA}{\ensuremath{-0.04}}          
\newcommand{\hatcurPPteffxxxxA}{\ensuremath{1014\pm14}}        
\newcommand{\hatcurPPthetaxxxxA}{\ensuremath{0.0231\pm0.0060}} 
\newcommand{\hatcurPPfluxperixxxxA}{\ensuremath{2.39\pm0.13}}  
\newcommand{\hatcurPPfluxperidimxxxxA}{\ensuremath{8}}         
\newcommand{\hatcurPPfluxapxxxxA}{\ensuremath{2.39\pm0.13}}    
\newcommand{\hatcurPPfluxapdimxxxxA}{\ensuremath{8}}           
\newcommand{\hatcurPPfluxavgxxxxA}{\ensuremath{2.39\pm0.13}}   
\newcommand{\hatcurPPfluxavgdimxxxxA}{\ensuremath{8}}          
\newcommand{\hatcurPPfluxavglogxxxxA}{\ensuremath{8.378\pm0.024}} 
\newcommand{\hatcurXsecphasexxxxA}{\ensuremath{0\pm0}}         
\newcommand{\hatcurXsecondaryxxxxA}{\ensuremath{2457629.50620\pm0.00028}} 
\newcommand{\hatcurXsecdurxxxxA}{\ensuremath{0.12441\pm0.00076}} 
\newcommand{\hatcurXsecingdurxxxxA}{\ensuremath{0.01644\pm0.00057}} 
\newcommand{\hatcurPPphiconjxxxxA}{\ensuremath{0\pm0}}         
\newcommand{\hatcurPPperixxxxA}{\ensuremath{2457626.21456\pm0.00028}} 
\newcommand{\hatcurPPaequivxxxxA}{\ensuremath{0.0756\pm0.0020}} 
\newcommand{\hatcurPPtcircxxxxA}{\ensuremath{148_{-35}^{+46}}} 
\newcommand{\hatcurPPtinfallxxxxA}{\ensuremath{49000_{-11000}^{+15000}}} 
\newcommand{\hatcurXdistxxxxA}{\ensuremath{354.4\pm9.0}}       
\newcommand{\hatcurXAvxxxxA}{\ensuremath{0.000\pm0.018}}       
\newcommand{\hatcurXdistredxxxxA}{\ensuremath{351.4\pm9.3}}    
\newcommand{\hatcurXEBVxxxxA}{\ensuremath{0.0000\pm0.0059}}    
\newcommand{\hatcurXmvisoredxxxxA}{\ensuremath{13.631\pm0.027}} 
\newcommand{\hatcurXmiisoredxxxxA}{\ensuremath{12.753\pm0.019}} 
\newcommand{\hatcurXmjisoredxxxxA}{\ensuremath{12.128\pm0.014}} 
\newcommand{\hatcurXmhisoredxxxxA}{\ensuremath{11.658\pm0.016}} 
\newcommand{\hatcurXmkisoredxxxxA}{\ensuremath{11.580\pm0.016}} 
\newcommand{\hatcurXviisoredxxxxA}{\ensuremath{0.878\pm0.012}} 
\newcommand{\hatcurXvkisoredxxxxA}{\ensuremath{2.051\pm0.032}} 
\newcommand{\hatcurXjhisoredxxxxA}{\ensuremath{0.4700\pm0.0061}} 
\newcommand{\hatcurXjkisoredxxxxA}{\ensuremath{0.5480\pm0.0072}} 
\newcommand{\hatcurCCpmraxxxxA}{\ensuremath{9.8\pm1.9}}        
\newcommand{\hatcurCCpmdecxxxxA}{\ensuremath{7.9\pm1.7}}       
\newcommand{\hatcurCCpmxxxxA}{\ensuremath{12.6\pm2.5}}         

\newcommand{\hatcurhtrxxxxB}{HATS599-006}                      
\newcommand{\hatcurfieldxxxxB}{\ensuremath{string}}            
\newcommand{\hatcurCCraxxxxB}{\ensuremath{05^{\mathrm h}37^{\mathrm m}18.41{\mathrm s}}}                     
\newcommand{\hatcurCCdecxxxxB}{\ensuremath{-27{\arcdeg}58{\arcmin}21.4{\arcsec}}}                    
\newcommand{\hatcurCCmagxxxxB}{14.428}                         
\newcommand{\hatcurCCtwomassxxxxB}{2MASS~05371842-2758214}     
\newcommand{\hatcurCCgscxxxxB}{GSC~6497-00040}                 
\newcommand{\hatcurCCtassmvxxxxB}{\ensuremath{14.428\pm0.010}} 
\newcommand{\hatcurCCtassmvshortxxxxB}{\ensuremath{14.4}}      
\newcommand{\hatcurCCtassmBxxxxB}{\ensuremath{15.487\pm0.020}} 
\newcommand{\hatcurCCtassmBshortxxxxB}{\ensuremath{15.5}}      
\newcommand{\hatcurCCtassmIxxxxB}{\ensuremath{nff\pmnff}}      
\newcommand{\hatcurCCtassmIshortxxxxB}{\ensuremath{0.0}}       
\newcommand{\hatcurCCtassmgxxxxB}{\ensuremath{14.933\pm0.010}} 
\newcommand{\hatcurCCtassmgshortxxxxB}{\ensuremath{14.9}}      
\newcommand{\hatcurCCtassmrxxxxB}{\ensuremath{14.086\pm0.010}} 
\newcommand{\hatcurCCtassmrshortxxxxB}{\ensuremath{14.1}}      
\newcommand{\hatcurCCtassmixxxxB}{\ensuremath{13.794\pm0.030}} 
\newcommand{\hatcurCCtassmishortxxxxB}{\ensuremath{13.8}}      
\newcommand{\hatcurCCtwomassJmagxxxxB}{\ensuremath{12.699\pm0.023}} 
\newcommand{\hatcurCCtwomassHmagxxxxB}{\ensuremath{12.234\pm0.022}} 
\newcommand{\hatcurCCtwomassKmagxxxxB}{\ensuremath{12.188\pm0.030}} 
\newcommand{\hatcurCCcitJmagxxxxB}{\ensuremath{12.708\pm0.024}} 
\newcommand{\hatcurCCcitHmagxxxxB}{\ensuremath{12.229\pm0.023}} 
\newcommand{\hatcurCCcitKmagxxxxB}{\ensuremath{12.212\pm0.030}} 
\newcommand{\hatcurCCbbJmagxxxxB}{\ensuremath{12.769\pm0.026}} 
\newcommand{\hatcurCCbbHmagxxxxB}{\ensuremath{12.250\pm0.023}} 
\newcommand{\hatcurCCbbKmagxxxxB}{\ensuremath{12.232\pm0.030}} 
\newcommand{\hatcurCCesoJmagxxxxB}{\ensuremath{12.773\pm0.028}} 
\newcommand{\hatcurCCesoHmagxxxxB}{\ensuremath{12.243\pm0.026}} 
\newcommand{\hatcurCCesoKmagxxxxB}{\ensuremath{12.230\pm0.031}} 
\newcommand{\hatcurCCesoJHmagxxxxB}{\ensuremath{0.529\pm0.035}} 
\newcommand{\hatcurCCesoJKmagxxxxB}{\ensuremath{0.544\pm0.041}} 
\newcommand{\hatcurCCesoHKmagxxxxB}{\ensuremath{0.014\pm0.041}} 
\newcommand{\hatcurLCdipxxxxB}{\ensuremath{13.0}}              
\newcommand{\hatcurLCrprstarxxxxB}{\ensuremath{0.129\pm0.010}} 
\newcommand{\hatcurLCbsqxxxxB}{\ensuremath{0.743_{-0.035}^{+0.051}}} 
\newcommand{\hatcurLCimpxxxxB}{\ensuremath{0.862_{-0.021}^{+0.029}}} 
\newcommand{\hatcurLCzetaxxxxB}{\ensuremath{42.0_{-1.7}^{+3.3}}} 
\newcommand{\hatcurLCdurxxxxB}{\ensuremath{0.0688\pm0.0017}}   
\newcommand{\hatcurLCdurshortxxxxB}{\ensuremath{0.0688}}       
\newcommand{\hatcurLCdurhrxxxxB}{\ensuremath{1.652\pm0.040}}   
\newcommand{\hatcurLCdurhrshortxxxxB}{\ensuremath{1.652}}      
\newcommand{\hatcurLCqxxxxB}{\ensuremath{0.02510\pm0.00062}}   
\newcommand{\hatcurLCqshortxxxxB}{\ensuremath{0.025}}          
\newcommand{\hatcurLCingdurxxxxB}{\ensuremath{0.029\pm0.034}}  
\newcommand{\hatcurLCPxxxxB}{\ensuremath{2.7439004\pm0.0000032}} 
\newcommand{\hatcurLCPprecxxxxB}{\ensuremath{2.7439004}}       
\newcommand{\hatcurLCPshortxxxxB}{\ensuremath{2.7439}}         
\newcommand{\hatcurLCTxxxxB}{\ensuremath{2456931.11384\pm0.00061}} 
\newcommand{\hatcurLCTAxxxxB}{\ensuremath{2455946.0536\pm0.0013}} 
\newcommand{\hatcurLCTBxxxxB}{\ensuremath{2457353.67449\pm0.00081}} 
\newcommand{\hatcurLCrhoxxxxB}{\ensuremath{1.59_{-0.28}^{+0.46}}} 
\newcommand{\hatcurSMEiteffxxxxB}{\ensuremath{5080\pm100}}     
\newcommand{\hatcurSMEizfehxxxxB}{\ensuremath{0.320\pm0.071}}  
\newcommand{\hatcurSMEizfehshortxxxxB}{\ensuremath{0.32}}      
\newcommand{\hatcurSMEiloggxxxxB}{\ensuremath{4.75\pm0.19}}    
\newcommand{\hatcurSMEivsinxxxxB}{\ensuremath{0.5\pm1.1}}      
\newcommand{\hatcurSMEivmacxxxxB}{\ensuremath{2.92\pm0.15}}            
\newcommand{\hatcurSMEivmicxxxxB}{\ensuremath{0.737\pm0.051}}            
\newcommand{\hatcurSMEiiteffxxxxB}{\ensuremath{4972\pm66}}     
\newcommand{\hatcurSMEiizfehxxxxB}{\ensuremath{0.320\pm0.051}} 
\newcommand{\hatcurSMEiizfehshortxxxxB}{\ensuremath{0.32}}     
\newcommand{\hatcurSMEiiloggxxxxB}{\ensuremath{4.485\pm0.045}} 
\newcommand{\hatcurSMEiivsinxxxxB}{\ensuremath{1.50\pm0.91}}   
\newcommand{\hatcurSMEiivmacxxxxB}{\ensuremath{2.75\pm0.10}}            
\newcommand{\hatcurSMEiivmicxxxxB}{\ensuremath{0.681\pm0.036}}            
\newcommand{\hatcurLBizxxxxB}{\ensuremath{0.3262}}             
\newcommand{\hatcurLBiizxxxxB}{\ensuremath{0.2723}}            
\newcommand{\hatcurLBiixxxxB}{\ensuremath{0.4195}}             
\newcommand{\hatcurLBiiixxxxB}{\ensuremath{0.2444}}            
\newcommand{\hatcurLBiIxxxxB}{\ensuremath{0.3882}}             
\newcommand{\hatcurLBiiIxxxxB}{\ensuremath{0.2535}}            
\newcommand{\hatcurLBigxxxxB}{\ensuremath{0.8140}}             
\newcommand{\hatcurLBiigxxxxB}{\ensuremath{0.0272}}            
\newcommand{\hatcurLBirxxxxB}{\ensuremath{0.5603}}             
\newcommand{\hatcurLBiirxxxxB}{\ensuremath{0.1933}}            
\newcommand{\hatcurLBiRxxxxB}{\ensuremath{0.5216}}             
\newcommand{\hatcurLBiiRxxxxB}{\ensuremath{0.2081}}            
\newcommand{\hatcurLBikepxxxxB}{\ensuremath{0.1000}}           
\newcommand{\hatcurLBiikepxxxxB}{\ensuremath{0.1000}}          
\newcommand{\hatcurISOmxxxxB}{\ensuremath{0.860\pm0.021}}      
\newcommand{\hatcurISOmshortxxxxB}{\ensuremath{0.86}}          
\newcommand{\hatcurISOmlongxxxxB}{\ensuremath{0.860\pm0.021}}  
\newcommand{\hatcurISOrxxxxB}{\ensuremath{0.847\pm0.036}}      
\newcommand{\hatcurISOrshortxxxxB}{\ensuremath{0.85}}          
\newcommand{\hatcurISOrlongxxxxB}{\ensuremath{0.847\pm0.036}}  
\newcommand{\hatcurISOrhoxxxxB}{\ensuremath{1.98\pm0.25}}      
\newcommand{\hatcurISOrholongxxxxB}{\ensuremath{1.98\pm0.25}}  
\newcommand{\hatcurISOloggxxxxB}{\ensuremath{4.514\pm0.036}}   
\newcommand{\hatcurISOlumxxxxB}{\ensuremath{0.400\pm0.051}}    
\newcommand{\hatcurISOlumshortxxxxB}{\ensuremath{0.40}}        
\newcommand{\hatcurISOmvxxxxB}{\ensuremath{6.01\pm0.16}}       
\newcommand{\hatcurISOvixxxxB}{\ensuremath{0.925\pm0.021}}     
\newcommand{\hatcurISOagexxxxB}{\ensuremath{9.7_{-4.0}^{+2.4}}} 
\newcommand{\hatcurISOsigmaxxxxB}{\ensuremath{0.00090\pm0.00024}} 
\newcommand{\hatcurISOMJxxxxB}{\ensuremath{4.41\pm0.12}}       
\newcommand{\hatcurISOMHxxxxB}{\ensuremath{3.93\pm0.11}}       
\newcommand{\hatcurISOMKxxxxB}{\ensuremath{3.85\pm0.11}}       
\newcommand{\hatcurISOJKxxxxB}{\ensuremath{0.41\pm0.25}}       
\newcommand{\hatcurISOspecxxxxB}{G}                            
\newcommand{\hatcurRVKxxxxB}{\ensuremath{90\pm17}}             
\newcommand{\hatcurRVrkxxxxB}{\ensuremath{0\pm0}}              
\newcommand{\hatcurRVrhxxxxB}{\ensuremath{0\pm0}}              
\newcommand{\hatcurRVkxxxxB}{\ensuremath{0\pm0}}               
\newcommand{\hatcurRVhxxxxB}{\ensuremath{0\pm0}}               
\newcommand{\hatcurRVtronexxxxB}{\ensuremath{0\pm0}}           
\newcommand{\hatcurRVtrtwoxxxxB}{\ensuremath{0\pm0}}           
\newcommand{\hatcurRVgammaxxxxB}{\ensuremath{44082\pm14}}      
\newcommand{\hatcurRVjitterxxxxB}{\ensuremath{49\pm11}}        
\newcommand{\hatcurRVjittertwosiglimxxxxB}{\ensuremath{<72.0}} 
\newcommand{\hatcurRVfitrmsxxxxB}{\ensuremath{.1fym}}          %
\newcommand{\hatcurRVeccenxxxxB}{\ensuremath{0\pm0}}           
\newcommand{\hatcurRVeccentwosiglimxxxxB}{\ensuremath{<0.000}} 
\newcommand{\hatcurRVomegaxxxxB}{\ensuremath{0\pm0}}           
\newcommand{\hatcurPPixxxxB}{\ensuremath{84.65\pm0.38}}        
\newcommand{\hatcurPPgxxxxB}{\ensuremath{11.9\pm3.3}}          
\newcommand{\hatcurPPloggxxxxB}{\ensuremath{3.08\pm0.13}}      
\newcommand{\hatcurPParxxxxB}{\ensuremath{9.24\pm0.38}}        
\newcommand{\hatcurPParelxxxxB}{\ensuremath{0.03649\pm0.00030}} 
\newcommand{\hatcurPPrhoxxxxB}{\ensuremath{0.56\pm0.19}}       
\newcommand{\hatcurPPmxxxxB}{\ensuremath{0.56\pm0.11}}         
\newcommand{\hatcurPPmshortxxxxB}{\ensuremath{0.56}}           
\newcommand{\hatcurPPmlongxxxxB}{\ensuremath{0.56\pm0.11}}     
\newcommand{\hatcurPPmexxxxB}{\ensuremath{179\pm34}}           
\newcommand{\hatcurPPmeshortxxxxB}{\ensuremath{179.3}}         
\newcommand{\hatcurPPmelongxxxxB}{\ensuremath{179\pm34}}       
\newcommand{\hatcurPPrxxxxB}{\ensuremath{1.067_{-0.071}^{+0.125}}} 
\newcommand{\hatcurPPrshortxxxxB}{\ensuremath{1.07}}           
\newcommand{\hatcurPPrlongxxxxB}{\ensuremath{1.067_{-0.071}^{+0.125}}} 
\newcommand{\hatcurPPrexxxxB}{\ensuremath{11.96_{-0.80}^{+1.40}}} 
\newcommand{\hatcurPPreshortxxxxB}{\ensuremath{12.0}}          
\newcommand{\hatcurPPrelongxxxxB}{\ensuremath{11.96_{-0.80}^{+1.40}}} 
\newcommand{\hatcurPPmrcorrxxxxB}{\ensuremath{0.01}}           
\newcommand{\hatcurPPteffxxxxB}{\ensuremath{1161\pm34}}        
\newcommand{\hatcurPPthetaxxxxB}{\ensuremath{0.0438\pm0.0095}} 
\newcommand{\hatcurPPfluxperixxxxB}{\ensuremath{4.10\pm0.50}}  
\newcommand{\hatcurPPfluxperidimxxxxB}{\ensuremath{8}}         
\newcommand{\hatcurPPfluxapxxxxB}{\ensuremath{4.10\pm0.50}}    
\newcommand{\hatcurPPfluxapdimxxxxB}{\ensuremath{8}}           
\newcommand{\hatcurPPfluxavgxxxxB}{\ensuremath{4.10\pm0.50}}   
\newcommand{\hatcurPPfluxavgdimxxxxB}{\ensuremath{8}}          
\newcommand{\hatcurPPfluxavglogxxxxB}{\ensuremath{8.613\pm0.051}} 
\newcommand{\hatcurXsecphasexxxxB}{\ensuremath{0\pm0}}         
\newcommand{\hatcurXsecondaryxxxxB}{\ensuremath{2456932.48579\pm0.00061}} 
\newcommand{\hatcurXsecdurxxxxB}{\ensuremath{0.0693\pm0.0040}} 
\newcommand{\hatcurXsecingdurxxxxB}{\ensuremath{0.0286\pm0.0063}} 
\newcommand{\hatcurPPphiconjxxxxB}{\ensuremath{0\pm0}}         
\newcommand{\hatcurPPperixxxxB}{\ensuremath{2456930.42786\pm0.00061}} 
\newcommand{\hatcurPPaequivxxxxB}{\ensuremath{0.0577\pm0.0034}} 
\newcommand{\hatcurPPtcircxxxxB}{\ensuremath{87\pm43}}         
\newcommand{\hatcurPPtinfallxxxxB}{\ensuremath{2710_{-630}^{+840}}} 
\newcommand{\hatcurXdistxxxxB}{\ensuremath{474\pm24}}          
\newcommand{\hatcurXAvxxxxB}{\ensuremath{0.095\pm0.064}}       
\newcommand{\hatcurXdistredxxxxB}{\ensuremath{463\pm23}}       
\newcommand{\hatcurXEBVxxxxB}{\ensuremath{0.031\pm0.021}}      
\newcommand{\hatcurXmvisoredxxxxB}{\ensuremath{14.428\pm0.010}} 
\newcommand{\hatcurXmiisoredxxxxB}{\ensuremath{13.453\pm0.016}} 
\newcommand{\hatcurXmjisoredxxxxB}{\ensuremath{12.763\pm0.015}} 
\newcommand{\hatcurXmhisoredxxxxB}{\ensuremath{12.277\pm0.019}} 
\newcommand{\hatcurXmkisoredxxxxB}{\ensuremath{12.189\pm0.020}} 
\newcommand{\hatcurXviisoredxxxxB}{\ensuremath{0.975\pm0.016}} 
\newcommand{\hatcurXvkisoredxxxxB}{\ensuremath{2.239\pm0.024}} 
\newcommand{\hatcurXjhisoredxxxxB}{\ensuremath{0.4860\pm0.0069}} 
\newcommand{\hatcurXjkisoredxxxxB}{\ensuremath{0.5750\pm0.0064}} 
\newcommand{\hatcurCCpmraxxxxB}{\ensuremath{-2.2\pm1.3}}       
\newcommand{\hatcurCCpmdecxxxxB}{\ensuremath{-3.1\pm1.6}}      
\newcommand{\hatcurCCpmxxxxB}{\ensuremath{3.8\pm2.1}}          

\newcommand{\hatcurhtrxxxxC}{HATS554-008}                      
\newcommand{\hatcurfieldxxxxC}{\ensuremath{string}}            
\newcommand{\hatcurCCraxxxxC}{\ensuremath{06^{\mathrm h}47^{\mathrm m}58.63{\mathrm s}}}                     
\newcommand{\hatcurCCdecxxxxC}{\ensuremath{-21{\arcdeg}54{\arcmin}38.5{\arcsec}}}                    
\newcommand{\hatcurCCmagxxxxC}{13.307}                         
\newcommand{\hatcurCCtwomassxxxxC}{2MASS~06475862-2154385}     
\newcommand{\hatcurCCgscxxxxC}{GSC~5961-02383}                 
\newcommand{\hatcurCCtassmvxxxxC}{\ensuremath{13.307\pm0.050}} 
\newcommand{\hatcurCCtassmvshortxxxxC}{\ensuremath{13.3}}      
\newcommand{\hatcurCCtassmBxxxxC}{\ensuremath{13.845\pm0.020}} 
\newcommand{\hatcurCCtassmBshortxxxxC}{\ensuremath{13.8}}      
\newcommand{\hatcurCCtassmIxxxxC}{\ensuremath{nff\pmnff}}      
\newcommand{\hatcurCCtassmIshortxxxxC}{\ensuremath{0.0}}       
\newcommand{\hatcurCCtassmgxxxxC}{\ensuremath{13.550\pm0.020}} 
\newcommand{\hatcurCCtassmgshortxxxxC}{\ensuremath{13.6}}      
\newcommand{\hatcurCCtassmrxxxxC}{\ensuremath{13.201\pm0.060}} 
\newcommand{\hatcurCCtassmrshortxxxxC}{\ensuremath{13.2}}      
\newcommand{\hatcurCCtassmixxxxC}{\ensuremath{13.162\pm0.040}} 
\newcommand{\hatcurCCtassmishortxxxxC}{\ensuremath{13.2}}      
\newcommand{\hatcurCCtwomassJmagxxxxC}{\ensuremath{12.364\pm0.024}} 
\newcommand{\hatcurCCtwomassHmagxxxxC}{\ensuremath{12.155\pm0.024}} 
\newcommand{\hatcurCCtwomassKmagxxxxC}{\ensuremath{12.137\pm0.021}} 
\newcommand{\hatcurCCcitJmagxxxxC}{\ensuremath{12.388\pm0.024}} 
\newcommand{\hatcurCCcitHmagxxxxC}{\ensuremath{12.151\pm0.024}} 
\newcommand{\hatcurCCcitKmagxxxxC}{\ensuremath{12.161\pm0.021}} 
\newcommand{\hatcurCCbbJmagxxxxC}{\ensuremath{12.426\pm0.026}} 
\newcommand{\hatcurCCbbHmagxxxxC}{\ensuremath{12.171\pm0.025}} 
\newcommand{\hatcurCCbbKmagxxxxC}{\ensuremath{12.181\pm0.021}} 
\newcommand{\hatcurCCesoJmagxxxxC}{\ensuremath{12.427\pm0.026}} 
\newcommand{\hatcurCCesoHmagxxxxC}{\ensuremath{12.164\pm0.027}} 
\newcommand{\hatcurCCesoKmagxxxxC}{\ensuremath{12.181\pm0.021}} 
\newcommand{\hatcurCCesoJHmagxxxxC}{\ensuremath{0.2630\pm0.0070}} 
\newcommand{\hatcurCCesoJKmagxxxxC}{\ensuremath{0.246\pm0.034}} 
\newcommand{\hatcurCCesoHKmagxxxxC}{\ensuremath{-0.017\pm0.035}} 
\newcommand{\hatcurLCdipxxxxC}{\ensuremath{9.7}}               
\newcommand{\hatcurLCrprstarxxxxC}{\ensuremath{0.1004\pm0.0042}} 
\newcommand{\hatcurLCbsqxxxxC}{\ensuremath{0.475_{-0.060}^{+0.049}}} 
\newcommand{\hatcurLCimpxxxxC}{\ensuremath{0.689_{-0.045}^{+0.034}}} 
\newcommand{\hatcurLCzetaxxxxC}{\ensuremath{18.65\pm0.25}}     
\newcommand{\hatcurLCdurxxxxC}{\ensuremath{0.1269\pm0.0021}}   
\newcommand{\hatcurLCdurshortxxxxC}{\ensuremath{0.1269}}       
\newcommand{\hatcurLCdurhrxxxxC}{\ensuremath{3.045\pm0.051}}   
\newcommand{\hatcurLCdurhrshortxxxxC}{\ensuremath{3.045}}      
\newcommand{\hatcurLCqxxxxC}{\ensuremath{0.03030\pm0.00051}}   
\newcommand{\hatcurLCqshortxxxxC}{\ensuremath{0.030}}          
\newcommand{\hatcurLCingdurxxxxC}{\ensuremath{0.0207\pm0.0023}} 
\newcommand{\hatcurLCPxxxxC}{\ensuremath{4.1876244\pm0.0000056}} 
\newcommand{\hatcurLCPprecxxxxC}{\ensuremath{4.1876244}}       
\newcommand{\hatcurLCPshortxxxxC}{\ensuremath{4.1876}}         
\newcommand{\hatcurLCTxxxxC}{\ensuremath{2456731.19533\pm0.00073}} 
\newcommand{\hatcurLCTAxxxxC}{\ensuremath{2455198.5248\pm0.0019}} 
\newcommand{\hatcurLCTBxxxxC}{\ensuremath{2457095.5187\pm0.0010}} 
\newcommand{\hatcurLChatnetmxxxxC}{\ensuremath{13.182730\pm0.000078}} 
\newcommand{\hatcurLCiblendxxxxC}{\ensuremath{0.810\pm0.075}}  
\newcommand{\hatcurLCrhoxxxxC}{\ensuremath{0.81_{-0.11}^{+0.16}}} 
\newcommand{\hatcurSMEiteffxxxxC}{\ensuremath{6740\pm150}}     
\newcommand{\hatcurSMEizfehxxxxC}{\ensuremath{0.190\pm0.087}}  
\newcommand{\hatcurSMEizfehshortxxxxC}{\ensuremath{0.19}}      
\newcommand{\hatcurSMEiloggxxxxC}{\ensuremath{4.52\pm0.23}}    
\newcommand{\hatcurSMEivsinxxxxC}{\ensuremath{9.89\pm0.38}}    
\newcommand{\hatcurSMEivmacxxxxC}{\ensuremath{5.5}}            
\newcommand{\hatcurSMEivmicxxxxC}{\ensuremath{2.1}}            
\newcommand{\hatcurSMEiiteffxxxxC}{\ensuremath{6450\pm110}}    
\newcommand{\hatcurSMEiizfehxxxxC}{\ensuremath{0.020\pm0.068}} 
\newcommand{\hatcurSMEiizfehshortxxxxC}{\ensuremath{0.02}}     
\newcommand{\hatcurSMEiiloggxxxxC}{\ensuremath{4.286\pm0.026}} 
\newcommand{\hatcurSMEiivsinxxxxC}{\ensuremath{9.90\pm0.40}}   
\newcommand{\hatcurSMEiivmacxxxxC}{\ensuremath{5.03}}          
\newcommand{\hatcurSMEiivmicxxxxC}{\ensuremath{1.69}}          
\newcommand{\hatcurLBizxxxxC}{\ensuremath{0.1286}}             
\newcommand{\hatcurLBiizxxxxC}{\ensuremath{0.3632}}            
\newcommand{\hatcurLBiixxxxC}{\ensuremath{0.1791}}             
\newcommand{\hatcurLBiiixxxxC}{\ensuremath{0.3719}}            
\newcommand{\hatcurLBiIxxxxC}{\ensuremath{0.1610}}             
\newcommand{\hatcurLBiiIxxxxC}{\ensuremath{0.3702}}            
\newcommand{\hatcurLBigxxxxC}{\ensuremath{0.4106}}             
\newcommand{\hatcurLBiigxxxxC}{\ensuremath{0.3344}}            
\newcommand{\hatcurLBirxxxxC}{\ensuremath{0.2511}}             
\newcommand{\hatcurLBiirxxxxC}{\ensuremath{0.3818}}            
\newcommand{\hatcurLBiRxxxxC}{\ensuremath{0.2306}}             
\newcommand{\hatcurLBiiRxxxxC}{\ensuremath{0.3807}}            
\newcommand{\hatcurLBikepxxxxC}{\ensuremath{0.1000}}           
\newcommand{\hatcurLBiikepxxxxC}{\ensuremath{0.1000}}          
\newcommand{\hatcurISOmxxxxC}{\ensuremath{1.272\pm0.048}}      
\newcommand{\hatcurISOmshortxxxxC}{\ensuremath{1.27}}          
\newcommand{\hatcurISOmlongxxxxC}{\ensuremath{1.272\pm0.048}}  
\newcommand{\hatcurISOrxxxxC}{\ensuremath{1.315\pm0.064}}      
\newcommand{\hatcurISOrshortxxxxC}{\ensuremath{1.32}}          
\newcommand{\hatcurISOrlongxxxxC}{\ensuremath{1.315\pm0.064}}  
\newcommand{\hatcurISOrhoxxxxC}{\ensuremath{0.79\pm0.10}}      
\newcommand{\hatcurISOrholongxxxxC}{\ensuremath{0.79\pm0.10}}  
\newcommand{\hatcurISOloggxxxxC}{\ensuremath{4.305\pm0.036}}   
\newcommand{\hatcurISOlumxxxxC}{\ensuremath{2.66\pm0.36}}      
\newcommand{\hatcurISOlumshortxxxxC}{\ensuremath{2.66}}        
\newcommand{\hatcurISOmvxxxxC}{\ensuremath{3.68\pm0.16}}       
\newcommand{\hatcurISOvixxxxC}{\ensuremath{0.515\pm0.027}}     
\newcommand{\hatcurISOagexxxxC}{\ensuremath{1.52\pm0.70}}      
\newcommand{\hatcurISOsigmaxxxxC}{\ensuremath{0.00060\pm0.00015}} 
\newcommand{\hatcurISOMJxxxxC}{\ensuremath{2.85\pm0.12}}       
\newcommand{\hatcurISOMHxxxxC}{\ensuremath{2.62\pm0.11}}       
\newcommand{\hatcurISOMKxxxxC}{\ensuremath{2.58\pm0.11}}       
\newcommand{\hatcurISOJKxxxxC}{\ensuremath{0.270\pm0.020}}     
\newcommand{\hatcurISOspecxxxxC}{F}                            
\newcommand{\hatcurRVKxxxxC}{\ensuremath{75\pm16}}             
\newcommand{\hatcurRVrkxxxxC}{\ensuremath{0\pm0}}              
\newcommand{\hatcurRVrhxxxxC}{\ensuremath{0\pm0}}              
\newcommand{\hatcurRVkxxxxC}{\ensuremath{0\pm0}}               
\newcommand{\hatcurRVhxxxxC}{\ensuremath{0\pm0}}               
\newcommand{\hatcurRVtronexxxxC}{\ensuremath{0\pm0}}           
\newcommand{\hatcurRVtrtwoxxxxC}{\ensuremath{0\pm0}}           
\newcommand{\hatcurRVgammaAxxxxC}{\ensuremath{19423\pm17}}     
\newcommand{\hatcurRVjitterAxxxxC}{\ensuremath{47\pm20}}       
\newcommand{\hatcurRVjittertwosiglimAxxxxC}{\ensuremath{<82.2}} 
\newcommand{\hatcurRVfitrmsAxxxxC}{\ensuremath{0.0}}           
\newcommand{\hatcurRVgammaBxxxxC}{\ensuremath{19372\pm14}}     
\newcommand{\hatcurRVjitterBxxxxC}{\ensuremath{0.0\pm5.4}}     
\newcommand{\hatcurRVjittertwosiglimBxxxxC}{\ensuremath{<12.4}} 
\newcommand{\hatcurRVfitrmsBxxxxC}{\ensuremath{0.0}}           
\newcommand{\hatcurRVeccenxxxxC}{\ensuremath{0\pm0}}           
\newcommand{\hatcurRVeccentwosiglimxxxxC}{\ensuremath{<0.000}} 
\newcommand{\hatcurRVomegaxxxxC}{\ensuremath{0\pm0}}           
\newcommand{\hatcurPPixxxxC}{\ensuremath{85.61\pm0.42}}        
\newcommand{\hatcurPPgxxxxC}{\ensuremath{10.6\pm2.9}}          
\newcommand{\hatcurPPloggxxxxC}{\ensuremath{3.02\pm0.12}}      
\newcommand{\hatcurPParxxxxC}{\ensuremath{9.02\pm0.39}}        
\newcommand{\hatcurPParelxxxxC}{\ensuremath{0.05511\pm0.00069}} 
\newcommand{\hatcurPPrhoxxxxC}{\ensuremath{0.41_{-0.11}^{+0.16}}} 
\newcommand{\hatcurPPmxxxxC}{\ensuremath{0.70\pm0.15}}         
\newcommand{\hatcurPPmshortxxxxC}{\ensuremath{0.70}}           
\newcommand{\hatcurPPmlongxxxxC}{\ensuremath{0.70\pm0.15}}     
\newcommand{\hatcurPPmexxxxC}{\ensuremath{223\pm49}}           
\newcommand{\hatcurPPmeshortxxxxC}{\ensuremath{222.8}}         
\newcommand{\hatcurPPmelongxxxxC}{\ensuremath{223\pm49}}       
\newcommand{\hatcurPPrxxxxC}{\ensuremath{1.286\pm0.093}}       
\newcommand{\hatcurPPrshortxxxxC}{\ensuremath{1.29}}           
\newcommand{\hatcurPPrlongxxxxC}{\ensuremath{1.286\pm0.093}}   
\newcommand{\hatcurPPrexxxxC}{\ensuremath{14.4\pm1.0}}         
\newcommand{\hatcurPPreshortxxxxC}{\ensuremath{14.4}}          
\newcommand{\hatcurPPrelongxxxxC}{\ensuremath{14.4\pm1.0}}     
\newcommand{\hatcurPPmrcorrxxxxC}{\ensuremath{0.01}}           
\newcommand{\hatcurPPteffxxxxC}{\ensuremath{1518\pm45}}        
\newcommand{\hatcurPPthetaxxxxC}{\ensuremath{0.047\pm0.011}}   
\newcommand{\hatcurPPfluxperixxxxC}{\ensuremath{1.20\pm0.14}}  
\newcommand{\hatcurPPfluxperidimxxxxC}{\ensuremath{9}}         
\newcommand{\hatcurPPfluxapxxxxC}{\ensuremath{1.20\pm0.14}}    
\newcommand{\hatcurPPfluxapdimxxxxC}{\ensuremath{9}}           
\newcommand{\hatcurPPfluxavgxxxxC}{\ensuremath{1.20\pm0.14}}   
\newcommand{\hatcurPPfluxavgdimxxxxC}{\ensuremath{9}}          
\newcommand{\hatcurPPfluxavglogxxxxC}{\ensuremath{9.079\pm0.051}} 
\newcommand{\hatcurXsecphasexxxxC}{\ensuremath{0\pm0}}         
\newcommand{\hatcurXsecondaryxxxxC}{\ensuremath{2456733.28915\pm0.00073}} 
\newcommand{\hatcurXsecdurxxxxC}{\ensuremath{0.1269\pm0.0021}} 
\newcommand{\hatcurXsecingdurxxxxC}{\ensuremath{0.0207\pm0.0023}} 
\newcommand{\hatcurPPphiconjxxxxC}{\ensuremath{0\pm0}}         
\newcommand{\hatcurPPperixxxxC}{\ensuremath{2456730.14843\pm0.00073}} 
\newcommand{\hatcurPPaequivxxxxC}{\ensuremath{0.0337\pm0.0020}} 
\newcommand{\hatcurPPtcircxxxxC}{\ensuremath{340_{-110}^{+210}}} 
\newcommand{\hatcurPPtinfallxxxxC}{\ensuremath{4400\pm1600}}   
\newcommand{\hatcurXdistxxxxC}{\ensuremath{831\pm43}}          
\newcommand{\hatcurXAvxxxxC}{\ensuremath{0.042_{-0.042}^{+0.106}}} 
\newcommand{\hatcurXdistredxxxxC}{\ensuremath{818\pm41}}       
\newcommand{\hatcurXEBVxxxxC}{\ensuremath{0.013_{-0.013}^{+0.035}}} 
\newcommand{\hatcurXmvisoredxxxxC}{\ensuremath{13.309\pm0.046}} 
\newcommand{\hatcurXmiisoredxxxxC}{\ensuremath{12.759\pm0.030}} 
\newcommand{\hatcurXmjisoredxxxxC}{\ensuremath{12.436\pm0.016}} 
\newcommand{\hatcurXmhisoredxxxxC}{\ensuremath{12.199\pm0.015}} 
\newcommand{\hatcurXmkisoredxxxxC}{\ensuremath{12.154\pm0.016}} 
\newcommand{\hatcurXviisoredxxxxC}{\ensuremath{0.547_{-0.022}^{+0.029}}} 
\newcommand{\hatcurXvkisoredxxxxC}{\ensuremath{1.155\pm0.052}} 
\newcommand{\hatcurXjhisoredxxxxC}{\ensuremath{0.235\pm0.016}} 
\newcommand{\hatcurXjkisoredxxxxC}{\ensuremath{0.281\pm0.016}} 
\newcommand{\hatcurCCpmraxxxxC}{\ensuremath{-5.1\pm2.5}}       
\newcommand{\hatcurCCpmdecxxxxC}{\ensuremath{2.8\pm1.6}}       
\newcommand{\hatcurCCpmxxxxC}{\ensuremath{5.8\pm3.0}}          

\newcommand{\hatcurhtrxxxxD}{HATS755-003}                            
\newcommand{\hatcurfieldxxxxD}{\ensuremath{string}}                  
\newcommand{\hatcurCCraxxxxD}{\ensuremath{00^{\mathrm h}26^{\mathrm m}48.58{\mathrm s}}}                           
\newcommand{\hatcurCCdecxxxxD}{\ensuremath{-56{\arcdeg}18{\arcmin}58.0{\arcsec}}}                          
\newcommand{\hatcurCCmagxxxxD}{13.634}                               
\newcommand{\hatcurCCtwomassxxxxD}{2MASS~00264858-5618580}           
\newcommand{\hatcurCCgscxxxxD}{GSC~8468-01248}                       
\newcommand{\hatcurCCtassmvxxxxD}{\ensuremath{13.634\pm0.050}}       
\newcommand{\hatcurCCtassmvshortxxxxD}{\ensuremath{13.6}}            
\newcommand{\hatcurCCtassmBxxxxD}{\ensuremath{14.421\pm0.010}}       
\newcommand{\hatcurCCtassmBshortxxxxD}{\ensuremath{14.4}}            
\newcommand{\hatcurCCtassmIxxxxD}{\ensuremath{nff\pmnff}}            
\newcommand{\hatcurCCtassmIshortxxxxD}{\ensuremath{0.0}}             
\newcommand{\hatcurCCtassmgxxxxD}{\ensuremath{14.018\pm0.010}}       
\newcommand{\hatcurCCtassmgshortxxxxD}{\ensuremath{14.0}}            
\newcommand{\hatcurCCtassmrxxxxD}{\ensuremath{13.487\pm0.020}}       
\newcommand{\hatcurCCtassmrshortxxxxD}{\ensuremath{13.5}}            
\newcommand{\hatcurCCtassmixxxxD}{\ensuremath{13.45\pm0.22}}         
\newcommand{\hatcurCCtassmishortxxxxD}{\ensuremath{13.4}}            
\newcommand{\hatcurCCtwomassJmagxxxxD}{\ensuremath{12.366\pm0.024}}  
\newcommand{\hatcurCCtwomassHmagxxxxD}{\ensuremath{11.993\pm0.022}}  
\newcommand{\hatcurCCtwomassKmagxxxxD}{\ensuremath{11.965\pm0.024}}  
\newcommand{\hatcurCCcitJmagxxxxD}{\ensuremath{12.381\pm0.025}}      
\newcommand{\hatcurCCcitHmagxxxxD}{\ensuremath{11.989\pm0.023}}      
\newcommand{\hatcurCCcitKmagxxxxD}{\ensuremath{11.989\pm0.024}}      
\newcommand{\hatcurCCbbJmagxxxxD}{\ensuremath{12.433\pm0.025}}       
\newcommand{\hatcurCCbbHmagxxxxD}{\ensuremath{12.009\pm0.023}}       
\newcommand{\hatcurCCbbKmagxxxxD}{\ensuremath{12.009\pm0.024}}       
\newcommand{\hatcurCCesoJmagxxxxD}{\ensuremath{12.436\pm0.027}}      
\newcommand{\hatcurCCesoHmagxxxxD}{\ensuremath{12.001\pm0.025}}      
\newcommand{\hatcurCCesoKmagxxxxD}{\ensuremath{12.008\pm0.025}}      
\newcommand{\hatcurCCesoJHmagxxxxD}{\ensuremath{0.434\pm0.036}}      
\newcommand{\hatcurCCesoJKmagxxxxD}{\ensuremath{0.428\pm0.010}}      
\newcommand{\hatcurCCesoHKmagxxxxD}{\ensuremath{-0.006\pm0.036}}     
\newcommand{\hatcurLCdipxxxxD}{\ensuremath{14.0}}                    
\newcommand{\hatcurLCrprstarxxxxD}{\ensuremath{0.1088\pm0.0027}}     
\newcommand{\hatcurLCbsqxxxxD}{\ensuremath{0.402_{-0.042}^{+0.055}}} 
\newcommand{\hatcurLCimpxxxxD}{\ensuremath{0.634_{-0.034}^{+0.042}}} 
\newcommand{\hatcurLCzetaxxxxD}{\ensuremath{23.21\pm0.28}}           
\newcommand{\hatcurLCdurxxxxD}{\ensuremath{0.1014\pm0.0019}}         
\newcommand{\hatcurLCdurshortxxxxD}{\ensuremath{0.1014}}             
\newcommand{\hatcurLCdurhrxxxxD}{\ensuremath{2.434\pm0.045}}         
\newcommand{\hatcurLCdurhrshortxxxxD}{\ensuremath{2.434}}            
\newcommand{\hatcurLCqxxxxD}{\ensuremath{0.02140\pm0.00039}}         
\newcommand{\hatcurLCqshortxxxxD}{\ensuremath{0.021}}                
\newcommand{\hatcurLCingdurxxxxD}{\ensuremath{0.0157\pm0.0017}}      
\newcommand{\hatcurLCPxxxxD}{\ensuremath{4.7423729\pm0.0000049}}     
\newcommand{\hatcurLCPprecxxxxD}{\ensuremath{4.7423729}}             
\newcommand{\hatcurLCPshortxxxxD}{\ensuremath{4.7424}}               
\newcommand{\hatcurLCTxxxxD}{\ensuremath{2457376.68539\pm0.00060}}   
\newcommand{\hatcurLCTAxxxxD}{\ensuremath{2455764.2786\pm0.0016}}    
\newcommand{\hatcurLCTBxxxxD}{\ensuremath{2457632.77353\pm0.00073}}  
\newcommand{\hatcurLChatnetmAxxxxD}{\ensuremath{13.524670\pm0.000076}} 
\newcommand{\hatcurLCiblendAxxxxD}{\ensuremath{0.877\pm0.061}}       
\newcommand{\hatcurLChatnetmBxxxxD}{\ensuremath{13.524720\pm0.000092}} 
\newcommand{\hatcurLCiblendBxxxxD}{\ensuremath{0.863\pm0.068}}       
\newcommand{\hatcurLCrhoxxxxD}{\ensuremath{3.20\pm0.72}}             
\newcommand{\hatcurSMEiteffxxxxD}{\ensuremath{5495\pm69}}            
\newcommand{\hatcurSMEizfehxxxxD}{\ensuremath{-0.060\pm0.046}}       
\newcommand{\hatcurSMEizfehshortxxxxD}{\ensuremath{-0.06}}           
\newcommand{\hatcurSMEiloggxxxxD}{\ensuremath{4.610\pm0.066}}        
\newcommand{\hatcurSMEivsinxxxxD}{\ensuremath{0.90\pm0.66}}          
\newcommand{\hatcurSMEivmacxxxxD}{\ensuremath{3.56\pm0.10}}                  
\newcommand{\hatcurSMEivmicxxxxD}{\ensuremath{0.932\pm0.033}}                  
\newcommand{\hatcurLBizxxxxD}{\ensuremath{0.2448}}                   
\newcommand{\hatcurLBiizxxxxD}{\ensuremath{0.3072}}                  
\newcommand{\hatcurLBiixxxxD}{\ensuremath{0.3112}}                   
\newcommand{\hatcurLBiiixxxxD}{\ensuremath{0.3042}}                  
\newcommand{\hatcurLBiIxxxxD}{\ensuremath{0.2892}}                   
\newcommand{\hatcurLBiiIxxxxD}{\ensuremath{0.3055}}                  
\newcommand{\hatcurLBigxxxxD}{\ensuremath{0.6142}}                   
\newcommand{\hatcurLBiigxxxxD}{\ensuremath{0.1896}}                  
\newcommand{\hatcurLBirxxxxD}{\ensuremath{0.4078}}                   
\newcommand{\hatcurLBiirxxxxD}{\ensuremath{0.2935}}                  
\newcommand{\hatcurLBiRxxxxD}{\ensuremath{0.3812}}                   
\newcommand{\hatcurLBiiRxxxxD}{\ensuremath{0.2972}}                  
\newcommand{\hatcurLBikepxxxxD}{\ensuremath{0.1000}}                 
\newcommand{\hatcurLBiikepxxxxD}{\ensuremath{0.1000}}                
\newcommand{\hatcurISOmxxxxD}{\ensuremath{0.917\pm0.027}}            
\newcommand{\hatcurISOmshortxxxxD}{\ensuremath{0.92}}                
\newcommand{\hatcurISOmlongxxxxD}{\ensuremath{0.917\pm0.027}}        
\newcommand{\hatcurISOrxxxxD}{\ensuremath{0.853_{-0.030}^{+0.040}}}  
\newcommand{\hatcurISOrshortxxxxD}{\ensuremath{0.85}}                
\newcommand{\hatcurISOrlongxxxxD}{\ensuremath{0.853_{-0.030}^{+0.040}}} 
\newcommand{\hatcurISOrhoxxxxD}{\ensuremath{2.10_{-0.29}^{+0.22}}}   
\newcommand{\hatcurISOrholongxxxxD}{\ensuremath{2.10_{-0.29}^{+0.22}}} 
\newcommand{\hatcurISOloggxxxxD}{\ensuremath{4.542\pm0.038}}         
\newcommand{\hatcurISOlumxxxxD}{\ensuremath{0.589\pm0.070}}          
\newcommand{\hatcurISOlumshortxxxxD}{\ensuremath{0.59}}              
\newcommand{\hatcurISOmvxxxxD}{\ensuremath{5.47\pm0.14}}             
\newcommand{\hatcurISOvixxxxD}{\ensuremath{0.784\pm0.019}}           
\newcommand{\hatcurISOagexxxxD}{\ensuremath{3.0_{-2.1}^{+3.4}}}      
\newcommand{\hatcurISOsigmaxxxxD}{\ensuremath{0.00040\pm0.00017}}    
\newcommand{\hatcurISOMJxxxxD}{\ensuremath{4.17\pm0.11}}             
\newcommand{\hatcurISOMHxxxxD}{\ensuremath{3.770\pm0.100}}           
\newcommand{\hatcurISOMKxxxxD}{\ensuremath{3.704\pm0.098}}           
\newcommand{\hatcurISOJKxxxxD}{\ensuremath{0.470\pm0.020}}           
\newcommand{\hatcurISOspecxxxxD}{G}                                  
\newcommand{\hatcurRVKxxxxD}{\ensuremath{22.1\pm8.0}}                
\newcommand{\hatcurRVrkxxxxD}{\ensuremath{0\pm0}}                    
\newcommand{\hatcurRVrhxxxxD}{\ensuremath{0\pm0}}                    
\newcommand{\hatcurRVkxxxxD}{\ensuremath{0\pm0}}                     
\newcommand{\hatcurRVhxxxxD}{\ensuremath{0\pm0}}                     
\newcommand{\hatcurRVtronexxxxD}{\ensuremath{0\pm0}}                 
\newcommand{\hatcurRVtrtwoxxxxD}{\ensuremath{0\pm0}}                 
\newcommand{\hatcurRVgammaAxxxxD}{\ensuremath{-30192.7\pm8.6}}       
\newcommand{\hatcurRVjitterAxxxxD}{\ensuremath{32.9\pm6.7}}          
\newcommand{\hatcurRVjittertwosiglimAxxxxD}{\ensuremath{<46.0}}      
\newcommand{\hatcurRVfitrmsAxxxxD}{\ensuremath{0.0}}                 
\newcommand{\hatcurRVgammaBxxxxD}{\ensuremath{-4.9\pm8.3}}           
\newcommand{\hatcurRVjitterBxxxxD}{\ensuremath{22.7\pm6.8}}          
\newcommand{\hatcurRVjittertwosiglimBxxxxD}{\ensuremath{<36.0}}      
\newcommand{\hatcurRVfitrmsBxxxxD}{\ensuremath{0.0}}                 
\newcommand{\hatcurRVeccenxxxxD}{\ensuremath{0\pm0}}                 
\newcommand{\hatcurRVeccentwosiglimxxxxD}{\ensuremath{<0.000}}       
\newcommand{\hatcurRVomegaxxxxD}{\ensuremath{0\pm0}}                 
\newcommand{\hatcurPPixxxxD}{\ensuremath{87.32_{-0.31}^{+0.22}}}     
\newcommand{\hatcurPPgxxxxD}{\ensuremath{5.2\pm2.0}}                 
\newcommand{\hatcurPPloggxxxxD}{\ensuremath{2.71_{-0.20}^{+0.14}}}   
\newcommand{\hatcurPParxxxxD}{\ensuremath{13.55_{-0.65}^{+0.45}}}    
\newcommand{\hatcurPParelxxxxD}{\ensuremath{0.05367\pm0.00053}}      
\newcommand{\hatcurPPrhoxxxxD}{\ensuremath{0.28\pm0.12}}             
\newcommand{\hatcurPPmxxxxD}{\ensuremath{0.173\pm0.062}}             
\newcommand{\hatcurPPmshortxxxxD}{\ensuremath{0.17}}                 
\newcommand{\hatcurPPmlongxxxxD}{\ensuremath{0.173\pm0.062}}         
\newcommand{\hatcurPPmexxxxD}{\ensuremath{55\pm20}}                  
\newcommand{\hatcurPPmeshortxxxxD}{\ensuremath{54.9}}                
\newcommand{\hatcurPPmelongxxxxD}{\ensuremath{55\pm20}}              
\newcommand{\hatcurPPrxxxxD}{\ensuremath{0.903_{-0.045}^{+0.058}}}   
\newcommand{\hatcurPPrshortxxxxD}{\ensuremath{0.90}}                 
\newcommand{\hatcurPPrlongxxxxD}{\ensuremath{0.903_{-0.045}^{+0.058}}} 
\newcommand{\hatcurPPrexxxxD}{\ensuremath{10.12_{-0.50}^{+0.65}}}    
\newcommand{\hatcurPPreshortxxxxD}{\ensuremath{10.1}}                
\newcommand{\hatcurPPrelongxxxxD}{\ensuremath{10.12_{-0.50}^{+0.65}}} 
\newcommand{\hatcurPPmrcorrxxxxD}{\ensuremath{-0.02}}                
\newcommand{\hatcurPPteffxxxxD}{\ensuremath{1054\pm29}}              
\newcommand{\hatcurPPthetaxxxxD}{\ensuremath{0.0222\pm0.0082}}       
\newcommand{\hatcurPPfluxperixxxxD}{\ensuremath{2.78\pm0.32}}        
\newcommand{\hatcurPPfluxperidimxxxxD}{\ensuremath{8}}               
\newcommand{\hatcurPPfluxapxxxxD}{\ensuremath{2.78\pm0.32}}          
\newcommand{\hatcurPPfluxapdimxxxxD}{\ensuremath{8}}                 
\newcommand{\hatcurPPfluxavgxxxxD}{\ensuremath{2.78\pm0.32}}         
\newcommand{\hatcurPPfluxavgdimxxxxD}{\ensuremath{8}}                
\newcommand{\hatcurPPfluxavglogxxxxD}{\ensuremath{8.445\pm0.048}}    
\newcommand{\hatcurXsecphasexxxxD}{\ensuremath{0\pm0}}               
\newcommand{\hatcurXsecondaryxxxxD}{\ensuremath{2457379.05658\pm0.00061}} 
\newcommand{\hatcurXsecdurxxxxD}{\ensuremath{0.1014\pm0.0019}}       
\newcommand{\hatcurXsecingdurxxxxD}{\ensuremath{0.0157\pm0.0017}}    
\newcommand{\hatcurPPphiconjxxxxD}{\ensuremath{0\pm0}}               
\newcommand{\hatcurPPperixxxxD}{\ensuremath{2457375.49980\pm0.00060}} 
\newcommand{\hatcurPPaequivxxxxD}{\ensuremath{0.0700\pm0.0038}}      
\newcommand{\hatcurPPtcircxxxxD}{\ensuremath{680\pm330}}             
\newcommand{\hatcurPPtinfallxxxxD}{\ensuremath{108000_{-34000}^{+66000}}} 
\newcommand{\hatcurXdistxxxxD}{\ensuremath{458\pm22}}                
\newcommand{\hatcurXAvxxxxD}{\ensuremath{0.000\pm0.013}}             
\newcommand{\hatcurXdistredxxxxD}{\ensuremath{448\pm22}}             
\newcommand{\hatcurXEBVxxxxD}{\ensuremath{0.0000\pm0.0041}}          
\newcommand{\hatcurXmvisoredxxxxD}{\ensuremath{13.730\pm0.046}}      
\newcommand{\hatcurXmiisoredxxxxD}{\ensuremath{12.945\pm0.030}}      
\newcommand{\hatcurXmjisoredxxxxD}{\ensuremath{12.433\pm0.016}}      
\newcommand{\hatcurXmhisoredxxxxD}{\ensuremath{12.027\pm0.016}}      
\newcommand{\hatcurXmkisoredxxxxD}{\ensuremath{11.960\pm0.017}}      
\newcommand{\hatcurXviisoredxxxxD}{\ensuremath{0.785\pm0.018}}       
\newcommand{\hatcurXvkisoredxxxxD}{\ensuremath{1.769\pm0.054}}       
\newcommand{\hatcurXjhisoredxxxxD}{\ensuremath{0.406\pm0.015}}       
\newcommand{\hatcurXjkisoredxxxxD}{\ensuremath{0.473\pm0.017}}       
\newcommand{\hatcurCCpmraxxxxD}{\ensuremath{21.3\pm1.7}}             
\newcommand{\hatcurCCpmdecxxxxD}{\ensuremath{5.0\pm1.9}}             
\newcommand{\hatcurCCpmxxxxD}{\ensuremath{21.9\pm2.5}}               

\newcommand{\hatcurCCbbHmag}[1]{\ifnum#1=43 %
\hatcurCCbbHmagxxxxA
\else
\ifnum#1=44 %
\hatcurCCbbHmagxxxxB
\else
\ifnum#1=45 %
\hatcurCCbbHmagxxxxC
\else
\ifnum#1=46 %
\hatcurCCbbHmagxxxxD
\else
??????\fi
\fi
\fi
\fi
}
\newcommand{\hatcurCCbbJmag}[1]{\ifnum#1=43 %
\hatcurCCbbJmagxxxxA
\else
\ifnum#1=44 %
\hatcurCCbbJmagxxxxB
\else
\ifnum#1=45 %
\hatcurCCbbJmagxxxxC
\else
\ifnum#1=46 %
\hatcurCCbbJmagxxxxD
\else
??????\fi
\fi
\fi
\fi
}
\newcommand{\hatcurCCbbKmag}[1]{\ifnum#1=43 %
\hatcurCCbbKmagxxxxA
\else
\ifnum#1=44 %
\hatcurCCbbKmagxxxxB
\else
\ifnum#1=45 %
\hatcurCCbbKmagxxxxC
\else
\ifnum#1=46 %
\hatcurCCbbKmagxxxxD
\else
??????\fi
\fi
\fi
\fi
}
\newcommand{\hatcurCCcitHmag}[1]{\ifnum#1=43 %
\hatcurCCcitHmagxxxxA
\else
\ifnum#1=44 %
\hatcurCCcitHmagxxxxB
\else
\ifnum#1=45 %
\hatcurCCcitHmagxxxxC
\else
\ifnum#1=46 %
\hatcurCCcitHmagxxxxD
\else
??????\fi
\fi
\fi
\fi
}
\newcommand{\hatcurCCcitJmag}[1]{\ifnum#1=43 %
\hatcurCCcitJmagxxxxA
\else
\ifnum#1=44 %
\hatcurCCcitJmagxxxxB
\else
\ifnum#1=45 %
\hatcurCCcitJmagxxxxC
\else
\ifnum#1=46 %
\hatcurCCcitJmagxxxxD
\else
??????\fi
\fi
\fi
\fi
}
\newcommand{\hatcurCCcitKmag}[1]{\ifnum#1=43 %
\hatcurCCcitKmagxxxxA
\else
\ifnum#1=44 %
\hatcurCCcitKmagxxxxB
\else
\ifnum#1=45 %
\hatcurCCcitKmagxxxxC
\else
\ifnum#1=46 %
\hatcurCCcitKmagxxxxD
\else
??????\fi
\fi
\fi
\fi
}
\newcommand{\hatcurCCdec}[1]{\ifnum#1=43 %
\hatcurCCdecxxxxA
\else
\ifnum#1=44 %
\hatcurCCdecxxxxB
\else
\ifnum#1=45 %
\hatcurCCdecxxxxC
\else
\ifnum#1=46 %
\hatcurCCdecxxxxD
\else
??????\fi
\fi
\fi
\fi
}
\newcommand{\hatcurCCesoHKmag}[1]{\ifnum#1=43 %
\hatcurCCesoHKmagxxxxA
\else
\ifnum#1=44 %
\hatcurCCesoHKmagxxxxB
\else
\ifnum#1=45 %
\hatcurCCesoHKmagxxxxC
\else
\ifnum#1=46 %
\hatcurCCesoHKmagxxxxD
\else
??????\fi
\fi
\fi
\fi
}
\newcommand{\hatcurCCesoHmag}[1]{\ifnum#1=43 %
\hatcurCCesoHmagxxxxA
\else
\ifnum#1=44 %
\hatcurCCesoHmagxxxxB
\else
\ifnum#1=45 %
\hatcurCCesoHmagxxxxC
\else
\ifnum#1=46 %
\hatcurCCesoHmagxxxxD
\else
??????\fi
\fi
\fi
\fi
}
\newcommand{\hatcurCCesoJHmag}[1]{\ifnum#1=43 %
\hatcurCCesoJHmagxxxxA
\else
\ifnum#1=44 %
\hatcurCCesoJHmagxxxxB
\else
\ifnum#1=45 %
\hatcurCCesoJHmagxxxxC
\else
\ifnum#1=46 %
\hatcurCCesoJHmagxxxxD
\else
??????\fi
\fi
\fi
\fi
}
\newcommand{\hatcurCCesoJKmag}[1]{\ifnum#1=43 %
\hatcurCCesoJKmagxxxxA
\else
\ifnum#1=44 %
\hatcurCCesoJKmagxxxxB
\else
\ifnum#1=45 %
\hatcurCCesoJKmagxxxxC
\else
\ifnum#1=46 %
\hatcurCCesoJKmagxxxxD
\else
??????\fi
\fi
\fi
\fi
}
\newcommand{\hatcurCCesoJmag}[1]{\ifnum#1=43 %
\hatcurCCesoJmagxxxxA
\else
\ifnum#1=44 %
\hatcurCCesoJmagxxxxB
\else
\ifnum#1=45 %
\hatcurCCesoJmagxxxxC
\else
\ifnum#1=46 %
\hatcurCCesoJmagxxxxD
\else
??????\fi
\fi
\fi
\fi
}
\newcommand{\hatcurCCesoKmag}[1]{\ifnum#1=43 %
\hatcurCCesoKmagxxxxA
\else
\ifnum#1=44 %
\hatcurCCesoKmagxxxxB
\else
\ifnum#1=45 %
\hatcurCCesoKmagxxxxC
\else
\ifnum#1=46 %
\hatcurCCesoKmagxxxxD
\else
??????\fi
\fi
\fi
\fi
}
\newcommand{\hatcurCCgsc}[1]{\ifnum#1=43 %
\hatcurCCgscxxxxA
\else
\ifnum#1=44 %
\hatcurCCgscxxxxB
\else
\ifnum#1=45 %
\hatcurCCgscxxxxC
\else
\ifnum#1=46 %
\hatcurCCgscxxxxD
\else
??????\fi
\fi
\fi
\fi
}
\newcommand{\hatcurCCmag}[1]{\ifnum#1=43 %
\hatcurCCmagxxxxA
\else
\ifnum#1=44 %
\hatcurCCmagxxxxB
\else
\ifnum#1=45 %
\hatcurCCmagxxxxC
\else
\ifnum#1=46 %
\hatcurCCmagxxxxD
\else
??????\fi
\fi
\fi
\fi
}
\newcommand{\hatcurCCpm}[1]{\ifnum#1=43 %
\hatcurCCpmxxxxA
\else
\ifnum#1=44 %
\hatcurCCpmxxxxB
\else
\ifnum#1=45 %
\hatcurCCpmxxxxC
\else
\ifnum#1=46 %
\hatcurCCpmxxxxD
\else
??????\fi
\fi
\fi
\fi
}
\newcommand{\hatcurCCpmdec}[1]{\ifnum#1=43 %
\hatcurCCpmdecxxxxA
\else
\ifnum#1=44 %
\hatcurCCpmdecxxxxB
\else
\ifnum#1=45 %
\hatcurCCpmdecxxxxC
\else
\ifnum#1=46 %
\hatcurCCpmdecxxxxD
\else
??????\fi
\fi
\fi
\fi
}
\newcommand{\hatcurCCpmra}[1]{\ifnum#1=43 %
\hatcurCCpmraxxxxA
\else
\ifnum#1=44 %
\hatcurCCpmraxxxxB
\else
\ifnum#1=45 %
\hatcurCCpmraxxxxC
\else
\ifnum#1=46 %
\hatcurCCpmraxxxxD
\else
??????\fi
\fi
\fi
\fi
}
\newcommand{\hatcurCCra}[1]{\ifnum#1=43 %
\hatcurCCraxxxxA
\else
\ifnum#1=44 %
\hatcurCCraxxxxB
\else
\ifnum#1=45 %
\hatcurCCraxxxxC
\else
\ifnum#1=46 %
\hatcurCCraxxxxD
\else
??????\fi
\fi
\fi
\fi
}
\newcommand{\hatcurCCtassmB}[1]{\ifnum#1=43 %
\hatcurCCtassmBxxxxA
\else
\ifnum#1=44 %
\hatcurCCtassmBxxxxB
\else
\ifnum#1=45 %
\hatcurCCtassmBxxxxC
\else
\ifnum#1=46 %
\hatcurCCtassmBxxxxD
\else
??????\fi
\fi
\fi
\fi
}
\newcommand{\hatcurCCtassmBshort}[1]{\ifnum#1=43 %
\hatcurCCtassmBshortxxxxA
\else
\ifnum#1=44 %
\hatcurCCtassmBshortxxxxB
\else
\ifnum#1=45 %
\hatcurCCtassmBshortxxxxC
\else
\ifnum#1=46 %
\hatcurCCtassmBshortxxxxD
\else
??????\fi
\fi
\fi
\fi
}
\newcommand{\hatcurCCtassmg}[1]{\ifnum#1=43 %
\hatcurCCtassmgxxxxA
\else
\ifnum#1=44 %
\hatcurCCtassmgxxxxB
\else
\ifnum#1=45 %
\hatcurCCtassmgxxxxC
\else
\ifnum#1=46 %
\hatcurCCtassmgxxxxD
\else
??????\fi
\fi
\fi
\fi
}
\newcommand{\hatcurCCtassmgshort}[1]{\ifnum#1=43 %
\hatcurCCtassmgshortxxxxA
\else
\ifnum#1=44 %
\hatcurCCtassmgshortxxxxB
\else
\ifnum#1=45 %
\hatcurCCtassmgshortxxxxC
\else
\ifnum#1=46 %
\hatcurCCtassmgshortxxxxD
\else
??????\fi
\fi
\fi
\fi
}
\newcommand{\hatcurCCtassmi}[1]{\ifnum#1=43 %
\hatcurCCtassmixxxxA
\else
\ifnum#1=44 %
\hatcurCCtassmixxxxB
\else
\ifnum#1=45 %
\hatcurCCtassmixxxxC
\else
\ifnum#1=46 %
\hatcurCCtassmixxxxD
\else
??????\fi
\fi
\fi
\fi
}
\newcommand{\hatcurCCtassmI}[1]{\ifnum#1=43 %
\hatcurCCtassmIxxxxA
\else
\ifnum#1=44 %
\hatcurCCtassmIxxxxB
\else
\ifnum#1=45 %
\hatcurCCtassmIxxxxC
\else
\ifnum#1=46 %
\hatcurCCtassmIxxxxD
\else
??????\fi
\fi
\fi
\fi
}
\newcommand{\hatcurCCtassmishort}[1]{\ifnum#1=43 %
\hatcurCCtassmishortxxxxA
\else
\ifnum#1=44 %
\hatcurCCtassmishortxxxxB
\else
\ifnum#1=45 %
\hatcurCCtassmishortxxxxC
\else
\ifnum#1=46 %
\hatcurCCtassmishortxxxxD
\else
??????\fi
\fi
\fi
\fi
}
\newcommand{\hatcurCCtassmIshort}[1]{\ifnum#1=43 %
\hatcurCCtassmIshortxxxxA
\else
\ifnum#1=44 %
\hatcurCCtassmIshortxxxxB
\else
\ifnum#1=45 %
\hatcurCCtassmIshortxxxxC
\else
\ifnum#1=46 %
\hatcurCCtassmIshortxxxxD
\else
??????\fi
\fi
\fi
\fi
}
\newcommand{\hatcurCCtassmr}[1]{\ifnum#1=43 %
\hatcurCCtassmrxxxxA
\else
\ifnum#1=44 %
\hatcurCCtassmrxxxxB
\else
\ifnum#1=45 %
\hatcurCCtassmrxxxxC
\else
\ifnum#1=46 %
\hatcurCCtassmrxxxxD
\else
??????\fi
\fi
\fi
\fi
}
\newcommand{\hatcurCCtassmrshort}[1]{\ifnum#1=43 %
\hatcurCCtassmrshortxxxxA
\else
\ifnum#1=44 %
\hatcurCCtassmrshortxxxxB
\else
\ifnum#1=45 %
\hatcurCCtassmrshortxxxxC
\else
\ifnum#1=46 %
\hatcurCCtassmrshortxxxxD
\else
??????\fi
\fi
\fi
\fi
}
\newcommand{\hatcurCCtassmv}[1]{\ifnum#1=43 %
\hatcurCCtassmvxxxxA
\else
\ifnum#1=44 %
\hatcurCCtassmvxxxxB
\else
\ifnum#1=45 %
\hatcurCCtassmvxxxxC
\else
\ifnum#1=46 %
\hatcurCCtassmvxxxxD
\else
??????\fi
\fi
\fi
\fi
}
\newcommand{\hatcurCCtassmvshort}[1]{\ifnum#1=43 %
\hatcurCCtassmvshortxxxxA
\else
\ifnum#1=44 %
\hatcurCCtassmvshortxxxxB
\else
\ifnum#1=45 %
\hatcurCCtassmvshortxxxxC
\else
\ifnum#1=46 %
\hatcurCCtassmvshortxxxxD
\else
??????\fi
\fi
\fi
\fi
}
\newcommand{\hatcurCCtwomass}[1]{\ifnum#1=43 %
\hatcurCCtwomassxxxxA
\else
\ifnum#1=44 %
\hatcurCCtwomassxxxxB
\else
\ifnum#1=45 %
\hatcurCCtwomassxxxxC
\else
\ifnum#1=46 %
\hatcurCCtwomassxxxxD
\else
??????\fi
\fi
\fi
\fi
}
\newcommand{\hatcurCCtwomassHmag}[1]{\ifnum#1=43 %
\hatcurCCtwomassHmagxxxxA
\else
\ifnum#1=44 %
\hatcurCCtwomassHmagxxxxB
\else
\ifnum#1=45 %
\hatcurCCtwomassHmagxxxxC
\else
\ifnum#1=46 %
\hatcurCCtwomassHmagxxxxD
\else
??????\fi
\fi
\fi
\fi
}
\newcommand{\hatcurCCtwomassJmag}[1]{\ifnum#1=43 %
\hatcurCCtwomassJmagxxxxA
\else
\ifnum#1=44 %
\hatcurCCtwomassJmagxxxxB
\else
\ifnum#1=45 %
\hatcurCCtwomassJmagxxxxC
\else
\ifnum#1=46 %
\hatcurCCtwomassJmagxxxxD
\else
??????\fi
\fi
\fi
\fi
}
\newcommand{\hatcurCCtwomassKmag}[1]{\ifnum#1=43 %
\hatcurCCtwomassKmagxxxxA
\else
\ifnum#1=44 %
\hatcurCCtwomassKmagxxxxB
\else
\ifnum#1=45 %
\hatcurCCtwomassKmagxxxxC
\else
\ifnum#1=46 %
\hatcurCCtwomassKmagxxxxD
\else
??????\fi
\fi
\fi
\fi
}
\newcommand{\hatcurfield}[1]{\ifnum#1=43 %
\hatcurfieldxxxxA
\else
\ifnum#1=44 %
\hatcurfieldxxxxB
\else
\ifnum#1=45 %
\hatcurfieldxxxxC
\else
\ifnum#1=46 %
\hatcurfieldxxxxD
\else
??????\fi
\fi
\fi
\fi
}
\newcommand{\hatcurhtr}[1]{\ifnum#1=43 %
\hatcurhtrxxxxA
\else
\ifnum#1=44 %
\hatcurhtrxxxxB
\else
\ifnum#1=45 %
\hatcurhtrxxxxC
\else
\ifnum#1=46 %
\hatcurhtrxxxxD
\else
??????\fi
\fi
\fi
\fi
}
\newcommand{\hatcurISOage}[1]{\ifnum#1=43 %
\hatcurISOagexxxxA
\else
\ifnum#1=44 %
\hatcurISOagexxxxB
\else
\ifnum#1=45 %
\hatcurISOagexxxxC
\else
\ifnum#1=46 %
\hatcurISOagexxxxD
\else
??????\fi
\fi
\fi
\fi
}
\newcommand{\hatcurISOJK}[1]{\ifnum#1=43 %
\hatcurISOJKxxxxA
\else
\ifnum#1=44 %
\hatcurISOJKxxxxB
\else
\ifnum#1=45 %
\hatcurISOJKxxxxC
\else
\ifnum#1=46 %
\hatcurISOJKxxxxD
\else
??????\fi
\fi
\fi
\fi
}
\newcommand{\hatcurISOlogg}[1]{\ifnum#1=43 %
\hatcurISOloggxxxxA
\else
\ifnum#1=44 %
\hatcurISOloggxxxxB
\else
\ifnum#1=45 %
\hatcurISOloggxxxxC
\else
\ifnum#1=46 %
\hatcurISOloggxxxxD
\else
??????\fi
\fi
\fi
\fi
}
\newcommand{\hatcurISOlum}[1]{\ifnum#1=43 %
\hatcurISOlumxxxxA
\else
\ifnum#1=44 %
\hatcurISOlumxxxxB
\else
\ifnum#1=45 %
\hatcurISOlumxxxxC
\else
\ifnum#1=46 %
\hatcurISOlumxxxxD
\else
??????\fi
\fi
\fi
\fi
}
\newcommand{\hatcurISOlumshort}[1]{\ifnum#1=43 %
\hatcurISOlumshortxxxxA
\else
\ifnum#1=44 %
\hatcurISOlumshortxxxxB
\else
\ifnum#1=45 %
\hatcurISOlumshortxxxxC
\else
\ifnum#1=46 %
\hatcurISOlumshortxxxxD
\else
??????\fi
\fi
\fi
\fi
}
\newcommand{\hatcurISOm}[1]{\ifnum#1=43 %
\hatcurISOmxxxxA
\else
\ifnum#1=44 %
\hatcurISOmxxxxB
\else
\ifnum#1=45 %
\hatcurISOmxxxxC
\else
\ifnum#1=46 %
\hatcurISOmxxxxD
\else
??????\fi
\fi
\fi
\fi
}
\newcommand{\hatcurISOMH}[1]{\ifnum#1=43 %
\hatcurISOMHxxxxA
\else
\ifnum#1=44 %
\hatcurISOMHxxxxB
\else
\ifnum#1=45 %
\hatcurISOMHxxxxC
\else
\ifnum#1=46 %
\hatcurISOMHxxxxD
\else
??????\fi
\fi
\fi
\fi
}
\newcommand{\hatcurISOMJ}[1]{\ifnum#1=43 %
\hatcurISOMJxxxxA
\else
\ifnum#1=44 %
\hatcurISOMJxxxxB
\else
\ifnum#1=45 %
\hatcurISOMJxxxxC
\else
\ifnum#1=46 %
\hatcurISOMJxxxxD
\else
??????\fi
\fi
\fi
\fi
}
\newcommand{\hatcurISOMK}[1]{\ifnum#1=43 %
\hatcurISOMKxxxxA
\else
\ifnum#1=44 %
\hatcurISOMKxxxxB
\else
\ifnum#1=45 %
\hatcurISOMKxxxxC
\else
\ifnum#1=46 %
\hatcurISOMKxxxxD
\else
??????\fi
\fi
\fi
\fi
}
\newcommand{\hatcurISOmlong}[1]{\ifnum#1=43 %
\hatcurISOmlongxxxxA
\else
\ifnum#1=44 %
\hatcurISOmlongxxxxB
\else
\ifnum#1=45 %
\hatcurISOmlongxxxxC
\else
\ifnum#1=46 %
\hatcurISOmlongxxxxD
\else
??????\fi
\fi
\fi
\fi
}
\newcommand{\hatcurISOmshort}[1]{\ifnum#1=43 %
\hatcurISOmshortxxxxA
\else
\ifnum#1=44 %
\hatcurISOmshortxxxxB
\else
\ifnum#1=45 %
\hatcurISOmshortxxxxC
\else
\ifnum#1=46 %
\hatcurISOmshortxxxxD
\else
??????\fi
\fi
\fi
\fi
}
\newcommand{\hatcurISOmv}[1]{\ifnum#1=43 %
\hatcurISOmvxxxxA
\else
\ifnum#1=44 %
\hatcurISOmvxxxxB
\else
\ifnum#1=45 %
\hatcurISOmvxxxxC
\else
\ifnum#1=46 %
\hatcurISOmvxxxxD
\else
??????\fi
\fi
\fi
\fi
}
\newcommand{\hatcurISOr}[1]{\ifnum#1=43 %
\hatcurISOrxxxxA
\else
\ifnum#1=44 %
\hatcurISOrxxxxB
\else
\ifnum#1=45 %
\hatcurISOrxxxxC
\else
\ifnum#1=46 %
\hatcurISOrxxxxD
\else
??????\fi
\fi
\fi
\fi
}
\newcommand{\hatcurISOrho}[1]{\ifnum#1=43 %
\hatcurISOrhoxxxxA
\else
\ifnum#1=44 %
\hatcurISOrhoxxxxB
\else
\ifnum#1=45 %
\hatcurISOrhoxxxxC
\else
\ifnum#1=46 %
\hatcurISOrhoxxxxD
\else
??????\fi
\fi
\fi
\fi
}
\newcommand{\hatcurISOrholong}[1]{\ifnum#1=43 %
\hatcurISOrholongxxxxA
\else
\ifnum#1=44 %
\hatcurISOrholongxxxxB
\else
\ifnum#1=45 %
\hatcurISOrholongxxxxC
\else
\ifnum#1=46 %
\hatcurISOrholongxxxxD
\else
??????\fi
\fi
\fi
\fi
}
\newcommand{\hatcurISOrlong}[1]{\ifnum#1=43 %
\hatcurISOrlongxxxxA
\else
\ifnum#1=44 %
\hatcurISOrlongxxxxB
\else
\ifnum#1=45 %
\hatcurISOrlongxxxxC
\else
\ifnum#1=46 %
\hatcurISOrlongxxxxD
\else
??????\fi
\fi
\fi
\fi
}
\newcommand{\hatcurISOrshort}[1]{\ifnum#1=43 %
\hatcurISOrshortxxxxA
\else
\ifnum#1=44 %
\hatcurISOrshortxxxxB
\else
\ifnum#1=45 %
\hatcurISOrshortxxxxC
\else
\ifnum#1=46 %
\hatcurISOrshortxxxxD
\else
??????\fi
\fi
\fi
\fi
}
\newcommand{\hatcurISOsigma}[1]{\ifnum#1=43 %
\hatcurISOsigmaxxxxA
\else
\ifnum#1=44 %
\hatcurISOsigmaxxxxB
\else
\ifnum#1=45 %
\hatcurISOsigmaxxxxC
\else
\ifnum#1=46 %
\hatcurISOsigmaxxxxD
\else
??????\fi
\fi
\fi
\fi
}
\newcommand{\hatcurISOspec}[1]{\ifnum#1=43 %
\hatcurISOspecxxxxA
\else
\ifnum#1=44 %
\hatcurISOspecxxxxB
\else
\ifnum#1=45 %
\hatcurISOspecxxxxC
\else
\ifnum#1=46 %
\hatcurISOspecxxxxD
\else
??????\fi
\fi
\fi
\fi
}
\newcommand{\hatcurISOvi}[1]{\ifnum#1=43 %
\hatcurISOvixxxxA
\else
\ifnum#1=44 %
\hatcurISOvixxxxB
\else
\ifnum#1=45 %
\hatcurISOvixxxxC
\else
\ifnum#1=46 %
\hatcurISOvixxxxD
\else
??????\fi
\fi
\fi
\fi
}
\newcommand{\hatcurLBig}[1]{\ifnum#1=43 %
\hatcurLBigxxxxA
\else
\ifnum#1=44 %
\hatcurLBigxxxxB
\else
\ifnum#1=45 %
\hatcurLBigxxxxC
\else
\ifnum#1=46 %
\hatcurLBigxxxxD
\else
??????\fi
\fi
\fi
\fi
}
\newcommand{\hatcurLBii}[1]{\ifnum#1=43 %
\hatcurLBiixxxxA
\else
\ifnum#1=44 %
\hatcurLBiixxxxB
\else
\ifnum#1=45 %
\hatcurLBiixxxxC
\else
\ifnum#1=46 %
\hatcurLBiixxxxD
\else
??????\fi
\fi
\fi
\fi
}
\newcommand{\hatcurLBiI}[1]{\ifnum#1=43 %
\hatcurLBiIxxxxA
\else
\ifnum#1=44 %
\hatcurLBiIxxxxB
\else
\ifnum#1=45 %
\hatcurLBiIxxxxC
\else
\ifnum#1=46 %
\hatcurLBiIxxxxD
\else
??????\fi
\fi
\fi
\fi
}
\newcommand{\hatcurLBiig}[1]{\ifnum#1=43 %
\hatcurLBiigxxxxA
\else
\ifnum#1=44 %
\hatcurLBiigxxxxB
\else
\ifnum#1=45 %
\hatcurLBiigxxxxC
\else
\ifnum#1=46 %
\hatcurLBiigxxxxD
\else
??????\fi
\fi
\fi
\fi
}
\newcommand{\hatcurLBiii}[1]{\ifnum#1=43 %
\hatcurLBiiixxxxA
\else
\ifnum#1=44 %
\hatcurLBiiixxxxB
\else
\ifnum#1=45 %
\hatcurLBiiixxxxC
\else
\ifnum#1=46 %
\hatcurLBiiixxxxD
\else
??????\fi
\fi
\fi
\fi
}
\newcommand{\hatcurLBiiI}[1]{\ifnum#1=43 %
\hatcurLBiiIxxxxA
\else
\ifnum#1=44 %
\hatcurLBiiIxxxxB
\else
\ifnum#1=45 %
\hatcurLBiiIxxxxC
\else
\ifnum#1=46 %
\hatcurLBiiIxxxxD
\else
??????\fi
\fi
\fi
\fi
}
\newcommand{\hatcurLBiikep}[1]{\ifnum#1=43 %
\hatcurLBiikepxxxxA
\else
\ifnum#1=44 %
\hatcurLBiikepxxxxB
\else
\ifnum#1=45 %
\hatcurLBiikepxxxxC
\else
\ifnum#1=46 %
\hatcurLBiikepxxxxD
\else
??????\fi
\fi
\fi
\fi
}
\newcommand{\hatcurLBiir}[1]{\ifnum#1=43 %
\hatcurLBiirxxxxA
\else
\ifnum#1=44 %
\hatcurLBiirxxxxB
\else
\ifnum#1=45 %
\hatcurLBiirxxxxC
\else
\ifnum#1=46 %
\hatcurLBiirxxxxD
\else
??????\fi
\fi
\fi
\fi
}
\newcommand{\hatcurLBiiR}[1]{\ifnum#1=43 %
\hatcurLBiiRxxxxA
\else
\ifnum#1=44 %
\hatcurLBiiRxxxxB
\else
\ifnum#1=45 %
\hatcurLBiiRxxxxC
\else
\ifnum#1=46 %
\hatcurLBiiRxxxxD
\else
??????\fi
\fi
\fi
\fi
}
\newcommand{\hatcurLBiiz}[1]{\ifnum#1=43 %
\hatcurLBiizxxxxA
\else
\ifnum#1=44 %
\hatcurLBiizxxxxB
\else
\ifnum#1=45 %
\hatcurLBiizxxxxC
\else
\ifnum#1=46 %
\hatcurLBiizxxxxD
\else
??????\fi
\fi
\fi
\fi
}
\newcommand{\hatcurLBikep}[1]{\ifnum#1=43 %
\hatcurLBikepxxxxA
\else
\ifnum#1=44 %
\hatcurLBikepxxxxB
\else
\ifnum#1=45 %
\hatcurLBikepxxxxC
\else
\ifnum#1=46 %
\hatcurLBikepxxxxD
\else
??????\fi
\fi
\fi
\fi
}
\newcommand{\hatcurLBir}[1]{\ifnum#1=43 %
\hatcurLBirxxxxA
\else
\ifnum#1=44 %
\hatcurLBirxxxxB
\else
\ifnum#1=45 %
\hatcurLBirxxxxC
\else
\ifnum#1=46 %
\hatcurLBirxxxxD
\else
??????\fi
\fi
\fi
\fi
}
\newcommand{\hatcurLBiR}[1]{\ifnum#1=43 %
\hatcurLBiRxxxxA
\else
\ifnum#1=44 %
\hatcurLBiRxxxxB
\else
\ifnum#1=45 %
\hatcurLBiRxxxxC
\else
\ifnum#1=46 %
\hatcurLBiRxxxxD
\else
??????\fi
\fi
\fi
\fi
}
\newcommand{\hatcurLBiz}[1]{\ifnum#1=43 %
\hatcurLBizxxxxA
\else
\ifnum#1=44 %
\hatcurLBizxxxxB
\else
\ifnum#1=45 %
\hatcurLBizxxxxC
\else
\ifnum#1=46 %
\hatcurLBizxxxxD
\else
??????\fi
\fi
\fi
\fi
}
\newcommand{\hatcurLCbsq}[1]{\ifnum#1=43 %
\hatcurLCbsqxxxxA
\else
\ifnum#1=44 %
\hatcurLCbsqxxxxB
\else
\ifnum#1=45 %
\hatcurLCbsqxxxxC
\else
\ifnum#1=46 %
\hatcurLCbsqxxxxD
\else
??????\fi
\fi
\fi
\fi
}
\newcommand{\hatcurLCdip}[1]{\ifnum#1=43 %
\hatcurLCdipxxxxA
\else
\ifnum#1=44 %
\hatcurLCdipxxxxB
\else
\ifnum#1=45 %
\hatcurLCdipxxxxC
\else
\ifnum#1=46 %
\hatcurLCdipxxxxD
\else
??????\fi
\fi
\fi
\fi
}
\newcommand{\hatcurLCdur}[1]{\ifnum#1=43 %
\hatcurLCdurxxxxA
\else
\ifnum#1=44 %
\hatcurLCdurxxxxB
\else
\ifnum#1=45 %
\hatcurLCdurxxxxC
\else
\ifnum#1=46 %
\hatcurLCdurxxxxD
\else
??????\fi
\fi
\fi
\fi
}
\newcommand{\hatcurLCdurhr}[1]{\ifnum#1=43 %
\hatcurLCdurhrxxxxA
\else
\ifnum#1=44 %
\hatcurLCdurhrxxxxB
\else
\ifnum#1=45 %
\hatcurLCdurhrxxxxC
\else
\ifnum#1=46 %
\hatcurLCdurhrxxxxD
\else
??????\fi
\fi
\fi
\fi
}
\newcommand{\hatcurLCdurhrshort}[1]{\ifnum#1=43 %
\hatcurLCdurhrshortxxxxA
\else
\ifnum#1=44 %
\hatcurLCdurhrshortxxxxB
\else
\ifnum#1=45 %
\hatcurLCdurhrshortxxxxC
\else
\ifnum#1=46 %
\hatcurLCdurhrshortxxxxD
\else
??????\fi
\fi
\fi
\fi
}
\newcommand{\hatcurLCdurshort}[1]{\ifnum#1=43 %
\hatcurLCdurshortxxxxA
\else
\ifnum#1=44 %
\hatcurLCdurshortxxxxB
\else
\ifnum#1=45 %
\hatcurLCdurshortxxxxC
\else
\ifnum#1=46 %
\hatcurLCdurshortxxxxD
\else
??????\fi
\fi
\fi
\fi
}
\newcommand{\hatcurLChatnetm}[1]{\ifnum#1=45 %
\hatcurLChatnetmxxxxC
\else
??????\fi
}
\newcommand{\hatcurLChatnetmA}[1]{\ifnum#1=46 %
\hatcurLChatnetmAxxxxD
\else
??????\fi
}
\newcommand{\hatcurLChatnetmB}[1]{\ifnum#1=46 %
\hatcurLChatnetmBxxxxD
\else
??????\fi
}
\newcommand{\hatcurLCiblend}[1]{\ifnum#1=45 %
\hatcurLCiblendxxxxC
\else
??????\fi
}
\newcommand{\hatcurLCiblendA}[1]{\ifnum#1=46 %
\hatcurLCiblendAxxxxD
\else
??????\fi
}
\newcommand{\hatcurLCiblendB}[1]{\ifnum#1=46 %
\hatcurLCiblendBxxxxD
\else
??????\fi
}
\newcommand{\hatcurLCimp}[1]{\ifnum#1=43 %
\hatcurLCimpxxxxA
\else
\ifnum#1=44 %
\hatcurLCimpxxxxB
\else
\ifnum#1=45 %
\hatcurLCimpxxxxC
\else
\ifnum#1=46 %
\hatcurLCimpxxxxD
\else
??????\fi
\fi
\fi
\fi
}
\newcommand{\hatcurLCingdur}[1]{\ifnum#1=43 %
\hatcurLCingdurxxxxA
\else
\ifnum#1=44 %
\hatcurLCingdurxxxxB
\else
\ifnum#1=45 %
\hatcurLCingdurxxxxC
\else
\ifnum#1=46 %
\hatcurLCingdurxxxxD
\else
??????\fi
\fi
\fi
\fi
}
\newcommand{\hatcurLCP}[1]{\ifnum#1=43 %
\hatcurLCPxxxxA
\else
\ifnum#1=44 %
\hatcurLCPxxxxB
\else
\ifnum#1=45 %
\hatcurLCPxxxxC
\else
\ifnum#1=46 %
\hatcurLCPxxxxD
\else
??????\fi
\fi
\fi
\fi
}
\newcommand{\hatcurLCPprec}[1]{\ifnum#1=43 %
\hatcurLCPprecxxxxA
\else
\ifnum#1=44 %
\hatcurLCPprecxxxxB
\else
\ifnum#1=45 %
\hatcurLCPprecxxxxC
\else
\ifnum#1=46 %
\hatcurLCPprecxxxxD
\else
??????\fi
\fi
\fi
\fi
}
\newcommand{\hatcurLCPshort}[1]{\ifnum#1=43 %
\hatcurLCPshortxxxxA
\else
\ifnum#1=44 %
\hatcurLCPshortxxxxB
\else
\ifnum#1=45 %
\hatcurLCPshortxxxxC
\else
\ifnum#1=46 %
\hatcurLCPshortxxxxD
\else
??????\fi
\fi
\fi
\fi
}
\newcommand{\hatcurLCq}[1]{\ifnum#1=43 %
\hatcurLCqxxxxA
\else
\ifnum#1=44 %
\hatcurLCqxxxxB
\else
\ifnum#1=45 %
\hatcurLCqxxxxC
\else
\ifnum#1=46 %
\hatcurLCqxxxxD
\else
??????\fi
\fi
\fi
\fi
}
\newcommand{\hatcurLCqshort}[1]{\ifnum#1=43 %
\hatcurLCqshortxxxxA
\else
\ifnum#1=44 %
\hatcurLCqshortxxxxB
\else
\ifnum#1=45 %
\hatcurLCqshortxxxxC
\else
\ifnum#1=46 %
\hatcurLCqshortxxxxD
\else
??????\fi
\fi
\fi
\fi
}
\newcommand{\hatcurLCrho}[1]{\ifnum#1=43 %
\hatcurLCrhoxxxxA
\else
\ifnum#1=44 %
\hatcurLCrhoxxxxB
\else
\ifnum#1=45 %
\hatcurLCrhoxxxxC
\else
\ifnum#1=46 %
\hatcurLCrhoxxxxD
\else
??????\fi
\fi
\fi
\fi
}
\newcommand{\hatcurLCrprstar}[1]{\ifnum#1=43 %
\hatcurLCrprstarxxxxA
\else
\ifnum#1=44 %
\hatcurLCrprstarxxxxB
\else
\ifnum#1=45 %
\hatcurLCrprstarxxxxC
\else
\ifnum#1=46 %
\hatcurLCrprstarxxxxD
\else
??????\fi
\fi
\fi
\fi
}
\newcommand{\hatcurLCT}[1]{\ifnum#1=43 %
\hatcurLCTxxxxA
\else
\ifnum#1=44 %
\hatcurLCTxxxxB
\else
\ifnum#1=45 %
\hatcurLCTxxxxC
\else
\ifnum#1=46 %
\hatcurLCTxxxxD
\else
??????\fi
\fi
\fi
\fi
}
\newcommand{\hatcurLCTA}[1]{\ifnum#1=43 %
\hatcurLCTAxxxxA
\else
\ifnum#1=44 %
\hatcurLCTAxxxxB
\else
\ifnum#1=45 %
\hatcurLCTAxxxxC
\else
\ifnum#1=46 %
\hatcurLCTAxxxxD
\else
??????\fi
\fi
\fi
\fi
}
\newcommand{\hatcurLCTB}[1]{\ifnum#1=43 %
\hatcurLCTBxxxxA
\else
\ifnum#1=44 %
\hatcurLCTBxxxxB
\else
\ifnum#1=45 %
\hatcurLCTBxxxxC
\else
\ifnum#1=46 %
\hatcurLCTBxxxxD
\else
??????\fi
\fi
\fi
\fi
}
\newcommand{\hatcurLCzeta}[1]{\ifnum#1=43 %
\hatcurLCzetaxxxxA
\else
\ifnum#1=44 %
\hatcurLCzetaxxxxB
\else
\ifnum#1=45 %
\hatcurLCzetaxxxxC
\else
\ifnum#1=46 %
\hatcurLCzetaxxxxD
\else
??????\fi
\fi
\fi
\fi
}
\newcommand{\hatcurPPaequiv}[1]{\ifnum#1=43 %
\hatcurPPaequivxxxxA
\else
\ifnum#1=44 %
\hatcurPPaequivxxxxB
\else
\ifnum#1=45 %
\hatcurPPaequivxxxxC
\else
\ifnum#1=46 %
\hatcurPPaequivxxxxD
\else
??????\fi
\fi
\fi
\fi
}
\newcommand{\hatcurPPar}[1]{\ifnum#1=43 %
\hatcurPParxxxxA
\else
\ifnum#1=44 %
\hatcurPParxxxxB
\else
\ifnum#1=45 %
\hatcurPParxxxxC
\else
\ifnum#1=46 %
\hatcurPParxxxxD
\else
??????\fi
\fi
\fi
\fi
}
\newcommand{\hatcurPParel}[1]{\ifnum#1=43 %
\hatcurPParelxxxxA
\else
\ifnum#1=44 %
\hatcurPParelxxxxB
\else
\ifnum#1=45 %
\hatcurPParelxxxxC
\else
\ifnum#1=46 %
\hatcurPParelxxxxD
\else
??????\fi
\fi
\fi
\fi
}
\newcommand{\hatcurPPfluxap}[1]{\ifnum#1=43 %
\hatcurPPfluxapxxxxA
\else
\ifnum#1=44 %
\hatcurPPfluxapxxxxB
\else
\ifnum#1=45 %
\hatcurPPfluxapxxxxC
\else
\ifnum#1=46 %
\hatcurPPfluxapxxxxD
\else
??????\fi
\fi
\fi
\fi
}
\newcommand{\hatcurPPfluxapdim}[1]{\ifnum#1=43 %
\hatcurPPfluxapdimxxxxA
\else
\ifnum#1=44 %
\hatcurPPfluxapdimxxxxB
\else
\ifnum#1=45 %
\hatcurPPfluxapdimxxxxC
\else
\ifnum#1=46 %
\hatcurPPfluxapdimxxxxD
\else
??????\fi
\fi
\fi
\fi
}
\newcommand{\hatcurPPfluxavg}[1]{\ifnum#1=43 %
\hatcurPPfluxavgxxxxA
\else
\ifnum#1=44 %
\hatcurPPfluxavgxxxxB
\else
\ifnum#1=45 %
\hatcurPPfluxavgxxxxC
\else
\ifnum#1=46 %
\hatcurPPfluxavgxxxxD
\else
??????\fi
\fi
\fi
\fi
}
\newcommand{\hatcurPPfluxavgdim}[1]{\ifnum#1=43 %
\hatcurPPfluxavgdimxxxxA
\else
\ifnum#1=44 %
\hatcurPPfluxavgdimxxxxB
\else
\ifnum#1=45 %
\hatcurPPfluxavgdimxxxxC
\else
\ifnum#1=46 %
\hatcurPPfluxavgdimxxxxD
\else
??????\fi
\fi
\fi
\fi
}
\newcommand{\hatcurPPfluxavglog}[1]{\ifnum#1=43 %
\hatcurPPfluxavglogxxxxA
\else
\ifnum#1=44 %
\hatcurPPfluxavglogxxxxB
\else
\ifnum#1=45 %
\hatcurPPfluxavglogxxxxC
\else
\ifnum#1=46 %
\hatcurPPfluxavglogxxxxD
\else
??????\fi
\fi
\fi
\fi
}
\newcommand{\hatcurPPfluxperi}[1]{\ifnum#1=43 %
\hatcurPPfluxperixxxxA
\else
\ifnum#1=44 %
\hatcurPPfluxperixxxxB
\else
\ifnum#1=45 %
\hatcurPPfluxperixxxxC
\else
\ifnum#1=46 %
\hatcurPPfluxperixxxxD
\else
??????\fi
\fi
\fi
\fi
}
\newcommand{\hatcurPPfluxperidim}[1]{\ifnum#1=43 %
\hatcurPPfluxperidimxxxxA
\else
\ifnum#1=44 %
\hatcurPPfluxperidimxxxxB
\else
\ifnum#1=45 %
\hatcurPPfluxperidimxxxxC
\else
\ifnum#1=46 %
\hatcurPPfluxperidimxxxxD
\else
??????\fi
\fi
\fi
\fi
}
\newcommand{\hatcurPPg}[1]{\ifnum#1=43 %
\hatcurPPgxxxxA
\else
\ifnum#1=44 %
\hatcurPPgxxxxB
\else
\ifnum#1=45 %
\hatcurPPgxxxxC
\else
\ifnum#1=46 %
\hatcurPPgxxxxD
\else
??????\fi
\fi
\fi
\fi
}
\newcommand{\hatcurPPi}[1]{\ifnum#1=43 %
\hatcurPPixxxxA
\else
\ifnum#1=44 %
\hatcurPPixxxxB
\else
\ifnum#1=45 %
\hatcurPPixxxxC
\else
\ifnum#1=46 %
\hatcurPPixxxxD
\else
??????\fi
\fi
\fi
\fi
}
\newcommand{\hatcurPPlogg}[1]{\ifnum#1=43 %
\hatcurPPloggxxxxA
\else
\ifnum#1=44 %
\hatcurPPloggxxxxB
\else
\ifnum#1=45 %
\hatcurPPloggxxxxC
\else
\ifnum#1=46 %
\hatcurPPloggxxxxD
\else
??????\fi
\fi
\fi
\fi
}
\newcommand{\hatcurPPm}[1]{\ifnum#1=43 %
\hatcurPPmxxxxA
\else
\ifnum#1=44 %
\hatcurPPmxxxxB
\else
\ifnum#1=45 %
\hatcurPPmxxxxC
\else
\ifnum#1=46 %
\hatcurPPmxxxxD
\else
??????\fi
\fi
\fi
\fi
}
\newcommand{\hatcurPPme}[1]{\ifnum#1=43 %
\hatcurPPmexxxxA
\else
\ifnum#1=44 %
\hatcurPPmexxxxB
\else
\ifnum#1=45 %
\hatcurPPmexxxxC
\else
\ifnum#1=46 %
\hatcurPPmexxxxD
\else
??????\fi
\fi
\fi
\fi
}
\newcommand{\hatcurPPmelong}[1]{\ifnum#1=43 %
\hatcurPPmelongxxxxA
\else
\ifnum#1=44 %
\hatcurPPmelongxxxxB
\else
\ifnum#1=45 %
\hatcurPPmelongxxxxC
\else
\ifnum#1=46 %
\hatcurPPmelongxxxxD
\else
??????\fi
\fi
\fi
\fi
}
\newcommand{\hatcurPPmeshort}[1]{\ifnum#1=43 %
\hatcurPPmeshortxxxxA
\else
\ifnum#1=44 %
\hatcurPPmeshortxxxxB
\else
\ifnum#1=45 %
\hatcurPPmeshortxxxxC
\else
\ifnum#1=46 %
\hatcurPPmeshortxxxxD
\else
??????\fi
\fi
\fi
\fi
}
\newcommand{\hatcurPPmlong}[1]{\ifnum#1=43 %
\hatcurPPmlongxxxxA
\else
\ifnum#1=44 %
\hatcurPPmlongxxxxB
\else
\ifnum#1=45 %
\hatcurPPmlongxxxxC
\else
\ifnum#1=46 %
\hatcurPPmlongxxxxD
\else
??????\fi
\fi
\fi
\fi
}
\newcommand{\hatcurPPmrcorr}[1]{\ifnum#1=43 %
\hatcurPPmrcorrxxxxA
\else
\ifnum#1=44 %
\hatcurPPmrcorrxxxxB
\else
\ifnum#1=45 %
\hatcurPPmrcorrxxxxC
\else
\ifnum#1=46 %
\hatcurPPmrcorrxxxxD
\else
??????\fi
\fi
\fi
\fi
}
\newcommand{\hatcurPPmshort}[1]{\ifnum#1=43 %
\hatcurPPmshortxxxxA
\else
\ifnum#1=44 %
\hatcurPPmshortxxxxB
\else
\ifnum#1=45 %
\hatcurPPmshortxxxxC
\else
\ifnum#1=46 %
\hatcurPPmshortxxxxD
\else
??????\fi
\fi
\fi
\fi
}
\newcommand{\hatcurPPperi}[1]{\ifnum#1=43 %
\hatcurPPperixxxxA
\else
\ifnum#1=44 %
\hatcurPPperixxxxB
\else
\ifnum#1=45 %
\hatcurPPperixxxxC
\else
\ifnum#1=46 %
\hatcurPPperixxxxD
\else
??????\fi
\fi
\fi
\fi
}
\newcommand{\hatcurPPphiconj}[1]{\ifnum#1=43 %
\hatcurPPphiconjxxxxA
\else
\ifnum#1=44 %
\hatcurPPphiconjxxxxB
\else
\ifnum#1=45 %
\hatcurPPphiconjxxxxC
\else
\ifnum#1=46 %
\hatcurPPphiconjxxxxD
\else
??????\fi
\fi
\fi
\fi
}
\newcommand{\hatcurPPr}[1]{\ifnum#1=43 %
\hatcurPPrxxxxA
\else
\ifnum#1=44 %
\hatcurPPrxxxxB
\else
\ifnum#1=45 %
\hatcurPPrxxxxC
\else
\ifnum#1=46 %
\hatcurPPrxxxxD
\else
??????\fi
\fi
\fi
\fi
}
\newcommand{\hatcurPPre}[1]{\ifnum#1=43 %
\hatcurPPrexxxxA
\else
\ifnum#1=44 %
\hatcurPPrexxxxB
\else
\ifnum#1=45 %
\hatcurPPrexxxxC
\else
\ifnum#1=46 %
\hatcurPPrexxxxD
\else
??????\fi
\fi
\fi
\fi
}
\newcommand{\hatcurPPrelong}[1]{\ifnum#1=43 %
\hatcurPPrelongxxxxA
\else
\ifnum#1=44 %
\hatcurPPrelongxxxxB
\else
\ifnum#1=45 %
\hatcurPPrelongxxxxC
\else
\ifnum#1=46 %
\hatcurPPrelongxxxxD
\else
??????\fi
\fi
\fi
\fi
}
\newcommand{\hatcurPPreshort}[1]{\ifnum#1=43 %
\hatcurPPreshortxxxxA
\else
\ifnum#1=44 %
\hatcurPPreshortxxxxB
\else
\ifnum#1=45 %
\hatcurPPreshortxxxxC
\else
\ifnum#1=46 %
\hatcurPPreshortxxxxD
\else
??????\fi
\fi
\fi
\fi
}
\newcommand{\hatcurPPrho}[1]{\ifnum#1=43 %
\hatcurPPrhoxxxxA
\else
\ifnum#1=44 %
\hatcurPPrhoxxxxB
\else
\ifnum#1=45 %
\hatcurPPrhoxxxxC
\else
\ifnum#1=46 %
\hatcurPPrhoxxxxD
\else
??????\fi
\fi
\fi
\fi
}
\newcommand{\hatcurPPrlong}[1]{\ifnum#1=43 %
\hatcurPPrlongxxxxA
\else
\ifnum#1=44 %
\hatcurPPrlongxxxxB
\else
\ifnum#1=45 %
\hatcurPPrlongxxxxC
\else
\ifnum#1=46 %
\hatcurPPrlongxxxxD
\else
??????\fi
\fi
\fi
\fi
}
\newcommand{\hatcurPPrshort}[1]{\ifnum#1=43 %
\hatcurPPrshortxxxxA
\else
\ifnum#1=44 %
\hatcurPPrshortxxxxB
\else
\ifnum#1=45 %
\hatcurPPrshortxxxxC
\else
\ifnum#1=46 %
\hatcurPPrshortxxxxD
\else
??????\fi
\fi
\fi
\fi
}
\newcommand{\hatcurPPtcirc}[1]{\ifnum#1=43 %
\hatcurPPtcircxxxxA
\else
\ifnum#1=44 %
\hatcurPPtcircxxxxB
\else
\ifnum#1=45 %
\hatcurPPtcircxxxxC
\else
\ifnum#1=46 %
\hatcurPPtcircxxxxD
\else
??????\fi
\fi
\fi
\fi
}
\newcommand{\hatcurPPteff}[1]{\ifnum#1=43 %
\hatcurPPteffxxxxA
\else
\ifnum#1=44 %
\hatcurPPteffxxxxB
\else
\ifnum#1=45 %
\hatcurPPteffxxxxC
\else
\ifnum#1=46 %
\hatcurPPteffxxxxD
\else
??????\fi
\fi
\fi
\fi
}
\newcommand{\hatcurPPtheta}[1]{\ifnum#1=43 %
\hatcurPPthetaxxxxA
\else
\ifnum#1=44 %
\hatcurPPthetaxxxxB
\else
\ifnum#1=45 %
\hatcurPPthetaxxxxC
\else
\ifnum#1=46 %
\hatcurPPthetaxxxxD
\else
??????\fi
\fi
\fi
\fi
}
\newcommand{\hatcurPPtinfall}[1]{\ifnum#1=43 %
\hatcurPPtinfallxxxxA
\else
\ifnum#1=44 %
\hatcurPPtinfallxxxxB
\else
\ifnum#1=45 %
\hatcurPPtinfallxxxxC
\else
\ifnum#1=46 %
\hatcurPPtinfallxxxxD
\else
??????\fi
\fi
\fi
\fi
}
\newcommand{\hatcurRVeccen}[1]{\ifnum#1=43 %
\hatcurRVeccenxxxxA
\else
\ifnum#1=44 %
\hatcurRVeccenxxxxB
\else
\ifnum#1=45 %
\hatcurRVeccenxxxxC
\else
\ifnum#1=46 %
\hatcurRVeccenxxxxD
\else
??????\fi
\fi
\fi
\fi
}
\newcommand{\hatcurRVeccentwosiglim}[1]{\ifnum#1=43 %
\hatcurRVeccentwosiglimxxxxA
\else
\ifnum#1=44 %
\hatcurRVeccentwosiglimxxxxB
\else
\ifnum#1=45 %
\hatcurRVeccentwosiglimxxxxC
\else
\ifnum#1=46 %
\hatcurRVeccentwosiglimxxxxD
\else
??????\fi
\fi
\fi
\fi
}
\newcommand{\hatcurRVfitrms}[1]{\ifnum#1=44 %
\hatcurRVfitrmsxxxxB
\else
??????\fi
}
\newcommand{\hatcurRVfitrmsA}[1]{\ifnum#1=43 %
\hatcurRVfitrmsAxxxxA
\else
\ifnum#1=45 %
\hatcurRVfitrmsAxxxxC
\else
\ifnum#1=46 %
\hatcurRVfitrmsAxxxxD
\else
??????\fi
\fi
\fi
}
\newcommand{\hatcurRVfitrmsB}[1]{\ifnum#1=43 %
\hatcurRVfitrmsBxxxxA
\else
\ifnum#1=45 %
\hatcurRVfitrmsBxxxxC
\else
\ifnum#1=46 %
\hatcurRVfitrmsBxxxxD
\else
??????\fi
\fi
\fi
}
\newcommand{\hatcurRVgamma}[1]{\ifnum#1=44 %
\hatcurRVgammaxxxxB
\else
??????\fi
}
\newcommand{\hatcurRVgammaA}[1]{\ifnum#1=43 %
\hatcurRVgammaAxxxxA
\else
\ifnum#1=45 %
\hatcurRVgammaAxxxxC
\else
\ifnum#1=46 %
\hatcurRVgammaAxxxxD
\else
??????\fi
\fi
\fi
}
\newcommand{\hatcurRVgammaB}[1]{\ifnum#1=43 %
\hatcurRVgammaBxxxxA
\else
\ifnum#1=45 %
\hatcurRVgammaBxxxxC
\else
\ifnum#1=46 %
\hatcurRVgammaBxxxxD
\else
??????\fi
\fi
\fi
}
\newcommand{\hatcurRVh}[1]{\ifnum#1=43 %
\hatcurRVhxxxxA
\else
\ifnum#1=44 %
\hatcurRVhxxxxB
\else
\ifnum#1=45 %
\hatcurRVhxxxxC
\else
\ifnum#1=46 %
\hatcurRVhxxxxD
\else
??????\fi
\fi
\fi
\fi
}
\newcommand{\hatcurRVjitter}[1]{\ifnum#1=44 %
\hatcurRVjitterxxxxB
\else
??????\fi
}
\newcommand{\hatcurRVjitterA}[1]{\ifnum#1=43 %
\hatcurRVjitterAxxxxA
\else
\ifnum#1=45 %
\hatcurRVjitterAxxxxC
\else
\ifnum#1=46 %
\hatcurRVjitterAxxxxD
\else
??????\fi
\fi
\fi
}
\newcommand{\hatcurRVjitterB}[1]{\ifnum#1=43 %
\hatcurRVjitterBxxxxA
\else
\ifnum#1=45 %
\hatcurRVjitterBxxxxC
\else
\ifnum#1=46 %
\hatcurRVjitterBxxxxD
\else
??????\fi
\fi
\fi
}
\newcommand{\hatcurRVjittertwosiglim}[1]{\ifnum#1=44 %
\hatcurRVjittertwosiglimxxxxB
\else
??????\fi
}
\newcommand{\hatcurRVjittertwosiglimA}[1]{\ifnum#1=43 %
\hatcurRVjittertwosiglimAxxxxA
\else
\ifnum#1=45 %
\hatcurRVjittertwosiglimAxxxxC
\else
\ifnum#1=46 %
\hatcurRVjittertwosiglimAxxxxD
\else
??????\fi
\fi
\fi
}
\newcommand{\hatcurRVjittertwosiglimB}[1]{\ifnum#1=43 %
\hatcurRVjittertwosiglimBxxxxA
\else
\ifnum#1=45 %
\hatcurRVjittertwosiglimBxxxxC
\else
\ifnum#1=46 %
\hatcurRVjittertwosiglimBxxxxD
\else
??????\fi
\fi
\fi
}
\newcommand{\hatcurRVk}[1]{\ifnum#1=43 %
\hatcurRVkxxxxA
\else
\ifnum#1=44 %
\hatcurRVkxxxxB
\else
\ifnum#1=45 %
\hatcurRVkxxxxC
\else
\ifnum#1=46 %
\hatcurRVkxxxxD
\else
??????\fi
\fi
\fi
\fi
}
\newcommand{\hatcurRVK}[1]{\ifnum#1=43 %
\hatcurRVKxxxxA
\else
\ifnum#1=44 %
\hatcurRVKxxxxB
\else
\ifnum#1=45 %
\hatcurRVKxxxxC
\else
\ifnum#1=46 %
\hatcurRVKxxxxD
\else
??????\fi
\fi
\fi
\fi
}
\newcommand{\hatcurRVomega}[1]{\ifnum#1=43 %
\hatcurRVomegaxxxxA
\else
\ifnum#1=44 %
\hatcurRVomegaxxxxB
\else
\ifnum#1=45 %
\hatcurRVomegaxxxxC
\else
\ifnum#1=46 %
\hatcurRVomegaxxxxD
\else
??????\fi
\fi
\fi
\fi
}
\newcommand{\hatcurRVrh}[1]{\ifnum#1=43 %
\hatcurRVrhxxxxA
\else
\ifnum#1=44 %
\hatcurRVrhxxxxB
\else
\ifnum#1=45 %
\hatcurRVrhxxxxC
\else
\ifnum#1=46 %
\hatcurRVrhxxxxD
\else
??????\fi
\fi
\fi
\fi
}
\newcommand{\hatcurRVrk}[1]{\ifnum#1=43 %
\hatcurRVrkxxxxA
\else
\ifnum#1=44 %
\hatcurRVrkxxxxB
\else
\ifnum#1=45 %
\hatcurRVrkxxxxC
\else
\ifnum#1=46 %
\hatcurRVrkxxxxD
\else
??????\fi
\fi
\fi
\fi
}
\newcommand{\hatcurRVtrone}[1]{\ifnum#1=43 %
\hatcurRVtronexxxxA
\else
\ifnum#1=44 %
\hatcurRVtronexxxxB
\else
\ifnum#1=45 %
\hatcurRVtronexxxxC
\else
\ifnum#1=46 %
\hatcurRVtronexxxxD
\else
??????\fi
\fi
\fi
\fi
}
\newcommand{\hatcurRVtrtwo}[1]{\ifnum#1=43 %
\hatcurRVtrtwoxxxxA
\else
\ifnum#1=44 %
\hatcurRVtrtwoxxxxB
\else
\ifnum#1=45 %
\hatcurRVtrtwoxxxxC
\else
\ifnum#1=46 %
\hatcurRVtrtwoxxxxD
\else
??????\fi
\fi
\fi
\fi
}
\newcommand{\hatcurSMEiilogg}[1]{\ifnum#1=43 %
\hatcurSMEiiloggxxxxA
\else
\ifnum#1=44 %
\hatcurSMEiiloggxxxxB
\else
\ifnum#1=45 %
\hatcurSMEiiloggxxxxC
\else
??????\fi
\fi
\fi
}
\newcommand{\hatcurSMEiiteff}[1]{\ifnum#1=43 %
\hatcurSMEiiteffxxxxA
\else
\ifnum#1=44 %
\hatcurSMEiiteffxxxxB
\else
\ifnum#1=45 %
\hatcurSMEiiteffxxxxC
\else
??????\fi
\fi
\fi
}
\newcommand{\hatcurSMEiivmac}[1]{\ifnum#1=43 %
\hatcurSMEiivmacxxxxA
\else
\ifnum#1=44 %
\hatcurSMEiivmacxxxxB
\else
\ifnum#1=45 %
\hatcurSMEiivmacxxxxC
\else
??????\fi
\fi
\fi
}
\newcommand{\hatcurSMEiivmic}[1]{\ifnum#1=43 %
\hatcurSMEiivmicxxxxA
\else
\ifnum#1=44 %
\hatcurSMEiivmicxxxxB
\else
\ifnum#1=45 %
\hatcurSMEiivmicxxxxC
\else
??????\fi
\fi
\fi
}
\newcommand{\hatcurSMEiivsin}[1]{\ifnum#1=43 %
\hatcurSMEiivsinxxxxA
\else
\ifnum#1=44 %
\hatcurSMEiivsinxxxxB
\else
\ifnum#1=45 %
\hatcurSMEiivsinxxxxC
\else
??????\fi
\fi
\fi
}
\newcommand{\hatcurSMEiizfeh}[1]{\ifnum#1=43 %
\hatcurSMEiizfehxxxxA
\else
\ifnum#1=44 %
\hatcurSMEiizfehxxxxB
\else
\ifnum#1=45 %
\hatcurSMEiizfehxxxxC
\else
??????\fi
\fi
\fi
}
\newcommand{\hatcurSMEiizfehshort}[1]{\ifnum#1=43 %
\hatcurSMEiizfehshortxxxxA
\else
\ifnum#1=44 %
\hatcurSMEiizfehshortxxxxB
\else
\ifnum#1=45 %
\hatcurSMEiizfehshortxxxxC
\else
??????\fi
\fi
\fi
}
\newcommand{\hatcurSMEilogg}[1]{\ifnum#1=43 %
\hatcurSMEiloggxxxxA
\else
\ifnum#1=44 %
\hatcurSMEiloggxxxxB
\else
\ifnum#1=45 %
\hatcurSMEiloggxxxxC
\else
\ifnum#1=46 %
\hatcurSMEiloggxxxxD
\else
??????\fi
\fi
\fi
\fi
}
\newcommand{\hatcurSMEiteff}[1]{\ifnum#1=43 %
\hatcurSMEiteffxxxxA
\else
\ifnum#1=44 %
\hatcurSMEiteffxxxxB
\else
\ifnum#1=45 %
\hatcurSMEiteffxxxxC
\else
\ifnum#1=46 %
\hatcurSMEiteffxxxxD
\else
??????\fi
\fi
\fi
\fi
}
\newcommand{\hatcurSMEivmac}[1]{\ifnum#1=43 %
\hatcurSMEivmacxxxxA
\else
\ifnum#1=44 %
\hatcurSMEivmacxxxxB
\else
\ifnum#1=45 %
\hatcurSMEivmacxxxxC
\else
\ifnum#1=46 %
\hatcurSMEivmacxxxxD
\else
??????\fi
\fi
\fi
\fi
}
\newcommand{\hatcurSMEivmic}[1]{\ifnum#1=43 %
\hatcurSMEivmicxxxxA
\else
\ifnum#1=44 %
\hatcurSMEivmicxxxxB
\else
\ifnum#1=45 %
\hatcurSMEivmicxxxxC
\else
\ifnum#1=46 %
\hatcurSMEivmicxxxxD
\else
??????\fi
\fi
\fi
\fi
}
\newcommand{\hatcurSMEivsin}[1]{\ifnum#1=43 %
\hatcurSMEivsinxxxxA
\else
\ifnum#1=44 %
\hatcurSMEivsinxxxxB
\else
\ifnum#1=45 %
\hatcurSMEivsinxxxxC
\else
\ifnum#1=46 %
\hatcurSMEivsinxxxxD
\else
??????\fi
\fi
\fi
\fi
}
\newcommand{\hatcurSMEizfeh}[1]{\ifnum#1=43 %
\hatcurSMEizfehxxxxA
\else
\ifnum#1=44 %
\hatcurSMEizfehxxxxB
\else
\ifnum#1=45 %
\hatcurSMEizfehxxxxC
\else
\ifnum#1=46 %
\hatcurSMEizfehxxxxD
\else
??????\fi
\fi
\fi
\fi
}
\newcommand{\hatcurSMEizfehshort}[1]{\ifnum#1=43 %
\hatcurSMEizfehshortxxxxA
\else
\ifnum#1=44 %
\hatcurSMEizfehshortxxxxB
\else
\ifnum#1=45 %
\hatcurSMEizfehshortxxxxC
\else
\ifnum#1=46 %
\hatcurSMEizfehshortxxxxD
\else
??????\fi
\fi
\fi
\fi
}
\newcommand{\hatcurXAv}[1]{\ifnum#1=43 %
\hatcurXAvxxxxA
\else
\ifnum#1=44 %
\hatcurXAvxxxxB
\else
\ifnum#1=45 %
\hatcurXAvxxxxC
\else
\ifnum#1=46 %
\hatcurXAvxxxxD
\else
??????\fi
\fi
\fi
\fi
}
\newcommand{\hatcurXdist}[1]{\ifnum#1=43 %
\hatcurXdistxxxxA
\else
\ifnum#1=44 %
\hatcurXdistxxxxB
\else
\ifnum#1=45 %
\hatcurXdistxxxxC
\else
\ifnum#1=46 %
\hatcurXdistxxxxD
\else
??????\fi
\fi
\fi
\fi
}
\newcommand{\hatcurXdistred}[1]{\ifnum#1=43 %
\hatcurXdistredxxxxA
\else
\ifnum#1=44 %
\hatcurXdistredxxxxB
\else
\ifnum#1=45 %
\hatcurXdistredxxxxC
\else
\ifnum#1=46 %
\hatcurXdistredxxxxD
\else
??????\fi
\fi
\fi
\fi
}
\newcommand{\hatcurXEBV}[1]{\ifnum#1=43 %
\hatcurXEBVxxxxA
\else
\ifnum#1=44 %
\hatcurXEBVxxxxB
\else
\ifnum#1=45 %
\hatcurXEBVxxxxC
\else
\ifnum#1=46 %
\hatcurXEBVxxxxD
\else
??????\fi
\fi
\fi
\fi
}
\newcommand{\hatcurXjhisored}[1]{\ifnum#1=43 %
\hatcurXjhisoredxxxxA
\else
\ifnum#1=44 %
\hatcurXjhisoredxxxxB
\else
\ifnum#1=45 %
\hatcurXjhisoredxxxxC
\else
\ifnum#1=46 %
\hatcurXjhisoredxxxxD
\else
??????\fi
\fi
\fi
\fi
}
\newcommand{\hatcurXjkisored}[1]{\ifnum#1=43 %
\hatcurXjkisoredxxxxA
\else
\ifnum#1=44 %
\hatcurXjkisoredxxxxB
\else
\ifnum#1=45 %
\hatcurXjkisoredxxxxC
\else
\ifnum#1=46 %
\hatcurXjkisoredxxxxD
\else
??????\fi
\fi
\fi
\fi
}
\newcommand{\hatcurXmhisored}[1]{\ifnum#1=43 %
\hatcurXmhisoredxxxxA
\else
\ifnum#1=44 %
\hatcurXmhisoredxxxxB
\else
\ifnum#1=45 %
\hatcurXmhisoredxxxxC
\else
\ifnum#1=46 %
\hatcurXmhisoredxxxxD
\else
??????\fi
\fi
\fi
\fi
}
\newcommand{\hatcurXmiisored}[1]{\ifnum#1=43 %
\hatcurXmiisoredxxxxA
\else
\ifnum#1=44 %
\hatcurXmiisoredxxxxB
\else
\ifnum#1=45 %
\hatcurXmiisoredxxxxC
\else
\ifnum#1=46 %
\hatcurXmiisoredxxxxD
\else
??????\fi
\fi
\fi
\fi
}
\newcommand{\hatcurXmjisored}[1]{\ifnum#1=43 %
\hatcurXmjisoredxxxxA
\else
\ifnum#1=44 %
\hatcurXmjisoredxxxxB
\else
\ifnum#1=45 %
\hatcurXmjisoredxxxxC
\else
\ifnum#1=46 %
\hatcurXmjisoredxxxxD
\else
??????\fi
\fi
\fi
\fi
}
\newcommand{\hatcurXmkisored}[1]{\ifnum#1=43 %
\hatcurXmkisoredxxxxA
\else
\ifnum#1=44 %
\hatcurXmkisoredxxxxB
\else
\ifnum#1=45 %
\hatcurXmkisoredxxxxC
\else
\ifnum#1=46 %
\hatcurXmkisoredxxxxD
\else
??????\fi
\fi
\fi
\fi
}
\newcommand{\hatcurXmvisored}[1]{\ifnum#1=43 %
\hatcurXmvisoredxxxxA
\else
\ifnum#1=44 %
\hatcurXmvisoredxxxxB
\else
\ifnum#1=45 %
\hatcurXmvisoredxxxxC
\else
\ifnum#1=46 %
\hatcurXmvisoredxxxxD
\else
??????\fi
\fi
\fi
\fi
}
\newcommand{\hatcurXsecdur}[1]{\ifnum#1=43 %
\hatcurXsecdurxxxxA
\else
\ifnum#1=44 %
\hatcurXsecdurxxxxB
\else
\ifnum#1=45 %
\hatcurXsecdurxxxxC
\else
\ifnum#1=46 %
\hatcurXsecdurxxxxD
\else
??????\fi
\fi
\fi
\fi
}
\newcommand{\hatcurXsecingdur}[1]{\ifnum#1=43 %
\hatcurXsecingdurxxxxA
\else
\ifnum#1=44 %
\hatcurXsecingdurxxxxB
\else
\ifnum#1=45 %
\hatcurXsecingdurxxxxC
\else
\ifnum#1=46 %
\hatcurXsecingdurxxxxD
\else
??????\fi
\fi
\fi
\fi
}
\newcommand{\hatcurXsecondary}[1]{\ifnum#1=43 %
\hatcurXsecondaryxxxxA
\else
\ifnum#1=44 %
\hatcurXsecondaryxxxxB
\else
\ifnum#1=45 %
\hatcurXsecondaryxxxxC
\else
\ifnum#1=46 %
\hatcurXsecondaryxxxxD
\else
??????\fi
\fi
\fi
\fi
}
\newcommand{\hatcurXsecphase}[1]{\ifnum#1=43 %
\hatcurXsecphasexxxxA
\else
\ifnum#1=44 %
\hatcurXsecphasexxxxB
\else
\ifnum#1=45 %
\hatcurXsecphasexxxxC
\else
\ifnum#1=46 %
\hatcurXsecphasexxxxD
\else
??????\fi
\fi
\fi
\fi
}
\newcommand{\hatcurXviisored}[1]{\ifnum#1=43 %
\hatcurXviisoredxxxxA
\else
\ifnum#1=44 %
\hatcurXviisoredxxxxB
\else
\ifnum#1=45 %
\hatcurXviisoredxxxxC
\else
\ifnum#1=46 %
\hatcurXviisoredxxxxD
\else
??????\fi
\fi
\fi
\fi
}
\newcommand{\hatcurXvkisored}[1]{\ifnum#1=43 %
\hatcurXvkisoredxxxxA
\else
\ifnum#1=44 %
\hatcurXvkisoredxxxxB
\else
\ifnum#1=45 %
\hatcurXvkisoredxxxxC
\else
\ifnum#1=46 %
\hatcurXvkisoredxxxxD
\else
??????\fi
\fi
\fi
\fi
}

\newcommand{\hatcurhtreccenxxxxA}{HATS598-015}                      
\newcommand{\hatcurfieldeccenxxxxA}{\ensuremath{string}}            
\newcommand{\hatcurCCraeccenxxxxA}{\ensuremath{05^{\mathrm h}22^{\mathrm m}09.16{\mathrm s}}}                     
\newcommand{\hatcurCCdececcenxxxxA}{\ensuremath{-30{\arcdeg}58{\arcmin}15.0{\arcsec}}}                    
\newcommand{\hatcurCCmageccenxxxxA}{13.593}                         
\newcommand{\hatcurCCtwomasseccenxxxxA}{2MASS~05220915-3058150}     
\newcommand{\hatcurCCgsceccenxxxxA}{GSC~7048-01851}                 
\newcommand{\hatcurCCtassmveccenxxxxA}{\ensuremath{13.593\pm0.030}} 
\newcommand{\hatcurCCtassmvshorteccenxxxxA}{\ensuremath{13.6}}      
\newcommand{\hatcurCCtassmBeccenxxxxA}{\ensuremath{14.471\pm0.050}} 
\newcommand{\hatcurCCtassmBshorteccenxxxxA}{\ensuremath{14.5}}      
\newcommand{\hatcurCCtassmIeccenxxxxA}{\ensuremath{nff\pmnff}}      
\newcommand{\hatcurCCtassmIshorteccenxxxxA}{\ensuremath{0.0}}       
\newcommand{\hatcurCCtassmgeccenxxxxA}{\ensuremath{13.973\pm0.030}} 
\newcommand{\hatcurCCtassmgshorteccenxxxxA}{\ensuremath{14.0}}      
\newcommand{\hatcurCCtassmreccenxxxxA}{\ensuremath{13.301\pm0.020}} 
\newcommand{\hatcurCCtassmrshorteccenxxxxA}{\ensuremath{13.3}}      
\newcommand{\hatcurCCtassmieccenxxxxA}{\ensuremath{13.094\pm0.070}} 
\newcommand{\hatcurCCtassmishorteccenxxxxA}{\ensuremath{13.1}}      
\newcommand{\hatcurCCtwomassJmageccenxxxxA}{\ensuremath{12.064\pm0.026}} 
\newcommand{\hatcurCCtwomassHmageccenxxxxA}{\ensuremath{11.646\pm0.022}} 
\newcommand{\hatcurCCtwomassKmageccenxxxxA}{\ensuremath{11.556\pm0.023}} 
\newcommand{\hatcurCCcitJmageccenxxxxA}{\ensuremath{12.073\pm0.026}} 
\newcommand{\hatcurCCcitHmageccenxxxxA}{\ensuremath{11.640\pm0.023}} 
\newcommand{\hatcurCCcitKmageccenxxxxA}{\ensuremath{11.580\pm0.023}} 
\newcommand{\hatcurCCbbJmageccenxxxxA}{\ensuremath{12.134\pm0.028}} 
\newcommand{\hatcurCCbbHmageccenxxxxA}{\ensuremath{11.662\pm0.023}} 
\newcommand{\hatcurCCbbKmageccenxxxxA}{\ensuremath{11.600\pm0.023}} 
\newcommand{\hatcurCCesoJmageccenxxxxA}{\ensuremath{12.138\pm0.030}} 
\newcommand{\hatcurCCesoHmageccenxxxxA}{\ensuremath{11.658\pm0.027}} 
\newcommand{\hatcurCCesoKmageccenxxxxA}{\ensuremath{11.598\pm0.024}} 
\newcommand{\hatcurCCesoJHmageccenxxxxA}{\ensuremath{0.480\pm0.038}} 
\newcommand{\hatcurCCesoJKmageccenxxxxA}{\ensuremath{0.540\pm0.038}} 
\newcommand{\hatcurCCesoHKmageccenxxxxA}{\ensuremath{0.060\pm0.036}} 
\newcommand{\hatcurLCdipeccenxxxxA}{\ensuremath{2.8}}               
\newcommand{\hatcurLCrprstareccenxxxxA}{\ensuremath{0.1492\pm0.0017}} 
\newcommand{\hatcurLCbsqeccenxxxxA}{\ensuremath{0.029_{-0.018}^{+0.046}}} 
\newcommand{\hatcurLCimpeccenxxxxA}{\ensuremath{0.172_{-0.066}^{+0.104}}} 
\newcommand{\hatcurLCzetaeccenxxxxA}{\ensuremath{18.519\pm0.075}}   
\newcommand{\hatcurLCdureccenxxxxA}{\ensuremath{0.12452\pm0.00090}} 
\newcommand{\hatcurLCdurshorteccenxxxxA}{\ensuremath{0.1245}}       
\newcommand{\hatcurLCdurhreccenxxxxA}{\ensuremath{2.988\pm0.022}}   
\newcommand{\hatcurLCdurhrshorteccenxxxxA}{\ensuremath{2.988}}      
\newcommand{\hatcurLCqeccenxxxxA}{\ensuremath{0.02840\pm0.00020}}   
\newcommand{\hatcurLCqshorteccenxxxxA}{\ensuremath{0.028}}          
\newcommand{\hatcurLCingdureccenxxxxA}{\ensuremath{0.01666\pm0.00070}} 
\newcommand{\hatcurLCPeccenxxxxA}{\ensuremath{4.3888497\pm0.0000059}} 
\newcommand{\hatcurLCPprececcenxxxxA}{\ensuremath{4.3888497}}       
\newcommand{\hatcurLCPshorteccenxxxxA}{\ensuremath{4.3888}}         
\newcommand{\hatcurLCTeccenxxxxA}{\ensuremath{2457636.08946\pm0.00025}} 
\newcommand{\hatcurLCTAeccenxxxxA}{\ensuremath{2456543.2659\pm0.0014}} 
\newcommand{\hatcurLCTBeccenxxxxA}{\ensuremath{2457688.75568\pm0.00028}} 
\newcommand{\hatcurLCrhoeccenxxxxA}{\ensuremath{1.96\pm1.00}}       
\newcommand{\hatcurSMEiteffeccenxxxxA}{\ensuremath{5268\pm86}}      
\newcommand{\hatcurSMEizfeheccenxxxxA}{\ensuremath{0.160\pm0.052}}  
\newcommand{\hatcurSMEizfehshorteccenxxxxA}{\ensuremath{0.16}}      
\newcommand{\hatcurSMEiloggeccenxxxxA}{\ensuremath{4.75\pm0.17}}    
\newcommand{\hatcurSMEivsineccenxxxxA}{\ensuremath{0.5\pm1.2}}      
\newcommand{\hatcurSMEivmaceccenxxxxA}{\ensuremath{3.21\pm0.13}}            
\newcommand{\hatcurSMEivmiceccenxxxxA}{\ensuremath{0.827\pm0.039}}            
\newcommand{\hatcurSMEiiteffeccenxxxxA}{\ensuremath{5099\pm61}}     
\newcommand{\hatcurSMEiizfeheccenxxxxA}{\ensuremath{0.050\pm0.041}} 
\newcommand{\hatcurSMEiizfehshorteccenxxxxA}{\ensuremath{0.05}}     
\newcommand{\hatcurSMEiiloggeccenxxxxA}{\ensuremath{4.527\pm0.018}} 
\newcommand{\hatcurSMEiivsineccenxxxxA}{\ensuremath{1.11\pm0.82}}   
\newcommand{\hatcurSMEiivmaceccenxxxxA}{\ensuremath{2.948\pm0.093}}            
\newcommand{\hatcurSMEiivmiceccenxxxxA}{\ensuremath{0.747\pm0.030}}            
\newcommand{\hatcurLBizeccenxxxxA}{\ensuremath{0.3050}}             
\newcommand{\hatcurLBiizeccenxxxxA}{\ensuremath{0.2769}}            
\newcommand{\hatcurLBiieccenxxxxA}{\ensuremath{0.3873}}             
\newcommand{\hatcurLBiiieccenxxxxA}{\ensuremath{0.2588}}            
\newcommand{\hatcurLBiIeccenxxxxA}{\ensuremath{0.3601}}             
\newcommand{\hatcurLBiiIeccenxxxxA}{\ensuremath{0.2648}}            
\newcommand{\hatcurLBigeccenxxxxA}{\ensuremath{0.7513}}             
\newcommand{\hatcurLBiigeccenxxxxA}{\ensuremath{0.0782}}            
\newcommand{\hatcurLBireccenxxxxA}{\ensuremath{0.5115}}             
\newcommand{\hatcurLBiireccenxxxxA}{\ensuremath{0.2239}}            
\newcommand{\hatcurLBiReccenxxxxA}{\ensuremath{0.4772}}             
\newcommand{\hatcurLBiiReccenxxxxA}{\ensuremath{0.2344}}            
\newcommand{\hatcurLBikepeccenxxxxA}{\ensuremath{0.1000}}           
\newcommand{\hatcurLBiikepeccenxxxxA}{\ensuremath{0.1000}}          
\newcommand{\hatcurISOmeccenxxxxA}{\ensuremath{0.837\pm0.023}}      
\newcommand{\hatcurISOmshorteccenxxxxA}{\ensuremath{0.84}}          
\newcommand{\hatcurISOmlongeccenxxxxA}{\ensuremath{0.837\pm0.023}}  
\newcommand{\hatcurISOreccenxxxxA}{\ensuremath{0.812\pm0.032}}      
\newcommand{\hatcurISOrshorteccenxxxxA}{\ensuremath{0.81}}          
\newcommand{\hatcurISOrlongeccenxxxxA}{\ensuremath{0.812\pm0.032}}  
\newcommand{\hatcurISOrhoeccenxxxxA}{\ensuremath{2.18_{-0.20}^{+0.36}}} 
\newcommand{\hatcurISOrholongeccenxxxxA}{\ensuremath{2.18_{-0.20}^{+0.36}}} 
\newcommand{\hatcurISOloggeccenxxxxA}{\ensuremath{4.539\pm0.036}}   
\newcommand{\hatcurISOlumeccenxxxxA}{\ensuremath{0.400\pm0.046}}    
\newcommand{\hatcurISOlumshorteccenxxxxA}{\ensuremath{0.40}}        
\newcommand{\hatcurISOmveccenxxxxA}{\ensuremath{5.97\pm0.14}}       
\newcommand{\hatcurISOvieccenxxxxA}{\ensuremath{0.878\pm0.015}}     
\newcommand{\hatcurISOageeccenxxxxA}{\ensuremath{8.6_{-4.8}^{+3.0}}} 
\newcommand{\hatcurISOsigmaeccenxxxxA}{\ensuremath{0.00050\pm0.00018}} 
\newcommand{\hatcurISOMJeccenxxxxA}{\ensuremath{4.46\pm0.11}}       
\newcommand{\hatcurISOMHeccenxxxxA}{\ensuremath{3.99\pm0.10}}       
\newcommand{\hatcurISOMKeccenxxxxA}{\ensuremath{3.91\pm0.10}}       
\newcommand{\hatcurISOJKeccenxxxxA}{\ensuremath{0.550\pm0.010}}     
\newcommand{\hatcurISOspececcenxxxxA}{G}                            
\newcommand{\hatcurRVKeccenxxxxA}{\ensuremath{37.5\pm8.0}}          
\newcommand{\hatcurRVrkeccenxxxxA}{\ensuremath{0.38_{-0.21}^{+0.11}}} 
\newcommand{\hatcurRVrheccenxxxxA}{\ensuremath{-0.141_{-0.079}^{+0.130}}} 
\newcommand{\hatcurRVkeccenxxxxA}{\ensuremath{0.159_{-0.121}^{+0.092}}} 
\newcommand{\hatcurRVheccenxxxxA}{\ensuremath{-0.060_{-0.035}^{+0.056}}} 
\newcommand{\hatcurRVtroneeccenxxxxA}{\ensuremath{0\pm0}}           
\newcommand{\hatcurRVtrtwoeccenxxxxA}{\ensuremath{0\pm0}}           
\newcommand{\hatcurRVgammaAeccenxxxxA}{\ensuremath{22071.5\pm7.6}}  
\newcommand{\hatcurRVjitterAeccenxxxxA}{\ensuremath{20.3\pm7.8}}    
\newcommand{\hatcurRVjittertwosiglimAeccenxxxxA}{\ensuremath{<34.2}} 
\newcommand{\hatcurRVfitrmsAeccenxxxxA}{\ensuremath{0.0}}           
\newcommand{\hatcurRVgammaBeccenxxxxA}{\ensuremath{22058.5\pm8.6}}  
\newcommand{\hatcurRVjitterBeccenxxxxA}{\ensuremath{0\pm10}}        
\newcommand{\hatcurRVjittertwosiglimBeccenxxxxA}{\ensuremath{<26.6}} 
\newcommand{\hatcurRVfitrmsBeccenxxxxA}{\ensuremath{0.0}}           
\newcommand{\hatcurRVecceneccenxxxxA}{\ensuremath{0.173\pm0.089}}   
\newcommand{\hatcurRVeccentwosiglimeccenxxxxA}{\ensuremath{<0.311}} 
\newcommand{\hatcurRVomegaeccenxxxxA}{\ensuremath{330\pm120}}       
\newcommand{\hatcurPPieccenxxxxA}{\ensuremath{89.24_{-0.41}^{+0.29}}} 
\newcommand{\hatcurPPgeccenxxxxA}{\ensuremath{4.7\pm1.1}}           
\newcommand{\hatcurPPloggeccenxxxxA}{\ensuremath{2.67\pm0.11}}      
\newcommand{\hatcurPPareccenxxxxA}{\ensuremath{13.04_{-0.41}^{+0.68}}} 
\newcommand{\hatcurPPareleccenxxxxA}{\ensuremath{0.04944\pm0.00046}} 
\newcommand{\hatcurPPrhoeccenxxxxA}{\ensuremath{0.191_{-0.038}^{+0.054}}} 
\newcommand{\hatcurPPmeccenxxxxA}{\ensuremath{0.261\pm0.054}}       
\newcommand{\hatcurPPmshorteccenxxxxA}{\ensuremath{0.26}}           
\newcommand{\hatcurPPmlongeccenxxxxA}{\ensuremath{0.261\pm0.054}}   
\newcommand{\hatcurPPmeeccenxxxxA}{\ensuremath{83\pm17}}            
\newcommand{\hatcurPPmeshorteccenxxxxA}{\ensuremath{82.9}}          
\newcommand{\hatcurPPmelongeccenxxxxA}{\ensuremath{83\pm17}}        
\newcommand{\hatcurPPreccenxxxxA}{\ensuremath{1.180\pm0.050}}       
\newcommand{\hatcurPPrshorteccenxxxxA}{\ensuremath{1.18}}           
\newcommand{\hatcurPPrlongeccenxxxxA}{\ensuremath{1.180\pm0.050}}   
\newcommand{\hatcurPPreeccenxxxxA}{\ensuremath{13.23\pm0.56}}       
\newcommand{\hatcurPPreshorteccenxxxxA}{\ensuremath{13.2}}          
\newcommand{\hatcurPPrelongeccenxxxxA}{\ensuremath{13.23\pm0.56}}   
\newcommand{\hatcurPPmrcorreccenxxxxA}{\ensuremath{-0.01}}          
\newcommand{\hatcurPPteffeccenxxxxA}{\ensuremath{1003\pm27}}        
\newcommand{\hatcurPPthetaeccenxxxxA}{\ensuremath{0.0265\pm0.0057}} 
\newcommand{\hatcurPPfluxperieccenxxxxA}{\ensuremath{3.22_{-0.68}^{+0.95}}} 
\newcommand{\hatcurPPfluxperidimeccenxxxxA}{\ensuremath{8}}         
\newcommand{\hatcurPPfluxapeccenxxxxA}{\ensuremath{1.61_{-0.25}^{+0.46}}} 
\newcommand{\hatcurPPfluxapdimeccenxxxxA}{\ensuremath{8}}           
\newcommand{\hatcurPPfluxavgeccenxxxxA}{\ensuremath{2.28\pm0.25}}   
\newcommand{\hatcurPPfluxavgdimeccenxxxxA}{\ensuremath{8}}          
\newcommand{\hatcurPPfluxavglogeccenxxxxA}{\ensuremath{8.359\pm0.047}} 
\newcommand{\hatcurXsecphaseeccenxxxxA}{\ensuremath{0.601\pm0.060}} 
\newcommand{\hatcurXsecondaryeccenxxxxA}{\ensuremath{2457638.73\pm0.26}} 
\newcommand{\hatcurXsecdureccenxxxxA}{\ensuremath{0.1114\pm0.0100}} 
\newcommand{\hatcurXsecingdureccenxxxxA}{\ensuremath{0.0152\pm0.0014}} 
\newcommand{\hatcurPPphiconjeccenxxxxA}{\ensuremath{0.234_{-0.049}^{+0.071}}} 
\newcommand{\hatcurPPperieccenxxxxA}{\ensuremath{2457635.06\pm0.70}} 
\newcommand{\hatcurPPaequiveccenxxxxA}{\ensuremath{0.0781\pm0.0042}} 
\newcommand{\hatcurPPtcirceccenxxxxA}{\ensuremath{146\pm52}}        
\newcommand{\hatcurPPtinfalleccenxxxxA}{\ensuremath{55000\pm29000}} 
\newcommand{\hatcurXdisteccenxxxxA}{\ensuremath{344\pm16}}          
\newcommand{\hatcurXAveccenxxxxA}{\ensuremath{0.000\pm0.018}}       
\newcommand{\hatcurXdistredeccenxxxxA}{\ensuremath{341\pm17}}       
\newcommand{\hatcurXEBVeccenxxxxA}{\ensuremath{0.0000\pm0.0058}}    
\newcommand{\hatcurXmvisoredeccenxxxxA}{\ensuremath{13.637\pm0.033}} 
\newcommand{\hatcurXmiisoredeccenxxxxA}{\ensuremath{12.757\pm0.022}} 
\newcommand{\hatcurXmjisoredeccenxxxxA}{\ensuremath{12.128\pm0.014}} 
\newcommand{\hatcurXmhisoredeccenxxxxA}{\ensuremath{11.656\pm0.017}} 
\newcommand{\hatcurXmkisoredeccenxxxxA}{\ensuremath{11.578\pm0.018}} 
\newcommand{\hatcurXviisoredeccenxxxxA}{\ensuremath{0.881\pm0.014}} 
\newcommand{\hatcurXvkisoredeccenxxxxA}{\ensuremath{2.059\pm0.040}} 
\newcommand{\hatcurXjhisoredeccenxxxxA}{\ensuremath{0.4720\pm0.0080}} 
\newcommand{\hatcurXjkisoredeccenxxxxA}{\ensuremath{0.5500\pm0.0094}} 
\newcommand{\hatcurCCpmraeccenxxxxA}{\ensuremath{9.8\pm1.9}}        
\newcommand{\hatcurCCpmdececcenxxxxA}{\ensuremath{7.9\pm1.7}}       
\newcommand{\hatcurCCpmeccenxxxxA}{\ensuremath{12.6\pm2.5}}         

\newcommand{\hatcurhtreccenxxxxB}{HATS599-006}                      
\newcommand{\hatcurfieldeccenxxxxB}{\ensuremath{string}}            
\newcommand{\hatcurCCraeccenxxxxB}{\ensuremath{05^{\mathrm h}37^{\mathrm m}18.41{\mathrm s}}}                     
\newcommand{\hatcurCCdececcenxxxxB}{\ensuremath{-27{\arcdeg}58{\arcmin}21.4{\arcsec}}}                    
\newcommand{\hatcurCCmageccenxxxxB}{14.428}                         
\newcommand{\hatcurCCtwomasseccenxxxxB}{2MASS~05371842-2758214}     
\newcommand{\hatcurCCgsceccenxxxxB}{GSC~6497-00040}                 
\newcommand{\hatcurCCtassmveccenxxxxB}{\ensuremath{14.428\pm0.010}} 
\newcommand{\hatcurCCtassmvshorteccenxxxxB}{\ensuremath{14.4}}      
\newcommand{\hatcurCCtassmBeccenxxxxB}{\ensuremath{15.487\pm0.020}} 
\newcommand{\hatcurCCtassmBshorteccenxxxxB}{\ensuremath{15.5}}      
\newcommand{\hatcurCCtassmIeccenxxxxB}{\ensuremath{nff\pmnff}}      
\newcommand{\hatcurCCtassmIshorteccenxxxxB}{\ensuremath{0.0}}       
\newcommand{\hatcurCCtassmgeccenxxxxB}{\ensuremath{14.933\pm0.010}} 
\newcommand{\hatcurCCtassmgshorteccenxxxxB}{\ensuremath{14.9}}      
\newcommand{\hatcurCCtassmreccenxxxxB}{\ensuremath{14.086\pm0.010}} 
\newcommand{\hatcurCCtassmrshorteccenxxxxB}{\ensuremath{14.1}}      
\newcommand{\hatcurCCtassmieccenxxxxB}{\ensuremath{13.794\pm0.030}} 
\newcommand{\hatcurCCtassmishorteccenxxxxB}{\ensuremath{13.8}}      
\newcommand{\hatcurCCtwomassJmageccenxxxxB}{\ensuremath{12.699\pm0.023}} 
\newcommand{\hatcurCCtwomassHmageccenxxxxB}{\ensuremath{12.234\pm0.022}} 
\newcommand{\hatcurCCtwomassKmageccenxxxxB}{\ensuremath{12.188\pm0.030}} 
\newcommand{\hatcurCCcitJmageccenxxxxB}{\ensuremath{12.708\pm0.024}} 
\newcommand{\hatcurCCcitHmageccenxxxxB}{\ensuremath{12.229\pm0.023}} 
\newcommand{\hatcurCCcitKmageccenxxxxB}{\ensuremath{12.212\pm0.030}} 
\newcommand{\hatcurCCbbJmageccenxxxxB}{\ensuremath{12.769\pm0.026}} 
\newcommand{\hatcurCCbbHmageccenxxxxB}{\ensuremath{12.250\pm0.023}} 
\newcommand{\hatcurCCbbKmageccenxxxxB}{\ensuremath{12.232\pm0.030}} 
\newcommand{\hatcurCCesoJmageccenxxxxB}{\ensuremath{12.773\pm0.028}} 
\newcommand{\hatcurCCesoHmageccenxxxxB}{\ensuremath{12.243\pm0.026}} 
\newcommand{\hatcurCCesoKmageccenxxxxB}{\ensuremath{12.230\pm0.031}} 
\newcommand{\hatcurCCesoJHmageccenxxxxB}{\ensuremath{0.529\pm0.035}} 
\newcommand{\hatcurCCesoJKmageccenxxxxB}{\ensuremath{0.544\pm0.041}} 
\newcommand{\hatcurCCesoHKmageccenxxxxB}{\ensuremath{0.014\pm0.041}} 
\newcommand{\hatcurLCdipeccenxxxxB}{\ensuremath{13.0}}              
\newcommand{\hatcurLCrprstareccenxxxxB}{\ensuremath{0.150\pm0.020}} 
\newcommand{\hatcurLCbsqeccenxxxxB}{\ensuremath{0.838_{-0.093}^{+0.088}}} 
\newcommand{\hatcurLCimpeccenxxxxB}{\ensuremath{0.916_{-0.052}^{+0.047}}} 
\newcommand{\hatcurLCzetaeccenxxxxB}{\ensuremath{48.4_{-7.2}^{+19.2}}} 
\newcommand{\hatcurLCdureccenxxxxB}{\ensuremath{0.0720\pm0.0030}}   
\newcommand{\hatcurLCdurshorteccenxxxxB}{\ensuremath{0.0720}}       
\newcommand{\hatcurLCdurhreccenxxxxB}{\ensuremath{1.728\pm0.072}}   
\newcommand{\hatcurLCdurhrshorteccenxxxxB}{\ensuremath{1.728}}      
\newcommand{\hatcurLCqeccenxxxxB}{\ensuremath{0.0262\pm0.0011}}     
\newcommand{\hatcurLCqshorteccenxxxxB}{\ensuremath{0.026}}          
\newcommand{\hatcurLCingdureccenxxxxB}{\ensuremath{0.087\pm0.028}}  
\newcommand{\hatcurLCPeccenxxxxB}{\ensuremath{2.7439009\pm0.0000034}} 
\newcommand{\hatcurLCPprececcenxxxxB}{\ensuremath{2.7439009}}       
\newcommand{\hatcurLCPshorteccenxxxxB}{\ensuremath{2.7439}}         
\newcommand{\hatcurLCTeccenxxxxB}{\ensuremath{2456813.12625\pm0.00063}} 
\newcommand{\hatcurLCTAeccenxxxxB}{\ensuremath{2455946.0536\pm0.0012}} 
\newcommand{\hatcurLCTBeccenxxxxB}{\ensuremath{2457353.67470\pm0.00094}} 
\newcommand{\hatcurLChatnetmAeccenxxxxB}{\ensuremath{14.25976\pm0.00014}} 
\newcommand{\hatcurLCiblendAeccenxxxxB}{\ensuremath{1\pm0}}         
\newcommand{\hatcurLChatnetmBeccenxxxxB}{\ensuremath{14.275870\pm0.000092}} 
\newcommand{\hatcurLCiblendBeccenxxxxB}{\ensuremath{1\pm0}}         
\newcommand{\hatcurLCrhoeccenxxxxB}{\ensuremath{4.7_{-3.0}^{+4.5}}} 
\newcommand{\hatcurSMEiteffeccenxxxxB}{\ensuremath{5080\pm100}}     
\newcommand{\hatcurSMEizfeheccenxxxxB}{\ensuremath{0.320\pm0.071}}  
\newcommand{\hatcurSMEizfehshorteccenxxxxB}{\ensuremath{0.32}}      
\newcommand{\hatcurSMEiloggeccenxxxxB}{\ensuremath{4.75\pm0.19}}    
\newcommand{\hatcurSMEivsineccenxxxxB}{\ensuremath{0.5\pm1.1}}      
\newcommand{\hatcurSMEivmaceccenxxxxB}{\ensuremath{2.92\pm0.15}}    
\newcommand{\hatcurSMEivmiceccenxxxxB}{\ensuremath{0.737\pm0.051}}  
\newcommand{\hatcurSMEiiteffeccenxxxxB}{\ensuremath{4972\pm66}}     
\newcommand{\hatcurSMEiizfeheccenxxxxB}{\ensuremath{0.320\pm0.051}} 
\newcommand{\hatcurSMEiizfehshorteccenxxxxB}{\ensuremath{0.32}}     
\newcommand{\hatcurSMEiiloggeccenxxxxB}{\ensuremath{4.485\pm0.045}} 
\newcommand{\hatcurSMEiivsineccenxxxxB}{\ensuremath{1.50\pm0.91}}   
\newcommand{\hatcurSMEiivmaceccenxxxxB}{\ensuremath{2.8}}           
\newcommand{\hatcurSMEiivmiceccenxxxxB}{\ensuremath{0.7}}           
\newcommand{\hatcurLBizeccenxxxxB}{\ensuremath{0.3262}}             
\newcommand{\hatcurLBiizeccenxxxxB}{\ensuremath{0.2723}}            
\newcommand{\hatcurLBiieccenxxxxB}{\ensuremath{0.4195}}             
\newcommand{\hatcurLBiiieccenxxxxB}{\ensuremath{0.2444}}            
\newcommand{\hatcurLBiIeccenxxxxB}{\ensuremath{0.3882}}             
\newcommand{\hatcurLBiiIeccenxxxxB}{\ensuremath{0.2535}}            
\newcommand{\hatcurLBigeccenxxxxB}{\ensuremath{0.8140}}             
\newcommand{\hatcurLBiigeccenxxxxB}{\ensuremath{0.0272}}            
\newcommand{\hatcurLBireccenxxxxB}{\ensuremath{0.5603}}             
\newcommand{\hatcurLBiireccenxxxxB}{\ensuremath{0.1933}}            
\newcommand{\hatcurLBiReccenxxxxB}{\ensuremath{0.5216}}             
\newcommand{\hatcurLBiiReccenxxxxB}{\ensuremath{0.2081}}            
\newcommand{\hatcurLBikepeccenxxxxB}{\ensuremath{0.1000}}           
\newcommand{\hatcurLBiikepeccenxxxxB}{\ensuremath{0.1000}}          
\newcommand{\hatcurISOmeccenxxxxB}{\ensuremath{0.862\pm0.025}}      
\newcommand{\hatcurISOmshorteccenxxxxB}{\ensuremath{0.86}}          
\newcommand{\hatcurISOmlongeccenxxxxB}{\ensuremath{0.862\pm0.025}}  
\newcommand{\hatcurISOreccenxxxxB}{\ensuremath{0.822\pm0.085}}      
\newcommand{\hatcurISOrshorteccenxxxxB}{\ensuremath{0.82}}          
\newcommand{\hatcurISOrlongeccenxxxxB}{\ensuremath{0.822\pm0.085}}  
\newcommand{\hatcurISOrhoeccenxxxxB}{\ensuremath{2.18\pm0.35}}      
\newcommand{\hatcurISOrholongeccenxxxxB}{\ensuremath{2.18\pm0.35}}  
\newcommand{\hatcurISOloggeccenxxxxB}{\ensuremath{4.543\pm0.063}}   
\newcommand{\hatcurISOlumeccenxxxxB}{\ensuremath{0.37\pm0.12}}      
\newcommand{\hatcurISOlumshorteccenxxxxB}{\ensuremath{0.37}}        
\newcommand{\hatcurISOmveccenxxxxB}{\ensuremath{6.09\pm0.22}}       
\newcommand{\hatcurISOvieccenxxxxB}{\ensuremath{0.932\pm0.024}}     
\newcommand{\hatcurISOageeccenxxxxB}{\ensuremath{7.2\pm3.9}}        
\newcommand{\hatcurISOsigmaeccenxxxxB}{\ensuremath{0.0021\pm0.0014}} 
\newcommand{\hatcurISOMJeccenxxxxB}{\ensuremath{4.48\pm0.19}}       
\newcommand{\hatcurISOMHeccenxxxxB}{\ensuremath{4.00\pm0.18}}       
\newcommand{\hatcurISOMKeccenxxxxB}{\ensuremath{3.92\pm0.18}}       
\newcommand{\hatcurISOJKeccenxxxxB}{\ensuremath{0.31\pm0.28}}       
\newcommand{\hatcurISOspececcenxxxxB}{G}                            
\newcommand{\hatcurRVKeccenxxxxB}{\ensuremath{96\pm18}}             
\newcommand{\hatcurRVrkeccenxxxxB}{\ensuremath{0.18\pm0.13}}        
\newcommand{\hatcurRVrheccenxxxxB}{\ensuremath{-0.31_{-0.12}^{+0.22}}} 
\newcommand{\hatcurRVkeccenxxxxB}{\ensuremath{0.067\pm0.057}}       
\newcommand{\hatcurRVheccenxxxxB}{\ensuremath{-0.117\pm0.095}}      
\newcommand{\hatcurRVtroneeccenxxxxB}{\ensuremath{0\pm0}}           
\newcommand{\hatcurRVtrtwoeccenxxxxB}{\ensuremath{0\pm0}}           
\newcommand{\hatcurRVgammaeccenxxxxB}{\ensuremath{44081\pm14}}      
\newcommand{\hatcurRVjittereccenxxxxB}{\ensuremath{45\pm12}}        
\newcommand{\hatcurRVjittertwosiglimeccenxxxxB}{\ensuremath{<68.1}} 
\newcommand{\hatcurRVfitrmseccenxxxxB}{\ensuremath{.1fym}}          %
\newcommand{\hatcurRVecceneccenxxxxB}{\ensuremath{0.152\pm0.081}}   
\newcommand{\hatcurRVeccentwosiglimeccenxxxxB}{\ensuremath{<0.279}} 
\newcommand{\hatcurRVomegaeccenxxxxB}{\ensuremath{296\pm78}}        
\newcommand{\hatcurPPieccenxxxxB}{\ensuremath{85.0\pm1.7}}          
\newcommand{\hatcurPPgeccenxxxxB}{\ensuremath{9.6_{-2.8}^{+4.1}}}   
\newcommand{\hatcurPPloggeccenxxxxB}{\ensuremath{2.98\pm0.16}}      
\newcommand{\hatcurPPareccenxxxxB}{\ensuremath{9.53\pm0.59}}        
\newcommand{\hatcurPPareleccenxxxxB}{\ensuremath{0.03651\pm0.00035}} 
\newcommand{\hatcurPPrhoeccenxxxxB}{\ensuremath{0.39_{-0.15}^{+0.26}}} 
\newcommand{\hatcurPPmeccenxxxxB}{\ensuremath{0.58\pm0.11}}         
\newcommand{\hatcurPPmshorteccenxxxxB}{\ensuremath{0.58}}           
\newcommand{\hatcurPPmlongeccenxxxxB}{\ensuremath{0.58\pm0.11}}     
\newcommand{\hatcurPPmeeccenxxxxB}{\ensuremath{185\pm34}}           
\newcommand{\hatcurPPmeshorteccenxxxxB}{\ensuremath{185.2}}         
\newcommand{\hatcurPPmelongeccenxxxxB}{\ensuremath{185\pm34}}       
\newcommand{\hatcurPPreccenxxxxB}{\ensuremath{1.21\pm0.21}}         
\newcommand{\hatcurPPrshorteccenxxxxB}{\ensuremath{1.21}}           
\newcommand{\hatcurPPrlongeccenxxxxB}{\ensuremath{1.21\pm0.21}}     
\newcommand{\hatcurPPreeccenxxxxB}{\ensuremath{13.6\pm2.4}}         
\newcommand{\hatcurPPreshorteccenxxxxB}{\ensuremath{13.6}}          
\newcommand{\hatcurPPrelongeccenxxxxB}{\ensuremath{13.6\pm2.4}}     
\newcommand{\hatcurPPmrcorreccenxxxxB}{\ensuremath{0.02}}           
\newcommand{\hatcurPPteffeccenxxxxB}{\ensuremath{1144\pm58}}        
\newcommand{\hatcurPPthetaeccenxxxxB}{\ensuremath{0.0404\pm0.0098}} 
\newcommand{\hatcurPPfluxperieccenxxxxB}{\ensuremath{5.4\pm6.9}}    
\newcommand{\hatcurPPfluxperidimeccenxxxxB}{\ensuremath{8}}         
\newcommand{\hatcurPPfluxapeccenxxxxB}{\ensuremath{2.90_{-0.60}^{+0.80}}} 
\newcommand{\hatcurPPfluxapdimeccenxxxxB}{\ensuremath{8}}           
\newcommand{\hatcurPPfluxavgeccenxxxxB}{\ensuremath{3.9\pm1.4}}     
\newcommand{\hatcurPPfluxavgdimeccenxxxxB}{\ensuremath{8}}          
\newcommand{\hatcurPPfluxavglogeccenxxxxB}{\ensuremath{8.587\pm0.078}} 
\newcommand{\hatcurXsecphaseeccenxxxxB}{\ensuremath{0.544\pm0.037}} 
\newcommand{\hatcurXsecondaryeccenxxxxB}{\ensuremath{2456814.62\pm0.10}} 
\newcommand{\hatcurXsecdureccenxxxxB}{\ensuremath{0.072\pm0.015}}   
\newcommand{\hatcurXsecingdureccenxxxxB}{\ensuremath{0.018\pm0.011}} 
\newcommand{\hatcurPPphiconjeccenxxxxB}{\ensuremath{0.371_{-0.165}^{+0.065}}} 
\newcommand{\hatcurPPperieccenxxxxB}{\ensuremath{2456812.11\pm0.68}} 
\newcommand{\hatcurPPaequiveccenxxxxB}{\ensuremath{0.0599\pm0.0046}} 
\newcommand{\hatcurPPtcirceccenxxxxB}{\ensuremath{39_{-24}^{+56}}}  
\newcommand{\hatcurPPtinfalleccenxxxxB}{\ensuremath{3100\pm1100}}   
\newcommand{\hatcurXdisteccenxxxxB}{\ensuremath{459\pm50}}          
\newcommand{\hatcurXAveccenxxxxB}{\ensuremath{0.071\pm0.063}}       
\newcommand{\hatcurXdistredeccenxxxxB}{\ensuremath{449\pm48}}       
\newcommand{\hatcurXEBVeccenxxxxB}{\ensuremath{0.023\pm0.020}}      
\newcommand{\hatcurXmvisoredeccenxxxxB}{\ensuremath{14.429\pm0.011}} 
\newcommand{\hatcurXmiisoredeccenxxxxB}{\ensuremath{13.457\pm0.016}} 
\newcommand{\hatcurXmjisoredeccenxxxxB}{\ensuremath{12.763\pm0.017}} 
\newcommand{\hatcurXmhisoredeccenxxxxB}{\ensuremath{12.274\pm0.023}} 
\newcommand{\hatcurXmkisoredeccenxxxxB}{\ensuremath{12.187\pm0.023}} 
\newcommand{\hatcurXviisoredeccenxxxxB}{\ensuremath{0.972\pm0.016}} 
\newcommand{\hatcurXvkisoredeccenxxxxB}{\ensuremath{2.242\pm0.029}} 
\newcommand{\hatcurXjhisoredeccenxxxxB}{\ensuremath{0.4880_{-0.0060}^{+0.0090}}} 
\newcommand{\hatcurXjkisoredeccenxxxxB}{\ensuremath{0.5760\pm0.0079}} 
\newcommand{\hatcurCCpmraeccenxxxxB}{\ensuremath{-2.2\pm1.3}}       
\newcommand{\hatcurCCpmdececcenxxxxB}{\ensuremath{-3.1\pm1.6}}      
\newcommand{\hatcurCCpmeccenxxxxB}{\ensuremath{3.8\pm2.1}}          

\newcommand{\hatcurhtreccenxxxxC}{HATS554-008}                      
\newcommand{\hatcurfieldeccenxxxxC}{\ensuremath{string}}            
\newcommand{\hatcurCCraeccenxxxxC}{\ensuremath{06^{\mathrm h}47^{\mathrm m}58.63{\mathrm s}}}                     
\newcommand{\hatcurCCdececcenxxxxC}{\ensuremath{-21{\arcdeg}54{\arcmin}38.5{\arcsec}}}                    
\newcommand{\hatcurCCmageccenxxxxC}{13.307}                         
\newcommand{\hatcurCCtwomasseccenxxxxC}{2MASS~06475862-2154385}     
\newcommand{\hatcurCCgsceccenxxxxC}{GSC~5961-02383}                 
\newcommand{\hatcurCCtassmveccenxxxxC}{\ensuremath{13.307\pm0.050}} 
\newcommand{\hatcurCCtassmvshorteccenxxxxC}{\ensuremath{13.3}}      
\newcommand{\hatcurCCtassmBeccenxxxxC}{\ensuremath{13.845\pm0.020}} 
\newcommand{\hatcurCCtassmBshorteccenxxxxC}{\ensuremath{13.8}}      
\newcommand{\hatcurCCtassmIeccenxxxxC}{\ensuremath{nff\pmnff}}      
\newcommand{\hatcurCCtassmIshorteccenxxxxC}{\ensuremath{0.0}}       
\newcommand{\hatcurCCtassmgeccenxxxxC}{\ensuremath{13.550\pm0.020}} 
\newcommand{\hatcurCCtassmgshorteccenxxxxC}{\ensuremath{13.6}}      
\newcommand{\hatcurCCtassmreccenxxxxC}{\ensuremath{13.201\pm0.060}} 
\newcommand{\hatcurCCtassmrshorteccenxxxxC}{\ensuremath{13.2}}      
\newcommand{\hatcurCCtassmieccenxxxxC}{\ensuremath{13.162\pm0.040}} 
\newcommand{\hatcurCCtassmishorteccenxxxxC}{\ensuremath{13.2}}      
\newcommand{\hatcurCCtwomassJmageccenxxxxC}{\ensuremath{12.364\pm0.024}} 
\newcommand{\hatcurCCtwomassHmageccenxxxxC}{\ensuremath{12.155\pm0.024}} 
\newcommand{\hatcurCCtwomassKmageccenxxxxC}{\ensuremath{12.137\pm0.021}} 
\newcommand{\hatcurCCcitJmageccenxxxxC}{\ensuremath{12.388\pm0.024}} 
\newcommand{\hatcurCCcitHmageccenxxxxC}{\ensuremath{12.151\pm0.024}} 
\newcommand{\hatcurCCcitKmageccenxxxxC}{\ensuremath{12.161\pm0.021}} 
\newcommand{\hatcurCCbbJmageccenxxxxC}{\ensuremath{12.426\pm0.026}} 
\newcommand{\hatcurCCbbHmageccenxxxxC}{\ensuremath{12.171\pm0.025}} 
\newcommand{\hatcurCCbbKmageccenxxxxC}{\ensuremath{12.181\pm0.021}} 
\newcommand{\hatcurCCesoJmageccenxxxxC}{\ensuremath{12.427\pm0.026}} 
\newcommand{\hatcurCCesoHmageccenxxxxC}{\ensuremath{12.164\pm0.027}} 
\newcommand{\hatcurCCesoKmageccenxxxxC}{\ensuremath{12.181\pm0.021}} 
\newcommand{\hatcurCCesoJHmageccenxxxxC}{\ensuremath{0.2630\pm0.0070}} 
\newcommand{\hatcurCCesoJKmageccenxxxxC}{\ensuremath{0.246\pm0.034}} 
\newcommand{\hatcurCCesoHKmageccenxxxxC}{\ensuremath{-0.017\pm0.035}} 
\newcommand{\hatcurLCdipeccenxxxxC}{\ensuremath{9.7}}               
\newcommand{\hatcurLCrprstareccenxxxxC}{\ensuremath{0.1007\pm0.0039}} 
\newcommand{\hatcurLCbsqeccenxxxxC}{\ensuremath{0.475_{-0.059}^{+0.065}}} 
\newcommand{\hatcurLCimpeccenxxxxC}{\ensuremath{0.689_{-0.045}^{+0.045}}} 
\newcommand{\hatcurLCzetaeccenxxxxC}{\ensuremath{18.64_{-0.18}^{+0.34}}} 
\newcommand{\hatcurLCdureccenxxxxC}{\ensuremath{0.1270\pm0.0022}}   
\newcommand{\hatcurLCdurshorteccenxxxxC}{\ensuremath{0.1270}}       
\newcommand{\hatcurLCdurhreccenxxxxC}{\ensuremath{3.049\pm0.054}}   
\newcommand{\hatcurLCdurhrshorteccenxxxxC}{\ensuremath{3.049}}      
\newcommand{\hatcurLCqeccenxxxxC}{\ensuremath{0.03030\pm0.00054}}   
\newcommand{\hatcurLCqshorteccenxxxxC}{\ensuremath{0.030}}          
\newcommand{\hatcurLCingdureccenxxxxC}{\ensuremath{0.0208\pm0.0028}} 
\newcommand{\hatcurLCPeccenxxxxC}{\ensuremath{4.1876255\pm0.0000059}} 
\newcommand{\hatcurLCPprececcenxxxxC}{\ensuremath{4.1876255}}       
\newcommand{\hatcurLCPshorteccenxxxxC}{\ensuremath{4.1876}}         
\newcommand{\hatcurLCTeccenxxxxC}{\ensuremath{2456756.32108\pm0.00074}} 
\newcommand{\hatcurLCTAeccenxxxxC}{\ensuremath{2455198.5243\pm0.0020}} 
\newcommand{\hatcurLCTBeccenxxxxC}{\ensuremath{2457095.5188\pm0.0010}} 
\newcommand{\hatcurLChatnetmeccenxxxxC}{\ensuremath{13.182720\pm0.000078}} 
\newcommand{\hatcurLCiblendeccenxxxxC}{\ensuremath{0.816\pm0.074}}  
\newcommand{\hatcurLCrhoeccenxxxxC}{\ensuremath{1.19_{-0.41}^{+1.34}}} 
\newcommand{\hatcurSMEiteffeccenxxxxC}{\ensuremath{6740\pm150}}     
\newcommand{\hatcurSMEizfeheccenxxxxC}{\ensuremath{0.190\pm0.087}}  
\newcommand{\hatcurSMEizfehshorteccenxxxxC}{\ensuremath{0.19}}      
\newcommand{\hatcurSMEiloggeccenxxxxC}{\ensuremath{4.52\pm0.23}}    
\newcommand{\hatcurSMEivsineccenxxxxC}{\ensuremath{9.89\pm0.38}}    
\newcommand{\hatcurSMEivmaceccenxxxxC}{\ensuremath{5.5}}            
\newcommand{\hatcurSMEivmiceccenxxxxC}{\ensuremath{2.1}}            
\newcommand{\hatcurSMEiiteffeccenxxxxC}{\ensuremath{6450\pm110}}    
\newcommand{\hatcurSMEiizfeheccenxxxxC}{\ensuremath{0.020\pm0.068}} 
\newcommand{\hatcurSMEiizfehshorteccenxxxxC}{\ensuremath{0.02}}     
\newcommand{\hatcurSMEiiloggeccenxxxxC}{\ensuremath{4.286\pm0.026}} 
\newcommand{\hatcurSMEiivsineccenxxxxC}{\ensuremath{9.90\pm0.40}}   
\newcommand{\hatcurSMEiivmaceccenxxxxC}{\ensuremath{5.03}}          
\newcommand{\hatcurSMEiivmiceccenxxxxC}{\ensuremath{1.69}}          
\newcommand{\hatcurLBizeccenxxxxC}{\ensuremath{0.1286}}             
\newcommand{\hatcurLBiizeccenxxxxC}{\ensuremath{0.3632}}            
\newcommand{\hatcurLBiieccenxxxxC}{\ensuremath{0.1791}}             
\newcommand{\hatcurLBiiieccenxxxxC}{\ensuremath{0.3719}}            
\newcommand{\hatcurLBiIeccenxxxxC}{\ensuremath{0.1610}}             
\newcommand{\hatcurLBiiIeccenxxxxC}{\ensuremath{0.3702}}            
\newcommand{\hatcurLBigeccenxxxxC}{\ensuremath{0.4106}}             
\newcommand{\hatcurLBiigeccenxxxxC}{\ensuremath{0.3344}}            
\newcommand{\hatcurLBireccenxxxxC}{\ensuremath{0.2511}}             
\newcommand{\hatcurLBiireccenxxxxC}{\ensuremath{0.3818}}            
\newcommand{\hatcurLBiReccenxxxxC}{\ensuremath{0.2306}}             
\newcommand{\hatcurLBiiReccenxxxxC}{\ensuremath{0.3807}}            
\newcommand{\hatcurLBikepeccenxxxxC}{\ensuremath{0.1000}}           
\newcommand{\hatcurLBiikepeccenxxxxC}{\ensuremath{0.1000}}          
\newcommand{\hatcurISOmeccenxxxxC}{\ensuremath{1.276\pm0.059}}      
\newcommand{\hatcurISOmshorteccenxxxxC}{\ensuremath{1.28}}          
\newcommand{\hatcurISOmlongeccenxxxxC}{\ensuremath{1.276\pm0.059}}  
\newcommand{\hatcurISOreccenxxxxC}{\ensuremath{1.324_{-0.097}^{+0.140}}} 
\newcommand{\hatcurISOrshorteccenxxxxC}{\ensuremath{1.32}}          
\newcommand{\hatcurISOrlongeccenxxxxC}{\ensuremath{1.324_{-0.097}^{+0.140}}} 
\newcommand{\hatcurISOrhoeccenxxxxC}{\ensuremath{0.78\pm0.18}}      
\newcommand{\hatcurISOrholongeccenxxxxC}{\ensuremath{0.78\pm0.18}}  
\newcommand{\hatcurISOloggeccenxxxxC}{\ensuremath{4.300\pm0.072}}   
\newcommand{\hatcurISOlumeccenxxxxC}{\ensuremath{2.71_{-0.49}^{+0.64}}} 
\newcommand{\hatcurISOlumshorteccenxxxxC}{\ensuremath{2.71}}        
\newcommand{\hatcurISOmveccenxxxxC}{\ensuremath{3.66\pm0.25}}       
\newcommand{\hatcurISOvieccenxxxxC}{\ensuremath{0.517\pm0.028}}     
\newcommand{\hatcurISOageeccenxxxxC}{\ensuremath{1.64\pm0.77}}      
\newcommand{\hatcurISOsigmaeccenxxxxC}{\ensuremath{0.00070\pm0.00082}} 
\newcommand{\hatcurISOMJeccenxxxxC}{\ensuremath{2.84\pm0.22}}       
\newcommand{\hatcurISOMHeccenxxxxC}{\ensuremath{2.61\pm0.22}}       
\newcommand{\hatcurISOMKeccenxxxxC}{\ensuremath{2.57\pm0.22}}       
\newcommand{\hatcurISOJKeccenxxxxC}{\ensuremath{0.270\pm0.020}}     
\newcommand{\hatcurISOspececcenxxxxC}{F}                            
\newcommand{\hatcurRVKeccenxxxxC}{\ensuremath{71\pm15}}             
\newcommand{\hatcurRVrkeccenxxxxC}{\ensuremath{-0.03\pm0.18}}       
\newcommand{\hatcurRVrheccenxxxxC}{\ensuremath{-0.02\pm0.22}}       
\newcommand{\hatcurRVkeccenxxxxC}{\ensuremath{-0.005\pm0.073}}      
\newcommand{\hatcurRVheccenxxxxC}{\ensuremath{-0.001\pm0.088}}      
\newcommand{\hatcurRVtroneeccenxxxxC}{\ensuremath{0\pm0}}           
\newcommand{\hatcurRVtrtwoeccenxxxxC}{\ensuremath{0\pm0}}           
\newcommand{\hatcurRVgammaAeccenxxxxC}{\ensuremath{19422\pm20}}     
\newcommand{\hatcurRVjitterAeccenxxxxC}{\ensuremath{47\pm18}}       
\newcommand{\hatcurRVjittertwosiglimAeccenxxxxC}{\ensuremath{<78.3}} 
\newcommand{\hatcurRVfitrmsAeccenxxxxC}{\ensuremath{0.0}}           
\newcommand{\hatcurRVgammaBeccenxxxxC}{\ensuremath{19370\pm14}}     
\newcommand{\hatcurRVjitterBeccenxxxxC}{\ensuremath{0.0\pm5.1}}     
\newcommand{\hatcurRVjittertwosiglimBeccenxxxxC}{\ensuremath{<10.4}} 
\newcommand{\hatcurRVfitrmsBeccenxxxxC}{\ensuremath{0.0}}           
\newcommand{\hatcurRVecceneccenxxxxC}{\ensuremath{0.067\pm0.079}}   
\newcommand{\hatcurRVeccentwosiglimeccenxxxxC}{\ensuremath{<0.240}} 
\newcommand{\hatcurRVomegaeccenxxxxC}{\ensuremath{190\pm93}}        
\newcommand{\hatcurPPieccenxxxxC}{\ensuremath{85.58_{-0.97}^{+0.58}}} 
\newcommand{\hatcurPPgeccenxxxxC}{\ensuremath{9.4\pm2.8}}           
\newcommand{\hatcurPPloggeccenxxxxC}{\ensuremath{2.98\pm0.13}}      
\newcommand{\hatcurPPareccenxxxxC}{\ensuremath{8.97_{-0.83}^{+0.63}}} 
\newcommand{\hatcurPPareleccenxxxxC}{\ensuremath{0.05516\pm0.00084}} 
\newcommand{\hatcurPPrhoeccenxxxxC}{\ensuremath{0.36\pm0.14}}       
\newcommand{\hatcurPPmeccenxxxxC}{\ensuremath{0.65\pm0.14}}         
\newcommand{\hatcurPPmshorteccenxxxxC}{\ensuremath{0.65}}           
\newcommand{\hatcurPPmlongeccenxxxxC}{\ensuremath{0.65\pm0.14}}     
\newcommand{\hatcurPPmeeccenxxxxC}{\ensuremath{208\pm45}}           
\newcommand{\hatcurPPmeshorteccenxxxxC}{\ensuremath{208.0}}         
\newcommand{\hatcurPPmelongeccenxxxxC}{\ensuremath{208\pm45}}       
\newcommand{\hatcurPPreccenxxxxC}{\ensuremath{1.31\pm0.15}}         
\newcommand{\hatcurPPrshorteccenxxxxC}{\ensuremath{1.31}}           
\newcommand{\hatcurPPrlongeccenxxxxC}{\ensuremath{1.31\pm0.15}}     
\newcommand{\hatcurPPreeccenxxxxC}{\ensuremath{14.6\pm1.7}}         
\newcommand{\hatcurPPreshorteccenxxxxC}{\ensuremath{14.6}}          
\newcommand{\hatcurPPrelongeccenxxxxC}{\ensuremath{14.6\pm1.7}}     
\newcommand{\hatcurPPmrcorreccenxxxxC}{\ensuremath{0.06}}           
\newcommand{\hatcurPPteffeccenxxxxC}{\ensuremath{1524\pm82}}        
\newcommand{\hatcurPPthetaeccenxxxxC}{\ensuremath{0.043\pm0.010}}   
\newcommand{\hatcurPPfluxperieccenxxxxC}{\ensuremath{1.37_{-0.22}^{+0.54}}} 
\newcommand{\hatcurPPfluxperidimeccenxxxxC}{\ensuremath{9}}         
\newcommand{\hatcurPPfluxapeccenxxxxC}{\ensuremath{1.08\pm0.19}}    
\newcommand{\hatcurPPfluxapdimeccenxxxxC}{\ensuremath{9}}           
\newcommand{\hatcurPPfluxavgeccenxxxxC}{\ensuremath{1.22_{-0.19}^{+0.27}}} 
\newcommand{\hatcurPPfluxavgdimeccenxxxxC}{\ensuremath{9}}          
\newcommand{\hatcurPPfluxavglogeccenxxxxC}{\ensuremath{9.086\pm0.090}} 
\newcommand{\hatcurXsecphaseeccenxxxxC}{\ensuremath{0.497\pm0.046}} 
\newcommand{\hatcurXsecondaryeccenxxxxC}{\ensuremath{2456758.40\pm0.19}} 
\newcommand{\hatcurXsecdureccenxxxxC}{\ensuremath{0.127\pm0.010}}   
\newcommand{\hatcurXsecingdureccenxxxxC}{\ensuremath{0.020\pm0.016}} 
\newcommand{\hatcurPPphiconjeccenxxxxC}{\ensuremath{-0.08\pm0.28}}  
\newcommand{\hatcurPPperieccenxxxxC}{\ensuremath{2456756.7\pm1.2}}  
\newcommand{\hatcurPPaequiveccenxxxxC}{\ensuremath{0.0335\pm0.0033}} 
\newcommand{\hatcurPPtcirceccenxxxxC}{\ensuremath{290\pm170}}       
\newcommand{\hatcurPPtinfalleccenxxxxC}{\ensuremath{4300_{-1500}^{+2300}}} 
\newcommand{\hatcurXdisteccenxxxxC}{\ensuremath{837_{-63}^{+87}}}   
\newcommand{\hatcurXAveccenxxxxC}{\ensuremath{0.033_{-0.033}^{+0.113}}} 
\newcommand{\hatcurXdistredeccenxxxxC}{\ensuremath{824_{-62}^{+86}}} 
\newcommand{\hatcurXEBVeccenxxxxC}{\ensuremath{0.011_{-0.011}^{+0.036}}} 
\newcommand{\hatcurXmvisoredeccenxxxxC}{\ensuremath{13.311\pm0.047}} 
\newcommand{\hatcurXmiisoredeccenxxxxC}{\ensuremath{12.761\pm0.031}} 
\newcommand{\hatcurXmjisoredeccenxxxxC}{\ensuremath{12.436\pm0.016}} 
\newcommand{\hatcurXmhisoredeccenxxxxC}{\ensuremath{12.199\pm0.016}} 
\newcommand{\hatcurXmkisoredeccenxxxxC}{\ensuremath{12.153\pm0.016}} 
\newcommand{\hatcurXviisoredeccenxxxxC}{\ensuremath{0.548\pm0.026}} 
\newcommand{\hatcurXvkisoredeccenxxxxC}{\ensuremath{1.158\pm0.053}} 
\newcommand{\hatcurXjhisoredeccenxxxxC}{\ensuremath{0.237\pm0.016}} 
\newcommand{\hatcurXjkisoredeccenxxxxC}{\ensuremath{0.282\pm0.016}} 
\newcommand{\hatcurCCpmraeccenxxxxC}{\ensuremath{-5.1\pm2.5}}       
\newcommand{\hatcurCCpmdececcenxxxxC}{\ensuremath{2.8\pm1.6}}       
\newcommand{\hatcurCCpmeccenxxxxC}{\ensuremath{5.8\pm3.0}}          

\newcommand{\hatcurhtreccenxxxxD}{HATS755-003}                      
\newcommand{\hatcurfieldeccenxxxxD}{\ensuremath{string}}            
\newcommand{\hatcurCCraeccenxxxxD}{\ensuremath{00^{\mathrm h}26^{\mathrm m}48.58{\mathrm s}}}                     
\newcommand{\hatcurCCdececcenxxxxD}{\ensuremath{-56{\arcdeg}18{\arcmin}58.0{\arcsec}}}                    
\newcommand{\hatcurCCmageccenxxxxD}{13.634}                         
\newcommand{\hatcurCCtwomasseccenxxxxD}{2MASS~00264858-5618580}     
\newcommand{\hatcurCCgsceccenxxxxD}{GSC~8468-01248}                 
\newcommand{\hatcurCCtassmveccenxxxxD}{\ensuremath{13.634\pm0.050}} 
\newcommand{\hatcurCCtassmvshorteccenxxxxD}{\ensuremath{13.6}}      
\newcommand{\hatcurCCtassmBeccenxxxxD}{\ensuremath{14.421\pm0.010}} 
\newcommand{\hatcurCCtassmBshorteccenxxxxD}{\ensuremath{14.4}}      
\newcommand{\hatcurCCtassmIeccenxxxxD}{\ensuremath{nff\pmnff}}      
\newcommand{\hatcurCCtassmIshorteccenxxxxD}{\ensuremath{0.0}}       
\newcommand{\hatcurCCtassmgeccenxxxxD}{\ensuremath{14.018\pm0.010}} 
\newcommand{\hatcurCCtassmgshorteccenxxxxD}{\ensuremath{14.0}}      
\newcommand{\hatcurCCtassmreccenxxxxD}{\ensuremath{13.487\pm0.020}} 
\newcommand{\hatcurCCtassmrshorteccenxxxxD}{\ensuremath{13.5}}      
\newcommand{\hatcurCCtassmieccenxxxxD}{\ensuremath{13.45\pm0.22}}   
\newcommand{\hatcurCCtassmishorteccenxxxxD}{\ensuremath{13.4}}      
\newcommand{\hatcurCCtwomassJmageccenxxxxD}{\ensuremath{12.366\pm0.024}} 
\newcommand{\hatcurCCtwomassHmageccenxxxxD}{\ensuremath{11.993\pm0.022}} 
\newcommand{\hatcurCCtwomassKmageccenxxxxD}{\ensuremath{11.965\pm0.024}} 
\newcommand{\hatcurCCcitJmageccenxxxxD}{\ensuremath{12.381\pm0.025}} 
\newcommand{\hatcurCCcitHmageccenxxxxD}{\ensuremath{11.989\pm0.023}} 
\newcommand{\hatcurCCcitKmageccenxxxxD}{\ensuremath{11.989\pm0.024}} 
\newcommand{\hatcurCCbbJmageccenxxxxD}{\ensuremath{12.433\pm0.025}} 
\newcommand{\hatcurCCbbHmageccenxxxxD}{\ensuremath{12.009\pm0.023}} 
\newcommand{\hatcurCCbbKmageccenxxxxD}{\ensuremath{12.009\pm0.024}} 
\newcommand{\hatcurCCesoJmageccenxxxxD}{\ensuremath{12.436\pm0.027}} 
\newcommand{\hatcurCCesoHmageccenxxxxD}{\ensuremath{12.001\pm0.025}} 
\newcommand{\hatcurCCesoKmageccenxxxxD}{\ensuremath{12.008\pm0.025}} 
\newcommand{\hatcurCCesoJHmageccenxxxxD}{\ensuremath{0.434\pm0.036}} 
\newcommand{\hatcurCCesoJKmageccenxxxxD}{\ensuremath{0.428\pm0.010}} 
\newcommand{\hatcurCCesoHKmageccenxxxxD}{\ensuremath{-0.006\pm0.036}} 
\newcommand{\hatcurLCdipeccenxxxxD}{\ensuremath{14.0}}              
\newcommand{\hatcurLCrprstareccenxxxxD}{\ensuremath{0.1048\pm0.0031}} 
\newcommand{\hatcurLCbsqeccenxxxxD}{\ensuremath{0.25_{-0.13}^{+0.12}}} 
\newcommand{\hatcurLCimpeccenxxxxD}{\ensuremath{0.50_{-0.15}^{+0.11}}} 
\newcommand{\hatcurLCzetaeccenxxxxD}{\ensuremath{23.41\pm0.30}}     
\newcommand{\hatcurLCdureccenxxxxD}{\ensuremath{0.0971\pm0.0026}}   
\newcommand{\hatcurLCdurshorteccenxxxxD}{\ensuremath{0.0971}}       
\newcommand{\hatcurLCdurhreccenxxxxD}{\ensuremath{2.331\pm0.063}}   
\newcommand{\hatcurLCdurhrshorteccenxxxxD}{\ensuremath{2.331}}      
\newcommand{\hatcurLCqeccenxxxxD}{\ensuremath{0.02050\pm0.00056}}   
\newcommand{\hatcurLCqshorteccenxxxxD}{\ensuremath{0.021}}          
\newcommand{\hatcurLCingdureccenxxxxD}{\ensuremath{0.0119\pm0.0023}} 
\newcommand{\hatcurLCPeccenxxxxD}{\ensuremath{4.7423726\pm0.0000047}} 
\newcommand{\hatcurLCPprececcenxxxxD}{\ensuremath{4.7423726}}       
\newcommand{\hatcurLCPshorteccenxxxxD}{\ensuremath{4.7424}}         
\newcommand{\hatcurLCTeccenxxxxD}{\ensuremath{2457357.71574\pm0.00053}} 
\newcommand{\hatcurLCTAeccenxxxxD}{\ensuremath{2455764.2786\pm0.0015}} 
\newcommand{\hatcurLCTBeccenxxxxD}{\ensuremath{2457632.77336\pm0.00065}} 
\newcommand{\hatcurLChatnetmAeccenxxxxD}{\ensuremath{13.524670\pm0.000073}} 
\newcommand{\hatcurLCiblendAeccenxxxxD}{\ensuremath{0.891\pm0.059}} 
\newcommand{\hatcurLChatnetmBeccenxxxxD}{\ensuremath{13.524720\pm0.000091}} 
\newcommand{\hatcurLCiblendBeccenxxxxD}{\ensuremath{0.878\pm0.068}} 
\newcommand{\hatcurLCrhoeccenxxxxD}{\ensuremath{2.8_{-1.4}^{+2.4}}} 
\newcommand{\hatcurSMEiteffeccenxxxxD}{\ensuremath{5495\pm69}}      
\newcommand{\hatcurSMEizfeheccenxxxxD}{\ensuremath{-0.060\pm0.046}} 
\newcommand{\hatcurSMEizfehshorteccenxxxxD}{\ensuremath{-0.06}}     
\newcommand{\hatcurSMEiloggeccenxxxxD}{\ensuremath{4.610\pm0.066}}  
\newcommand{\hatcurSMEivsineccenxxxxD}{\ensuremath{0.90\pm0.66}}    
\newcommand{\hatcurSMEivmaceccenxxxxD}{\ensuremath{3.56\pm0.10}}            
\newcommand{\hatcurSMEivmiceccenxxxxD}{\ensuremath{0.932\pm0.033}}            
\newcommand{\hatcurLBizeccenxxxxD}{\ensuremath{0.2448}}             
\newcommand{\hatcurLBiizeccenxxxxD}{\ensuremath{0.3072}}            
\newcommand{\hatcurLBiieccenxxxxD}{\ensuremath{0.3112}}             
\newcommand{\hatcurLBiiieccenxxxxD}{\ensuremath{0.3042}}            
\newcommand{\hatcurLBiIeccenxxxxD}{\ensuremath{0.2892}}             
\newcommand{\hatcurLBiiIeccenxxxxD}{\ensuremath{0.3055}}            
\newcommand{\hatcurLBigeccenxxxxD}{\ensuremath{0.6142}}             
\newcommand{\hatcurLBiigeccenxxxxD}{\ensuremath{0.1896}}            
\newcommand{\hatcurLBireccenxxxxD}{\ensuremath{0.4078}}             
\newcommand{\hatcurLBiireccenxxxxD}{\ensuremath{0.2935}}            
\newcommand{\hatcurLBiReccenxxxxD}{\ensuremath{0.3812}}             
\newcommand{\hatcurLBiiReccenxxxxD}{\ensuremath{0.2972}}            
\newcommand{\hatcurLBikepeccenxxxxD}{\ensuremath{0.1000}}           
\newcommand{\hatcurLBiikepeccenxxxxD}{\ensuremath{0.1000}}          
\newcommand{\hatcurISOmeccenxxxxD}{\ensuremath{0.911\pm0.033}}      
\newcommand{\hatcurISOmshorteccenxxxxD}{\ensuremath{0.91}}          
\newcommand{\hatcurISOmlongeccenxxxxD}{\ensuremath{0.911\pm0.033}}  
\newcommand{\hatcurISOreccenxxxxD}{\ensuremath{0.911_{-0.062}^{+0.121}}} 
\newcommand{\hatcurISOrshorteccenxxxxD}{\ensuremath{0.91}}          
\newcommand{\hatcurISOrlongeccenxxxxD}{\ensuremath{0.911_{-0.062}^{+0.121}}} 
\newcommand{\hatcurISOrhoeccenxxxxD}{\ensuremath{1.69\pm0.50}}      
\newcommand{\hatcurISOrholongeccenxxxxD}{\ensuremath{1.69\pm0.50}}  
\newcommand{\hatcurISOloggeccenxxxxD}{\ensuremath{4.47\pm0.12}}     
\newcommand{\hatcurISOlumeccenxxxxD}{\ensuremath{0.68_{-0.11}^{+0.22}}} 
\newcommand{\hatcurISOlumshorteccenxxxxD}{\ensuremath{0.68}}        
\newcommand{\hatcurISOmveccenxxxxD}{\ensuremath{5.30\pm0.35}}       
\newcommand{\hatcurISOvieccenxxxxD}{\ensuremath{0.779\pm0.019}}     
\newcommand{\hatcurISOageeccenxxxxD}{\ensuremath{7.4\pm3.8}}        
\newcommand{\hatcurISOsigmaeccenxxxxD}{\ensuremath{0.0004\pm0.0028}} 
\newcommand{\hatcurISOMJeccenxxxxD}{\ensuremath{4.02\pm0.33}}       
\newcommand{\hatcurISOMHeccenxxxxD}{\ensuremath{3.62\pm0.33}}       
\newcommand{\hatcurISOMKeccenxxxxD}{\ensuremath{3.56\pm0.32}}       
\newcommand{\hatcurISOJKeccenxxxxD}{\ensuremath{0.43\pm0.13}}       
\newcommand{\hatcurISOspececcenxxxxD}{G}                            
\newcommand{\hatcurRVKeccenxxxxD}{\ensuremath{23.8\pm9.4}}          
\newcommand{\hatcurRVrkeccenxxxxD}{\ensuremath{-0.06\pm0.32}}       
\newcommand{\hatcurRVrheccenxxxxD}{\ensuremath{0.38\pm0.20}}        
\newcommand{\hatcurRVkeccenxxxxD}{\ensuremath{-0.02\pm0.19}}        
\newcommand{\hatcurRVheccenxxxxD}{\ensuremath{0.18\pm0.14}}         
\newcommand{\hatcurRVtroneeccenxxxxD}{\ensuremath{0\pm0}}           
\newcommand{\hatcurRVtrtwoeccenxxxxD}{\ensuremath{0\pm0}}           
\newcommand{\hatcurRVgammaAeccenxxxxD}{\ensuremath{-30191.6\pm8.7}} 
\newcommand{\hatcurRVjitterAeccenxxxxD}{\ensuremath{33.4\pm7.2}}    
\newcommand{\hatcurRVjittertwosiglimAeccenxxxxD}{\ensuremath{<47.7}} 
\newcommand{\hatcurRVfitrmsAeccenxxxxD}{\ensuremath{0.0}}           
\newcommand{\hatcurRVgammaBeccenxxxxD}{\ensuremath{-5.5\pm8.7}}     
\newcommand{\hatcurRVjitterBeccenxxxxD}{\ensuremath{23.0\pm7.1}}    
\newcommand{\hatcurRVjittertwosiglimBeccenxxxxD}{\ensuremath{<36.7}} 
\newcommand{\hatcurRVfitrmsBeccenxxxxD}{\ensuremath{0.0}}           
\newcommand{\hatcurRVecceneccenxxxxD}{\ensuremath{0.26\pm0.15}}     
\newcommand{\hatcurRVeccentwosiglimeccenxxxxD}{\ensuremath{<0.559}} 
\newcommand{\hatcurRVomegaeccenxxxxD}{\ensuremath{100\pm58}}        
\newcommand{\hatcurPPieccenxxxxD}{\ensuremath{87.24_{-0.96}^{+0.65}}} 
\newcommand{\hatcurPPgeccenxxxxD}{\ensuremath{4.6\pm2.2}}           
\newcommand{\hatcurPPloggeccenxxxxD}{\ensuremath{2.66_{-0.27}^{+0.18}}} 
\newcommand{\hatcurPPareccenxxxxD}{\ensuremath{12.61_{-1.48}^{+1.00}}} 
\newcommand{\hatcurPPareleccenxxxxD}{\ensuremath{0.05355\pm0.00063}} 
\newcommand{\hatcurPPrhoeccenxxxxD}{\ensuremath{0.24\pm0.13}}       
\newcommand{\hatcurPPmeccenxxxxD}{\ensuremath{0.168\pm0.067}}       
\newcommand{\hatcurPPmshorteccenxxxxD}{\ensuremath{0.17}}           
\newcommand{\hatcurPPmlongeccenxxxxD}{\ensuremath{0.168\pm0.067}}   
\newcommand{\hatcurPPmeeccenxxxxD}{\ensuremath{53\pm21}}            
\newcommand{\hatcurPPmeshorteccenxxxxD}{\ensuremath{53.3}}          
\newcommand{\hatcurPPmelongeccenxxxxD}{\ensuremath{53\pm21}}        
\newcommand{\hatcurPPreccenxxxxD}{\ensuremath{0.932_{-0.069}^{+0.127}}} 
\newcommand{\hatcurPPrshorteccenxxxxD}{\ensuremath{0.93}}           
\newcommand{\hatcurPPrlongeccenxxxxD}{\ensuremath{0.932_{-0.069}^{+0.127}}} 
\newcommand{\hatcurPPreeccenxxxxD}{\ensuremath{10.45_{-0.77}^{+1.42}}} 
\newcommand{\hatcurPPreshorteccenxxxxD}{\ensuremath{10.4}}          
\newcommand{\hatcurPPrelongeccenxxxxD}{\ensuremath{10.45_{-0.77}^{+1.42}}} 
\newcommand{\hatcurPPmrcorreccenxxxxD}{\ensuremath{-0.16}}          
\newcommand{\hatcurPPteffeccenxxxxD}{\ensuremath{1105_{-50}^{+90}}} 
\newcommand{\hatcurPPthetaeccenxxxxD}{\ensuremath{0.0214\pm0.0091}} 
\newcommand{\hatcurPPfluxperieccenxxxxD}{\ensuremath{5.9_{-2.1}^{+5.5}}} 
\newcommand{\hatcurPPfluxperidimeccenxxxxD}{\ensuremath{8}}         
\newcommand{\hatcurPPfluxapeccenxxxxD}{\ensuremath{2.18_{-0.37}^{+0.49}}} 
\newcommand{\hatcurPPfluxapdimeccenxxxxD}{\ensuremath{8}}           
\newcommand{\hatcurPPfluxavgeccenxxxxD}{\ensuremath{3.37_{-0.58}^{+1.24}}} 
\newcommand{\hatcurPPfluxavgdimeccenxxxxD}{\ensuremath{8}}          
\newcommand{\hatcurPPfluxavglogeccenxxxxD}{\ensuremath{8.528_{-0.082}^{+0.136}}} 
\newcommand{\hatcurXsecphaseeccenxxxxD}{\ensuremath{0.49\pm0.12}}   
\newcommand{\hatcurXsecondaryeccenxxxxD}{\ensuremath{2457360.02\pm0.58}} 
\newcommand{\hatcurXsecdureccenxxxxD}{\ensuremath{0.118\pm0.036}}   
\newcommand{\hatcurXsecingdureccenxxxxD}{\ensuremath{0.020\pm0.027}} 
\newcommand{\hatcurPPphiconjeccenxxxxD}{\ensuremath{-0.010\pm0.099}} 
\newcommand{\hatcurPPperieccenxxxxD}{\ensuremath{2457357.76\pm0.47}} 
\newcommand{\hatcurPPaequiveccenxxxxD}{\ensuremath{0.0648_{-0.0084}^{+0.0058}}} 
\newcommand{\hatcurPPtcirceccenxxxxD}{\ensuremath{340_{-260}^{+400}}} 
\newcommand{\hatcurPPtinfalleccenxxxxD}{\ensuremath{76000_{-38000}^{+64000}}} 
\newcommand{\hatcurXdisteccenxxxxD}{\ensuremath{490_{-35}^{+67}}}   
\newcommand{\hatcurXAveccenxxxxD}{\ensuremath{0.000\pm0.018}}       
\newcommand{\hatcurXdistredeccenxxxxD}{\ensuremath{480_{-35}^{+67}}} 
\newcommand{\hatcurXEBVeccenxxxxD}{\ensuremath{0.0000\pm0.0058}}    
\newcommand{\hatcurXmvisoredeccenxxxxD}{\ensuremath{13.714\pm0.046}} 
\newcommand{\hatcurXmiisoredeccenxxxxD}{\ensuremath{12.934\pm0.030}} 
\newcommand{\hatcurXmjisoredeccenxxxxD}{\ensuremath{12.430\pm0.016}} 
\newcommand{\hatcurXmhisoredeccenxxxxD}{\ensuremath{12.030\pm0.016}} 
\newcommand{\hatcurXmkisoredeccenxxxxD}{\ensuremath{11.964\pm0.017}} 
\newcommand{\hatcurXviisoredeccenxxxxD}{\ensuremath{0.781\pm0.018}} 
\newcommand{\hatcurXvkisoredeccenxxxxD}{\ensuremath{1.750\pm0.053}} 
\newcommand{\hatcurXjhisoredeccenxxxxD}{\ensuremath{0.400\pm0.015}} 
\newcommand{\hatcurXjkisoredeccenxxxxD}{\ensuremath{0.466\pm0.018}} 
\newcommand{\hatcurCCpmraeccenxxxxD}{\ensuremath{21.3\pm1.7}}       
\newcommand{\hatcurCCpmdececcenxxxxD}{\ensuremath{5.0\pm1.9}}       
\newcommand{\hatcurCCpmeccenxxxxD}{\ensuremath{21.9\pm2.5}}         

\newcommand{\hatcurCCbbHmageccen}[1]{\ifnum#1=43 %
\hatcurCCbbHmageccenxxxxA
\else
\ifnum#1=44 %
\hatcurCCbbHmageccenxxxxB
\else
\ifnum#1=45 %
\hatcurCCbbHmageccenxxxxC
\else
\ifnum#1=46 %
\hatcurCCbbHmageccenxxxxD
\else
??????\fi
\fi
\fi
\fi
}
\newcommand{\hatcurCCbbJmageccen}[1]{\ifnum#1=43 %
\hatcurCCbbJmageccenxxxxA
\else
\ifnum#1=44 %
\hatcurCCbbJmageccenxxxxB
\else
\ifnum#1=45 %
\hatcurCCbbJmageccenxxxxC
\else
\ifnum#1=46 %
\hatcurCCbbJmageccenxxxxD
\else
??????\fi
\fi
\fi
\fi
}
\newcommand{\hatcurCCbbKmageccen}[1]{\ifnum#1=43 %
\hatcurCCbbKmageccenxxxxA
\else
\ifnum#1=44 %
\hatcurCCbbKmageccenxxxxB
\else
\ifnum#1=45 %
\hatcurCCbbKmageccenxxxxC
\else
\ifnum#1=46 %
\hatcurCCbbKmageccenxxxxD
\else
??????\fi
\fi
\fi
\fi
}
\newcommand{\hatcurCCcitHmageccen}[1]{\ifnum#1=43 %
\hatcurCCcitHmageccenxxxxA
\else
\ifnum#1=44 %
\hatcurCCcitHmageccenxxxxB
\else
\ifnum#1=45 %
\hatcurCCcitHmageccenxxxxC
\else
\ifnum#1=46 %
\hatcurCCcitHmageccenxxxxD
\else
??????\fi
\fi
\fi
\fi
}
\newcommand{\hatcurCCcitJmageccen}[1]{\ifnum#1=43 %
\hatcurCCcitJmageccenxxxxA
\else
\ifnum#1=44 %
\hatcurCCcitJmageccenxxxxB
\else
\ifnum#1=45 %
\hatcurCCcitJmageccenxxxxC
\else
\ifnum#1=46 %
\hatcurCCcitJmageccenxxxxD
\else
??????\fi
\fi
\fi
\fi
}
\newcommand{\hatcurCCcitKmageccen}[1]{\ifnum#1=43 %
\hatcurCCcitKmageccenxxxxA
\else
\ifnum#1=44 %
\hatcurCCcitKmageccenxxxxB
\else
\ifnum#1=45 %
\hatcurCCcitKmageccenxxxxC
\else
\ifnum#1=46 %
\hatcurCCcitKmageccenxxxxD
\else
??????\fi
\fi
\fi
\fi
}
\newcommand{\hatcurCCdececcen}[1]{\ifnum#1=43 %
\hatcurCCdececcenxxxxA
\else
\ifnum#1=44 %
\hatcurCCdececcenxxxxB
\else
\ifnum#1=45 %
\hatcurCCdececcenxxxxC
\else
\ifnum#1=46 %
\hatcurCCdececcenxxxxD
\else
??????\fi
\fi
\fi
\fi
}
\newcommand{\hatcurCCesoHKmageccen}[1]{\ifnum#1=43 %
\hatcurCCesoHKmageccenxxxxA
\else
\ifnum#1=44 %
\hatcurCCesoHKmageccenxxxxB
\else
\ifnum#1=45 %
\hatcurCCesoHKmageccenxxxxC
\else
\ifnum#1=46 %
\hatcurCCesoHKmageccenxxxxD
\else
??????\fi
\fi
\fi
\fi
}
\newcommand{\hatcurCCesoHmageccen}[1]{\ifnum#1=43 %
\hatcurCCesoHmageccenxxxxA
\else
\ifnum#1=44 %
\hatcurCCesoHmageccenxxxxB
\else
\ifnum#1=45 %
\hatcurCCesoHmageccenxxxxC
\else
\ifnum#1=46 %
\hatcurCCesoHmageccenxxxxD
\else
??????\fi
\fi
\fi
\fi
}
\newcommand{\hatcurCCesoJHmageccen}[1]{\ifnum#1=43 %
\hatcurCCesoJHmageccenxxxxA
\else
\ifnum#1=44 %
\hatcurCCesoJHmageccenxxxxB
\else
\ifnum#1=45 %
\hatcurCCesoJHmageccenxxxxC
\else
\ifnum#1=46 %
\hatcurCCesoJHmageccenxxxxD
\else
??????\fi
\fi
\fi
\fi
}
\newcommand{\hatcurCCesoJKmageccen}[1]{\ifnum#1=43 %
\hatcurCCesoJKmageccenxxxxA
\else
\ifnum#1=44 %
\hatcurCCesoJKmageccenxxxxB
\else
\ifnum#1=45 %
\hatcurCCesoJKmageccenxxxxC
\else
\ifnum#1=46 %
\hatcurCCesoJKmageccenxxxxD
\else
??????\fi
\fi
\fi
\fi
}
\newcommand{\hatcurCCesoJmageccen}[1]{\ifnum#1=43 %
\hatcurCCesoJmageccenxxxxA
\else
\ifnum#1=44 %
\hatcurCCesoJmageccenxxxxB
\else
\ifnum#1=45 %
\hatcurCCesoJmageccenxxxxC
\else
\ifnum#1=46 %
\hatcurCCesoJmageccenxxxxD
\else
??????\fi
\fi
\fi
\fi
}
\newcommand{\hatcurCCesoKmageccen}[1]{\ifnum#1=43 %
\hatcurCCesoKmageccenxxxxA
\else
\ifnum#1=44 %
\hatcurCCesoKmageccenxxxxB
\else
\ifnum#1=45 %
\hatcurCCesoKmageccenxxxxC
\else
\ifnum#1=46 %
\hatcurCCesoKmageccenxxxxD
\else
??????\fi
\fi
\fi
\fi
}
\newcommand{\hatcurCCgsceccen}[1]{\ifnum#1=43 %
\hatcurCCgsceccenxxxxA
\else
\ifnum#1=44 %
\hatcurCCgsceccenxxxxB
\else
\ifnum#1=45 %
\hatcurCCgsceccenxxxxC
\else
\ifnum#1=46 %
\hatcurCCgsceccenxxxxD
\else
??????\fi
\fi
\fi
\fi
}
\newcommand{\hatcurCCmageccen}[1]{\ifnum#1=43 %
\hatcurCCmageccenxxxxA
\else
\ifnum#1=44 %
\hatcurCCmageccenxxxxB
\else
\ifnum#1=45 %
\hatcurCCmageccenxxxxC
\else
\ifnum#1=46 %
\hatcurCCmageccenxxxxD
\else
??????\fi
\fi
\fi
\fi
}
\newcommand{\hatcurCCpmdececcen}[1]{\ifnum#1=43 %
\hatcurCCpmdececcenxxxxA
\else
\ifnum#1=44 %
\hatcurCCpmdececcenxxxxB
\else
\ifnum#1=45 %
\hatcurCCpmdececcenxxxxC
\else
\ifnum#1=46 %
\hatcurCCpmdececcenxxxxD
\else
??????\fi
\fi
\fi
\fi
}
\newcommand{\hatcurCCpmeccen}[1]{\ifnum#1=43 %
\hatcurCCpmeccenxxxxA
\else
\ifnum#1=44 %
\hatcurCCpmeccenxxxxB
\else
\ifnum#1=45 %
\hatcurCCpmeccenxxxxC
\else
\ifnum#1=46 %
\hatcurCCpmeccenxxxxD
\else
??????\fi
\fi
\fi
\fi
}
\newcommand{\hatcurCCpmraeccen}[1]{\ifnum#1=43 %
\hatcurCCpmraeccenxxxxA
\else
\ifnum#1=44 %
\hatcurCCpmraeccenxxxxB
\else
\ifnum#1=45 %
\hatcurCCpmraeccenxxxxC
\else
\ifnum#1=46 %
\hatcurCCpmraeccenxxxxD
\else
??????\fi
\fi
\fi
\fi
}
\newcommand{\hatcurCCraeccen}[1]{\ifnum#1=43 %
\hatcurCCraeccenxxxxA
\else
\ifnum#1=44 %
\hatcurCCraeccenxxxxB
\else
\ifnum#1=45 %
\hatcurCCraeccenxxxxC
\else
\ifnum#1=46 %
\hatcurCCraeccenxxxxD
\else
??????\fi
\fi
\fi
\fi
}
\newcommand{\hatcurCCtassmBeccen}[1]{\ifnum#1=43 %
\hatcurCCtassmBeccenxxxxA
\else
\ifnum#1=44 %
\hatcurCCtassmBeccenxxxxB
\else
\ifnum#1=45 %
\hatcurCCtassmBeccenxxxxC
\else
\ifnum#1=46 %
\hatcurCCtassmBeccenxxxxD
\else
??????\fi
\fi
\fi
\fi
}
\newcommand{\hatcurCCtassmBshorteccen}[1]{\ifnum#1=43 %
\hatcurCCtassmBshorteccenxxxxA
\else
\ifnum#1=44 %
\hatcurCCtassmBshorteccenxxxxB
\else
\ifnum#1=45 %
\hatcurCCtassmBshorteccenxxxxC
\else
\ifnum#1=46 %
\hatcurCCtassmBshorteccenxxxxD
\else
??????\fi
\fi
\fi
\fi
}
\newcommand{\hatcurCCtassmgeccen}[1]{\ifnum#1=43 %
\hatcurCCtassmgeccenxxxxA
\else
\ifnum#1=44 %
\hatcurCCtassmgeccenxxxxB
\else
\ifnum#1=45 %
\hatcurCCtassmgeccenxxxxC
\else
\ifnum#1=46 %
\hatcurCCtassmgeccenxxxxD
\else
??????\fi
\fi
\fi
\fi
}
\newcommand{\hatcurCCtassmgshorteccen}[1]{\ifnum#1=43 %
\hatcurCCtassmgshorteccenxxxxA
\else
\ifnum#1=44 %
\hatcurCCtassmgshorteccenxxxxB
\else
\ifnum#1=45 %
\hatcurCCtassmgshorteccenxxxxC
\else
\ifnum#1=46 %
\hatcurCCtassmgshorteccenxxxxD
\else
??????\fi
\fi
\fi
\fi
}
\newcommand{\hatcurCCtassmieccen}[1]{\ifnum#1=43 %
\hatcurCCtassmieccenxxxxA
\else
\ifnum#1=44 %
\hatcurCCtassmieccenxxxxB
\else
\ifnum#1=45 %
\hatcurCCtassmieccenxxxxC
\else
\ifnum#1=46 %
\hatcurCCtassmieccenxxxxD
\else
??????\fi
\fi
\fi
\fi
}
\newcommand{\hatcurCCtassmIeccen}[1]{\ifnum#1=43 %
\hatcurCCtassmIeccenxxxxA
\else
\ifnum#1=44 %
\hatcurCCtassmIeccenxxxxB
\else
\ifnum#1=45 %
\hatcurCCtassmIeccenxxxxC
\else
\ifnum#1=46 %
\hatcurCCtassmIeccenxxxxD
\else
??????\fi
\fi
\fi
\fi
}
\newcommand{\hatcurCCtassmishorteccen}[1]{\ifnum#1=43 %
\hatcurCCtassmishorteccenxxxxA
\else
\ifnum#1=44 %
\hatcurCCtassmishorteccenxxxxB
\else
\ifnum#1=45 %
\hatcurCCtassmishorteccenxxxxC
\else
\ifnum#1=46 %
\hatcurCCtassmishorteccenxxxxD
\else
??????\fi
\fi
\fi
\fi
}
\newcommand{\hatcurCCtassmIshorteccen}[1]{\ifnum#1=43 %
\hatcurCCtassmIshorteccenxxxxA
\else
\ifnum#1=44 %
\hatcurCCtassmIshorteccenxxxxB
\else
\ifnum#1=45 %
\hatcurCCtassmIshorteccenxxxxC
\else
\ifnum#1=46 %
\hatcurCCtassmIshorteccenxxxxD
\else
??????\fi
\fi
\fi
\fi
}
\newcommand{\hatcurCCtassmreccen}[1]{\ifnum#1=43 %
\hatcurCCtassmreccenxxxxA
\else
\ifnum#1=44 %
\hatcurCCtassmreccenxxxxB
\else
\ifnum#1=45 %
\hatcurCCtassmreccenxxxxC
\else
\ifnum#1=46 %
\hatcurCCtassmreccenxxxxD
\else
??????\fi
\fi
\fi
\fi
}
\newcommand{\hatcurCCtassmrshorteccen}[1]{\ifnum#1=43 %
\hatcurCCtassmrshorteccenxxxxA
\else
\ifnum#1=44 %
\hatcurCCtassmrshorteccenxxxxB
\else
\ifnum#1=45 %
\hatcurCCtassmrshorteccenxxxxC
\else
\ifnum#1=46 %
\hatcurCCtassmrshorteccenxxxxD
\else
??????\fi
\fi
\fi
\fi
}
\newcommand{\hatcurCCtassmveccen}[1]{\ifnum#1=43 %
\hatcurCCtassmveccenxxxxA
\else
\ifnum#1=44 %
\hatcurCCtassmveccenxxxxB
\else
\ifnum#1=45 %
\hatcurCCtassmveccenxxxxC
\else
\ifnum#1=46 %
\hatcurCCtassmveccenxxxxD
\else
??????\fi
\fi
\fi
\fi
}
\newcommand{\hatcurCCtassmvshorteccen}[1]{\ifnum#1=43 %
\hatcurCCtassmvshorteccenxxxxA
\else
\ifnum#1=44 %
\hatcurCCtassmvshorteccenxxxxB
\else
\ifnum#1=45 %
\hatcurCCtassmvshorteccenxxxxC
\else
\ifnum#1=46 %
\hatcurCCtassmvshorteccenxxxxD
\else
??????\fi
\fi
\fi
\fi
}
\newcommand{\hatcurCCtwomasseccen}[1]{\ifnum#1=43 %
\hatcurCCtwomasseccenxxxxA
\else
\ifnum#1=44 %
\hatcurCCtwomasseccenxxxxB
\else
\ifnum#1=45 %
\hatcurCCtwomasseccenxxxxC
\else
\ifnum#1=46 %
\hatcurCCtwomasseccenxxxxD
\else
??????\fi
\fi
\fi
\fi
}
\newcommand{\hatcurCCtwomassHmageccen}[1]{\ifnum#1=43 %
\hatcurCCtwomassHmageccenxxxxA
\else
\ifnum#1=44 %
\hatcurCCtwomassHmageccenxxxxB
\else
\ifnum#1=45 %
\hatcurCCtwomassHmageccenxxxxC
\else
\ifnum#1=46 %
\hatcurCCtwomassHmageccenxxxxD
\else
??????\fi
\fi
\fi
\fi
}
\newcommand{\hatcurCCtwomassJmageccen}[1]{\ifnum#1=43 %
\hatcurCCtwomassJmageccenxxxxA
\else
\ifnum#1=44 %
\hatcurCCtwomassJmageccenxxxxB
\else
\ifnum#1=45 %
\hatcurCCtwomassJmageccenxxxxC
\else
\ifnum#1=46 %
\hatcurCCtwomassJmageccenxxxxD
\else
??????\fi
\fi
\fi
\fi
}
\newcommand{\hatcurCCtwomassKmageccen}[1]{\ifnum#1=43 %
\hatcurCCtwomassKmageccenxxxxA
\else
\ifnum#1=44 %
\hatcurCCtwomassKmageccenxxxxB
\else
\ifnum#1=45 %
\hatcurCCtwomassKmageccenxxxxC
\else
\ifnum#1=46 %
\hatcurCCtwomassKmageccenxxxxD
\else
??????\fi
\fi
\fi
\fi
}
\newcommand{\hatcurfieldeccen}[1]{\ifnum#1=43 %
\hatcurfieldeccenxxxxA
\else
\ifnum#1=44 %
\hatcurfieldeccenxxxxB
\else
\ifnum#1=45 %
\hatcurfieldeccenxxxxC
\else
\ifnum#1=46 %
\hatcurfieldeccenxxxxD
\else
??????\fi
\fi
\fi
\fi
}
\newcommand{\hatcurhtreccen}[1]{\ifnum#1=43 %
\hatcurhtreccenxxxxA
\else
\ifnum#1=44 %
\hatcurhtreccenxxxxB
\else
\ifnum#1=45 %
\hatcurhtreccenxxxxC
\else
\ifnum#1=46 %
\hatcurhtreccenxxxxD
\else
??????\fi
\fi
\fi
\fi
}
\newcommand{\hatcurISOageeccen}[1]{\ifnum#1=43 %
\hatcurISOageeccenxxxxA
\else
\ifnum#1=44 %
\hatcurISOageeccenxxxxB
\else
\ifnum#1=45 %
\hatcurISOageeccenxxxxC
\else
\ifnum#1=46 %
\hatcurISOageeccenxxxxD
\else
??????\fi
\fi
\fi
\fi
}
\newcommand{\hatcurISOJKeccen}[1]{\ifnum#1=43 %
\hatcurISOJKeccenxxxxA
\else
\ifnum#1=44 %
\hatcurISOJKeccenxxxxB
\else
\ifnum#1=45 %
\hatcurISOJKeccenxxxxC
\else
\ifnum#1=46 %
\hatcurISOJKeccenxxxxD
\else
??????\fi
\fi
\fi
\fi
}
\newcommand{\hatcurISOloggeccen}[1]{\ifnum#1=43 %
\hatcurISOloggeccenxxxxA
\else
\ifnum#1=44 %
\hatcurISOloggeccenxxxxB
\else
\ifnum#1=45 %
\hatcurISOloggeccenxxxxC
\else
\ifnum#1=46 %
\hatcurISOloggeccenxxxxD
\else
??????\fi
\fi
\fi
\fi
}
\newcommand{\hatcurISOlumeccen}[1]{\ifnum#1=43 %
\hatcurISOlumeccenxxxxA
\else
\ifnum#1=44 %
\hatcurISOlumeccenxxxxB
\else
\ifnum#1=45 %
\hatcurISOlumeccenxxxxC
\else
\ifnum#1=46 %
\hatcurISOlumeccenxxxxD
\else
??????\fi
\fi
\fi
\fi
}
\newcommand{\hatcurISOlumshorteccen}[1]{\ifnum#1=43 %
\hatcurISOlumshorteccenxxxxA
\else
\ifnum#1=44 %
\hatcurISOlumshorteccenxxxxB
\else
\ifnum#1=45 %
\hatcurISOlumshorteccenxxxxC
\else
\ifnum#1=46 %
\hatcurISOlumshorteccenxxxxD
\else
??????\fi
\fi
\fi
\fi
}
\newcommand{\hatcurISOmeccen}[1]{\ifnum#1=43 %
\hatcurISOmeccenxxxxA
\else
\ifnum#1=44 %
\hatcurISOmeccenxxxxB
\else
\ifnum#1=45 %
\hatcurISOmeccenxxxxC
\else
\ifnum#1=46 %
\hatcurISOmeccenxxxxD
\else
??????\fi
\fi
\fi
\fi
}
\newcommand{\hatcurISOMHeccen}[1]{\ifnum#1=43 %
\hatcurISOMHeccenxxxxA
\else
\ifnum#1=44 %
\hatcurISOMHeccenxxxxB
\else
\ifnum#1=45 %
\hatcurISOMHeccenxxxxC
\else
\ifnum#1=46 %
\hatcurISOMHeccenxxxxD
\else
??????\fi
\fi
\fi
\fi
}
\newcommand{\hatcurISOMJeccen}[1]{\ifnum#1=43 %
\hatcurISOMJeccenxxxxA
\else
\ifnum#1=44 %
\hatcurISOMJeccenxxxxB
\else
\ifnum#1=45 %
\hatcurISOMJeccenxxxxC
\else
\ifnum#1=46 %
\hatcurISOMJeccenxxxxD
\else
??????\fi
\fi
\fi
\fi
}
\newcommand{\hatcurISOMKeccen}[1]{\ifnum#1=43 %
\hatcurISOMKeccenxxxxA
\else
\ifnum#1=44 %
\hatcurISOMKeccenxxxxB
\else
\ifnum#1=45 %
\hatcurISOMKeccenxxxxC
\else
\ifnum#1=46 %
\hatcurISOMKeccenxxxxD
\else
??????\fi
\fi
\fi
\fi
}
\newcommand{\hatcurISOmlongeccen}[1]{\ifnum#1=43 %
\hatcurISOmlongeccenxxxxA
\else
\ifnum#1=44 %
\hatcurISOmlongeccenxxxxB
\else
\ifnum#1=45 %
\hatcurISOmlongeccenxxxxC
\else
\ifnum#1=46 %
\hatcurISOmlongeccenxxxxD
\else
??????\fi
\fi
\fi
\fi
}
\newcommand{\hatcurISOmshorteccen}[1]{\ifnum#1=43 %
\hatcurISOmshorteccenxxxxA
\else
\ifnum#1=44 %
\hatcurISOmshorteccenxxxxB
\else
\ifnum#1=45 %
\hatcurISOmshorteccenxxxxC
\else
\ifnum#1=46 %
\hatcurISOmshorteccenxxxxD
\else
??????\fi
\fi
\fi
\fi
}
\newcommand{\hatcurISOmveccen}[1]{\ifnum#1=43 %
\hatcurISOmveccenxxxxA
\else
\ifnum#1=44 %
\hatcurISOmveccenxxxxB
\else
\ifnum#1=45 %
\hatcurISOmveccenxxxxC
\else
\ifnum#1=46 %
\hatcurISOmveccenxxxxD
\else
??????\fi
\fi
\fi
\fi
}
\newcommand{\hatcurISOreccen}[1]{\ifnum#1=43 %
\hatcurISOreccenxxxxA
\else
\ifnum#1=44 %
\hatcurISOreccenxxxxB
\else
\ifnum#1=45 %
\hatcurISOreccenxxxxC
\else
\ifnum#1=46 %
\hatcurISOreccenxxxxD
\else
??????\fi
\fi
\fi
\fi
}
\newcommand{\hatcurISOrhoeccen}[1]{\ifnum#1=43 %
\hatcurISOrhoeccenxxxxA
\else
\ifnum#1=44 %
\hatcurISOrhoeccenxxxxB
\else
\ifnum#1=45 %
\hatcurISOrhoeccenxxxxC
\else
\ifnum#1=46 %
\hatcurISOrhoeccenxxxxD
\else
??????\fi
\fi
\fi
\fi
}
\newcommand{\hatcurISOrholongeccen}[1]{\ifnum#1=43 %
\hatcurISOrholongeccenxxxxA
\else
\ifnum#1=44 %
\hatcurISOrholongeccenxxxxB
\else
\ifnum#1=45 %
\hatcurISOrholongeccenxxxxC
\else
\ifnum#1=46 %
\hatcurISOrholongeccenxxxxD
\else
??????\fi
\fi
\fi
\fi
}
\newcommand{\hatcurISOrlongeccen}[1]{\ifnum#1=43 %
\hatcurISOrlongeccenxxxxA
\else
\ifnum#1=44 %
\hatcurISOrlongeccenxxxxB
\else
\ifnum#1=45 %
\hatcurISOrlongeccenxxxxC
\else
\ifnum#1=46 %
\hatcurISOrlongeccenxxxxD
\else
??????\fi
\fi
\fi
\fi
}
\newcommand{\hatcurISOrshorteccen}[1]{\ifnum#1=43 %
\hatcurISOrshorteccenxxxxA
\else
\ifnum#1=44 %
\hatcurISOrshorteccenxxxxB
\else
\ifnum#1=45 %
\hatcurISOrshorteccenxxxxC
\else
\ifnum#1=46 %
\hatcurISOrshorteccenxxxxD
\else
??????\fi
\fi
\fi
\fi
}
\newcommand{\hatcurISOsigmaeccen}[1]{\ifnum#1=43 %
\hatcurISOsigmaeccenxxxxA
\else
\ifnum#1=44 %
\hatcurISOsigmaeccenxxxxB
\else
\ifnum#1=45 %
\hatcurISOsigmaeccenxxxxC
\else
\ifnum#1=46 %
\hatcurISOsigmaeccenxxxxD
\else
??????\fi
\fi
\fi
\fi
}
\newcommand{\hatcurISOspececcen}[1]{\ifnum#1=43 %
\hatcurISOspececcenxxxxA
\else
\ifnum#1=44 %
\hatcurISOspececcenxxxxB
\else
\ifnum#1=45 %
\hatcurISOspececcenxxxxC
\else
\ifnum#1=46 %
\hatcurISOspececcenxxxxD
\else
??????\fi
\fi
\fi
\fi
}
\newcommand{\hatcurISOvieccen}[1]{\ifnum#1=43 %
\hatcurISOvieccenxxxxA
\else
\ifnum#1=44 %
\hatcurISOvieccenxxxxB
\else
\ifnum#1=45 %
\hatcurISOvieccenxxxxC
\else
\ifnum#1=46 %
\hatcurISOvieccenxxxxD
\else
??????\fi
\fi
\fi
\fi
}
\newcommand{\hatcurLBigeccen}[1]{\ifnum#1=43 %
\hatcurLBigeccenxxxxA
\else
\ifnum#1=44 %
\hatcurLBigeccenxxxxB
\else
\ifnum#1=45 %
\hatcurLBigeccenxxxxC
\else
\ifnum#1=46 %
\hatcurLBigeccenxxxxD
\else
??????\fi
\fi
\fi
\fi
}
\newcommand{\hatcurLBiieccen}[1]{\ifnum#1=43 %
\hatcurLBiieccenxxxxA
\else
\ifnum#1=44 %
\hatcurLBiieccenxxxxB
\else
\ifnum#1=45 %
\hatcurLBiieccenxxxxC
\else
\ifnum#1=46 %
\hatcurLBiieccenxxxxD
\else
??????\fi
\fi
\fi
\fi
}
\newcommand{\hatcurLBiIeccen}[1]{\ifnum#1=43 %
\hatcurLBiIeccenxxxxA
\else
\ifnum#1=44 %
\hatcurLBiIeccenxxxxB
\else
\ifnum#1=45 %
\hatcurLBiIeccenxxxxC
\else
\ifnum#1=46 %
\hatcurLBiIeccenxxxxD
\else
??????\fi
\fi
\fi
\fi
}
\newcommand{\hatcurLBiigeccen}[1]{\ifnum#1=43 %
\hatcurLBiigeccenxxxxA
\else
\ifnum#1=44 %
\hatcurLBiigeccenxxxxB
\else
\ifnum#1=45 %
\hatcurLBiigeccenxxxxC
\else
\ifnum#1=46 %
\hatcurLBiigeccenxxxxD
\else
??????\fi
\fi
\fi
\fi
}
\newcommand{\hatcurLBiiieccen}[1]{\ifnum#1=43 %
\hatcurLBiiieccenxxxxA
\else
\ifnum#1=44 %
\hatcurLBiiieccenxxxxB
\else
\ifnum#1=45 %
\hatcurLBiiieccenxxxxC
\else
\ifnum#1=46 %
\hatcurLBiiieccenxxxxD
\else
??????\fi
\fi
\fi
\fi
}
\newcommand{\hatcurLBiiIeccen}[1]{\ifnum#1=43 %
\hatcurLBiiIeccenxxxxA
\else
\ifnum#1=44 %
\hatcurLBiiIeccenxxxxB
\else
\ifnum#1=45 %
\hatcurLBiiIeccenxxxxC
\else
\ifnum#1=46 %
\hatcurLBiiIeccenxxxxD
\else
??????\fi
\fi
\fi
\fi
}
\newcommand{\hatcurLBiikepeccen}[1]{\ifnum#1=43 %
\hatcurLBiikepeccenxxxxA
\else
\ifnum#1=44 %
\hatcurLBiikepeccenxxxxB
\else
\ifnum#1=45 %
\hatcurLBiikepeccenxxxxC
\else
\ifnum#1=46 %
\hatcurLBiikepeccenxxxxD
\else
??????\fi
\fi
\fi
\fi
}
\newcommand{\hatcurLBiireccen}[1]{\ifnum#1=43 %
\hatcurLBiireccenxxxxA
\else
\ifnum#1=44 %
\hatcurLBiireccenxxxxB
\else
\ifnum#1=45 %
\hatcurLBiireccenxxxxC
\else
\ifnum#1=46 %
\hatcurLBiireccenxxxxD
\else
??????\fi
\fi
\fi
\fi
}
\newcommand{\hatcurLBiiReccen}[1]{\ifnum#1=43 %
\hatcurLBiiReccenxxxxA
\else
\ifnum#1=44 %
\hatcurLBiiReccenxxxxB
\else
\ifnum#1=45 %
\hatcurLBiiReccenxxxxC
\else
\ifnum#1=46 %
\hatcurLBiiReccenxxxxD
\else
??????\fi
\fi
\fi
\fi
}
\newcommand{\hatcurLBiizeccen}[1]{\ifnum#1=43 %
\hatcurLBiizeccenxxxxA
\else
\ifnum#1=44 %
\hatcurLBiizeccenxxxxB
\else
\ifnum#1=45 %
\hatcurLBiizeccenxxxxC
\else
\ifnum#1=46 %
\hatcurLBiizeccenxxxxD
\else
??????\fi
\fi
\fi
\fi
}
\newcommand{\hatcurLBikepeccen}[1]{\ifnum#1=43 %
\hatcurLBikepeccenxxxxA
\else
\ifnum#1=44 %
\hatcurLBikepeccenxxxxB
\else
\ifnum#1=45 %
\hatcurLBikepeccenxxxxC
\else
\ifnum#1=46 %
\hatcurLBikepeccenxxxxD
\else
??????\fi
\fi
\fi
\fi
}
\newcommand{\hatcurLBireccen}[1]{\ifnum#1=43 %
\hatcurLBireccenxxxxA
\else
\ifnum#1=44 %
\hatcurLBireccenxxxxB
\else
\ifnum#1=45 %
\hatcurLBireccenxxxxC
\else
\ifnum#1=46 %
\hatcurLBireccenxxxxD
\else
??????\fi
\fi
\fi
\fi
}
\newcommand{\hatcurLBiReccen}[1]{\ifnum#1=43 %
\hatcurLBiReccenxxxxA
\else
\ifnum#1=44 %
\hatcurLBiReccenxxxxB
\else
\ifnum#1=45 %
\hatcurLBiReccenxxxxC
\else
\ifnum#1=46 %
\hatcurLBiReccenxxxxD
\else
??????\fi
\fi
\fi
\fi
}
\newcommand{\hatcurLBizeccen}[1]{\ifnum#1=43 %
\hatcurLBizeccenxxxxA
\else
\ifnum#1=44 %
\hatcurLBizeccenxxxxB
\else
\ifnum#1=45 %
\hatcurLBizeccenxxxxC
\else
\ifnum#1=46 %
\hatcurLBizeccenxxxxD
\else
??????\fi
\fi
\fi
\fi
}
\newcommand{\hatcurLCbsqeccen}[1]{\ifnum#1=43 %
\hatcurLCbsqeccenxxxxA
\else
\ifnum#1=44 %
\hatcurLCbsqeccenxxxxB
\else
\ifnum#1=45 %
\hatcurLCbsqeccenxxxxC
\else
\ifnum#1=46 %
\hatcurLCbsqeccenxxxxD
\else
??????\fi
\fi
\fi
\fi
}
\newcommand{\hatcurLCdipeccen}[1]{\ifnum#1=43 %
\hatcurLCdipeccenxxxxA
\else
\ifnum#1=44 %
\hatcurLCdipeccenxxxxB
\else
\ifnum#1=45 %
\hatcurLCdipeccenxxxxC
\else
\ifnum#1=46 %
\hatcurLCdipeccenxxxxD
\else
??????\fi
\fi
\fi
\fi
}
\newcommand{\hatcurLCdureccen}[1]{\ifnum#1=43 %
\hatcurLCdureccenxxxxA
\else
\ifnum#1=44 %
\hatcurLCdureccenxxxxB
\else
\ifnum#1=45 %
\hatcurLCdureccenxxxxC
\else
\ifnum#1=46 %
\hatcurLCdureccenxxxxD
\else
??????\fi
\fi
\fi
\fi
}
\newcommand{\hatcurLCdurhreccen}[1]{\ifnum#1=43 %
\hatcurLCdurhreccenxxxxA
\else
\ifnum#1=44 %
\hatcurLCdurhreccenxxxxB
\else
\ifnum#1=45 %
\hatcurLCdurhreccenxxxxC
\else
\ifnum#1=46 %
\hatcurLCdurhreccenxxxxD
\else
??????\fi
\fi
\fi
\fi
}
\newcommand{\hatcurLCdurhrshorteccen}[1]{\ifnum#1=43 %
\hatcurLCdurhrshorteccenxxxxA
\else
\ifnum#1=44 %
\hatcurLCdurhrshorteccenxxxxB
\else
\ifnum#1=45 %
\hatcurLCdurhrshorteccenxxxxC
\else
\ifnum#1=46 %
\hatcurLCdurhrshorteccenxxxxD
\else
??????\fi
\fi
\fi
\fi
}
\newcommand{\hatcurLCdurshorteccen}[1]{\ifnum#1=43 %
\hatcurLCdurshorteccenxxxxA
\else
\ifnum#1=44 %
\hatcurLCdurshorteccenxxxxB
\else
\ifnum#1=45 %
\hatcurLCdurshorteccenxxxxC
\else
\ifnum#1=46 %
\hatcurLCdurshorteccenxxxxD
\else
??????\fi
\fi
\fi
\fi
}
\newcommand{\hatcurLChatnetmAeccen}[1]{\ifnum#1=44 %
\hatcurLChatnetmAeccenxxxxB
\else
\ifnum#1=46 %
\hatcurLChatnetmAeccenxxxxD
\else
??????\fi
\fi
}
\newcommand{\hatcurLChatnetmBeccen}[1]{\ifnum#1=44 %
\hatcurLChatnetmBeccenxxxxB
\else
\ifnum#1=46 %
\hatcurLChatnetmBeccenxxxxD
\else
??????\fi
\fi
}
\newcommand{\hatcurLChatnetmeccen}[1]{\ifnum#1=45 %
\hatcurLChatnetmeccenxxxxC
\else
??????\fi
}
\newcommand{\hatcurLCiblendAeccen}[1]{\ifnum#1=44 %
\hatcurLCiblendAeccenxxxxB
\else
\ifnum#1=46 %
\hatcurLCiblendAeccenxxxxD
\else
??????\fi
\fi
}
\newcommand{\hatcurLCiblendBeccen}[1]{\ifnum#1=44 %
\hatcurLCiblendBeccenxxxxB
\else
\ifnum#1=46 %
\hatcurLCiblendBeccenxxxxD
\else
??????\fi
\fi
}
\newcommand{\hatcurLCiblendeccen}[1]{\ifnum#1=45 %
\hatcurLCiblendeccenxxxxC
\else
??????\fi
}
\newcommand{\hatcurLCimpeccen}[1]{\ifnum#1=43 %
\hatcurLCimpeccenxxxxA
\else
\ifnum#1=44 %
\hatcurLCimpeccenxxxxB
\else
\ifnum#1=45 %
\hatcurLCimpeccenxxxxC
\else
\ifnum#1=46 %
\hatcurLCimpeccenxxxxD
\else
??????\fi
\fi
\fi
\fi
}
\newcommand{\hatcurLCingdureccen}[1]{\ifnum#1=43 %
\hatcurLCingdureccenxxxxA
\else
\ifnum#1=44 %
\hatcurLCingdureccenxxxxB
\else
\ifnum#1=45 %
\hatcurLCingdureccenxxxxC
\else
\ifnum#1=46 %
\hatcurLCingdureccenxxxxD
\else
??????\fi
\fi
\fi
\fi
}
\newcommand{\hatcurLCPeccen}[1]{\ifnum#1=43 %
\hatcurLCPeccenxxxxA
\else
\ifnum#1=44 %
\hatcurLCPeccenxxxxB
\else
\ifnum#1=45 %
\hatcurLCPeccenxxxxC
\else
\ifnum#1=46 %
\hatcurLCPeccenxxxxD
\else
??????\fi
\fi
\fi
\fi
}
\newcommand{\hatcurLCPprececcen}[1]{\ifnum#1=43 %
\hatcurLCPprececcenxxxxA
\else
\ifnum#1=44 %
\hatcurLCPprececcenxxxxB
\else
\ifnum#1=45 %
\hatcurLCPprececcenxxxxC
\else
\ifnum#1=46 %
\hatcurLCPprececcenxxxxD
\else
??????\fi
\fi
\fi
\fi
}
\newcommand{\hatcurLCPshorteccen}[1]{\ifnum#1=43 %
\hatcurLCPshorteccenxxxxA
\else
\ifnum#1=44 %
\hatcurLCPshorteccenxxxxB
\else
\ifnum#1=45 %
\hatcurLCPshorteccenxxxxC
\else
\ifnum#1=46 %
\hatcurLCPshorteccenxxxxD
\else
??????\fi
\fi
\fi
\fi
}
\newcommand{\hatcurLCqeccen}[1]{\ifnum#1=43 %
\hatcurLCqeccenxxxxA
\else
\ifnum#1=44 %
\hatcurLCqeccenxxxxB
\else
\ifnum#1=45 %
\hatcurLCqeccenxxxxC
\else
\ifnum#1=46 %
\hatcurLCqeccenxxxxD
\else
??????\fi
\fi
\fi
\fi
}
\newcommand{\hatcurLCqshorteccen}[1]{\ifnum#1=43 %
\hatcurLCqshorteccenxxxxA
\else
\ifnum#1=44 %
\hatcurLCqshorteccenxxxxB
\else
\ifnum#1=45 %
\hatcurLCqshorteccenxxxxC
\else
\ifnum#1=46 %
\hatcurLCqshorteccenxxxxD
\else
??????\fi
\fi
\fi
\fi
}
\newcommand{\hatcurLCrhoeccen}[1]{\ifnum#1=43 %
\hatcurLCrhoeccenxxxxA
\else
\ifnum#1=44 %
\hatcurLCrhoeccenxxxxB
\else
\ifnum#1=45 %
\hatcurLCrhoeccenxxxxC
\else
\ifnum#1=46 %
\hatcurLCrhoeccenxxxxD
\else
??????\fi
\fi
\fi
\fi
}
\newcommand{\hatcurLCrprstareccen}[1]{\ifnum#1=43 %
\hatcurLCrprstareccenxxxxA
\else
\ifnum#1=44 %
\hatcurLCrprstareccenxxxxB
\else
\ifnum#1=45 %
\hatcurLCrprstareccenxxxxC
\else
\ifnum#1=46 %
\hatcurLCrprstareccenxxxxD
\else
??????\fi
\fi
\fi
\fi
}
\newcommand{\hatcurLCTAeccen}[1]{\ifnum#1=43 %
\hatcurLCTAeccenxxxxA
\else
\ifnum#1=44 %
\hatcurLCTAeccenxxxxB
\else
\ifnum#1=45 %
\hatcurLCTAeccenxxxxC
\else
\ifnum#1=46 %
\hatcurLCTAeccenxxxxD
\else
??????\fi
\fi
\fi
\fi
}
\newcommand{\hatcurLCTBeccen}[1]{\ifnum#1=43 %
\hatcurLCTBeccenxxxxA
\else
\ifnum#1=44 %
\hatcurLCTBeccenxxxxB
\else
\ifnum#1=45 %
\hatcurLCTBeccenxxxxC
\else
\ifnum#1=46 %
\hatcurLCTBeccenxxxxD
\else
??????\fi
\fi
\fi
\fi
}
\newcommand{\hatcurLCTeccen}[1]{\ifnum#1=43 %
\hatcurLCTeccenxxxxA
\else
\ifnum#1=44 %
\hatcurLCTeccenxxxxB
\else
\ifnum#1=45 %
\hatcurLCTeccenxxxxC
\else
\ifnum#1=46 %
\hatcurLCTeccenxxxxD
\else
??????\fi
\fi
\fi
\fi
}
\newcommand{\hatcurLCzetaeccen}[1]{\ifnum#1=43 %
\hatcurLCzetaeccenxxxxA
\else
\ifnum#1=44 %
\hatcurLCzetaeccenxxxxB
\else
\ifnum#1=45 %
\hatcurLCzetaeccenxxxxC
\else
\ifnum#1=46 %
\hatcurLCzetaeccenxxxxD
\else
??????\fi
\fi
\fi
\fi
}
\newcommand{\hatcurPPaequiveccen}[1]{\ifnum#1=43 %
\hatcurPPaequiveccenxxxxA
\else
\ifnum#1=44 %
\hatcurPPaequiveccenxxxxB
\else
\ifnum#1=45 %
\hatcurPPaequiveccenxxxxC
\else
\ifnum#1=46 %
\hatcurPPaequiveccenxxxxD
\else
??????\fi
\fi
\fi
\fi
}
\newcommand{\hatcurPPareccen}[1]{\ifnum#1=43 %
\hatcurPPareccenxxxxA
\else
\ifnum#1=44 %
\hatcurPPareccenxxxxB
\else
\ifnum#1=45 %
\hatcurPPareccenxxxxC
\else
\ifnum#1=46 %
\hatcurPPareccenxxxxD
\else
??????\fi
\fi
\fi
\fi
}
\newcommand{\hatcurPPareleccen}[1]{\ifnum#1=43 %
\hatcurPPareleccenxxxxA
\else
\ifnum#1=44 %
\hatcurPPareleccenxxxxB
\else
\ifnum#1=45 %
\hatcurPPareleccenxxxxC
\else
\ifnum#1=46 %
\hatcurPPareleccenxxxxD
\else
??????\fi
\fi
\fi
\fi
}
\newcommand{\hatcurPPfluxapdimeccen}[1]{\ifnum#1=43 %
\hatcurPPfluxapdimeccenxxxxA
\else
\ifnum#1=44 %
\hatcurPPfluxapdimeccenxxxxB
\else
\ifnum#1=45 %
\hatcurPPfluxapdimeccenxxxxC
\else
\ifnum#1=46 %
\hatcurPPfluxapdimeccenxxxxD
\else
??????\fi
\fi
\fi
\fi
}
\newcommand{\hatcurPPfluxapeccen}[1]{\ifnum#1=43 %
\hatcurPPfluxapeccenxxxxA
\else
\ifnum#1=44 %
\hatcurPPfluxapeccenxxxxB
\else
\ifnum#1=45 %
\hatcurPPfluxapeccenxxxxC
\else
\ifnum#1=46 %
\hatcurPPfluxapeccenxxxxD
\else
??????\fi
\fi
\fi
\fi
}
\newcommand{\hatcurPPfluxavgdimeccen}[1]{\ifnum#1=43 %
\hatcurPPfluxavgdimeccenxxxxA
\else
\ifnum#1=44 %
\hatcurPPfluxavgdimeccenxxxxB
\else
\ifnum#1=45 %
\hatcurPPfluxavgdimeccenxxxxC
\else
\ifnum#1=46 %
\hatcurPPfluxavgdimeccenxxxxD
\else
??????\fi
\fi
\fi
\fi
}
\newcommand{\hatcurPPfluxavgeccen}[1]{\ifnum#1=43 %
\hatcurPPfluxavgeccenxxxxA
\else
\ifnum#1=44 %
\hatcurPPfluxavgeccenxxxxB
\else
\ifnum#1=45 %
\hatcurPPfluxavgeccenxxxxC
\else
\ifnum#1=46 %
\hatcurPPfluxavgeccenxxxxD
\else
??????\fi
\fi
\fi
\fi
}
\newcommand{\hatcurPPfluxavglogeccen}[1]{\ifnum#1=43 %
\hatcurPPfluxavglogeccenxxxxA
\else
\ifnum#1=44 %
\hatcurPPfluxavglogeccenxxxxB
\else
\ifnum#1=45 %
\hatcurPPfluxavglogeccenxxxxC
\else
\ifnum#1=46 %
\hatcurPPfluxavglogeccenxxxxD
\else
??????\fi
\fi
\fi
\fi
}
\newcommand{\hatcurPPfluxperidimeccen}[1]{\ifnum#1=43 %
\hatcurPPfluxperidimeccenxxxxA
\else
\ifnum#1=44 %
\hatcurPPfluxperidimeccenxxxxB
\else
\ifnum#1=45 %
\hatcurPPfluxperidimeccenxxxxC
\else
\ifnum#1=46 %
\hatcurPPfluxperidimeccenxxxxD
\else
??????\fi
\fi
\fi
\fi
}
\newcommand{\hatcurPPfluxperieccen}[1]{\ifnum#1=43 %
\hatcurPPfluxperieccenxxxxA
\else
\ifnum#1=44 %
\hatcurPPfluxperieccenxxxxB
\else
\ifnum#1=45 %
\hatcurPPfluxperieccenxxxxC
\else
\ifnum#1=46 %
\hatcurPPfluxperieccenxxxxD
\else
??????\fi
\fi
\fi
\fi
}
\newcommand{\hatcurPPgeccen}[1]{\ifnum#1=43 %
\hatcurPPgeccenxxxxA
\else
\ifnum#1=44 %
\hatcurPPgeccenxxxxB
\else
\ifnum#1=45 %
\hatcurPPgeccenxxxxC
\else
\ifnum#1=46 %
\hatcurPPgeccenxxxxD
\else
??????\fi
\fi
\fi
\fi
}
\newcommand{\hatcurPPieccen}[1]{\ifnum#1=43 %
\hatcurPPieccenxxxxA
\else
\ifnum#1=44 %
\hatcurPPieccenxxxxB
\else
\ifnum#1=45 %
\hatcurPPieccenxxxxC
\else
\ifnum#1=46 %
\hatcurPPieccenxxxxD
\else
??????\fi
\fi
\fi
\fi
}
\newcommand{\hatcurPPloggeccen}[1]{\ifnum#1=43 %
\hatcurPPloggeccenxxxxA
\else
\ifnum#1=44 %
\hatcurPPloggeccenxxxxB
\else
\ifnum#1=45 %
\hatcurPPloggeccenxxxxC
\else
\ifnum#1=46 %
\hatcurPPloggeccenxxxxD
\else
??????\fi
\fi
\fi
\fi
}
\newcommand{\hatcurPPmeccen}[1]{\ifnum#1=43 %
\hatcurPPmeccenxxxxA
\else
\ifnum#1=44 %
\hatcurPPmeccenxxxxB
\else
\ifnum#1=45 %
\hatcurPPmeccenxxxxC
\else
\ifnum#1=46 %
\hatcurPPmeccenxxxxD
\else
??????\fi
\fi
\fi
\fi
}
\newcommand{\hatcurPPmeeccen}[1]{\ifnum#1=43 %
\hatcurPPmeeccenxxxxA
\else
\ifnum#1=44 %
\hatcurPPmeeccenxxxxB
\else
\ifnum#1=45 %
\hatcurPPmeeccenxxxxC
\else
\ifnum#1=46 %
\hatcurPPmeeccenxxxxD
\else
??????\fi
\fi
\fi
\fi
}
\newcommand{\hatcurPPmelongeccen}[1]{\ifnum#1=43 %
\hatcurPPmelongeccenxxxxA
\else
\ifnum#1=44 %
\hatcurPPmelongeccenxxxxB
\else
\ifnum#1=45 %
\hatcurPPmelongeccenxxxxC
\else
\ifnum#1=46 %
\hatcurPPmelongeccenxxxxD
\else
??????\fi
\fi
\fi
\fi
}
\newcommand{\hatcurPPmeshorteccen}[1]{\ifnum#1=43 %
\hatcurPPmeshorteccenxxxxA
\else
\ifnum#1=44 %
\hatcurPPmeshorteccenxxxxB
\else
\ifnum#1=45 %
\hatcurPPmeshorteccenxxxxC
\else
\ifnum#1=46 %
\hatcurPPmeshorteccenxxxxD
\else
??????\fi
\fi
\fi
\fi
}
\newcommand{\hatcurPPmlongeccen}[1]{\ifnum#1=43 %
\hatcurPPmlongeccenxxxxA
\else
\ifnum#1=44 %
\hatcurPPmlongeccenxxxxB
\else
\ifnum#1=45 %
\hatcurPPmlongeccenxxxxC
\else
\ifnum#1=46 %
\hatcurPPmlongeccenxxxxD
\else
??????\fi
\fi
\fi
\fi
}
\newcommand{\hatcurPPmrcorreccen}[1]{\ifnum#1=43 %
\hatcurPPmrcorreccenxxxxA
\else
\ifnum#1=44 %
\hatcurPPmrcorreccenxxxxB
\else
\ifnum#1=45 %
\hatcurPPmrcorreccenxxxxC
\else
\ifnum#1=46 %
\hatcurPPmrcorreccenxxxxD
\else
??????\fi
\fi
\fi
\fi
}
\newcommand{\hatcurPPmshorteccen}[1]{\ifnum#1=43 %
\hatcurPPmshorteccenxxxxA
\else
\ifnum#1=44 %
\hatcurPPmshorteccenxxxxB
\else
\ifnum#1=45 %
\hatcurPPmshorteccenxxxxC
\else
\ifnum#1=46 %
\hatcurPPmshorteccenxxxxD
\else
??????\fi
\fi
\fi
\fi
}
\newcommand{\hatcurPPperieccen}[1]{\ifnum#1=43 %
\hatcurPPperieccenxxxxA
\else
\ifnum#1=44 %
\hatcurPPperieccenxxxxB
\else
\ifnum#1=45 %
\hatcurPPperieccenxxxxC
\else
\ifnum#1=46 %
\hatcurPPperieccenxxxxD
\else
??????\fi
\fi
\fi
\fi
}
\newcommand{\hatcurPPphiconjeccen}[1]{\ifnum#1=43 %
\hatcurPPphiconjeccenxxxxA
\else
\ifnum#1=44 %
\hatcurPPphiconjeccenxxxxB
\else
\ifnum#1=45 %
\hatcurPPphiconjeccenxxxxC
\else
\ifnum#1=46 %
\hatcurPPphiconjeccenxxxxD
\else
??????\fi
\fi
\fi
\fi
}
\newcommand{\hatcurPPreccen}[1]{\ifnum#1=43 %
\hatcurPPreccenxxxxA
\else
\ifnum#1=44 %
\hatcurPPreccenxxxxB
\else
\ifnum#1=45 %
\hatcurPPreccenxxxxC
\else
\ifnum#1=46 %
\hatcurPPreccenxxxxD
\else
??????\fi
\fi
\fi
\fi
}
\newcommand{\hatcurPPreeccen}[1]{\ifnum#1=43 %
\hatcurPPreeccenxxxxA
\else
\ifnum#1=44 %
\hatcurPPreeccenxxxxB
\else
\ifnum#1=45 %
\hatcurPPreeccenxxxxC
\else
\ifnum#1=46 %
\hatcurPPreeccenxxxxD
\else
??????\fi
\fi
\fi
\fi
}
\newcommand{\hatcurPPrelongeccen}[1]{\ifnum#1=43 %
\hatcurPPrelongeccenxxxxA
\else
\ifnum#1=44 %
\hatcurPPrelongeccenxxxxB
\else
\ifnum#1=45 %
\hatcurPPrelongeccenxxxxC
\else
\ifnum#1=46 %
\hatcurPPrelongeccenxxxxD
\else
??????\fi
\fi
\fi
\fi
}
\newcommand{\hatcurPPreshorteccen}[1]{\ifnum#1=43 %
\hatcurPPreshorteccenxxxxA
\else
\ifnum#1=44 %
\hatcurPPreshorteccenxxxxB
\else
\ifnum#1=45 %
\hatcurPPreshorteccenxxxxC
\else
\ifnum#1=46 %
\hatcurPPreshorteccenxxxxD
\else
??????\fi
\fi
\fi
\fi
}
\newcommand{\hatcurPPrhoeccen}[1]{\ifnum#1=43 %
\hatcurPPrhoeccenxxxxA
\else
\ifnum#1=44 %
\hatcurPPrhoeccenxxxxB
\else
\ifnum#1=45 %
\hatcurPPrhoeccenxxxxC
\else
\ifnum#1=46 %
\hatcurPPrhoeccenxxxxD
\else
??????\fi
\fi
\fi
\fi
}
\newcommand{\hatcurPPrlongeccen}[1]{\ifnum#1=43 %
\hatcurPPrlongeccenxxxxA
\else
\ifnum#1=44 %
\hatcurPPrlongeccenxxxxB
\else
\ifnum#1=45 %
\hatcurPPrlongeccenxxxxC
\else
\ifnum#1=46 %
\hatcurPPrlongeccenxxxxD
\else
??????\fi
\fi
\fi
\fi
}
\newcommand{\hatcurPPrshorteccen}[1]{\ifnum#1=43 %
\hatcurPPrshorteccenxxxxA
\else
\ifnum#1=44 %
\hatcurPPrshorteccenxxxxB
\else
\ifnum#1=45 %
\hatcurPPrshorteccenxxxxC
\else
\ifnum#1=46 %
\hatcurPPrshorteccenxxxxD
\else
??????\fi
\fi
\fi
\fi
}
\newcommand{\hatcurPPtcirceccen}[1]{\ifnum#1=43 %
\hatcurPPtcirceccenxxxxA
\else
\ifnum#1=44 %
\hatcurPPtcirceccenxxxxB
\else
\ifnum#1=45 %
\hatcurPPtcirceccenxxxxC
\else
\ifnum#1=46 %
\hatcurPPtcirceccenxxxxD
\else
??????\fi
\fi
\fi
\fi
}
\newcommand{\hatcurPPteffeccen}[1]{\ifnum#1=43 %
\hatcurPPteffeccenxxxxA
\else
\ifnum#1=44 %
\hatcurPPteffeccenxxxxB
\else
\ifnum#1=45 %
\hatcurPPteffeccenxxxxC
\else
\ifnum#1=46 %
\hatcurPPteffeccenxxxxD
\else
??????\fi
\fi
\fi
\fi
}
\newcommand{\hatcurPPthetaeccen}[1]{\ifnum#1=43 %
\hatcurPPthetaeccenxxxxA
\else
\ifnum#1=44 %
\hatcurPPthetaeccenxxxxB
\else
\ifnum#1=45 %
\hatcurPPthetaeccenxxxxC
\else
\ifnum#1=46 %
\hatcurPPthetaeccenxxxxD
\else
??????\fi
\fi
\fi
\fi
}
\newcommand{\hatcurPPtinfalleccen}[1]{\ifnum#1=43 %
\hatcurPPtinfalleccenxxxxA
\else
\ifnum#1=44 %
\hatcurPPtinfalleccenxxxxB
\else
\ifnum#1=45 %
\hatcurPPtinfalleccenxxxxC
\else
\ifnum#1=46 %
\hatcurPPtinfalleccenxxxxD
\else
??????\fi
\fi
\fi
\fi
}
\newcommand{\hatcurRVecceneccen}[1]{\ifnum#1=43 %
\hatcurRVecceneccenxxxxA
\else
\ifnum#1=44 %
\hatcurRVecceneccenxxxxB
\else
\ifnum#1=45 %
\hatcurRVecceneccenxxxxC
\else
\ifnum#1=46 %
\hatcurRVecceneccenxxxxD
\else
??????\fi
\fi
\fi
\fi
}
\newcommand{\hatcurRVeccentwosiglimeccen}[1]{\ifnum#1=43 %
\hatcurRVeccentwosiglimeccenxxxxA
\else
\ifnum#1=44 %
\hatcurRVeccentwosiglimeccenxxxxB
\else
\ifnum#1=45 %
\hatcurRVeccentwosiglimeccenxxxxC
\else
\ifnum#1=46 %
\hatcurRVeccentwosiglimeccenxxxxD
\else
??????\fi
\fi
\fi
\fi
}
\newcommand{\hatcurRVfitrmsAeccen}[1]{\ifnum#1=43 %
\hatcurRVfitrmsAeccenxxxxA
\else
\ifnum#1=45 %
\hatcurRVfitrmsAeccenxxxxC
\else
\ifnum#1=46 %
\hatcurRVfitrmsAeccenxxxxD
\else
??????\fi
\fi
\fi
}
\newcommand{\hatcurRVfitrmsBeccen}[1]{\ifnum#1=43 %
\hatcurRVfitrmsBeccenxxxxA
\else
\ifnum#1=45 %
\hatcurRVfitrmsBeccenxxxxC
\else
\ifnum#1=46 %
\hatcurRVfitrmsBeccenxxxxD
\else
??????\fi
\fi
\fi
}
\newcommand{\hatcurRVfitrmseccen}[1]{\ifnum#1=44 %
\hatcurRVfitrmseccenxxxxB
\else
??????\fi
}
\newcommand{\hatcurRVgammaAeccen}[1]{\ifnum#1=43 %
\hatcurRVgammaAeccenxxxxA
\else
\ifnum#1=45 %
\hatcurRVgammaAeccenxxxxC
\else
\ifnum#1=46 %
\hatcurRVgammaAeccenxxxxD
\else
??????\fi
\fi
\fi
}
\newcommand{\hatcurRVgammaBeccen}[1]{\ifnum#1=43 %
\hatcurRVgammaBeccenxxxxA
\else
\ifnum#1=45 %
\hatcurRVgammaBeccenxxxxC
\else
\ifnum#1=46 %
\hatcurRVgammaBeccenxxxxD
\else
??????\fi
\fi
\fi
}
\newcommand{\hatcurRVgammaeccen}[1]{\ifnum#1=44 %
\hatcurRVgammaeccenxxxxB
\else
??????\fi
}
\newcommand{\hatcurRVheccen}[1]{\ifnum#1=43 %
\hatcurRVheccenxxxxA
\else
\ifnum#1=44 %
\hatcurRVheccenxxxxB
\else
\ifnum#1=45 %
\hatcurRVheccenxxxxC
\else
\ifnum#1=46 %
\hatcurRVheccenxxxxD
\else
??????\fi
\fi
\fi
\fi
}
\newcommand{\hatcurRVjitterAeccen}[1]{\ifnum#1=43 %
\hatcurRVjitterAeccenxxxxA
\else
\ifnum#1=45 %
\hatcurRVjitterAeccenxxxxC
\else
\ifnum#1=46 %
\hatcurRVjitterAeccenxxxxD
\else
??????\fi
\fi
\fi
}
\newcommand{\hatcurRVjitterBeccen}[1]{\ifnum#1=43 %
\hatcurRVjitterBeccenxxxxA
\else
\ifnum#1=45 %
\hatcurRVjitterBeccenxxxxC
\else
\ifnum#1=46 %
\hatcurRVjitterBeccenxxxxD
\else
??????\fi
\fi
\fi
}
\newcommand{\hatcurRVjittereccen}[1]{\ifnum#1=44 %
\hatcurRVjittereccenxxxxB
\else
??????\fi
}
\newcommand{\hatcurRVjittertwosiglimAeccen}[1]{\ifnum#1=43 %
\hatcurRVjittertwosiglimAeccenxxxxA
\else
\ifnum#1=45 %
\hatcurRVjittertwosiglimAeccenxxxxC
\else
\ifnum#1=46 %
\hatcurRVjittertwosiglimAeccenxxxxD
\else
??????\fi
\fi
\fi
}
\newcommand{\hatcurRVjittertwosiglimBeccen}[1]{\ifnum#1=43 %
\hatcurRVjittertwosiglimBeccenxxxxA
\else
\ifnum#1=45 %
\hatcurRVjittertwosiglimBeccenxxxxC
\else
\ifnum#1=46 %
\hatcurRVjittertwosiglimBeccenxxxxD
\else
??????\fi
\fi
\fi
}
\newcommand{\hatcurRVjittertwosiglimeccen}[1]{\ifnum#1=44 %
\hatcurRVjittertwosiglimeccenxxxxB
\else
??????\fi
}
\newcommand{\hatcurRVkeccen}[1]{\ifnum#1=43 %
\hatcurRVkeccenxxxxA
\else
\ifnum#1=44 %
\hatcurRVkeccenxxxxB
\else
\ifnum#1=45 %
\hatcurRVkeccenxxxxC
\else
\ifnum#1=46 %
\hatcurRVkeccenxxxxD
\else
??????\fi
\fi
\fi
\fi
}
\newcommand{\hatcurRVKeccen}[1]{\ifnum#1=43 %
\hatcurRVKeccenxxxxA
\else
\ifnum#1=44 %
\hatcurRVKeccenxxxxB
\else
\ifnum#1=45 %
\hatcurRVKeccenxxxxC
\else
\ifnum#1=46 %
\hatcurRVKeccenxxxxD
\else
??????\fi
\fi
\fi
\fi
}
\newcommand{\hatcurRVomegaeccen}[1]{\ifnum#1=43 %
\hatcurRVomegaeccenxxxxA
\else
\ifnum#1=44 %
\hatcurRVomegaeccenxxxxB
\else
\ifnum#1=45 %
\hatcurRVomegaeccenxxxxC
\else
\ifnum#1=46 %
\hatcurRVomegaeccenxxxxD
\else
??????\fi
\fi
\fi
\fi
}
\newcommand{\hatcurRVrheccen}[1]{\ifnum#1=43 %
\hatcurRVrheccenxxxxA
\else
\ifnum#1=44 %
\hatcurRVrheccenxxxxB
\else
\ifnum#1=45 %
\hatcurRVrheccenxxxxC
\else
\ifnum#1=46 %
\hatcurRVrheccenxxxxD
\else
??????\fi
\fi
\fi
\fi
}
\newcommand{\hatcurRVrkeccen}[1]{\ifnum#1=43 %
\hatcurRVrkeccenxxxxA
\else
\ifnum#1=44 %
\hatcurRVrkeccenxxxxB
\else
\ifnum#1=45 %
\hatcurRVrkeccenxxxxC
\else
\ifnum#1=46 %
\hatcurRVrkeccenxxxxD
\else
??????\fi
\fi
\fi
\fi
}
\newcommand{\hatcurRVtroneeccen}[1]{\ifnum#1=43 %
\hatcurRVtroneeccenxxxxA
\else
\ifnum#1=44 %
\hatcurRVtroneeccenxxxxB
\else
\ifnum#1=45 %
\hatcurRVtroneeccenxxxxC
\else
\ifnum#1=46 %
\hatcurRVtroneeccenxxxxD
\else
??????\fi
\fi
\fi
\fi
}
\newcommand{\hatcurRVtrtwoeccen}[1]{\ifnum#1=43 %
\hatcurRVtrtwoeccenxxxxA
\else
\ifnum#1=44 %
\hatcurRVtrtwoeccenxxxxB
\else
\ifnum#1=45 %
\hatcurRVtrtwoeccenxxxxC
\else
\ifnum#1=46 %
\hatcurRVtrtwoeccenxxxxD
\else
??????\fi
\fi
\fi
\fi
}
\newcommand{\hatcurSMEiiloggeccen}[1]{\ifnum#1=43 %
\hatcurSMEiiloggeccenxxxxA
\else
\ifnum#1=44 %
\hatcurSMEiiloggeccenxxxxB
\else
\ifnum#1=45 %
\hatcurSMEiiloggeccenxxxxC
\else
??????\fi
\fi
\fi
}
\newcommand{\hatcurSMEiiteffeccen}[1]{\ifnum#1=43 %
\hatcurSMEiiteffeccenxxxxA
\else
\ifnum#1=44 %
\hatcurSMEiiteffeccenxxxxB
\else
\ifnum#1=45 %
\hatcurSMEiiteffeccenxxxxC
\else
??????\fi
\fi
\fi
}
\newcommand{\hatcurSMEiivmaceccen}[1]{\ifnum#1=43 %
\hatcurSMEiivmaceccenxxxxA
\else
\ifnum#1=44 %
\hatcurSMEiivmaceccenxxxxB
\else
\ifnum#1=45 %
\hatcurSMEiivmaceccenxxxxC
\else
??????\fi
\fi
\fi
}
\newcommand{\hatcurSMEiivmiceccen}[1]{\ifnum#1=43 %
\hatcurSMEiivmiceccenxxxxA
\else
\ifnum#1=44 %
\hatcurSMEiivmiceccenxxxxB
\else
\ifnum#1=45 %
\hatcurSMEiivmiceccenxxxxC
\else
??????\fi
\fi
\fi
}
\newcommand{\hatcurSMEiivsineccen}[1]{\ifnum#1=43 %
\hatcurSMEiivsineccenxxxxA
\else
\ifnum#1=44 %
\hatcurSMEiivsineccenxxxxB
\else
\ifnum#1=45 %
\hatcurSMEiivsineccenxxxxC
\else
??????\fi
\fi
\fi
}
\newcommand{\hatcurSMEiizfeheccen}[1]{\ifnum#1=43 %
\hatcurSMEiizfeheccenxxxxA
\else
\ifnum#1=44 %
\hatcurSMEiizfeheccenxxxxB
\else
\ifnum#1=45 %
\hatcurSMEiizfeheccenxxxxC
\else
??????\fi
\fi
\fi
}
\newcommand{\hatcurSMEiizfehshorteccen}[1]{\ifnum#1=43 %
\hatcurSMEiizfehshorteccenxxxxA
\else
\ifnum#1=44 %
\hatcurSMEiizfehshorteccenxxxxB
\else
\ifnum#1=45 %
\hatcurSMEiizfehshorteccenxxxxC
\else
??????\fi
\fi
\fi
}
\newcommand{\hatcurSMEiloggeccen}[1]{\ifnum#1=43 %
\hatcurSMEiloggeccenxxxxA
\else
\ifnum#1=44 %
\hatcurSMEiloggeccenxxxxB
\else
\ifnum#1=45 %
\hatcurSMEiloggeccenxxxxC
\else
\ifnum#1=46 %
\hatcurSMEiloggeccenxxxxD
\else
??????\fi
\fi
\fi
\fi
}
\newcommand{\hatcurSMEiteffeccen}[1]{\ifnum#1=43 %
\hatcurSMEiteffeccenxxxxA
\else
\ifnum#1=44 %
\hatcurSMEiteffeccenxxxxB
\else
\ifnum#1=45 %
\hatcurSMEiteffeccenxxxxC
\else
\ifnum#1=46 %
\hatcurSMEiteffeccenxxxxD
\else
??????\fi
\fi
\fi
\fi
}
\newcommand{\hatcurSMEivmaceccen}[1]{\ifnum#1=43 %
\hatcurSMEivmaceccenxxxxA
\else
\ifnum#1=44 %
\hatcurSMEivmaceccenxxxxB
\else
\ifnum#1=45 %
\hatcurSMEivmaceccenxxxxC
\else
\ifnum#1=46 %
\hatcurSMEivmaceccenxxxxD
\else
??????\fi
\fi
\fi
\fi
}
\newcommand{\hatcurSMEivmiceccen}[1]{\ifnum#1=43 %
\hatcurSMEivmiceccenxxxxA
\else
\ifnum#1=44 %
\hatcurSMEivmiceccenxxxxB
\else
\ifnum#1=45 %
\hatcurSMEivmiceccenxxxxC
\else
\ifnum#1=46 %
\hatcurSMEivmiceccenxxxxD
\else
??????\fi
\fi
\fi
\fi
}
\newcommand{\hatcurSMEivsineccen}[1]{\ifnum#1=43 %
\hatcurSMEivsineccenxxxxA
\else
\ifnum#1=44 %
\hatcurSMEivsineccenxxxxB
\else
\ifnum#1=45 %
\hatcurSMEivsineccenxxxxC
\else
\ifnum#1=46 %
\hatcurSMEivsineccenxxxxD
\else
??????\fi
\fi
\fi
\fi
}
\newcommand{\hatcurSMEizfeheccen}[1]{\ifnum#1=43 %
\hatcurSMEizfeheccenxxxxA
\else
\ifnum#1=44 %
\hatcurSMEizfeheccenxxxxB
\else
\ifnum#1=45 %
\hatcurSMEizfeheccenxxxxC
\else
\ifnum#1=46 %
\hatcurSMEizfeheccenxxxxD
\else
??????\fi
\fi
\fi
\fi
}
\newcommand{\hatcurSMEizfehshorteccen}[1]{\ifnum#1=43 %
\hatcurSMEizfehshorteccenxxxxA
\else
\ifnum#1=44 %
\hatcurSMEizfehshorteccenxxxxB
\else
\ifnum#1=45 %
\hatcurSMEizfehshorteccenxxxxC
\else
\ifnum#1=46 %
\hatcurSMEizfehshorteccenxxxxD
\else
??????\fi
\fi
\fi
\fi
}
\newcommand{\hatcurXAveccen}[1]{\ifnum#1=43 %
\hatcurXAveccenxxxxA
\else
\ifnum#1=44 %
\hatcurXAveccenxxxxB
\else
\ifnum#1=45 %
\hatcurXAveccenxxxxC
\else
\ifnum#1=46 %
\hatcurXAveccenxxxxD
\else
??????\fi
\fi
\fi
\fi
}
\newcommand{\hatcurXdisteccen}[1]{\ifnum#1=43 %
\hatcurXdisteccenxxxxA
\else
\ifnum#1=44 %
\hatcurXdisteccenxxxxB
\else
\ifnum#1=45 %
\hatcurXdisteccenxxxxC
\else
\ifnum#1=46 %
\hatcurXdisteccenxxxxD
\else
??????\fi
\fi
\fi
\fi
}
\newcommand{\hatcurXdistredeccen}[1]{\ifnum#1=43 %
\hatcurXdistredeccenxxxxA
\else
\ifnum#1=44 %
\hatcurXdistredeccenxxxxB
\else
\ifnum#1=45 %
\hatcurXdistredeccenxxxxC
\else
\ifnum#1=46 %
\hatcurXdistredeccenxxxxD
\else
??????\fi
\fi
\fi
\fi
}
\newcommand{\hatcurXEBVeccen}[1]{\ifnum#1=43 %
\hatcurXEBVeccenxxxxA
\else
\ifnum#1=44 %
\hatcurXEBVeccenxxxxB
\else
\ifnum#1=45 %
\hatcurXEBVeccenxxxxC
\else
\ifnum#1=46 %
\hatcurXEBVeccenxxxxD
\else
??????\fi
\fi
\fi
\fi
}
\newcommand{\hatcurXjhisoredeccen}[1]{\ifnum#1=43 %
\hatcurXjhisoredeccenxxxxA
\else
\ifnum#1=44 %
\hatcurXjhisoredeccenxxxxB
\else
\ifnum#1=45 %
\hatcurXjhisoredeccenxxxxC
\else
\ifnum#1=46 %
\hatcurXjhisoredeccenxxxxD
\else
??????\fi
\fi
\fi
\fi
}
\newcommand{\hatcurXjkisoredeccen}[1]{\ifnum#1=43 %
\hatcurXjkisoredeccenxxxxA
\else
\ifnum#1=44 %
\hatcurXjkisoredeccenxxxxB
\else
\ifnum#1=45 %
\hatcurXjkisoredeccenxxxxC
\else
\ifnum#1=46 %
\hatcurXjkisoredeccenxxxxD
\else
??????\fi
\fi
\fi
\fi
}
\newcommand{\hatcurXmhisoredeccen}[1]{\ifnum#1=43 %
\hatcurXmhisoredeccenxxxxA
\else
\ifnum#1=44 %
\hatcurXmhisoredeccenxxxxB
\else
\ifnum#1=45 %
\hatcurXmhisoredeccenxxxxC
\else
\ifnum#1=46 %
\hatcurXmhisoredeccenxxxxD
\else
??????\fi
\fi
\fi
\fi
}
\newcommand{\hatcurXmiisoredeccen}[1]{\ifnum#1=43 %
\hatcurXmiisoredeccenxxxxA
\else
\ifnum#1=44 %
\hatcurXmiisoredeccenxxxxB
\else
\ifnum#1=45 %
\hatcurXmiisoredeccenxxxxC
\else
\ifnum#1=46 %
\hatcurXmiisoredeccenxxxxD
\else
??????\fi
\fi
\fi
\fi
}
\newcommand{\hatcurXmjisoredeccen}[1]{\ifnum#1=43 %
\hatcurXmjisoredeccenxxxxA
\else
\ifnum#1=44 %
\hatcurXmjisoredeccenxxxxB
\else
\ifnum#1=45 %
\hatcurXmjisoredeccenxxxxC
\else
\ifnum#1=46 %
\hatcurXmjisoredeccenxxxxD
\else
??????\fi
\fi
\fi
\fi
}
\newcommand{\hatcurXmkisoredeccen}[1]{\ifnum#1=43 %
\hatcurXmkisoredeccenxxxxA
\else
\ifnum#1=44 %
\hatcurXmkisoredeccenxxxxB
\else
\ifnum#1=45 %
\hatcurXmkisoredeccenxxxxC
\else
\ifnum#1=46 %
\hatcurXmkisoredeccenxxxxD
\else
??????\fi
\fi
\fi
\fi
}
\newcommand{\hatcurXmvisoredeccen}[1]{\ifnum#1=43 %
\hatcurXmvisoredeccenxxxxA
\else
\ifnum#1=44 %
\hatcurXmvisoredeccenxxxxB
\else
\ifnum#1=45 %
\hatcurXmvisoredeccenxxxxC
\else
\ifnum#1=46 %
\hatcurXmvisoredeccenxxxxD
\else
??????\fi
\fi
\fi
\fi
}
\newcommand{\hatcurXsecdureccen}[1]{\ifnum#1=43 %
\hatcurXsecdureccenxxxxA
\else
\ifnum#1=44 %
\hatcurXsecdureccenxxxxB
\else
\ifnum#1=45 %
\hatcurXsecdureccenxxxxC
\else
\ifnum#1=46 %
\hatcurXsecdureccenxxxxD
\else
??????\fi
\fi
\fi
\fi
}
\newcommand{\hatcurXsecingdureccen}[1]{\ifnum#1=43 %
\hatcurXsecingdureccenxxxxA
\else
\ifnum#1=44 %
\hatcurXsecingdureccenxxxxB
\else
\ifnum#1=45 %
\hatcurXsecingdureccenxxxxC
\else
\ifnum#1=46 %
\hatcurXsecingdureccenxxxxD
\else
??????\fi
\fi
\fi
\fi
}
\newcommand{\hatcurXsecondaryeccen}[1]{\ifnum#1=43 %
\hatcurXsecondaryeccenxxxxA
\else
\ifnum#1=44 %
\hatcurXsecondaryeccenxxxxB
\else
\ifnum#1=45 %
\hatcurXsecondaryeccenxxxxC
\else
\ifnum#1=46 %
\hatcurXsecondaryeccenxxxxD
\else
??????\fi
\fi
\fi
\fi
}
\newcommand{\hatcurXsecphaseeccen}[1]{\ifnum#1=43 %
\hatcurXsecphaseeccenxxxxA
\else
\ifnum#1=44 %
\hatcurXsecphaseeccenxxxxB
\else
\ifnum#1=45 %
\hatcurXsecphaseeccenxxxxC
\else
\ifnum#1=46 %
\hatcurXsecphaseeccenxxxxD
\else
??????\fi
\fi
\fi
\fi
}
\newcommand{\hatcurXviisoredeccen}[1]{\ifnum#1=43 %
\hatcurXviisoredeccenxxxxA
\else
\ifnum#1=44 %
\hatcurXviisoredeccenxxxxB
\else
\ifnum#1=45 %
\hatcurXviisoredeccenxxxxC
\else
\ifnum#1=46 %
\hatcurXviisoredeccenxxxxD
\else
??????\fi
\fi
\fi
\fi
}
\newcommand{\hatcurXvkisoredeccen}[1]{\ifnum#1=43 %
\hatcurXvkisoredeccenxxxxA
\else
\ifnum#1=44 %
\hatcurXvkisoredeccenxxxxB
\else
\ifnum#1=45 %
\hatcurXvkisoredeccenxxxxC
\else
\ifnum#1=46 %
\hatcurXvkisoredeccenxxxxD
\else
??????\fi
\fi
\fi
\fi
}
\newcommand{\hatcurxxxxC}{HATS-45}
\newcommand{\hatcurbxxxxC}{HATS-45b}
\newcommand{\hatcurcxxxxC}{HATS-45c}
\newcommand{\hatcurplanetnumxxxxC}{45}
\newcommand{\hatcurCCtwomassshortxxxxC}{06475862-2154385}
\newcommand{\hatcurRVgammaabsxxxxC}{\hatcurRVgammaA{\hatcurplanetnumxxxxC}}                           
\newcommand{\hatcurRVgammarelxxxxC}{\hatcurRVgammaA{\hatcurplanetnumxxxxC}}                           
\newcommand{\hatcurCCtassvixxxxC}{\ensuremath{NULL\pm NULL}}                  
\newcommand{\hatcurSMEversionxxxxC}{ii}                                       
\newcommand{\hatcurisoshortxxxxC}{YY}
\newcommand{\hatcurisofullxxxxC}{Yonsei-Yale (YY)}
\newcommand{\hatcurisocitexxxxC}{yi:2001}
\newcommand{\hatcurlumindxxxxC}{\arstar}
\newcommand{\hatcurjhkfilsetxxxxC}{ESO}
\newcommand{\hatcurSMEteffxxxxC}{\ifthenelse{\equal{\hatcurSMEversionxxxxC}{i}}{\hatcurSMEiteff{\hatcurplanetnumxxxxC}}{\hatcurSMEiiteff{\hatcurplanetnumxxxxC}}}
\newcommand{\hatcurSMEzfehxxxxC}{\ifthenelse{\equal{\hatcurSMEversionxxxxC}{i}}{\hatcurSMEizfeh{\hatcurplanetnumxxxxC}}{\hatcurSMEiizfeh{\hatcurplanetnumxxxxC}}}
\newcommand{\hatcurSMEzfehshortxxxxC}{\ifthenelse{\equal{\hatcurSMEversionxxxxC}{i}}{\hatcurSMEizfehshort{\hatcurplanetnumxxxxC}}{\hatcurSMEiizfehshort{\hatcurplanetnumxxxxC}}}
\newcommand{\hatcurSMEloggxxxxC}{\ifthenelse{\equal{\hatcurSMEversionxxxxC}{i}}{\hatcurSMEilogg{\hatcurplanetnumxxxxC}}{\hatcurSMEiilogg{\hatcurplanetnumxxxxC}}}
\newcommand{\hatcurSMEvsinxxxxC}{\ifthenelse{\equal{\hatcurSMEversionxxxxC}{i}}{\hatcurSMEivsin{\hatcurplanetnumxxxxC}}{\hatcurSMEiivsin{\hatcurplanetnumxxxxC}}}
\newcommand{\hatcurSMEvmacxxxxC}{\ifthenelse{\equal{\hatcurSMEversionxxxxC}{i}}{\hatcurSMEivmac{\hatcurplanetnumxxxxC}}{\hatcurSMEiivmac{\hatcurplanetnumxxxxC}}}
\newcommand{\hatcurSMEvmicxxxxC}{\ifthenelse{\equal{\hatcurSMEversionxxxxC}{i}}{\hatcurSMEivmic{\hatcurplanetnumxxxxC}}{\hatcurSMEiivmic{\hatcurplanetnumxxxxC}}}

\newcommand{\hatcurxxxxA}{HATS-43}
\newcommand{\hatcurbxxxxA}{HATS-43b}
\newcommand{\hatcurcxxxxA}{HATS-43c}
\newcommand{\hatcurplanetnumxxxxA}{43}
\newcommand{\hatcurCCtwomassshortxxxxA}{05220915-3058150}
\newcommand{\hatcurRVgammaabsxxxxA}{\hatcurRVgammaA{\hatcurplanetnumxxxxA}}                           
\newcommand{\hatcurRVgammarelxxxxA}{\hatcurRVgammaA{\hatcurplanetnumxxxxA}}                           
\newcommand{\hatcurCCtassvixxxxA}{\ensuremath{NULL\pm NULL}}                  
\newcommand{\hatcurSMEversionxxxxA}{ii}                                       
\newcommand{\hatcurisoshortxxxxA}{YY}
\newcommand{\hatcurisofullxxxxA}{Yonsei-Yale (YY)}
\newcommand{\hatcurisocitexxxxA}{yi:2001}
\newcommand{\hatcurlumindxxxxA}{\arstar}
\newcommand{\hatcurjhkfilsetxxxxA}{ESO}
\newcommand{\hatcurSMEteffxxxxA}{\ifthenelse{\equal{\hatcurSMEversionxxxxA}{i}}{\hatcurSMEiteff{\hatcurplanetnumxxxxA}}{\hatcurSMEiiteff{\hatcurplanetnumxxxxA}}}
\newcommand{\hatcurSMEzfehxxxxA}{\ifthenelse{\equal{\hatcurSMEversionxxxxA}{i}}{\hatcurSMEizfeh{\hatcurplanetnumxxxxA}}{\hatcurSMEiizfeh{\hatcurplanetnumxxxxA}}}
\newcommand{\hatcurSMEzfehshortxxxxA}{\ifthenelse{\equal{\hatcurSMEversionxxxxA}{i}}{\hatcurSMEizfehshort{\hatcurplanetnumxxxxA}}{\hatcurSMEiizfehshort{\hatcurplanetnumxxxxA}}}
\newcommand{\hatcurSMEloggxxxxA}{\ifthenelse{\equal{\hatcurSMEversionxxxxA}{i}}{\hatcurSMEilogg{\hatcurplanetnumxxxxA}}{\hatcurSMEiilogg{\hatcurplanetnumxxxxA}}}
\newcommand{\hatcurSMEvsinxxxxA}{\ifthenelse{\equal{\hatcurSMEversionxxxxA}{i}}{\hatcurSMEivsin{\hatcurplanetnumxxxxA}}{\hatcurSMEiivsin{\hatcurplanetnumxxxxA}}}
\newcommand{\hatcurSMEvmacxxxxA}{\ifthenelse{\equal{\hatcurSMEversionxxxxA}{i}}{\hatcurSMEivmac{\hatcurplanetnumxxxxA}}{\hatcurSMEiivmac{\hatcurplanetnumxxxxA}}}
\newcommand{\hatcurSMEvmicxxxxA}{\ifthenelse{\equal{\hatcurSMEversionxxxxA}{i}}{\hatcurSMEivmic{\hatcurplanetnumxxxxA}}{\hatcurSMEiivmic{\hatcurplanetnumxxxxA}}}

\newcommand{\hatcurxxxxB}{HATS-44}
\newcommand{\hatcurbxxxxB}{HATS-44b}
\newcommand{\hatcurcxxxxB}{HATS-44c}
\newcommand{\hatcurplanetnumxxxxB}{44}
\newcommand{\hatcurCCtwomassshortxxxxB}{05371842-2758214}
\newcommand{\hatcurRVgammaabsxxxxB}{\hatcurRVgamma{\hatcurplanetnumxxxxB}}                           
\newcommand{\hatcurRVgammarelxxxxB}{\hatcurRVgamma{\hatcurplanetnumxxxxB}}                           
\newcommand{\hatcurCCtassvixxxxB}{\ensuremath{NULL\pm NULL}}                  
\newcommand{\hatcurSMEversionxxxxB}{i}                                       
\newcommand{\hatcurisoshortxxxxB}{YY}
\newcommand{\hatcurisofullxxxxB}{Yonsei-Yale (YY)}
\newcommand{\hatcurisocitexxxxB}{yi:2001}
\newcommand{\hatcurlumindxxxxB}{\arstar}
\newcommand{\hatcurjhkfilsetxxxxB}{ESO}
\newcommand{\hatcurSMEteffxxxxB}{\ifthenelse{\equal{\hatcurSMEversionxxxxB}{i}}{\hatcurSMEiteff{\hatcurplanetnumxxxxB}}{\hatcurSMEiiteff{\hatcurplanetnumxxxxB}}}
\newcommand{\hatcurSMEzfehxxxxB}{\ifthenelse{\equal{\hatcurSMEversionxxxxB}{i}}{\hatcurSMEizfeh{\hatcurplanetnumxxxxB}}{\hatcurSMEiizfeh{\hatcurplanetnumxxxxB}}}
\newcommand{\hatcurSMEzfehshortxxxxB}{\ifthenelse{\equal{\hatcurSMEversionxxxxB}{i}}{\hatcurSMEizfehshort{\hatcurplanetnumxxxxB}}{\hatcurSMEiizfehshort{\hatcurplanetnumxxxxB}}}
\newcommand{\hatcurSMEloggxxxxB}{\ifthenelse{\equal{\hatcurSMEversionxxxxB}{i}}{\hatcurSMEilogg{\hatcurplanetnumxxxxB}}{\hatcurSMEiilogg{\hatcurplanetnumxxxxB}}}
\newcommand{\hatcurSMEvsinxxxxB}{\ifthenelse{\equal{\hatcurSMEversionxxxxB}{i}}{\hatcurSMEivsin{\hatcurplanetnumxxxxB}}{\hatcurSMEiivsin{\hatcurplanetnumxxxxB}}}
\newcommand{\hatcurSMEvmacxxxxB}{\ifthenelse{\equal{\hatcurSMEversionxxxxB}{i}}{\hatcurSMEivmac{\hatcurplanetnumxxxxB}}{\hatcurSMEiivmac{\hatcurplanetnumxxxxB}}}
\newcommand{\hatcurSMEvmicxxxxB}{\ifthenelse{\equal{\hatcurSMEversionxxxxB}{i}}{\hatcurSMEivmic{\hatcurplanetnumxxxxB}}{\hatcurSMEiivmic{\hatcurplanetnumxxxxB}}}

\newcommand{\hatcurxxxxD}{HATS-46}
\newcommand{\hatcurbxxxxD}{HATS-46b}
\newcommand{\hatcurcxxxxD}{HATS-46c}
\newcommand{\hatcurplanetnumxxxxD}{46}
\newcommand{\hatcurCCtwomassshortxxxxD}{00264858-5618580}
\newcommand{\hatcurRVgammaabsxxxxD}{\hatcurRVgammaA{\hatcurplanetnumxxxxD}}                           
\newcommand{\hatcurRVgammarelxxxxD}{\hatcurRVgammaA{\hatcurplanetnumxxxxD}}                           
\newcommand{\hatcurCCtassvixxxxD}{\ensuremath{NULL\pm NULL}}                  
\newcommand{\hatcurSMEversionxxxxD}{i}                                       
\newcommand{\hatcurisoshortxxxxD}{YY}
\newcommand{\hatcurisofullxxxxD}{Yonsei-Yale (YY)}
\newcommand{\hatcurisocitexxxxD}{yi:2001}
\newcommand{\hatcurlumindxxxxD}{\arstar}
\newcommand{\hatcurjhkfilsetxxxxD}{ESO}
\newcommand{\hatcurSMEteffxxxxD}{\ifthenelse{\equal{\hatcurSMEversionxxxxD}{i}}{\hatcurSMEiteff{\hatcurplanetnumxxxxD}}{\hatcurSMEiiteff{\hatcurplanetnumxxxxD}}}
\newcommand{\hatcurSMEzfehxxxxD}{\ifthenelse{\equal{\hatcurSMEversionxxxxD}{i}}{\hatcurSMEizfeh{\hatcurplanetnumxxxxD}}{\hatcurSMEiizfeh{\hatcurplanetnumxxxxD}}}
\newcommand{\hatcurSMEzfehshortxxxxD}{\ifthenelse{\equal{\hatcurSMEversionxxxxD}{i}}{\hatcurSMEizfehshort{\hatcurplanetnumxxxxD}}{\hatcurSMEiizfehshort{\hatcurplanetnumxxxxD}}}
\newcommand{\hatcurSMEloggxxxxD}{\ifthenelse{\equal{\hatcurSMEversionxxxxD}{i}}{\hatcurSMEilogg{\hatcurplanetnumxxxxD}}{\hatcurSMEiilogg{\hatcurplanetnumxxxxD}}}
\newcommand{\hatcurSMEvsinxxxxD}{\ifthenelse{\equal{\hatcurSMEversionxxxxD}{i}}{\hatcurSMEivsin{\hatcurplanetnumxxxxD}}{\hatcurSMEiivsin{\hatcurplanetnumxxxxD}}}
\newcommand{\hatcurSMEvmacxxxxD}{\ifthenelse{\equal{\hatcurSMEversionxxxxD}{i}}{\hatcurSMEivmac{\hatcurplanetnumxxxxD}}{\hatcurSMEiivmac{\hatcurplanetnumxxxxD}}}
\newcommand{\hatcurSMEvmicxxxxD}{\ifthenelse{\equal{\hatcurSMEversionxxxxD}{i}}{\hatcurSMEivmic{\hatcurplanetnumxxxxD}}{\hatcurSMEiivmic{\hatcurplanetnumxxxxD}}}

\newcommand{\hatcur}[1]{\ifnum#1=43 %
\hatcurxxxxA
\else
\ifnum#1=44 %
\hatcurxxxxB
\else
\ifnum#1=45 %
\hatcurxxxxC
\else
\ifnum#1=46 %
\hatcurxxxxD
\else
??????\fi
\fi
\fi
\fi
}
\newcommand{\hatcurb}[1]{\ifnum#1=43 %
\hatcurbxxxxA
\else
\ifnum#1=44 %
\hatcurbxxxxB
\else
\ifnum#1=45 %
\hatcurbxxxxC
\else
\ifnum#1=46 %
\hatcurbxxxxD
\else
??????\fi
\fi
\fi
\fi
}
\newcommand{\hatcurc}[1]{\ifnum#1=43 %
\hatcurcxxxxA
\else
\ifnum#1=44 %
\hatcurcxxxxB
\else
\ifnum#1=45 %
\hatcurcxxxxC
\else
\ifnum#1=46 %
\hatcurcxxxxD
\else
??????\fi
\fi
\fi
\fi
}
\newcommand{\hatcurCCtassvi}[1]{\ifnum#1=43 %
\hatcurCCtassvixxxxA
\else
\ifnum#1=44 %
\hatcurCCtassvixxxxB
\else
\ifnum#1=45 %
\hatcurCCtassvixxxxC
\else
\ifnum#1=46 %
\hatcurCCtassvixxxxD
\else
??????\fi
\fi
\fi
\fi
}
\newcommand{\hatcurCCtwomassshort}[1]{\ifnum#1=43 %
\hatcurCCtwomassshortxxxxA
\else
\ifnum#1=44 %
\hatcurCCtwomassshortxxxxB
\else
\ifnum#1=45 %
\hatcurCCtwomassshortxxxxC
\else
\ifnum#1=46 %
\hatcurCCtwomassshortxxxxD
\else
??????\fi
\fi
\fi
\fi
}
\newcommand{\hatcurisocite}[1]{\ifnum#1=43 %
\hatcurisocitexxxxA
\else
\ifnum#1=44 %
\hatcurisocitexxxxB
\else
\ifnum#1=45 %
\hatcurisocitexxxxC
\else
\ifnum#1=46 %
\hatcurisocitexxxxD
\else
??????\fi
\fi
\fi
\fi
}
\newcommand{\hatcurisofull}[1]{\ifnum#1=43 %
\hatcurisofullxxxxA
\else
\ifnum#1=44 %
\hatcurisofullxxxxB
\else
\ifnum#1=45 %
\hatcurisofullxxxxC
\else
\ifnum#1=46 %
\hatcurisofullxxxxD
\else
??????\fi
\fi
\fi
\fi
}
\newcommand{\hatcurisoshort}[1]{\ifnum#1=43 %
\hatcurisoshortxxxxA
\else
\ifnum#1=44 %
\hatcurisoshortxxxxB
\else
\ifnum#1=45 %
\hatcurisoshortxxxxC
\else
\ifnum#1=46 %
\hatcurisoshortxxxxD
\else
??????\fi
\fi
\fi
\fi
}
\newcommand{\hatcurjhkfilset}[1]{\ifnum#1=43 %
\hatcurjhkfilsetxxxxA
\else
\ifnum#1=44 %
\hatcurjhkfilsetxxxxB
\else
\ifnum#1=45 %
\hatcurjhkfilsetxxxxC
\else
\ifnum#1=46 %
\hatcurjhkfilsetxxxxD
\else
??????\fi
\fi
\fi
\fi
}
\newcommand{\hatcurlumind}[1]{\ifnum#1=43 %
\hatcurlumindxxxxA
\else
\ifnum#1=44 %
\hatcurlumindxxxxB
\else
\ifnum#1=45 %
\hatcurlumindxxxxC
\else
\ifnum#1=46 %
\hatcurlumindxxxxD
\else
??????\fi
\fi
\fi
\fi
}
\newcommand{\hatcurplanetnum}[1]{\ifnum#1=43 %
\hatcurplanetnumxxxxA
\else
\ifnum#1=44 %
\hatcurplanetnumxxxxB
\else
\ifnum#1=45 %
\hatcurplanetnumxxxxC
\else
\ifnum#1=46 %
\hatcurplanetnumxxxxD
\else
??????\fi
\fi
\fi
\fi
}
\newcommand{\hatcurRVgammaabs}[1]{\ifnum#1=43 %
\hatcurRVgammaabsxxxxA
\else
\ifnum#1=44 %
\hatcurRVgammaabsxxxxB
\else
\ifnum#1=45 %
\hatcurRVgammaabsxxxxC
\else
\ifnum#1=46 %
\hatcurRVgammaabsxxxxD
\else
??????\fi
\fi
\fi
\fi
}
\newcommand{\hatcurRVgammarel}[1]{\ifnum#1=43 %
\hatcurRVgammarelxxxxA
\else
\ifnum#1=44 %
\hatcurRVgammarelxxxxB
\else
\ifnum#1=45 %
\hatcurRVgammarelxxxxC
\else
\ifnum#1=46 %
\hatcurRVgammarelxxxxD
\else
??????\fi
\fi
\fi
\fi
}
\newcommand{\hatcurSMElogg}[1]{\ifnum#1=43 %
\hatcurSMEloggxxxxA
\else
\ifnum#1=44 %
\hatcurSMEloggxxxxB
\else
\ifnum#1=45 %
\hatcurSMEloggxxxxC
\else
\ifnum#1=46 %
\hatcurSMEloggxxxxD
\else
??????\fi
\fi
\fi
\fi
}
\newcommand{\hatcurSMEteff}[1]{\ifnum#1=43 %
\hatcurSMEteffxxxxA
\else
\ifnum#1=44 %
\hatcurSMEteffxxxxB
\else
\ifnum#1=45 %
\hatcurSMEteffxxxxC
\else
\ifnum#1=46 %
\hatcurSMEteffxxxxD
\else
??????\fi
\fi
\fi
\fi
}
\newcommand{\hatcurSMEversion}[1]{\ifnum#1=43 %
\hatcurSMEversionxxxxA
\else
\ifnum#1=44 %
\hatcurSMEversionxxxxB
\else
\ifnum#1=45 %
\hatcurSMEversionxxxxC
\else
\ifnum#1=46 %
\hatcurSMEversionxxxxD
\else
??????\fi
\fi
\fi
\fi
}
\newcommand{\hatcurSMEvmac}[1]{\ifnum#1=43 %
\hatcurSMEvmacxxxxA
\else
\ifnum#1=44 %
\hatcurSMEvmacxxxxB
\else
\ifnum#1=45 %
\hatcurSMEvmacxxxxC
\else
\ifnum#1=46 %
\hatcurSMEvmacxxxxD
\else
??????\fi
\fi
\fi
\fi
}
\newcommand{\hatcurSMEvmic}[1]{\ifnum#1=43 %
\hatcurSMEvmicxxxxA
\else
\ifnum#1=44 %
\hatcurSMEvmicxxxxB
\else
\ifnum#1=45 %
\hatcurSMEvmicxxxxC
\else
\ifnum#1=46 %
\hatcurSMEvmicxxxxD
\else
??????\fi
\fi
\fi
\fi
}
\newcommand{\hatcurSMEvsin}[1]{\ifnum#1=43 %
\hatcurSMEvsinxxxxA
\else
\ifnum#1=44 %
\hatcurSMEvsinxxxxB
\else
\ifnum#1=45 %
\hatcurSMEvsinxxxxC
\else
\ifnum#1=46 %
\hatcurSMEvsinxxxxD
\else
??????\fi
\fi
\fi
\fi
}
\newcommand{\hatcurSMEzfeh}[1]{\ifnum#1=43 %
\hatcurSMEzfehxxxxA
\else
\ifnum#1=44 %
\hatcurSMEzfehxxxxB
\else
\ifnum#1=45 %
\hatcurSMEzfehxxxxC
\else
\ifnum#1=46 %
\hatcurSMEzfehxxxxD
\else
??????\fi
\fi
\fi
\fi
}
\newcommand{\hatcurSMEzfehshort}[1]{\ifnum#1=43 %
\hatcurSMEzfehshortxxxxA
\else
\ifnum#1=44 %
\hatcurSMEzfehshortxxxxB
\else
\ifnum#1=45 %
\hatcurSMEzfehshortxxxxC
\else
\ifnum#1=46 %
\hatcurSMEzfehshortxxxxD
\else
??????\fi
\fi
\fi
\fi
}

\newcounter{planetcounter}

\newboolean{emulateapj}
\setboolean{emulateapj}{true}

\newboolean{rvtablelong}
\setboolean{rvtablelong}{true}

\newboolean{astroph}
\setboolean{astroph}{true}

\shortauthors{Brahm et al.}
\shorttitle{\hatcur{43}\lowercase{b}--\hatcur{46}\lowercase{b}}
\ifthenelse{\boolean{emulateapj}}{
    \newcommand{\titledag}{$\dagger$}
}{
    \newcommand{\titledag}{\dagger}
}

\begin{document}

\title{
\hatcur{43}\lowercase{b}, \hatcur{44}\lowercase{b}, \hatcur{45}\lowercase{b}, and \hatcur{46}\lowercase{b}: Four Short Period Transiting Giant Planets in the Neptune-Jupiter Mass Range\altaffilmark{\titledag}
}

\author{
    R.~Brahm\altaffilmark{1,2},
    J.~D.~Hartman\altaffilmark{3},
    A.~Jord\'an\altaffilmark{2,1,4},
    G.~\'A.~Bakos\altaffilmark{3,$\star$,$\star\star$},
    N.~Espinoza\altaffilmark{2,1},
    M.~Rabus\altaffilmark{2,4},
    W.~Bhatti\altaffilmark{3},
    K.~Penev\altaffilmark{3},
    P.~Sarkis\altaffilmark{4,5},
    V.~Suc\altaffilmark{2},
    Z.~Csubry\altaffilmark{3},
    D.~Bayliss\altaffilmark{6},
    J.~Bento\altaffilmark{7},
    G.~Zhou\altaffilmark{8},
    L.~Mancini\altaffilmark{4,9,10},
    T.~Henning\altaffilmark{4},
    S.~Ciceri\altaffilmark{11},
    M.~de~Val-Borro\altaffilmark{3},
    S. Shectman\altaffilmark{12},
    J.~D.~Crane\altaffilmark{12},
    P. Arriagada\altaffilmark{13},
    P. Butler\altaffilmark{13},
    J. Teske\altaffilmark{12, 13},
    I. Thompson\altaffilmark{12},
    D. Osip\altaffilmark{14},
    M. D\'iaz\altaffilmark{15},
    B.~Schmidt\altaffilmark{7},
    J.~L\'az\'ar\altaffilmark{16},
    I.~Papp\altaffilmark{16},
    P.~S\'ari\altaffilmark{16}
}
\altaffiltext{1}{Millennium Institute of Astrophysics, Santiago, Chile; rbrahm@astro.puc.cl}
\altaffiltext{2}{Instituto de Astrof\'isica, Facultad de F\'isica, Pontificia Universidad Cat\'olica de Chile, Av. Vicu\~na Mackenna 4860, 7820436 Macul, Santiago, Chile.}
\altaffiltext{3}{Department of Astrophysical Sciences, Princeton University, NJ 08544, USA.}
\altaffiltext{4}{Max-Planck-Institut f\"ur Astronomie, K\"onigstuhl 17, 69117 Heidelberg, Germany.}
\altaffiltext{5}{Physikalisches Institut, Universit\"{a}t Bern, Gesellschaftstrasse 6, 3012 Bern, Switzerland}
\altaffiltext{6}{Observatoire Astronomique de l'Universit\'e de Gen\`eve, 51 ch. des Maillettes, 1290 Versoix, Switzerland.}
\altaffiltext{7}{Research School of Astronomy and Astrophysics, Australian National University, Canberra, ACT 2611, Australia.}
\altaffiltext{8}{Harvard-Smithsonian Center for Astrophysics, 60 Garden
St., Cambridge, MA 02138, USA.}
\altaffiltext{9}{Dipartimento di Fisica, Universita di Roma Tor Vergata, Via
della Ricerca Scientifica 1, 00133 – Roma, Italy}
\altaffiltext{10}{INAF -- Osservatorio Astrofisico di Torino, Via Osservatorio 20, 10025 -- Pino Torinese, Italy}
\altaffiltext{11}{Department of Astronomy, Stockholm University, 11419, Stockholm, Sweden}
\altaffiltext{12}{The Observatories of the Carnegie Institution for Science, 813 Santa Barbara Street, Pasadena, CA 91101, USA}
\altaffiltext{13}{Department of Terrestrial Magnetism, The Carnegie Institution for Science, NW Washington, DC 20015-1305, USA}
\altaffiltext{14}{Las Campanas Observatory, The Carnegie Institution for Science, Colina el Pino, Casilla 601 La Serena, Chile}
\altaffiltext{15}{Departamento de Astronom\'ia, Universidad de Chile, Camino el Observatorio 1515, Casilla 36-D, Las Condes, Santiago, Chile}
\altaffiltext{16}{Hungarian Astronomical Association, 1451 Budapest, Hungary}
\altaffiltext{$\star$}{Alfred P.~Sloan Research Fellow}
\altaffiltext{$\star\star$}{Packard Fellow}
\altaffiltext{$\dagger$}{
 The HATSouth network is operated by a collaboration consisting of
Princeton University (PU), the Max Planck Institute f\"ur Astronomie
(MPIA), the Australian National University (ANU), and the Pontificia
Universidad Cat\'olica de Chile (PUC).  The station at Las Campanas
Observatory (LCO) of the Carnegie Institute is operated by PU in
conjunction with PUC, the station at the High Energy Spectroscopic
Survey (H.E.S.S.) site is operated in conjunction with MPIA, and the
station at Siding Spring Observatory (SSO) is operated jointly with
ANU.
This  paper  includes  data  gathered  with  the  MPG  2.2 m  and
ESO 3.6 m telescopes at the ESO Observatory in La Silla.
This paper includes data gathered with the 6.5~meter Magellan Telescopes located at Las Campanas Observatory, Chile.
}


\begin{abstract}

\setcounter{footnote}{10}
We report the discovery of four short period extrasolar planets transiting moderately bright stars 
from photometric measurements of the HATSouth network coupled to additional spectroscopic
and photometric follow-up observations.
While the planet masses range from 0.26 to 0.90 \mjup, the radii are all approximately a Jupiter radii, resulting in a wide range of bulk densities.  The orbital period of the planets range from 2.7d to 4.7d, with HATS-43b having an orbit that appears to be marginally non-circular (e$=0.173 \pm 0.089$).  HATS-44 is notable for a high metallicity (\feh$ = 0.320\pm0.071$). The host stars spectral types range from late F to early K, and all of them are moderately bright (13.3$<$V$<$14.4), allowing the execution of future detailed follow-up observations.
\hatcurb{43} and \hatcurb{46}, with expected transmission signals of 2350 ppm and 1500 ppm, respectively,  are particularly well suited targets for atmospheric characterisation via transmission spectroscopy.

\setcounter{footnote}{0}
\end{abstract}

\keywords{
    planetary systems ---
    stars: individual (
\setcounter{planetcounter}{1}
\hatcur{43},
\hatcurCCgsc{43}\loopcommanoperiod
\setcounter{planetcounter}{2}
\hatcur{44},
\hatcurCCgsc{44}\loopcommanoperiod
\setcounter{planetcounter}{3}
\hatcur{45},
\hatcurCCgsc{45}\loopcommanoperiod
\setcounter{planetcounter}{3}
\hatcur{46},
\hatcurCCgsc{46}\loopcommanoperiod
\setcounter{planetcounter}{4}
) 
    techniques: spectroscopic, photometric
}


\section{Introduction}
\label{sec:introduction}

By measuring their masses and radii, transiting planets orbiting moderately bright stars offer the unique opportunity of studying in depth
the physical structure of planets other than those present in the solar system. This property makes transiting planets one of the
most valuable resources for testing current models of planet formation and evolution \citep[e.g.,][]{mordasini:2012}. In addition, detailed follow-up observations of these systems allowed us to study the structure and composition of their atmospheres \citep[e.g.,][]{fraine:2014}, and to refine their
orbital configuration and evolution by measuring the misalignment angle between the orbit and the spin axis of their host stars \citep[e.g.,][]{zhou:2015}.

Ground-based photometric surveys like HATNet \citep{bakos:2004:hatnet} and SuperWASP \citep{pollacco:2006} have been a key and cost-efficient resource in contributing to the current number
of $\sim$300 discovered transiting systems with masses and radii determined with a precision better than 30\%. Even though the population of discovered systems is highly biased towards
the detection of giant planets at short distances from their stars (hot Jupiters), the increase of discoveries in this narrow region of the parameter space has proven to be fundamental for
achieving statistically significant results that are helping to solve some of the theoretical challenges present in the field. For example, \citet{hartman:2016}, using the full population of well
characterised transiting systems, found that the radii of close-in planets increase as the parent stars
evolve. This fact supports the theories that propose that hot Jupiters are inflated due to energy deposited
deep into the planet interior \citep[e.g. ][]{batygin:2010} and not due to a delayed cooling \citep{burrows:2007}. Another recent example concerning the using of the full population of well characterised transiting systems is that of \citet{hellier:2016}, who find that there
is no difference within the so-called Jupiter bulge between the planetary period distributions around greater-than versus less-than-solar metallicity host stars. 
This calls into question the idea that the lack of hot Jupiters discovered by the $Kepler$ mission is due to a bias
in the metallicity of the $Kepler$ sample \citep{dawson:2013}.

Despite the importance of continuing the discovery of short period gas giants, most of the efforts of the community are being put into the detection of well characterised transiting planets located in sparsely populated
regions of the parameter space. Planets with periods longer than ten days and/or sub-Saturn mass planets are particularly interesting.
While these types of planets are among the prime discovery targets of the ongoing $Kepler$ K2 mission \citep{howell:2014}, and will be also efficiently discovered by TESS \citep{ricker:2014}, several new ground-based photometric surveys were
designed to dig into these planetary regimes. Specifically, the HATSouth survey of robotic telescopes \citep{bakos:2013:hatsouth}, having three stations in three well separated locations in the southern hemisphere, has an increased efficiency
for detecting both longer period giant planets (warm Jupiters), and planets with in the Neptune-Saturn mass range, if compared to typical single site surveys. The ability of the HATSouth network to
discover these type of systems has been already demonstrated with the  discoveries of two super-Neptunes \citep{bakos:2015, bayliss:2015}, and of a warm Jupiter with an orbital period of $\approx16$ days \citep{brahm:2016:hs17}.

In this paper we present the discovery of four new well characterised short-period transiting planets from the HATSouth survey.
In Section \ref{sec:obs} we summarise the detection of the photometric planetary signal, and spectroscopic and photometric follow up observations.
In Section \ref{sec:analysis} we describe the analysis that was carried out in order to rule out false positives, and to derive the
parameters of the planets and host stars. Finally, in Section \ref{sec:discussion} we summarise the properties of each system and we discuss our
findings in the context of the full population of discovered transiting systems.

\section{Observations}
\label{sec:obs}

\subsection{Photometric detection}
\label{sec:detection}
The discovery of the periodic planetary-like photometric signals for the four systems
presented in this study were obtained from the images registered by the three
stations of the HATSouth network (the HS1 and HS2 instruments in Chile,
HS3 and HS4 in Namibia, and HS5 and HS6 in Australia).
Observations were performed with a typical cadence of 5 minutes
using a Sloan $r$ photometric filter. The specific properties of the observations
that allowed the discovery of the planets are summarised in Table \ref{tab:photobs}.
As can be noticed in this table, the total number of images obtained per
station varies from 700 to 9000 images, and were strongly dependent on the weather
conditions and technical issues present on each  site. In addition, the number of 
images obtained for different objects also depends on the adopted observing strategy
and the position of the target in the field, because targets located in the edges of
fields are usually also monitored when observing contiguous fields, as is the case in this work for \hatcur{44} and \hatcur{46}.

The original images were  reduced to photometric
light curves by following the procedures described in \citet{penev:2013:hats1}.
The light curves thus generated were corrected for systematic signals using the Trend Filtering
Algorithm \citep[TFA,][]{kovacs:2005:TFA} and then the Box Least Squares
\citep[BLS,][]{kovacs:2002:BLS} method was applied to identify periodic transit-like
features on them. Figure~\ref{fig:hatsouth} shows the phase-folded light curves for
the four systems analysed in this study, which presented periodic dimmings in their fluxes with depths (10--30 mmag) and durations (1.6--3.2 h) compatible with being short-period transiting giant planets. Due to these  properties, these systems
were added to the HATSouth database of planetary candidates.
The light curve data for these four systems are presented in Table~\ref{tab:phfu}.

%
%
\ifthenelse{\boolean{emulateapj}}{
    \begin{figure*}[!ht]
}{
    \begin{figure}[!ht]
}
\plottwo{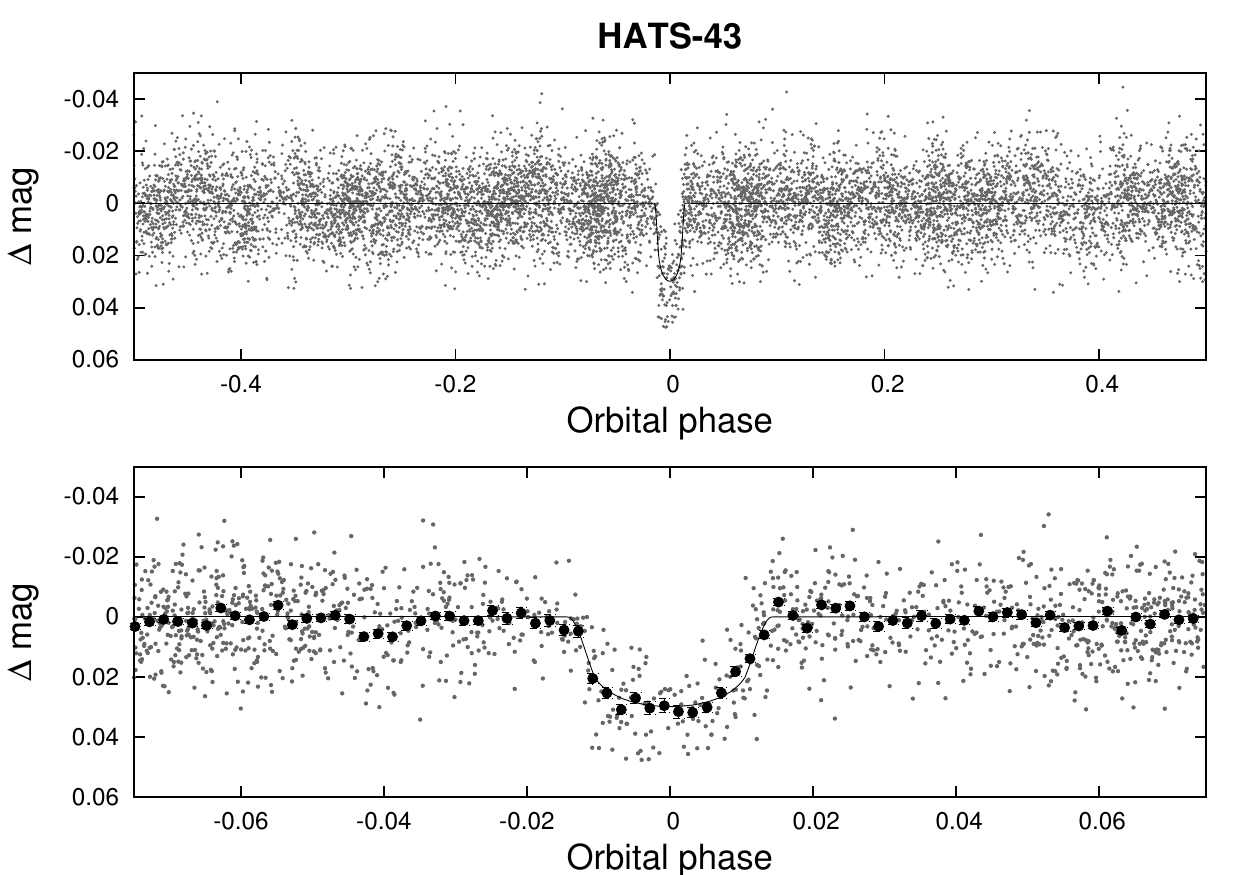}{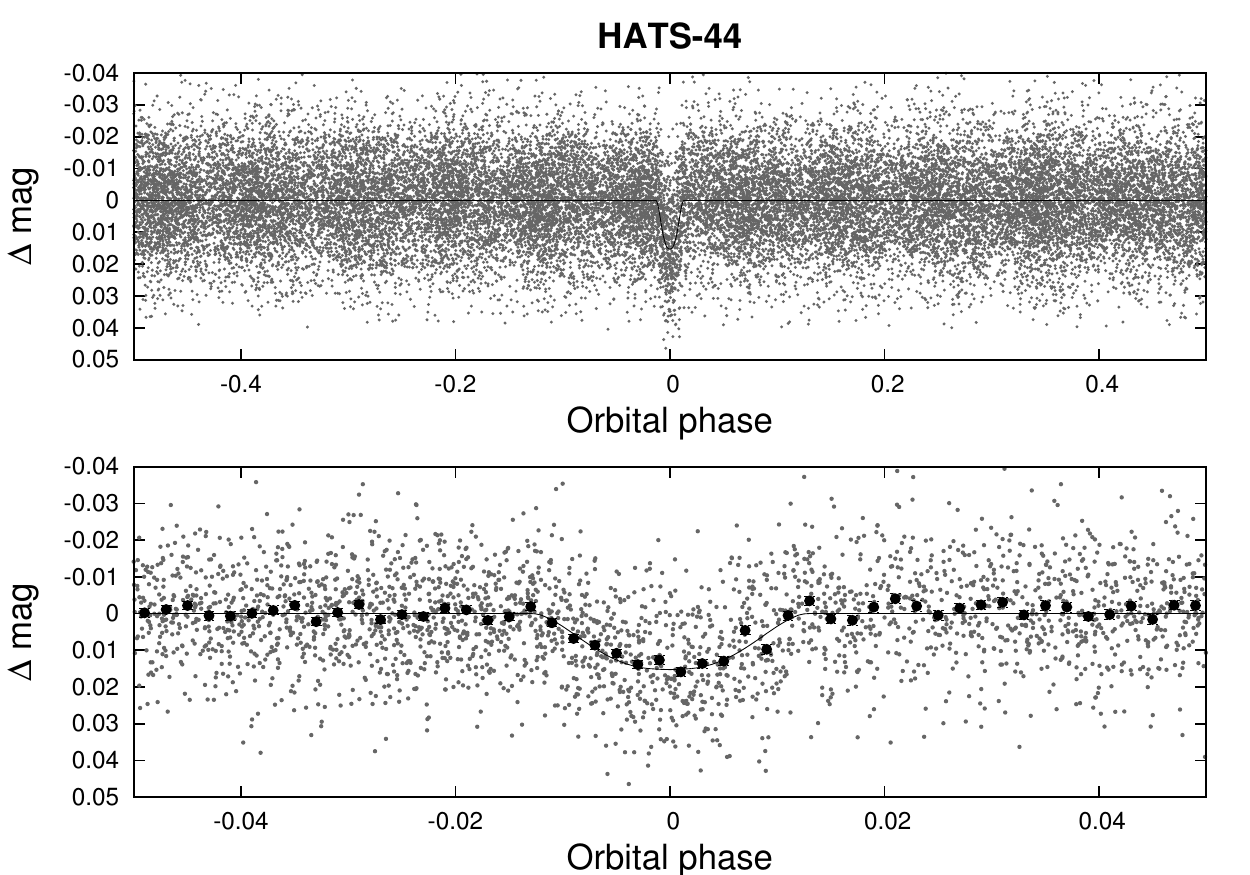}
\plottwo{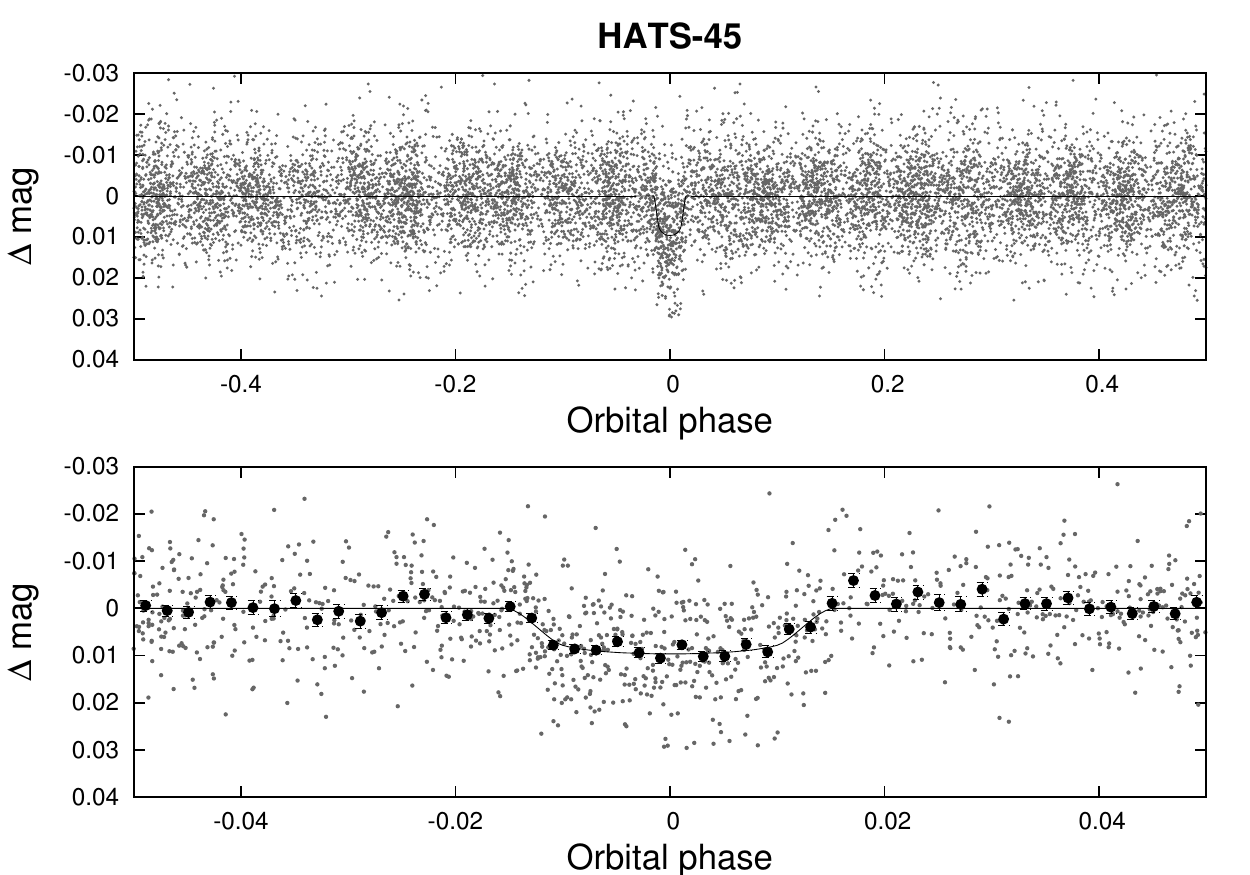}{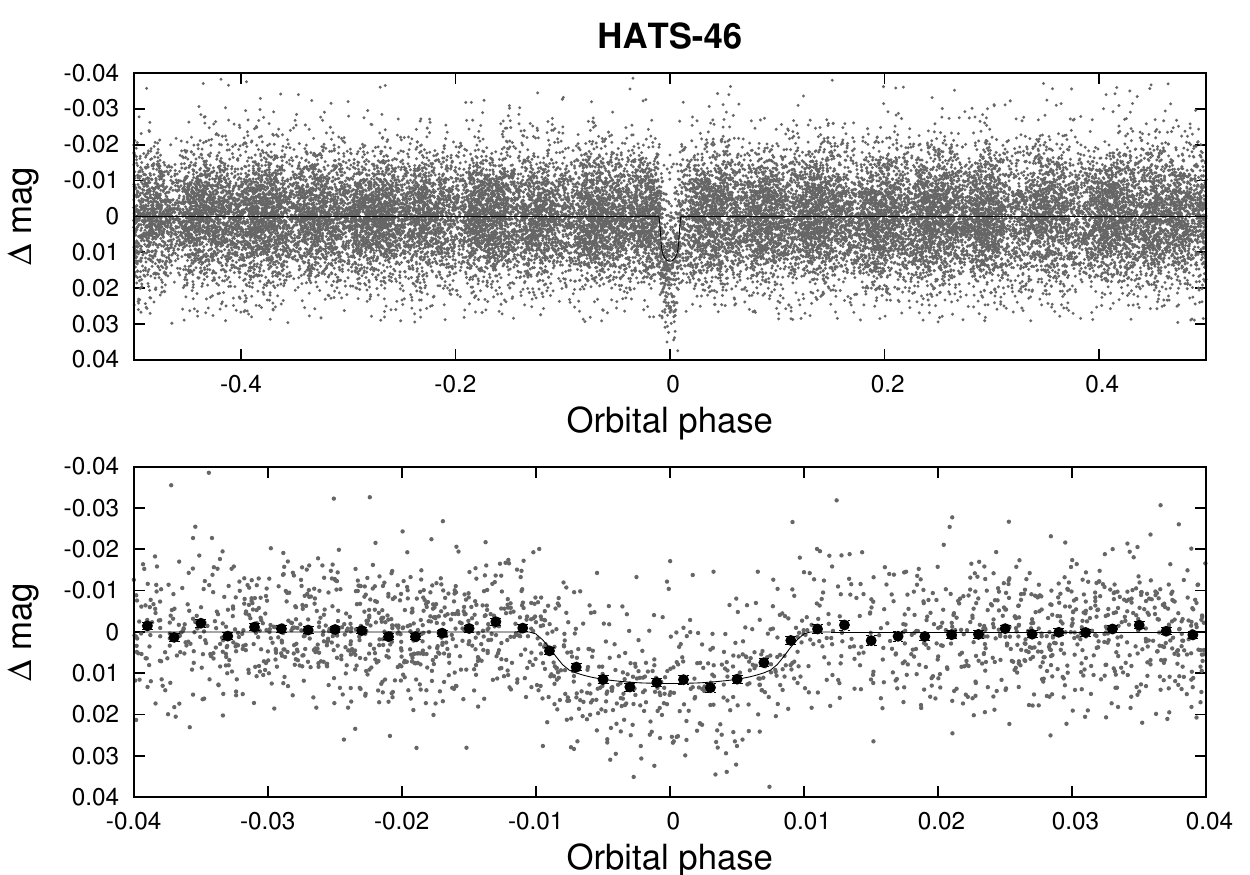}
\caption[]{
    Phase-folded unbinned HATSouth light curves for \hatcur{43} (upper left), \hatcur{44} (upper right), \hatcur{45} (lower left) and \hatcur{46} (lower right). In each case we show two panels. The
    top panel shows the full light curve, while the bottom panel shows
    the light curve zoomed-in on the transit. The solid lines show the
    model fits to the light curves. The dark filled circles in the
    bottom panels show the light curves binned in phase with a bin
    size of 0.002.
\label{fig:hatsouth}}
\ifthenelse{\boolean{emulateapj}}{
    \end{figure*}
}{
    \end{figure}
}

\ifthenelse{\boolean{emulateapj}}{
    \begin{deluxetable*}{llrrrr}
}{
    \begin{deluxetable}{llrrrr}
}
\tablewidth{0pc}
\tabletypesize{\scriptsize}
\tablecaption{
    Summary of photometric observations
    \label{tab:photobs}
}
\tablehead{
    \multicolumn{1}{c}{Instrument/Field\tablenotemark{a}} &
    \multicolumn{1}{c}{Date(s)} &
    \multicolumn{1}{c}{\# Images} &
    \multicolumn{1}{c}{Cadence\tablenotemark{b}} &
    \multicolumn{1}{c}{Filter} &
    \multicolumn{1}{c}{Precision\tablenotemark{c}} \\
    \multicolumn{1}{c}{} &
    \multicolumn{1}{c}{} &
    \multicolumn{1}{c}{} &
    \multicolumn{1}{c}{(sec)} &
    \multicolumn{1}{c}{} &
    \multicolumn{1}{c}{(mmag)}
}
\startdata
\sidehead{\textbf{\hatcur{43}}}
~~~~HS-2.4/G598 & 2013 Sep--2014 Mar & 739 & 285 & $r$ & 13.4 \\
~~~~HS-4.4/G598 & 2013 Sep--2014 Feb & 4154 & 345 & $r$ & 11.4 \\
~~~~HS-6.4/G598 & 2013 Sep--2014 Mar & 3865 & 357 & $r$ & 10.9 \\
~~~~LCOGT~1\,m+CTIO/SBIG & 2016 Aug 21 & 44 & 219 & $i$ & 1.3 \\
~~~~LCOGT~1\,m+CTIO/sinistro & 2016 Oct 26 & 69 & 219 & $i$ & 1.1 \\
\sidehead{\textbf{\hatcur{44}}}
~~~~HS-2.3/G598 & 2013 Sep--2014 Mar & 745 & 285 & $r$ & 12.4 \\
~~~~HS-4.3/G598 & 2013 Sep--2014 Feb & 4143 & 345 & $r$ & 13.5 \\
~~~~HS-6.3/G598 & 2013 Sep--2014 Mar & 3836 & 357 & $r$ & 13.3 \\
~~~~HS-1.2/G599 & 2012 Jan--2013 Apr & 9325 & 292 & $r$ & 12.1 \\
~~~~HS-3.2/G599 & 2012 Jan--2013 Apr & 3174 & 288 & $r$ & 12.9 \\
~~~~HS-5.2/G599 & 2012 Jan--2013 Apr & 5004 & 288 & $r$ & 12.7 \\
~~~~LCOGT~1\,m+SAAO/SBIG & 2015 Nov 07 & 63 & 194 & $i$ & 2.2 \\
~~~~LCOGT~1\,m+CTIO/sinistro & 2015 Nov 15 & 55 & 219 & $i$ & 5.4 \\
~~~~LCOGT~1\,m+CTIO/SBIG & 2015 Nov 26 & 41 & 194 & $g$ & 6.9 \\
\sidehead{\textbf{\hatcur{45}}}
~~~~HS-2.2/G554 & 2009 Dec--2011 May & 6414 & 296 & $r$ & 8.0 \\
~~~~HS-4.2/G554 & 2009 Dec--2011 Mar & 953 & 383 & $r$ & 10.2 \\
~~~~HS-6.2/G554 & 2010 Dec--2011 May & 2097 & 300 & $r$ & 8.5 \\
~~~~Swope~1\,m/e2v & 2014 Mar 22 & 81 & 160 & $i$ & 1.6 \\
~~~~CTIO~0.9\,m & 2014 Oct 13 & 36 & 181 & $i$ & 1.9 \\
~~~~LCOGT~1\,m+SAAO/SBIG & 2015 Mar 09 & 25 & 200 & $i$ & 1.4 \\
~~~~LCOGT~1\,m+CTIO/sinistro & 2015 Mar 14 & 51 & 226 & $i$ & 1.5 \\
\sidehead{\textbf{\hatcur{46}}}
~~~~HS-2.3/G754 & 2012 Sep--2012 Dec & 3875 & 282 & $r$ & 9.2 \\
~~~~HS-4.3/G754 & 2012 Sep--2013 Jan & 3191 & 292 & $r$ & 9.9 \\
~~~~HS-6.3/G754 & 2012 Sep--2012 Dec & 2994 & 285 & $r$ & 9.7 \\
~~~~HS-1.2/G755 & 2011 Jul--2012 Oct & 5265 & 292 & $r$ & 9.6 \\
~~~~HS-3.2/G755 & 2011 Jul--2012 Oct & 4851 & 287 & $r$ & 10.3 \\
~~~~HS-5.2/G755 & 2011 Jul--2012 Oct & 6018 & 296 & $r$ & 9.3 \\
~~~~Swope~1\,m/e2v & 2014 Nov 30 & 64 & 169 & $i$ & 1.7 \\
~~~~LCOGT~1\,m+CTIO/sinistro & 2016 Aug 31 & 55 & 219 & $i$ & 1.1 \\
\enddata
\tablenotetext{a}{
    For HATSouth data we list the HATSouth unit, CCD and field name
    from which the observations are taken. HS-1 and -2 are located at
    Las Campanas Observatory in Chile, HS-3 and -4 are located at the
    H.E.S.S. site in Namibia, and HS-5 and -6 are located at Siding
    Spring Observatory in Australia. Each unit has 4 ccds. Each field
    corresponds to one of 838 fixed pointings used to cover the full
    4$\pi$ celestial sphere. All data from a given HATSouth field and
    CCD number are reduced together, while detrending through External
    Parameter Decorrelation (EPD) is done independently for each
    unique unit+CCD+field combination.
}
\tablenotetext{b}{
    The median time between consecutive images rounded to the nearest
    second. Due to factors such as weather, the day--night cycle,
    guiding and focus corrections the cadence is only approximately
    uniform over short timescales.
}
\tablenotetext{c}{
    The RMS of the residuals from the best-fit model.
} \ifthenelse{\boolean{emulateapj}}{
    \end{deluxetable*}
}{
    \end{deluxetable}
}


\subsection{Spectroscopic Observations}
\label{sec:obsspec}

To confirm the planetary nature of the candidates discovered by the HATSouth network,
they are subject of an extensive follow-up campaign that involves the use of different facilities
containing spectroscopic instruments with a wide range of capabilities. All the spectrographs
used for the discovery of the four HATS planets presented in this work are listed in Table~\ref{tab:specobs} along with the
general properties of the observations.

\subsubsection{Reconnaissance Spectroscopy}

After the initial photometric detection, the four candidates presented in the previous section were first
observed with the WiFeS spectrograph \citep{dopita:2007} installed at the ANU 2.3m telescope.
As described in \citet{bayliss:2013:hats3}, observations are performed with two instrument configurations.
A single R=3000 spectra is obtained for each candidate in order to perform a spectral classification of
the star by determining its effective temperature (\teff), surface gravity (\logg) and metallicity ([Fe/H])
by using a library of synthetic spectra with the principal goal of identifying giant stars for which
the observed transit depth could not be produced by planetary companions. In this way, \hatcur{43} and \hatcur{44} 
were identified as a K-dwarfs with \teff = 4900 $\pm$ 300 K, \logg = 5.0 $\pm$ 0.3 dex, and 
 \teff = 4750 $\pm$ 300 K, \logg = 4.3 $\pm$ 0.3 dex, respectively, while \hatcur{45}  and
 \hatcur{46} were typed as a F-dwarf (6250 $\pm$ 300 K, 3.7 $\pm$ 0.3) and a G-dwarf
 (5800 $\pm$ 300 K, 4.5 $\pm$ 0.3), respectively.
 Additionally multiple spectra for each candidate are obtained with a resolving power of R=7000,
 with the goal of measuring radial velocity (RV) variations with a precision of $\sim$2 km s$^{-1}$ for identifying
 systems in which the transit-like signal is produced by stellar mass companions. The number of velocity points
 obtained for each candidate were 2, 4, 3 and 3 for HATS-43 -44 -45 and -46, respectively and they
 were focused on phases $\sim$0.25 and $\sim$0.75 where the maximum velocity difference is expected. No significant
 velocity variations at the level of $\sim$4 \kms were observed for any of the four candidates.
 
 The lack of high-amplitude velocity variations and the dwarf status of the host stars for the
 four HATS candidates provided the first evidences in favour of the planetary origin of the
 photometric signals described in Section~\ref{sec:detection}
 
\subsubsection{Precision Radial Velocities}

Precision radial velocities are required to confirm the planetary nature of a transiting companion
by providing the means to estimate its mass and orbital parameters. For this purpose, we used the FEROS
spectrograph \citep{kaufer:1998} installed at the MPG 2.2m telescope. The high efficiency of this
instrument coupled to its high resolving power of R=50000 allows us to achieve a long term radial
velocity precision in the range of 10 -- 50 m s$^{-1}$ for our V$>$13 candidates by obtaining spectra with
$\sim$1800s of exposure time using the simultaneous  wavelength calibration technique \citep{baranne:1996}.
In the case of HATS-43, -44, and -45, we obtained on the order of 15 FEROS spectra, while for HATS-46,
31 FEROS spectra were acquired. All FEROS spectra were reduced, extracted and analysed
using the CERES pipeline \citep{brahm:2017:ceres}, where after applying an optimal extraction algorithm,
each spectrum is wavelength calibrated and corrected by instrumental drifts before calculating
the radial velocity by cross-correlation with a G2-like binary mask. Bisector spans are also measured for each spectrum
by CERES.
Figure~\ref{fig:rvbis} shows that the FEROS velocities obtained for the four systems have a time correlated variation, which is in phase with the photometric
ephemerides and has an amplitude consistent with planetary-mass objects.
However, only in the case of HATS-44, the FEROS velocities were enough for measuring the semi-amplitude
with a precision better than 25\% (K=90$\pm$17 m s$^{-1}$). For the remaining candidates
additional observations were required as we now describe.

\hatcur{43} and \hatcur{45} were observed with the HARPS instrument \citep{mayor:2003} mounted on the
ESO 3.6m telescope, which is located at the ESO La Silla Observatory.
Observations were performed with the Object+Sky
mode because the daily internal drifts of the instrument are smaller than 1 ms$^{-1}$
which is significantly smaller than the radial velocity precision that we require.
HARPS data was also reduced and analysed with CERES. By using these additional velocities,
the semi-amplitudes of the orbits for \hatcur{43} and \hatcur{45} were determined with a precision better than 25$\%$.
While the combined RVs for \hatcur{45} suggest that the star is orbited by a typical hot-Jupiter, the velocities for \hatcur{43} are consistent with
the presence of a Saturn-mass planet with a non-negligible eccentricity.
For these two planets we also obtained
a single spectrum using the Coralie Spectrograph mounted on the 1.2m Euler telescope installed at the ESO La Silla Observatory.
Both spectra were processed with CERES but were not used in the analysis because no velocity variations could
be computed from a single radial velocity epoch.

Finally, \hatcur{46} was observed with the Planet Finder Spectrograph \citep[PFS,][]{crane:2010} mounted on 
the Magellan/Clay 6.5m telescope at Las Campanas Observatory. 
As has been described in previous HATSouth discoveries \citep{jordan:2014:hats4,zhou:2014:hats5,hartman:2015:hats6},
we obtained a template spectrum by using the 0.5\arcsec\ slit, which was then used as reference for
computing the radial velocities at different epochs by obtaining spectra with a I2-cell. The
11 spectra that were acquired with the I2-cell were processed as described in \citet{butler:1996}. The mean RV precision achieved was $\sim$20 ms$^{-1}$ and 
was principally limited by the faintness of the star. 
By combining the
velocities measured by FEROS and PFS for \hatcur{46} we were able to confirm the planetary
nature of this candidate and infer for it a sub-Saturn mass.

The radial velocities and bisector spans obtained with FEROS, Coralie, and HARPS  for the
four discovered planets are presented in Table~\ref{tab:rvs} at the end of the paper. We investigated
if there is a significant degree of correlation between the radial velocities and the bisector span measurements
that could suggest that the observed velocity variations are produced by a blended stellar companion.
Specifically, we followed the procedure adopted in \citet{bhatti:2016} and for each of our four systems
we computed the error weighted distribution for the Pearson correlation coefficient by applying a bootstrap
method. The derived 95\% confidence intervals for the correlation coefficient are
[-0.57,  0.30], [-0.59,  0.02], [-0.35,  0.46], and [-0.22,  0.65], for \hatcur{43}, \hatcur{44}, \hatcur{45}, and \hatcur{46},
respectively, implying that there is no significant correlation and thus supporting the planetary hypothesis as the
cause of the observed velocity variations. Figure~\ref{fig:rvbs} shows the radial velocities vs. bisector spans diagrams 
for each of the four systems.

\ifthenelse{\boolean{emulateapj}}{
    \begin{figure*}[!ht]
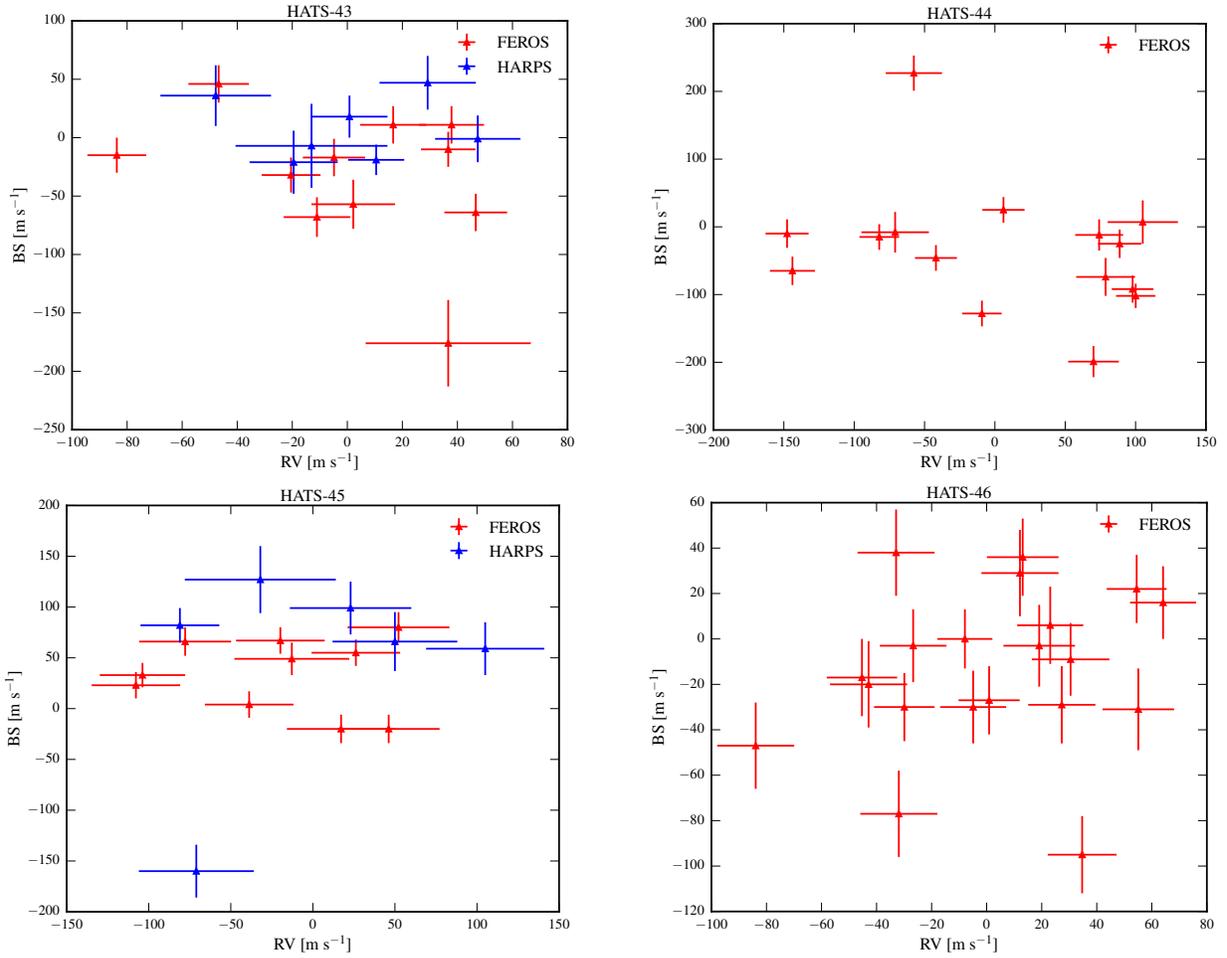

}{
    \begin{figure}[!ht]
}
\plottwo{\hatcur{43}-bs-rv-correl.pdf}{\hatcur{44}-bs-rv-correl.pdf}
\plottwo{\hatcur{45}-bs-rv-correl.pdf}{\hatcur{46}-bs-rv-correl.pdf}
\caption[]{
    RV vs bisector span measurements for \hatcur{43} (upper left), \hatcur{44} (upper right), \hatcur{45} (lower left) and \hatcur{46} (lower right).
    No significant correlation at the 95\% level was identified, which indicates that the RV variations are probably produces by planetary mass
    orbital companions.
\label{fig:rvbs}}
\ifthenelse{\boolean{emulateapj}}{
    \end{figure*}
}{
    \end{figure}
}

\ifthenelse{\boolean{emulateapj}}{
    \begin{deluxetable*}{llrrrrr}
}{
    \begin{deluxetable}{llrrrrrrrr}
}
\tablewidth{0pc}
\tabletypesize{\scriptsize}
\tablecaption{
    Summary of spectroscopy observations 
    \label{tab:specobs}
}
\tablehead{
    \multicolumn{1}{c}{Instrument}          &
    \multicolumn{1}{c}{UT Date(s)}             &
    \multicolumn{1}{c}{\# Spec.}   &
    \multicolumn{1}{c}{Res.}          &
    \multicolumn{1}{c}{S/N Range\tablenotemark{a}}           &
    \multicolumn{1}{c}{$\gamma_{\rm RV}$\tablenotemark{b}} &
    \multicolumn{1}{c}{RV Precision\tablenotemark{c}} \\
    &
    &
    &
    \multicolumn{1}{c}{$\Delta \lambda$/$\lambda$/1000} &
    &
    \multicolumn{1}{c}{(\kms)}              &
    \multicolumn{1}{c}{(\ms)}
}
\startdata
%
%
\sidehead{\textbf{\hatcur{43}}}\\
ANU~2.3\,m/WiFeS & 2015 Feb 6 & 1 & 3 & 68 & $\cdots$ & $\cdots$ \\
ANU~2.3\,m/WiFeS & 2015 Feb 6--8 & 2 & 7 & 57--82 & 23.2 & 4000 \\
MPG~2.2\,m/FEROS & 2015 Oct--2016 Dec \tablenotemark{d} & 12 & 48 & 15--50 & 22.078 & 25 \\
ESO~3.6\,m/HARPS & 2016 Apr--Nov & 7 & 115 & 10--24 & 22.053 & 24 \\
Euler~1.2\,m/Coralie \tablenotemark{d} & 2016 Aug 10 & 1 & 60 & 12 & 22.002 & $\cdots$ \\
\sidehead{\textbf{\hatcur{44}}}\\
ANU~2.3\,m/WiFeS & 2015 Jan 1 & 1 & 3 & 54 & $\cdots$ & $\cdots$ \\
ANU~2.3\,m/WiFeS & 2015 Jan--Aug & 4 & 7 & 50--101 & 47.4 & 4000 \\
MPG~2.2\,m/FEROS & 2015 Oct--2016 Dec & 15 & 48 & 19--36 & 44.082 & 50 \\
\sidehead{\textbf{\hatcur{45}}}\\
ANU~2.3\,m/WiFeS & 2013 Sep 27 & 1 & 3 & 87 & $\cdots$ & $\cdots$ \\
ANU~2.3\,m/WiFeS & 2014 Feb 17--23 & 3 & 7 & 44--70 &  20.8 & 4000 \\
Euler~1.2\,m/Coralie \tablenotemark{d} & 2014 Sep 12 & 1 & 60 & 12 & 19.19 & $\cdots$ \\
MPG~2.2\,m/FEROS \tablenotemark{d} & 2015 Jan--2016 Feb & 13 & 48 & 22--70 & 19.423 & 50 \\
ESO~3.6\,m/HARPS & 2015 Feb 14--19 & 6 & 115 & 14--24 & 19.372 & 28 \\
\sidehead{\textbf{\hatcur{46}}}\\
ANU~2.3\,m/WiFeS & 2014 Oct 4 & 1 & 3 & 64 & $\cdots$ & $\cdots$ \\
ANU~2.3\,m/WiFeS & 2014 Oct 4--7 & 3 & 7 & 36--80 & -30.5 & 4000 \\
MPG~2.2\,m/FEROS \tablenotemark{d} & 2015 Jun--2016 Dec & 31 & 48 & 16--57 & -30.193 & 35 \\
Magellan~6.5\,m/PFS+I$_{2}$ & 2015 Jun--2016 Dec & 11 & 76 & 45--55 & $\cdots$ & 24 \\
Magellan~6.5\,m/PFS & 2015 Jun & 3 & 76 & 59--61 & $\cdots$ & $\cdots$ \\
\enddata 
\tablenotetext{a}{
    S/N per resolution element near 5180\,\AA.
}
\tablenotetext{b}{
    For high-precision RV observations included in the orbit determination this is the zero-point RV from the best-fit orbit. For other instruments it is the mean value. We do not provide this quantity for the lower resolution WiFeS observations which were only used to measure stellar atmospheric parameters.
}
\tablenotetext{c}{
    For high-precision RV observations included in the orbit
    determination this is the scatter in the RV residuals from the
    best-fit orbit (which may include astrophysical jitter), for other
    instruments this is either an estimate of the precision (not
    including jitter), or the measured standard deviation. We do not
    provide this quantity for low-resolution observations from the
    ANU~2.3\,m/WiFeS.
}
\tablenotetext{d}{
    We excluded from the analysis the single Coralie observations of \hatcur{43} and \hatcur{45}. We also excluded from the analysis one FEROS observation of \hatcur{43} obtained on UT 2015 Oct 30, and one FEROS observation of \hatcur{45} obtained on UT 2015 Feb 2, both of which were affected by significant sky contamination. For \hatcur{46}, which has a very low amplitude RV orbital wobble, and for which even slight sky contamination can obscure the signal, we excluded 11 FEROS observations due to evidence of sky contamination as seen in the computed CCFs. The excluded observations are from UT 2015 Jun 10, and 21, Jul 20, Oct 2, 4, 26, 27 and 30, and Nov 3, and 2016 Jul 26 and Sep 14.
}
\ifthenelse{\boolean{emulateapj}}{
    \end{deluxetable*}
}{
    \end{deluxetable}
}

%
\setcounter{planetcounter}{1}
%
\ifthenelse{\boolean{emulateapj}}{
    \begin{figure*} [ht]
}{
    \begin{figure}[ht]
}
\plottwo{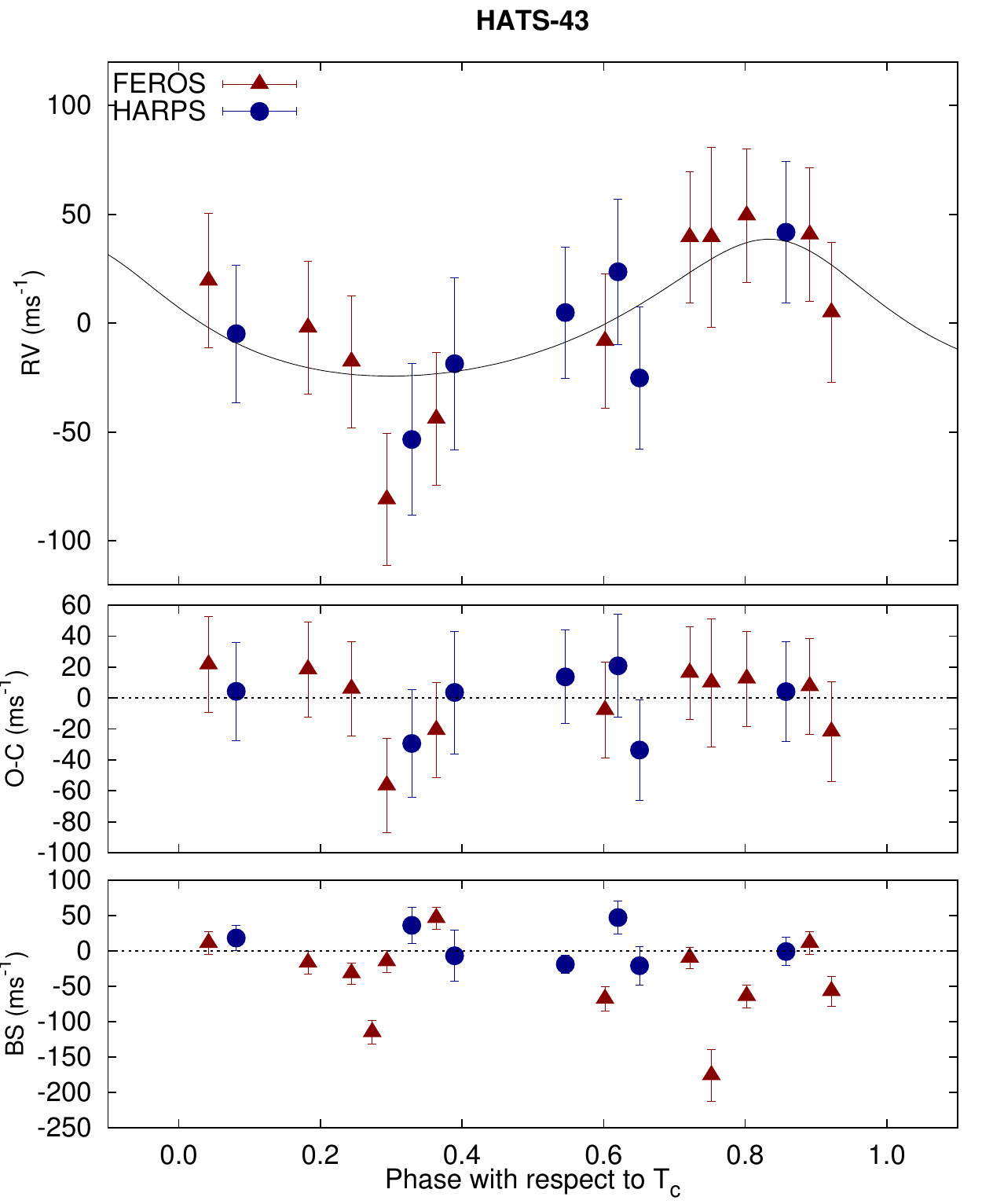}{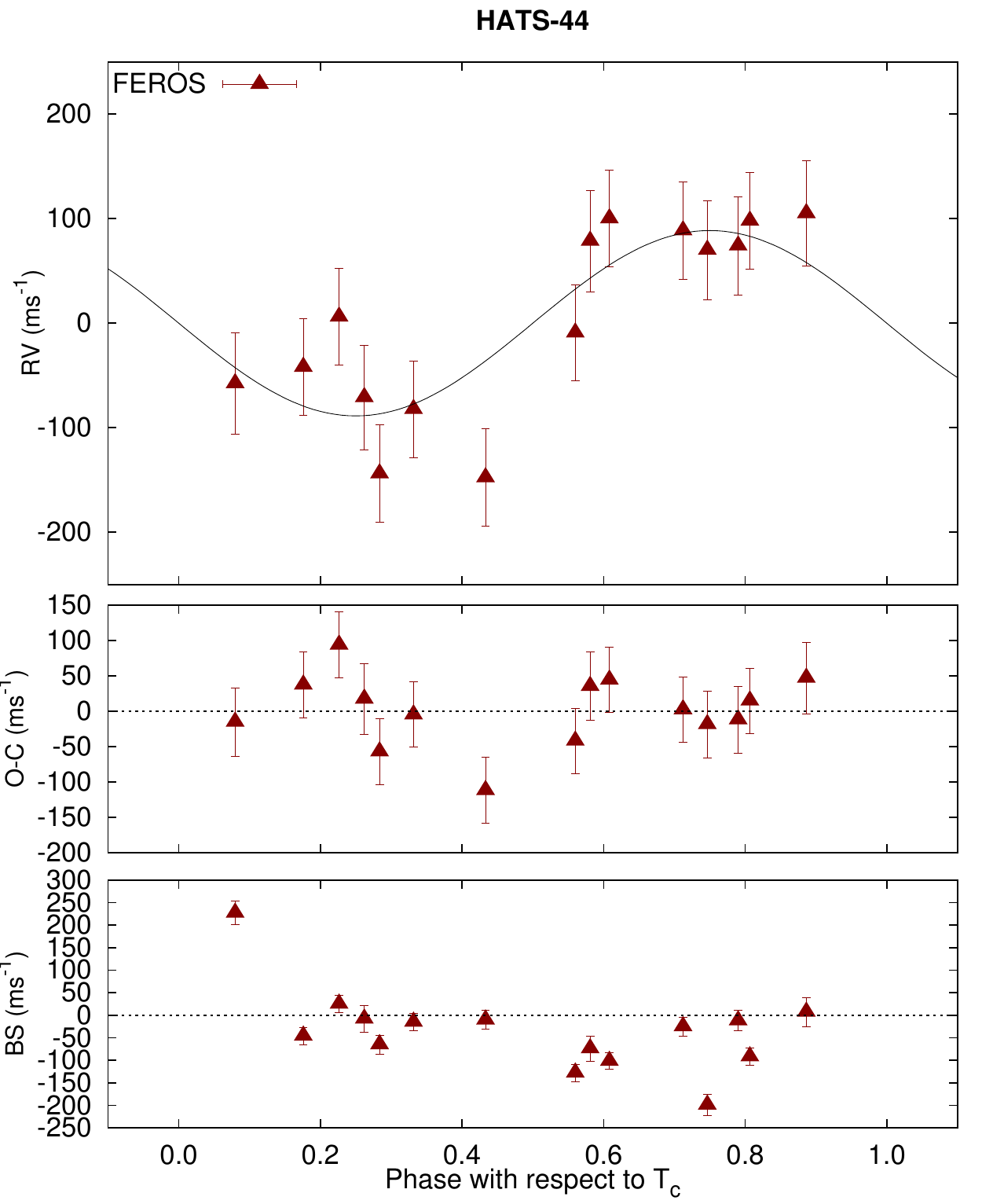}
\plottwo{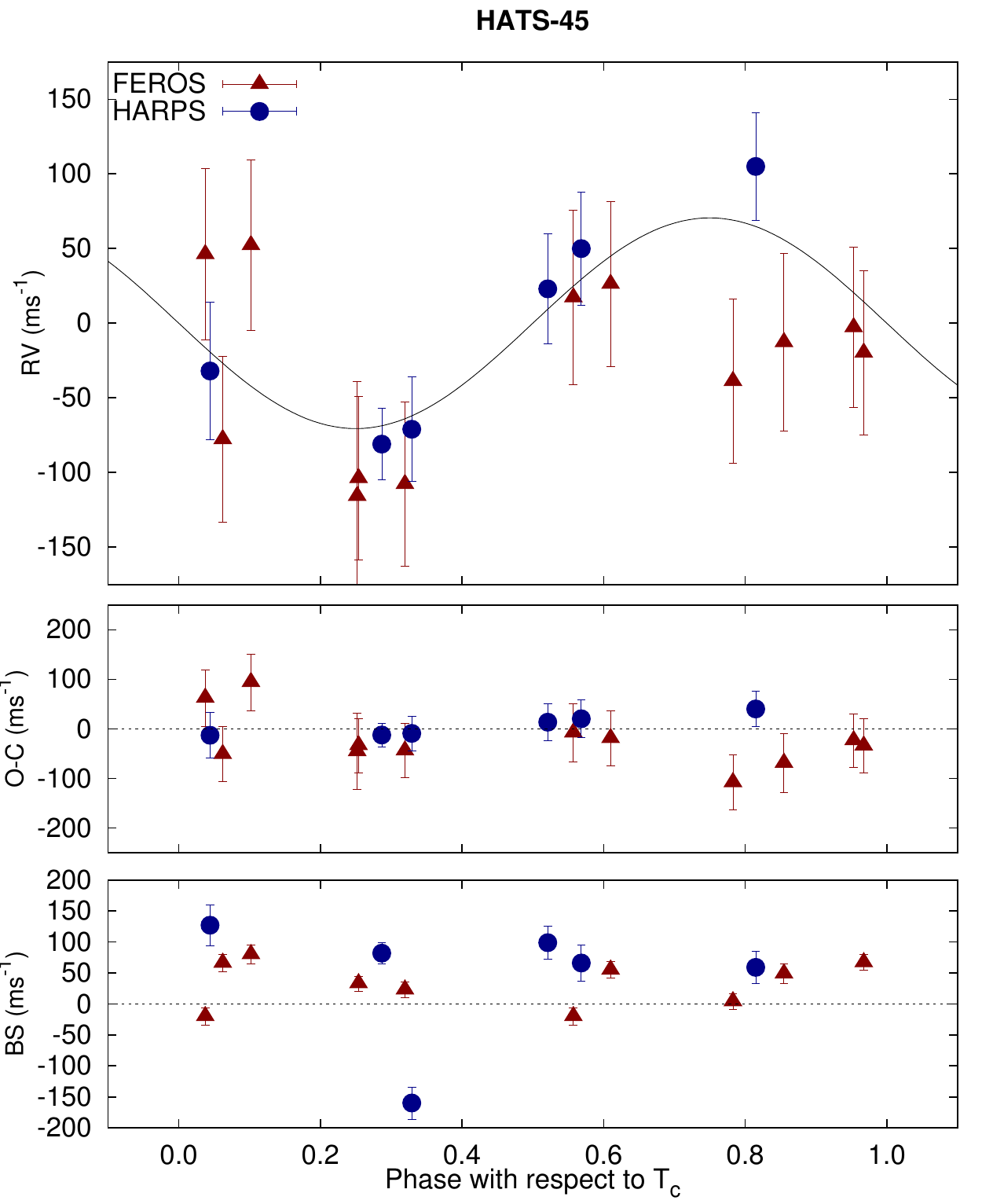}{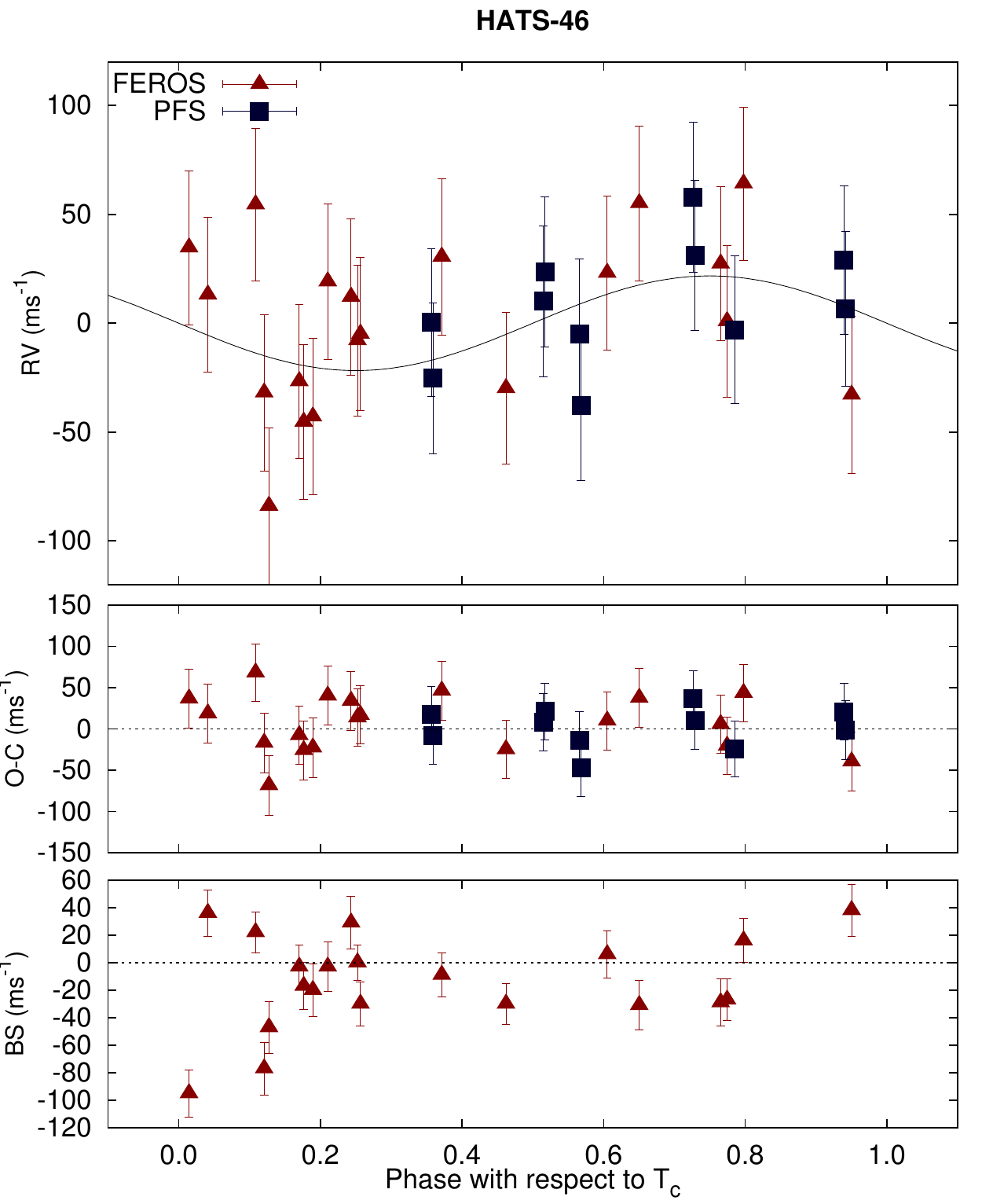}
\caption{
    Phased high-precision RV measurements for \hbox{\hatcur{43}{}} (upper left), \hbox{\hatcur{44}{}} (upper right), \hbox{\hatcur{45}{}} (lower left) and \hbox{\hatcur{46}{}} (lower right). The instruments used are labeled in the plots. In each case we show three panels. The top panel shows the phased measurements together with our best-fit model (see \reftabl{planetparam}) for each system. Zero-phase corresponds to the time of mid-transit. The center-of-mass velocity has been subtracted. The second panel shows the velocity $O\!-\!C$ residuals from the best fit. The error bars include the jitter terms listed in \reftabl{planetparam} added in quadrature to the formal errors for each instrument. The third panel shows the bisector spans (BS). Note the different vertical scales of the panels.
}
\label{fig:rvbis}
\ifthenelse{\boolean{emulateapj}}{
    \end{figure*}
}{
    \end{figure}
}


\subsection{Photometric follow-up observations}
\label{sec:phot}

In addition to the HATSouth discovery light curves, we observed transits for the four
discovered planets using telescopes with larger apertures in order to: i) confirm that the photometric signals are real, ii) refine the ephemerides of the systems, and iii) measure
the transiting parameters with a higher precision; an accurate determination
of $R_P/R_{\star}$ and $a/R_{\star}$ is particularly important for obtaining a reliable
estimation of the planetary physical parameters. The basic configurations used in these
observations are listed in \reftabl{photobs}, while  the light curve data is presented in
\reftabl{phfu}.

\begin{figure*}[!ht]
\plotone{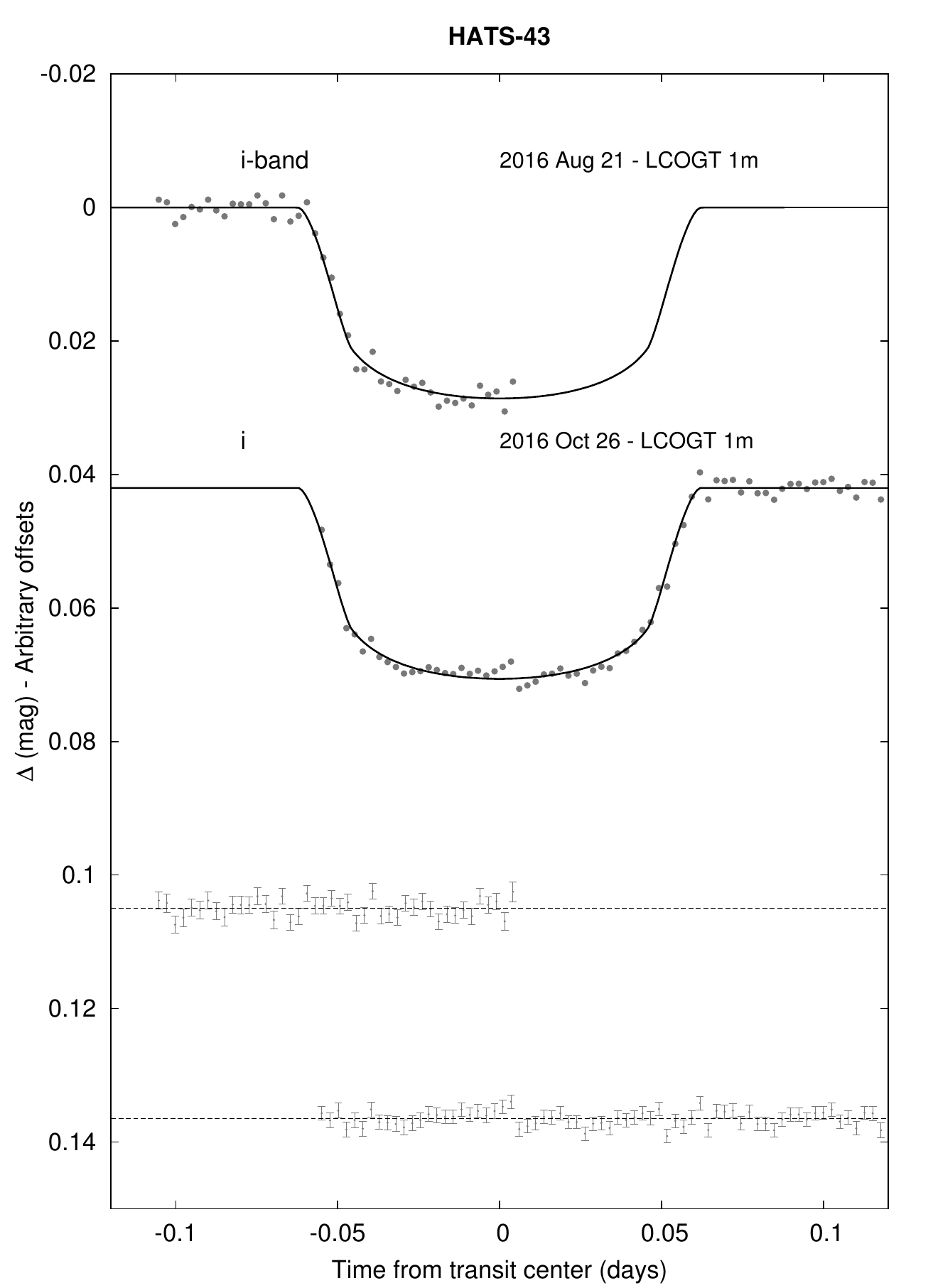}
\caption{
    Unbinned transit \lcs{} for \hatcur{43}.  The light curves have been
    corrected for quadratic trends in time, and linear trends with up
    to three parameters characterizing the shape of the PSF, fitted
    simultaneously with the transit model.
    The dates of the events, filters and instruments used are
    indicated.  Light curves following the first are displaced
    vertically for clarity.  Our best fit from the global modeling
    described in \refsecl{globmod} is shown by the solid lines. The
    residuals from the best-fit model are shown below in the same
    order as the original light curves.  The error bars represent the
    photon and background shot noise, plus the readout noise.
}
\label{fig:lc:43}
\end{figure*}

As shown in Figure~\ref{fig:lc:43}, an ingress and an almost full transit including a complete
egress of \hatcurb{43} were observed using the 1\,m telescope of of the Las Cumbres Observatory Global Telescope (LCOGT) network \citep{brown:2013:lcogt}
located at the Cerro Tololo International Observatory (CTIO).
Both observations were performed during the second semester of 2016, approximately 3
years after the original HATSouth photometry was obtained. Even though both light curves were
obtained with the Sloan $i$ band, the one containing the egress was registered by a SBIG
camera, while the full transit was registered using the Sinistro camera. In both cases the
per-point photometric precision was of $\sim$1.5 mmag with a cadence of $\approx 219$ sec.

\begin{figure*}[!ht]
\plotone{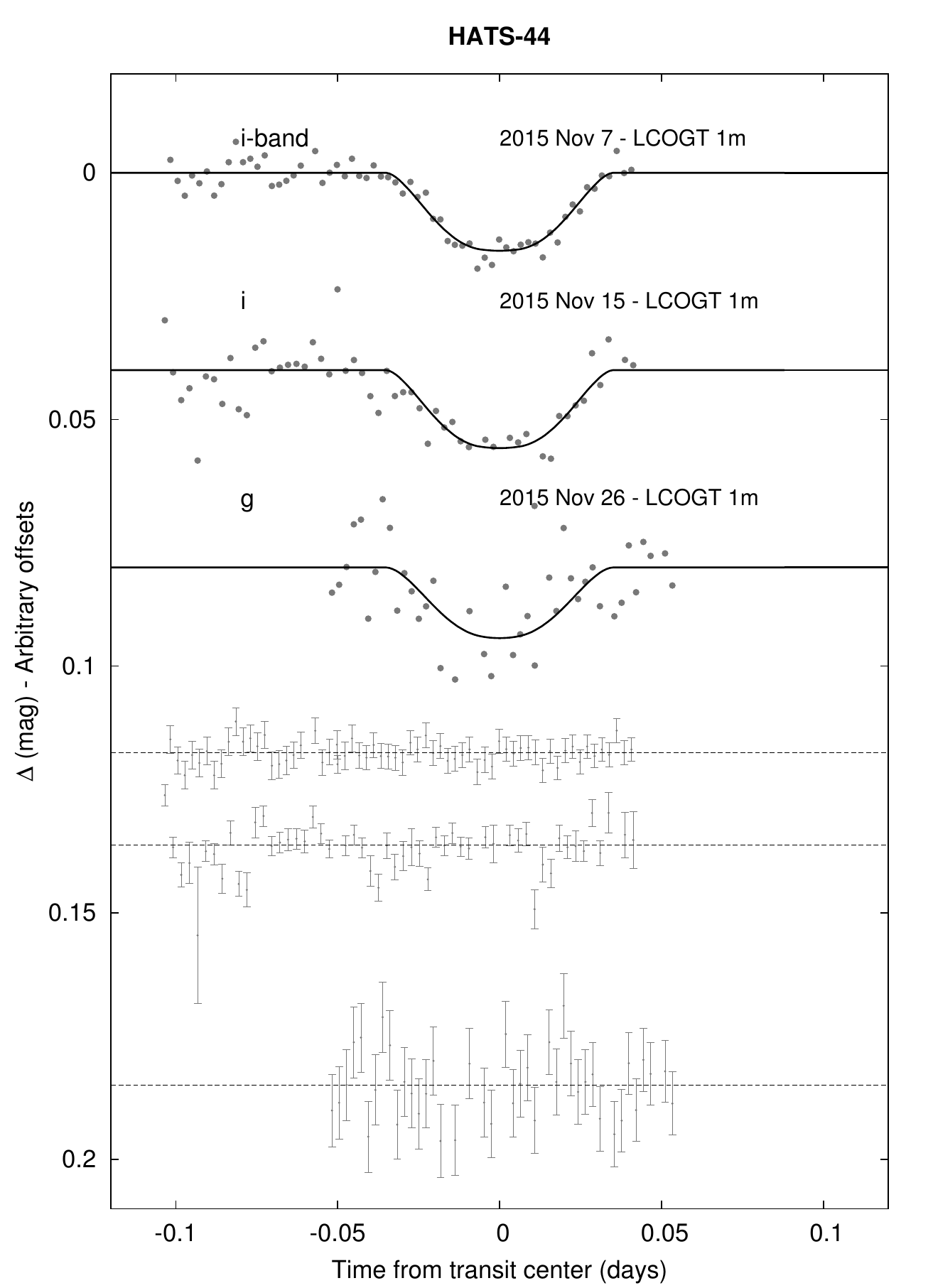}
\caption{
    Similar to Fig.~\ref{fig:lc:43}, here we show \lcs{} for \hatcur{44}. In this case the residuals are plotted on the right-hand-side of the figure, in the same order as the original light curves on the left-hand-side.
}
\label{fig:lc:44}
\end{figure*}

Three full transits of \hatcurb{44} were observed in November 2015 using the LCOGT 1m network
(see Figure \ref{fig:lc:44}).
The first one was obtained from the South African Astronomical Observatory (SAAO) using the Sloan $i$ filter and a SBIG camera, achieving
a photometric precision of $\approx 2$ mmag with a cadence of $\approx 200$ sec. The second transit was observed from CTIO with the same
filter but using a Sinistro camera. The observing conditions were not optimal which resulted in
the photometric precision being only $\sim$5 mmag with a cadence of $\approx 200$ sec. The last transit was also observed
from CTIO but this time the Sloan $g$ band was used with the goal of checking that there was no
colour dependence of the transit depth such as would be produced by a blended eclipsing binary system,
given the slightly triangular shape of this transit. Even
though the precision obtained for this transit was relatively low ($\sim$7 mmag with a cadence of $\approx 200$ sec), it was enough
to confirm that there was no significant variation in the transit depth between the $r$ and $g$ filters.

\begin{figure*}[!ht]
\plotone{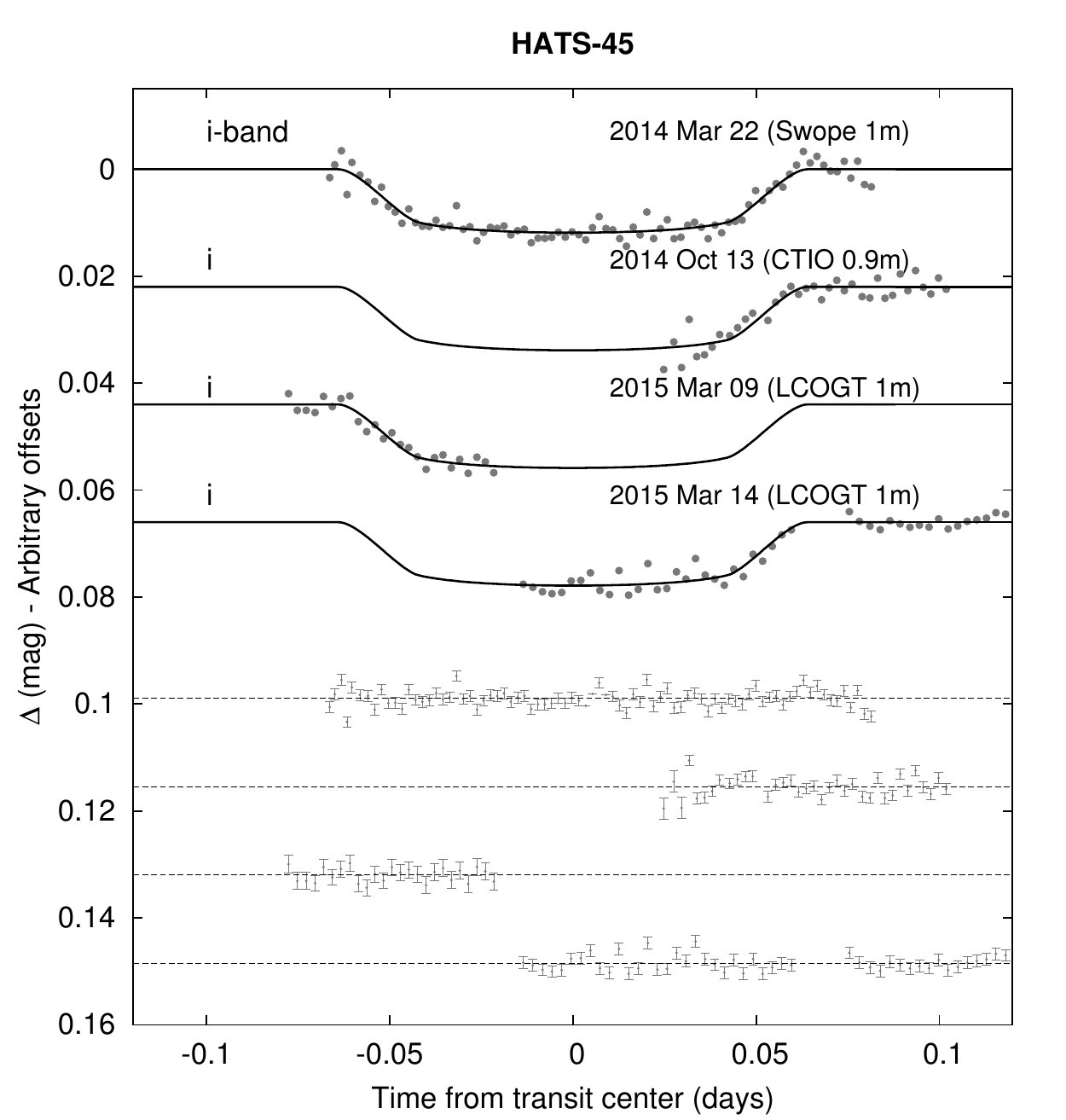}
\caption{
    Same as Fig.~\ref{fig:lc:43}, here we show \lcs{} for \hatcur{45}.
}
\label{fig:lc:45}
\end{figure*}

For \hatcurb{45} we obtained four $i$-band follow-up light curves, which are shown in Figure~\ref{fig:lc:45}.
The first light curve was obtained in March 2014 and registered a full transit using the 1m Swope telescope
and e2v CCD camera. The second light curve was obtained on October
2014 using the 0.9m Telescope of CTIO and registered only an egress. The last two light curves were obtained
on March 2015 with the LCOGT 1m network, with an ingress  observed from SAAO and an egress
from CTIO. The photometric precision for these four light curves was in the 1 -- 2 mmag range with cadences $\approx 200$ sec.

\begin{figure*}[!ht]
\plotone{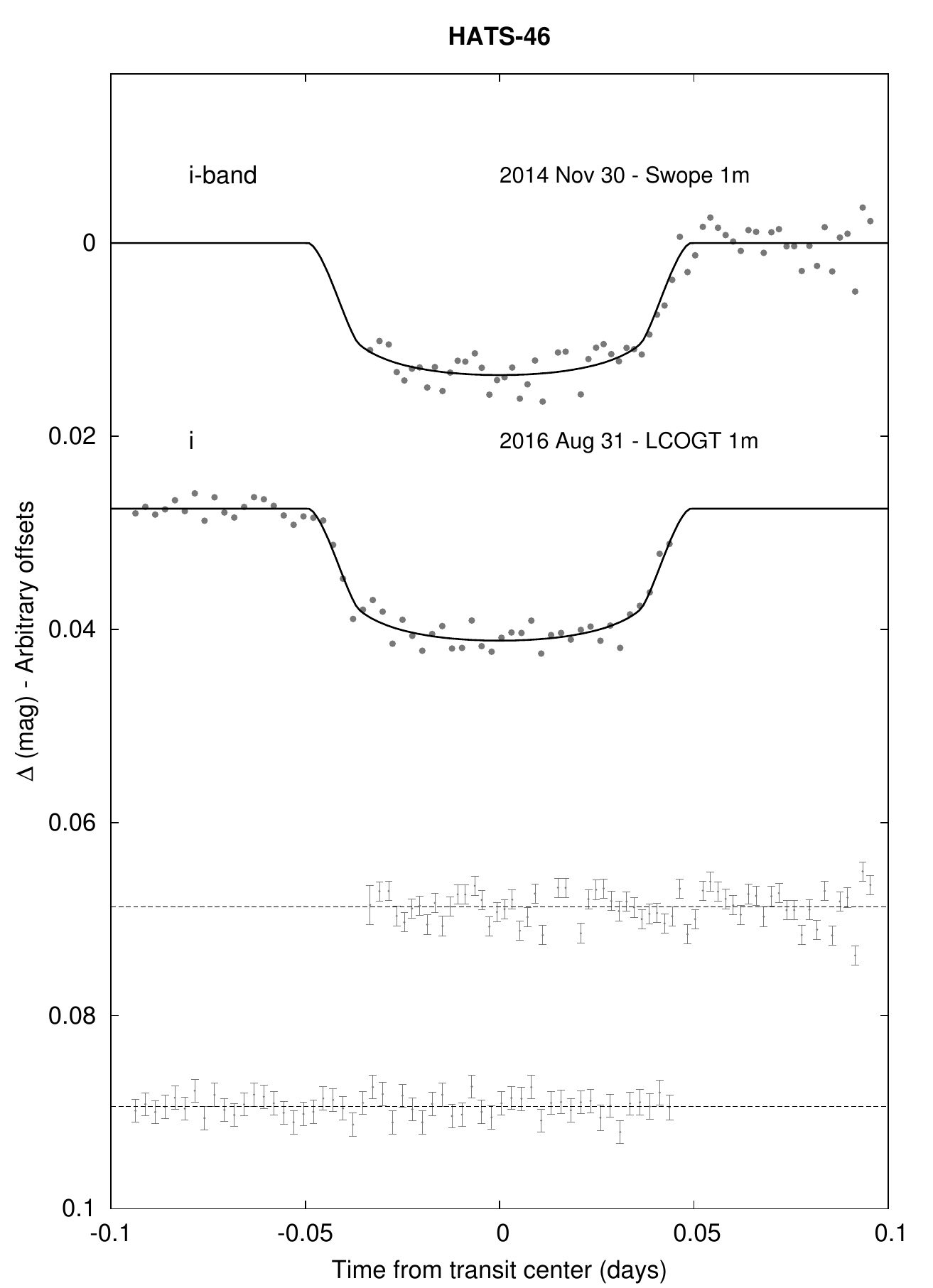}
\caption{
    Same as Fig.~\ref{fig:lc:43}, here we show \lcs{} for \hatcur{46}.
}
\label{fig:lc:46}
\end{figure*}

Finally, two $i$-band transits were observed for \hatcurb{46}, which are shown in Figure \ref{fig:lc:46}.
In November 2014 a partial transit containing an egress was observed with the Swope 1m telescope,
while in August 2016 we registered a full transit with the LCOGT 1m telescope installed at CTIO using
a Sinistro camera. Both light curves achieved a photometric precision below 2 mmag at $\approx 200$ sec cadence.

The instrument specifications, observing strategies and reduction procedures that we apply in the case
of the three instruments that were used to obtain photometry for our four planets have been previously
discussed in \citet{bayliss:2015}, \citet{penev:2013:hats1}, and \citet{hartman:2015:hats6}, for
LCOGT, Swope 1m, and CTIO 0.9m, respectively.

%
\setcounter{planetcounter}{1}
%

\clearpage

%
%
\ifthenelse{\boolean{emulateapj}}{
    \begin{deluxetable*}{llrrrrl}
}{
    \begin{deluxetable}{llrrrrl}
}
\tablewidth{0pc}
\tablecaption{
    Light curve data for \hatcur{43}, \hatcur{44}, \hatcur{45} and \hatcur{46}\label{tab:phfu}.
}
\tablehead{
    \colhead{Object\tablenotemark{a}} &
    \colhead{BJD\tablenotemark{b}} & 
    \colhead{Mag\tablenotemark{c}} & 
    \colhead{\ensuremath{\sigma_{\rm Mag}}} &
    \colhead{Mag(orig)\tablenotemark{d}} & 
    \colhead{Filter} &
    \colhead{Instrument} \\
    \colhead{} &
    \colhead{\hbox{~~~~(2,400,000$+$)~~~~}} & 
    \colhead{} & 
    \colhead{} &
    \colhead{} & 
    \colhead{} &
    \colhead{}
}
\startdata

   HATS-43 & $ 56659.57177 $ & $  -0.02617 $ & $   0.00629 $ & $ \cdots $ & $ r$ &         HS\\
   HATS-43 & $ 56602.51755 $ & $  -0.00594 $ & $   0.00565 $ & $ \cdots $ & $ r$ &         HS\\
   HATS-43 & $ 56663.96208 $ & $  -0.01437 $ & $   0.00523 $ & $ \cdots $ & $ r$ &         HS\\
   HATS-43 & $ 56620.07431 $ & $   0.01737 $ & $   0.00516 $ & $ \cdots $ & $ r$ &         HS\\
   HATS-43 & $ 56716.63096 $ & $   0.00000 $ & $   0.00584 $ & $ \cdots $ & $ r$ &         HS\\
   HATS-43 & $ 56602.52112 $ & $  -0.00586 $ & $   0.00563 $ & $ \cdots $ & $ r$ &         HS\\
   HATS-43 & $ 56659.57617 $ & $  -0.00933 $ & $   0.00646 $ & $ \cdots $ & $ r$ &         HS\\
   HATS-43 & $ 56620.07762 $ & $  -0.00014 $ & $   0.00519 $ & $ \cdots $ & $ r$ &         HS\\
   HATS-43 & $ 56663.96632 $ & $  -0.00456 $ & $   0.00531 $ & $ \cdots $ & $ r$ &         HS\\
   HATS-43 & $ 56580.57936 $ & $  -0.03117 $ & $   0.00689 $ & $ \cdots $ & $ r$ &         HS\\

\enddata
\tablenotetext{a}{
    Either \hatcur{43}, \hatcur{44}, \hatcur{45} or \hatcur{46}.
}
\tablenotetext{b}{
    Barycentric Julian Date is computed directly from the UTC time
    without correction for leap seconds.
}
\tablenotetext{c}{
    The out-of-transit level has been subtracted. For observations
    made with the HATSouth instruments (identified by ``HS'' in the
    ``Instrument'' column) these magnitudes have been corrected for
    trends using the EPD and TFA procedures applied {\em prior} to
    fitting the transit model. This procedure may lead to an
    artificial dilution in the transit depths. The blend factors for
    the HATSouth light curves are listed in
    Table~\ref{tab:planetparam}. For
    observations made with follow-up instruments (anything other than
    ``HS'' in the ``Instrument'' column), the magnitudes have been
    corrected for a quadratic trend in time, and for variations
    correlated with up to three PSF shape parameters, fit simultaneously
    with the transit.
}
\tablenotetext{d}{
    Raw magnitude values without correction for the quadratic trend in
    time, or for trends correlated with the seeing. These are only
    reported for the follow-up observations.
}
\tablecomments{
    This table is available in a machine-readable form in the online
    journal.  A portion is shown here for guidance regarding its form
    and content.
}
\ifthenelse{\boolean{emulateapj}}{
    \end{deluxetable*}
}{
    \end{deluxetable}
}

\subsection{High Spatial Resolution Imaging}
\label{sec:highresimaging}

As part of our follow-up campaign we also obtain high resolution “lucky” imaging in order to identify close stellar companions to our candidates that could be affecting the depth of the transits. In this context HATS-45 was observed with the Astralux Sur camera \citep{hippler:2009} mounted on the New Technology Telescope (NTT) at La Silla Observatory, in Chile on December 22, 2015 in the sloan $z'$ band. Instrument specifications, observing strategy, and reductions of Astralux data are described in \citet{espinoza:2016}. The only change in this work is that we instead use the plate scale derived in \citet{Janson:2017} of 15.2 mas/pixel, which a better estimate that the one we estimated in our previous work. No evident companion can be identified in the neighborhood of HATS-45 at the achieved resolution limit of FWHM$_{eff} = 40 \pm 4.6$ mas, which is within the expected telescope diffraction limit of \citep[$\sim 50$ mas][]{hippler:2009}. Figure 8 presents the contrast curve generated form the HATS-45 Astralux observations.

\ifthenelse{\boolean{emulateapj}}{
    \begin{figure}[!ht]
}{
    \begin{figure}[!ht]
}
\plotone{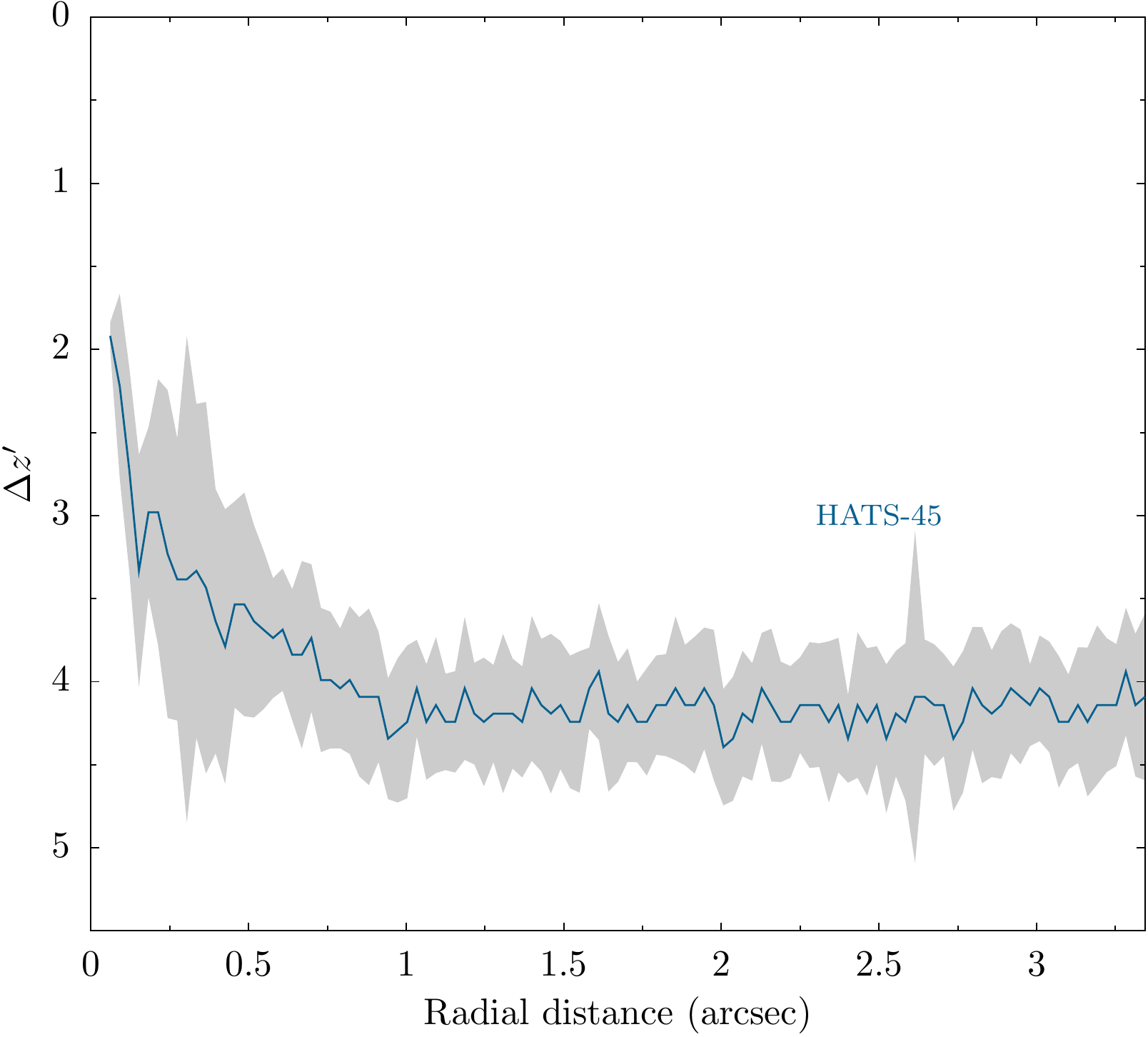}
\caption{
   Contrast curve for \hatcur{45} constructed from the $z'$ band Astralux images. Gray bands show the uncertainty
given by the scatter in the contrast in the azimuthal direction at a given radius.
}

\label{fig:astralux}

\ifthenelse{\boolean{emulateapj}}{
    \end{figure}
}{
    \end{figure}
}

\section{Analysis}
\label{sec:analysis}

\subsection{Properties of the parent star}
\label{sec:stelparam}

An initial estimation of the atmospheric parameters (\teff, \logg, \feh, and \vsini) for the four host stars was
computed using the Zonal Atmospheric Parameters Estimator \citep[ZASPE,][]{brahm:2017:zaspe} applied
to the FEROS follow-up spectra presented in Section~\ref{sec:obsspec}.  Due to the moderately low signal-to-noise ratio (SNR)
of each individual spectrum, they were shifted to a common rest frame and co-added to construct a spectral
template with SNR $\approx$ 50 for each star. ZASPE determines the atmospheric parameters by comparing
the observed spectra with a grid of synthetic models in the spectral regions most sensitive to changes in the
parameters. Additionally, reliable errors are obtained by performing Monte Carlo simulations where the
synthetic models are randomly modified in the sensitive regions by values obtained from the systematic
mismatch between the observed spectra and the best fit model.

To determine the physical and evolutionary parameters of the host star (M$_{\star}$, R$_{\star}$, age)
we use the Yonsei-Yale \citep[Y2;][]{yi:2001} stellar isochrones to search for the mass and age of the
model that produces the temperature and luminosity indicators closest to the observed ones.
While the spectroscopic \teff\ determined with ZASPE can be used as a direct temperature indicator,
the uncertainty in the spectroscopic \logg\ is usually too large to use this parameter as a reliable luminosity tracer.
As is a common procedure now, the stellar luminosity indicator is obtained from the transiting light-curve,
via the parameter $a/R_{\star}$ which as described in Section \ref{sec:globmod} is directly related to the stellar density \citep{sozzetti:2007}.
However, given that the modelling of the transiting light curve partially depends of the stellar parameters
by the selection of the \citet{claret:2004} limb darkening coefficients, we follow an iterative procedure
containing the following steps: i) determination of the ZASPE parameters, ii) global modelling (Section \ref{sec:globmod}),
and iii) isochronal fitting.
For the four transiting systems presented in this study, only two iterations were required.

Table~\ref{tab:stellar} presents the final atmospheric and physical parameters adopted for the four host stars,
while  Figure~\ref{fig:iso} displays their evolutionary states in the \teff -- \rhostar\ space, along with a set of different YY isochrones
for the specific spectroscopically derived metallicities. All four stars are currently on the main sequence. \hatcur{43} and \hatcur{44} have relatively low masses of 
M$_{\star}$ $\approx$ 0.85 M$_{\odot}$, characteristic of early K-dwarf stars. On the other
hand, \hatcur{45}, as expected from its higher spectroscopic derived \teff = \hatcurSMEteff{45}\,K,
is a relatively massive star with an isochronal derived mass of \mstar = \hatcurISOmlong{45} \msun.
Finally, the derived properties of \hatcur{46} are similar, but slightly smaller than the ones of the
sun (\mstar = \hatcurISOmlong{46} \msun, \rstar = \hatcurISOrlong{46} \rsun).
Only \hatcur{44}, with \feh =   \hatcurSMEzfeh{44},  presents a significant deviation from solar metallicity.

\ifthenelse{\boolean{emulateapj}}{
    \begin{figure*}[!ht]
}{
    \begin{figure}[!ht]
}
\plottwo{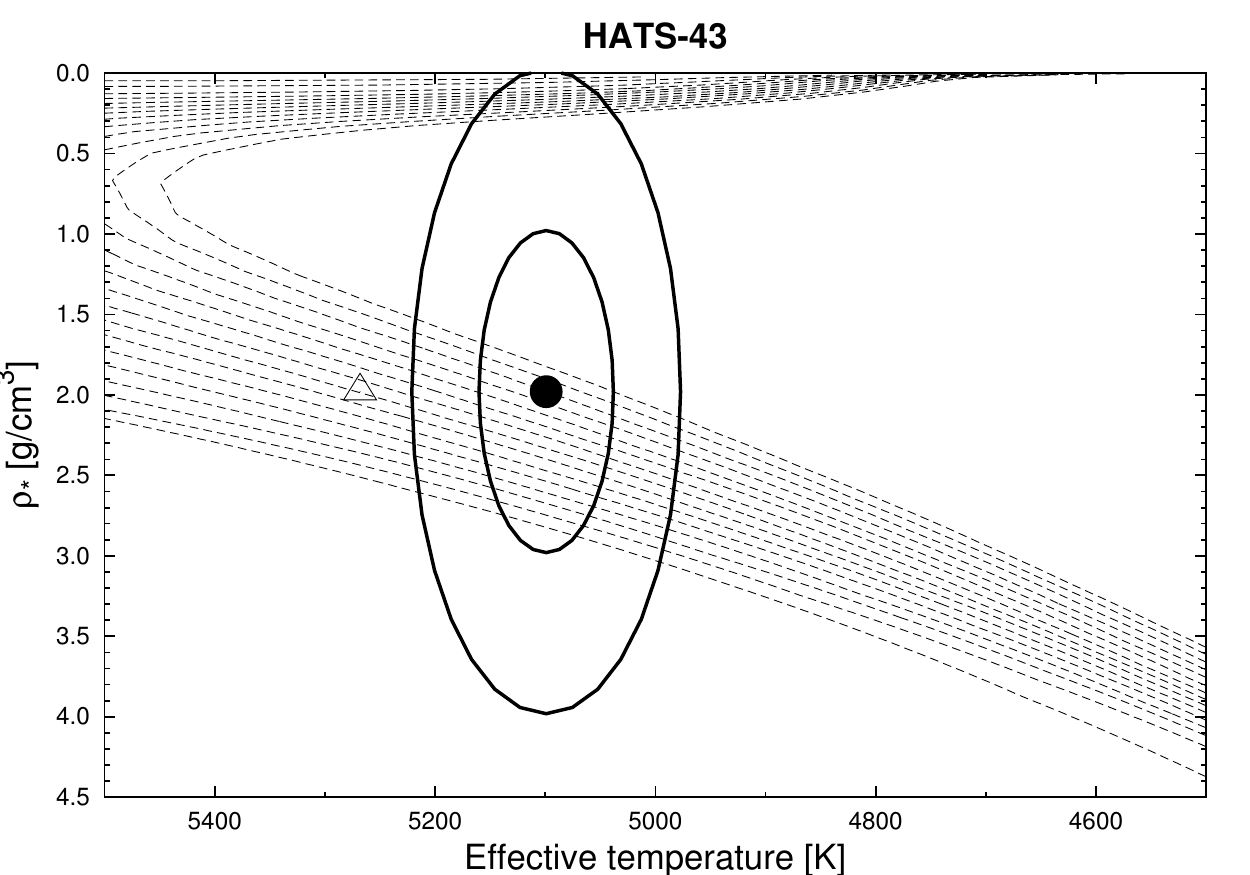}{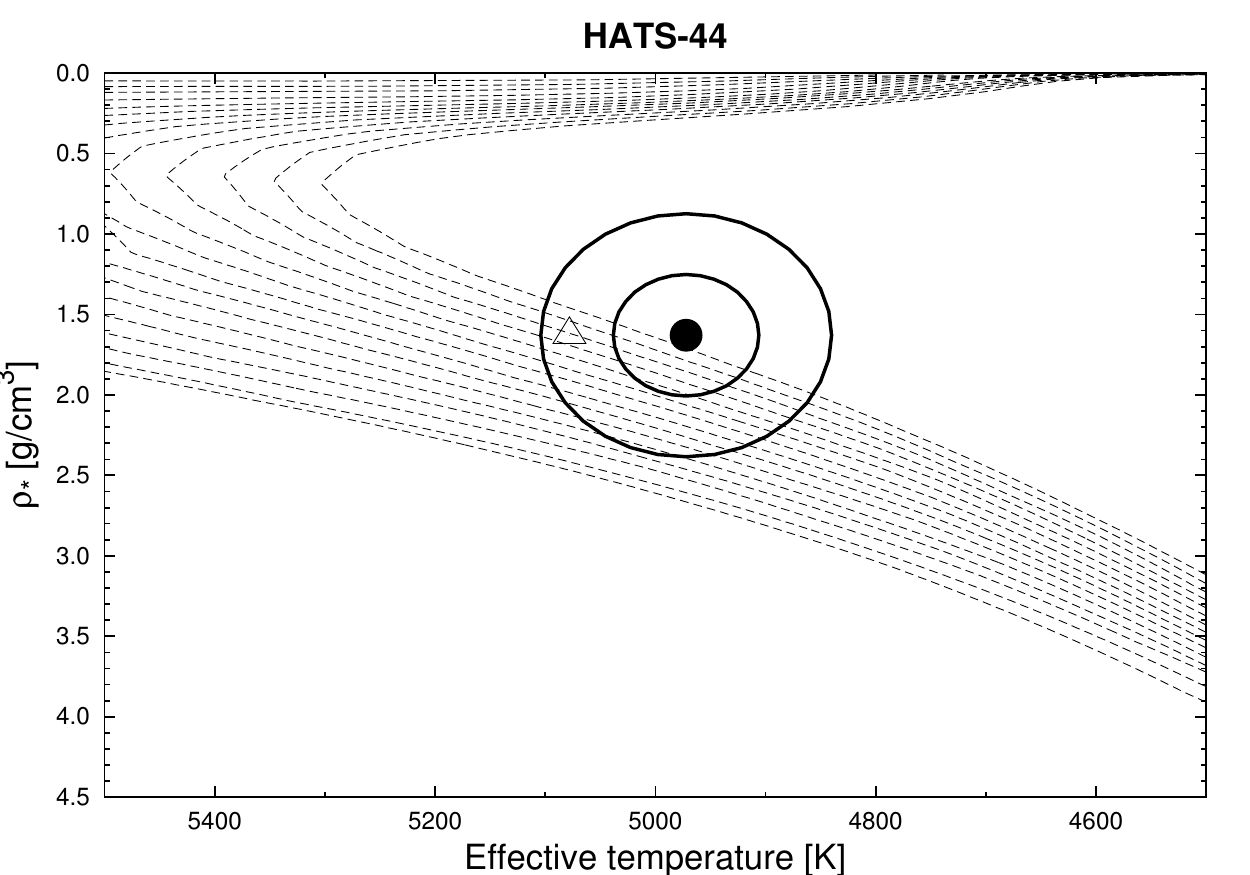}
\plottwo{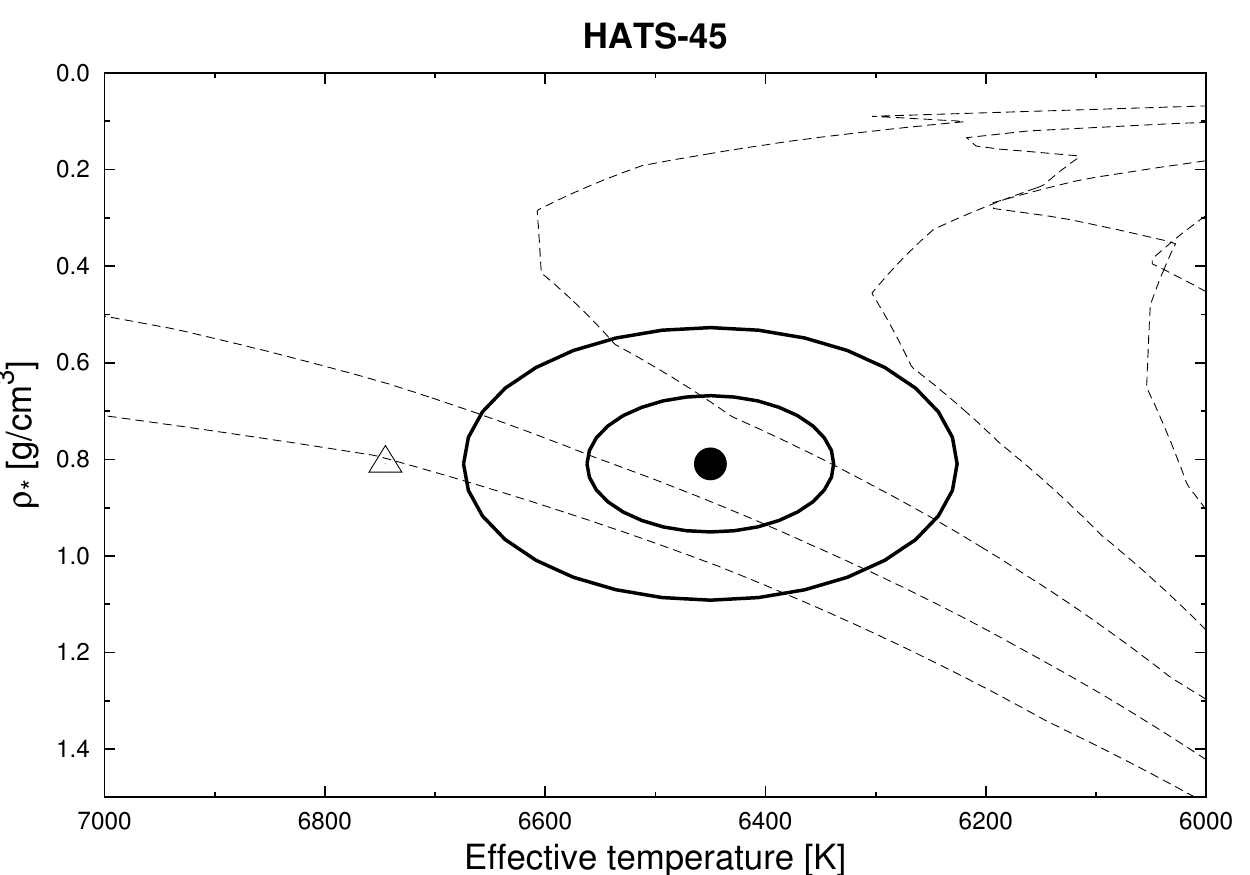}{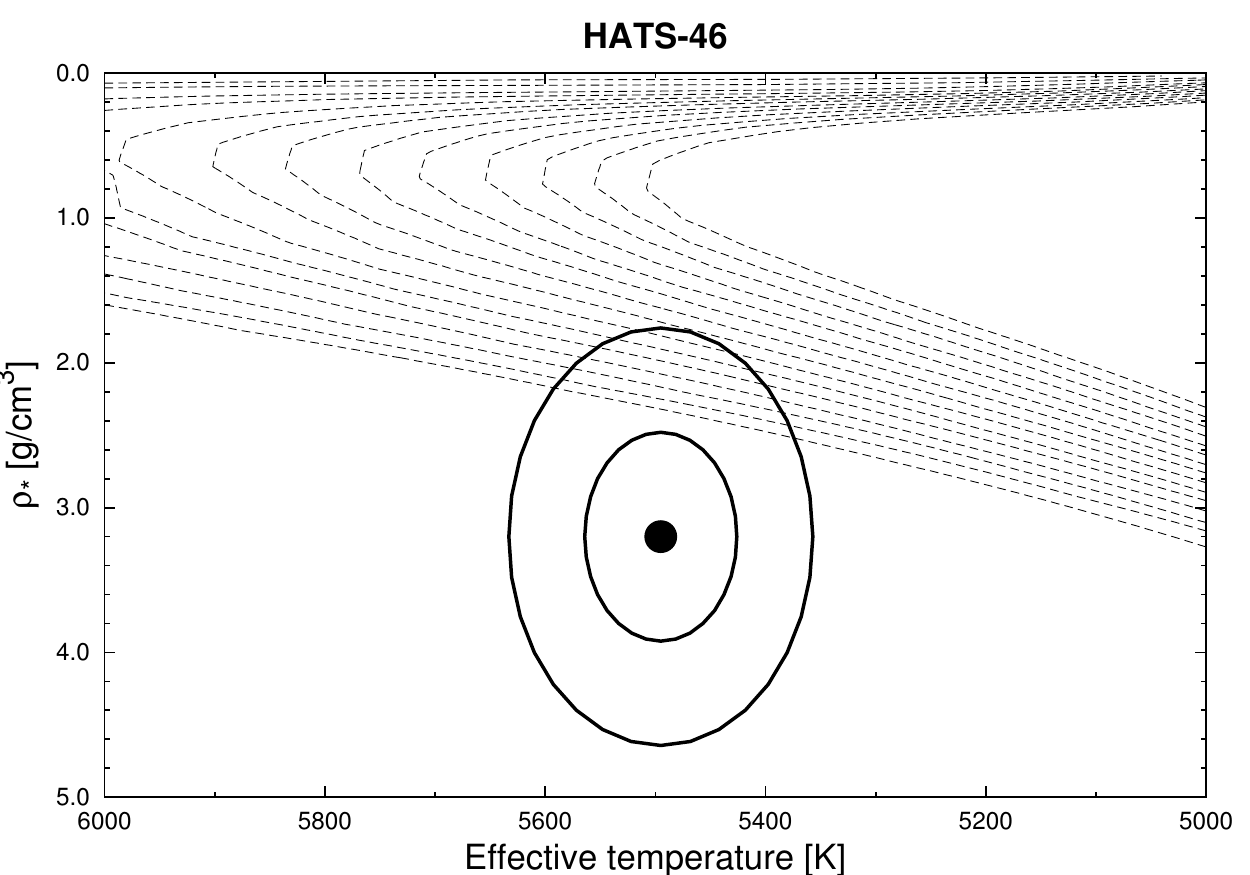}
\caption{
    Model isochrones from \cite{\hatcurisocite{43}} for the measured
    metallicities of \hatcur{43} (upper left), \hatcur{44} (upper right), \hatcur{45} (lower left) and \hatcur{46} (lower right). We show models for ages of 0.2\,Gyr and 1.0 to 14.0\,Gyr in 1.0\,Gyr increments (ages increasing from left to right). The
    adopted values of $\teffstar$ and \rhostar\ are shown together with
    their 1$\sigma$ and 2$\sigma$ confidence ellipsoids.  The initial
    values of \teffstar\ and \rhostar\ from the first ZASPE and \lc\
    analyses are represented with a triangle.
}
\label{fig:iso}
\ifthenelse{\boolean{emulateapj}}{
    \end{figure*}
}{
    \end{figure}
}

%
%
\ifthenelse{\boolean{emulateapj}}{
    \begin{deluxetable*}{lccccl}
}{
    \begin{deluxetable}{lccccl}
}
\tablewidth{0pc}
\tabletypesize{\footnotesize}
\tablecaption{
    Stellar parameters for \hatcur{43}--\hatcur{46}
    \label{tab:stellar}
}
\tablehead{
    \multicolumn{1}{c}{} &
    \multicolumn{1}{c}{\bf HATS-43} &
    \multicolumn{1}{c}{\bf HATS-44} &
    \multicolumn{1}{c}{\bf HATS-45} &
    \multicolumn{1}{c}{\bf HATS-46} &
    \multicolumn{1}{c}{} \\
    \multicolumn{1}{c}{~~~~~~~~Parameter~~~~~~~~} &
    \multicolumn{1}{c}{Value}                     &
    \multicolumn{1}{c}{Value}                     &
    \multicolumn{1}{c}{Value}                     &
    \multicolumn{1}{c}{Value}                     &
    \multicolumn{1}{c}{Source}
}
\startdata
\noalign{\vskip -3pt}
\sidehead{Astrometric properties and cross-identifications}
~~~~2MASS-ID\dotfill               & \hatcurCCtwomassshort{43}  & \hatcurCCtwomassshort{44} & \hatcurCCtwomassshort{45} & \hatcurCCtwomassshort{46} & \\
~~~~GSC-ID\dotfill                 & \hatcurCCgsceccen{43}      & \hatcurCCgsc{44}     & \hatcurCCgsc{45}     & \hatcurCCgsc{46}     & \\
~~~~R.A. (J2000)\dotfill            & \hatcurCCraeccen{43}       & \hatcurCCra{44}    & \hatcurCCra{45}    & \hatcurCCra{46}    & 2MASS\\
~~~~Dec. (J2000)\dotfill            & \hatcurCCdececcen{43}      & \hatcurCCdec{44}   & \hatcurCCdec{45}   & \hatcurCCdec{46}   & 2MASS\\
~~~~$\mu_{\rm R.A.}$ (\masy)              & \hatcurCCpmraeccen{43}     & \hatcurCCpmra{44} & \hatcurCCpmra{45} & \hatcurCCpmra{46} & UCAC4\\
~~~~$\mu_{\rm Dec.}$ (\masy)              & \hatcurCCpmdececcen{43}    & \hatcurCCpmdec{44} & \hatcurCCpmdec{45} & \hatcurCCpmdec{46} & UCAC4\\
\sidehead{Spectroscopic properties}
~~~~$\teffstar$ (K)\dotfill         &  \hatcurSMEteff{43}   & \hatcurSMEteff{44} & \hatcurSMEteff{45} & \hatcurSMEteff{46} & ZASPE\tablenotemark{a}\\
~~~~$\feh$\dotfill                  &  \hatcurSMEzfeh{43}   & \hatcurSMEzfeh{44} & \hatcurSMEzfeh{45} & \hatcurSMEzfeh{46} & ZASPE               \\
~~~~$\vsini$ (\kms)\dotfill         &  \hatcurSMEvsin{43}   & \hatcurSMEvsin{44} & \hatcurSMEvsin{45} & \hatcurSMEvsin{46} & ZASPE                \\
~~~~$\vmac$ (\kms)\dotfill          &  $\hatcurSMEvmac{43}$   & $\hatcurSMEvmac{44}$ & $\hatcurSMEvmac{45}$ & $\hatcurSMEvmac{46}$ & Assumed              \\
~~~~$\vmic$ (\kms)\dotfill          &  $\hatcurSMEvmic{43}$   & $\hatcurSMEvmic{44}$ & $\hatcurSMEvmic{45}$ & $\hatcurSMEvmic{46}$ & Assumed              \\
~~~~$\gamma_{\rm RV}$ (\ms)\dotfill&  \hatcurRVgammaabs{43}  & \hatcurRVgammaabs{44} & \hatcurRVgammaabs{45} & \hatcurRVgammaabs{46} & FEROS\tablenotemark{b}  \\
\sidehead{Photometric properties}
~~~~$B$ (mag)\dotfill               &  \hatcurCCtassmBeccen{43}  & \hatcurCCtassmB{44} & \hatcurCCtassmB{45} & \hatcurCCtassmB{46} & APASS\tablenotemark{c} \\
~~~~$V$ (mag)\dotfill               &  \hatcurCCtassmveccen{43}  & \hatcurCCtassmv{44} & \hatcurCCtassmv{45} & \hatcurCCtassmv{46} & APASS\tablenotemark{c} \\
~~~~$g$ (mag)\dotfill               &  \hatcurCCtassmgeccen{43}  & \hatcurCCtassmg{44} & \hatcurCCtassmg{45} & \hatcurCCtassmg{46} & APASS\tablenotemark{c} \\
~~~~$r$ (mag)\dotfill               &  \hatcurCCtassmreccen{43}  & \hatcurCCtassmr{44} & \hatcurCCtassmr{45} & \hatcurCCtassmr{46} & APASS\tablenotemark{c} \\
~~~~$i$ (mag)\dotfill               &  \hatcurCCtassmieccen{43}  & \hatcurCCtassmi{44} & \hatcurCCtassmi{45} & \hatcurCCtassmi{46} & APASS\tablenotemark{c} \\
~~~~$J$ (mag)\dotfill               &  \hatcurCCtwomassJmageccen{43} & \hatcurCCtwomassJmag{44} & \hatcurCCtwomassJmag{45} & \hatcurCCtwomassJmag{46} & 2MASS           \\
~~~~$H$ (mag)\dotfill               &  \hatcurCCtwomassHmageccen{43} & \hatcurCCtwomassHmag{44} & \hatcurCCtwomassHmag{45} & \hatcurCCtwomassHmag{46} & 2MASS           \\
~~~~$K_s$ (mag)\dotfill             &  \hatcurCCtwomassKmageccen{43} & \hatcurCCtwomassKmag{44} & \hatcurCCtwomassKmag{45} & \hatcurCCtwomassKmag{46} & 2MASS           \\
\sidehead{Derived properties}
~~~~$\mstar$ ($\msun$)\dotfill      &  \hatcurISOmlongeccen{43}   & \hatcurISOmlong{44} & \hatcurISOmlong{45} & \hatcurISOmlong{46} & YY+$\rhostar$+ZASPE \tablenotemark{d}\\
~~~~$\rstar$ ($\rsun$)\dotfill      &  \hatcurISOrlongeccen{43}   & \hatcurISOrlong{44} & \hatcurISOrlong{45} & \hatcurISOrlong{46} & YY+$\rhostar$+ZASPE         \\
~~~~$\loggstar$ (cgs)\dotfill       &  \hatcurISOloggeccen{43}    & \hatcurISOlogg{44} & \hatcurISOlogg{45} & \hatcurISOlogg{46} & YY+$\rhostar$+ZASPE         \\
~~~~$\rhostar$ (\gcmc) \tablenotemark{e}\dotfill       &  \hatcurLCrhoeccen{43}    & \hatcurLCrho{44} & \hatcurLCrho{45} & \hatcurLCrho{46} & Light curves         \\
~~~~$\rhostar$ (\gcmc) \tablenotemark{e}\dotfill       &  \hatcurISOrhoeccen{43}    & \hatcurISOrho{44} & \hatcurISOrho{45} & \hatcurISOrho{46} & YY+Light curves+ZASPE         \\
~~~~$\lstar$ ($\lsun$)\dotfill      &  \hatcurISOlumeccen{43}     & \hatcurISOlum{44} & \hatcurISOlum{45} & \hatcurISOlum{46} & YY+$\rhostar$+ZASPE         \\
~~~~$M_V$ (mag)\dotfill             &  \hatcurISOmveccen{43}      & \hatcurISOmv{44} & \hatcurISOmv{45} & \hatcurISOmv{46} & YY+$\rhostar$+ZASPE         \\
~~~~$M_K$ (mag,\hatcurjhkfilset{43})\dotfill &  \hatcurISOMKeccen{43} & \hatcurISOMK{44} & \hatcurISOMK{45} & \hatcurISOMK{46} & YY+$\rhostar$+ZASPE         \\
~~~~Age (Gyr)\dotfill               &  \hatcurISOageeccen{43}     & \hatcurISOage{44} & \hatcurISOage{45} & \hatcurISOage{46} & YY+$\rhostar$+ZASPE         \\
~~~~$A_{V}$ (mag)\dotfill               &  \hatcurXAveccen{43}     & \hatcurXAv{44} & \hatcurXAv{45} & \hatcurXAv{46} & YY+$\rhostar$+ZASPE         \\
~~~~Distance (pc)\dotfill           &  \hatcurXdistredeccen{43}\phn  & \hatcurXdistred{44} & \hatcurXdistred{45} & \hatcurXdistred{46} & YY+$\rhostar$+ZASPE\\ [-1.5ex]
\enddata
\tablecomments{
For \hatcur{43} we adopt a model in which the eccentricity is allowed to vary. For the other three systems we adopt a model in which the orbit is assumed to be circular. See the discussion in Section~\ref{sec:globmod}.
}
\tablenotetext{a}{
    ZASPE = Zonal Atmospherical Stellar Parameter Estimator routine
    for the analysis of high-resolution spectra
    \citep{brahm:2017:zaspe}, applied to the FEROS spectra of each
    system. These parameters rely primarily on ZASPE, but have a small
    dependence also on the iterative analysis incorporating the
    isochrone search and global modeling of the data.
}
\tablenotetext{b}{
    The error on $\gamma_{\rm RV}$ is determined from the orbital fit
    to the RV measurements, and does not include the systematic
    uncertainty in transforming the velocities to the IAU standard
    system. The velocities have not been corrected for gravitational
    redshifts.
} \tablenotetext{c}{
    From APASS DR6 for as
    listed in the UCAC 4 catalog \citep{zacharias:2012:ucac4}.  
}
\tablenotetext{d}{
    \hatcurisoshort{43}+\rhostar+ZASPE = Based on the \hatcurisoshort{43}
    isochrones \citep{\hatcurisocite{43}}, \rhostar\ as a luminosity
    indicator, and the ZASPE results.
}
\tablenotetext{e}{
    In the case of $\rhostar$ we list two values. The first value is
    determined from the global fit to the light curves and RV data,
    without imposing a constraint that the parameters match the
    stellar evolution models. The second value results from
    restricting the posterior distribution to combinations of
    $\rhostar$+$\teffstar$+$\feh$ that match to a \hatcurisoshort{43}
    stellar model.
}
\ifthenelse{\boolean{emulateapj}}{
    \end{deluxetable*}
}{
    \end{deluxetable}
}

\subsection{Excluding blend scenarios}
\label{sec:blend}

To exclude blend scenarios we carried out an analysis
following \citet{hartman:2012:hat39hat41}. We attempt to model the
available photometric data (including light curves and catalog
broad-band photometric measurements) for each object as a blend
between an eclipsing binary star system and a third star along the
line of sight. The physical properties of the stars are constrained
using the Padova isochrones \citep{girardi:2000}, while we also
require that the brightest of the three stars in the blend have
atmospheric parameters consistent with those measured with ZASPE. We
also simulate composite cross-correlation functions (CCFs) and use
them to predict radial velocities and bisector spans for each blend scenario considered. The results for each system are as follows:

\begin{itemize}
\item {\em \hatcur{43}} -- all blend scenarios tested provide a poorer fit to the photometric data than a model consisting of a single star with a planet. Based on this all blend models can be rejected with at least $3\sigma$ confidence. Moreover, blend models that come closest to fitting the photometric data have obvious double peaks in their CCFs and would produce many \kms\ BS and RV variations that we do not detect.
\item {\em \hatcur{44}} -- similar to \hatcur{43}, except here we can only reject blend models at $2.3\sigma$ confidence based on the photometry.  The blend models which provide the best fit to the photometry (i.e., those that can be rejected with greater than $2.3\sigma$ confidence based on the photometry, but which cannot be rejected with greater than $5\sigma$ confidence) have simulated RV measurements that do not resemble the observed sinusoidal RV variation. The best fit blend model has $\Delta \chi^2 = 13.4$ compared to the adopted planetary orbit model, when including jitter in the uncertainties, and, based on an F-test, can be rejected with 99.7\% confidence. Combining the RVs and photometry, all blend models can be rejected with greater than $4\sigma$ confidence.
\item {\em \hatcur{45}} -- similar to \hatcur{43}, except here we can only reject blend models at $1.4\sigma$ confidence based on the photometry alone. However, for blend models that cannot be rejected with at least $5\sigma$ confidence based on the photometry, both the simulated BSs and RVs vary by more than 500\,\ms, and in most cases by well over 1\,\kms (compared to the measured scatter of 36\,\ms\, and 61\,\ms\ for the FEROS BS and RV values --including the planetary signal-- of this target, respectively, and compared to the measured scatter of 104\,\ms\ and 73\,\ms\ for the HARPS BS and RV values, respectively). 
\item {\em \hatcur{46}} -- similar to \hatcur{43}, in this case all blend models tested can be rejected with $2.4\sigma$ confidence based solely on the photometry. For blend models that cannot be rejected with at least $4\sigma$ confidence based on the photometry, both the simulated BSs and RVs vary by more than 200\,\ms\ (compared to the measured scatter of 35\,\ms\, and 39\,\ms\ for the FEROS BS and RV values of this target, respectively).
\end{itemize}

\subsection{Global modeling of the data}
\label{sec:globmod}

In order to obtain the orbital and physical parameters of the planets we
simultaneously modelled for each system the HATSouth photometry,
the follow-up photometry, and the high-precision RV measurements
following \citet{pal:2008:hat7,bakos:2010:hat11,hartman:2012:hat39hat41}. 

Photometric light curves are modelled using the \citet{mandel:2002} models.
For HATSouth light curves, we consider a dilution factor for the transit depth
that compensates the blending effect produced by the presence of neighboring stars,
and also the possible over-correction introduced by the trend-filtered algorithm.
In the case of the follow-up light curves, systematic trends for each event are
corrected by including a time-dependent quadratic signal to the transit
model, and a linear signal with up to three parameters describing the shape
of the PSF.

Radial velocity data are modeled using Keplerian orbits, where we consider
independent zero-points and RV jitter factors for each instrument, which are
allowed to vary in the fit.
We fitted the four systems by considering two possible cases, the eccentricity as a free parameter,
and also by forcing circular orbits. For each system we estimated
the Bayesian evidence of each scenario by using the method presented
in \citet{weinberg:2013}.
We find that for \hatcur{43} the free-eccentricity model with $e = \hatcurRVecceneccen{43}$
has a significantly higher evidence compared to the model with fixed eccentricity.

For \hatcur{44} and \hatcur{45} the Bayesian evidence for the
free-eccentricity models are slightly higher than for the fixed
circular orbit models, however in both cases these results
are generated by outlier radial velocity points.
For both of these systems we therefore adopt
the fixed circular orbit solutions, but note that the eccentricities
are poorly constrained by the observations, with 95\% confidence upper
limits of $e\hatcurRVeccentwosiglimeccen{44}$, and
$e\hatcurRVeccentwosiglimeccen{45}$, respectively.
For \hatcur{46} we find that the
fixed circular orbit model has higher Bayesian evidence, and we adopt
the parameters from that model for this system as well. The 95\%
confidence upper limit on the eccentricity for \hatcur{46} is $e
\hatcurRVeccentwosiglimeccen{46}$.

We used a Differential Evolution Markov Chain Monte Carlo
procedure to explore the fitness landscape and to determine the
posterior distribution of the parameters.
The resulting parameters and uncertainties for each system are listed in
\reftabl{planetparam} and summarised below:
\begin{itemize}
\item \hatcurb{43} has a Saturn-like mass of \mpl = $\hatcurPPmlongeccen{43}$ \mjup, and a radius of \rpl = $\hatcurPPrlongeccen{43}$ \rjup,
which results in relatively low density of \rhopl = $\hatcurPPrhoeccen{43}$ \gcmc. Its orbit is moderately eccentric and
due to the low luminosity of its K-type host star \hatcurb{43} has a rather warm equilibrium temperature of 
$T_{\rm eq}$ =  $\hatcurPPteffeccen{43}$ K.

\item \hatcurb{44} has a sub-Jupiter mass of \mpl = $\hatcurPPmlong{44}$ \mjup, and a radius of \rpl = $\hatcurPPrlong{44}$ \rjup,
which results in a density of \rhopl = $\hatcurPPrho{44}$ \gcmc. Even though the luminosities of \hatcur{43} and  \hatcur{44} are
similar, the smaller semi-major axis of \hatcurb{44} results in a higher equilibrium temperature of $T_{\rm eq}$ =   $\hatcurPPteff{44}$ K.

\item \hatcurb{45} has also a sub-Jupiter mass of \mpl = $\hatcurPPmlong{45}$ \mjup, and an inflated radius of \rpl = $\hatcurPPrlong{45}$ \rjup,
which results in a density of \rhopl = $\hatcurPPrho{45}$ \gcmc. \hatcurb{45} suffers from moderately strong irradiation from its F-type host star,
which produces a high equilibrium temperature of $T_{\rm eq}$ =   $\hatcurPPteff{45}$ K.

\item \hatcurb{46} has a mass of  \mpl = $\hatcurPPmlong{46}$ \mjup\ which lies in the Neptune-Saturn mass range. We measured
a radius of \rpl = $\hatcurPPrlong{45}$ \rjup\ for \hatcurb{46} which combined with the mass gives a density of \rhopl = $\hatcurPPrho{46}$ \gcmc.
The low luminosity of the host star produces a relatively low equilibrium temperature of  $T_{\rm eq}$ =   $\hatcurPPteff{46}$ K for \hatcurb{46}.

\end{itemize}

%
\ifthenelse{\boolean{emulateapj}}{
    \begin{deluxetable*}{lcccc}
}{
    \begin{deluxetable}{lcccc}
}
\tabletypesize{\scriptsize}
\tablecaption{Orbital and planetary parameters for \hatcurb{43}--\hatcurb{46}\label{tab:planetparam}}
\tablehead{
    \multicolumn{1}{c}{} &
    \multicolumn{1}{c}{\bf HATS-43b} &
    \multicolumn{1}{c}{\bf HATS-44b} &
    \multicolumn{1}{c}{\bf HATS-45b} &
    \multicolumn{1}{c}{\bf HATS-46b} \\ 
    \multicolumn{1}{c}{~~~~~~~~~~~~~~~Parameter~~~~~~~~~~~~~~~} &
    \multicolumn{1}{c}{Value} &
    \multicolumn{1}{c}{Value} &
    \multicolumn{1}{c}{Value} &
    \multicolumn{1}{c}{Value}
}
\startdata
\noalign{\vskip -3pt}
\sidehead{\Lc{} parameters}
~~~$P$ (days)             \dotfill    & $\hatcurLCPeccen{43}$ & $\hatcurLCP{44}$ & $\hatcurLCP{45}$ & $\hatcurLCP{46}$ \\
~~~$T_c$ (${\rm BJD}$)    
      \tablenotemark{a}   \dotfill    & $\hatcurLCTeccen{43}$ & $\hatcurLCT{44}$ & $\hatcurLCT{45}$ & $\hatcurLCT{46}$ \\
~~~$T_{14}$ (days)
      \tablenotemark{a}   \dotfill    & $\hatcurLCdureccen{43}$ & $\hatcurLCdur{44}$ & $\hatcurLCdur{45}$ & $\hatcurLCdur{46}$ \\
~~~$T_{12} = T_{34}$ (days)
      \tablenotemark{a}   \dotfill    & $\hatcurLCingdureccen{43}$ & $\hatcurLCingdur{44}$ & $\hatcurLCingdur{45}$ & $\hatcurLCingdur{46}$ \\
~~~$\arstar$              \dotfill    & $\hatcurPPareccen{43}$ & $\hatcurPPar{44}$ & $\hatcurPPar{45}$ & $\hatcurPPar{46}$ \\
~~~$\zrstar$ \tablenotemark{b}             \dotfill    & $\hatcurLCzetaeccen{43}$\phn & $\hatcurLCzeta{44}$\phn & $\hatcurLCzeta{45}$\phn & $\hatcurLCzeta{46}$\phn \\
~~~$\rpl/\rstar$          \dotfill    & $\hatcurLCrprstareccen{43}$ & $\hatcurLCrprstar{44}$ & $\hatcurLCrprstar{45}$ & $\hatcurLCrprstar{46}$ \\
~~~$b^2$                  \dotfill    & $\hatcurLCbsqeccen{43}$ & $\hatcurLCbsq{44}$ & $\hatcurLCbsq{45}$ & $\hatcurLCbsq{46}$ \\
~~~$b \equiv a \cos i/\rstar$
                          \dotfill    & $\hatcurLCimpeccen{43}$ & $\hatcurLCimp{44}$ & $\hatcurLCimp{45}$ & $\hatcurLCimp{46}$ \\
~~~$i$ (deg)              \dotfill    & $\hatcurPPieccen{43}$\phn & $\hatcurPPi{44}$\phn & $\hatcurPPi{45}$\phn & $\hatcurPPi{46}$\phn \\

\sidehead{HATSouth blend factors \tablenotemark{d}}
~~~Blend factor \dotfill & $1.0$ & $1.0$ & $\hatcurLCiblend{45}$ & $\hatcurLCiblendA{46}$, $\hatcurLCiblendB{46}$ \\

\sidehead{Limb-darkening coefficients \tablenotemark{e}}
~~~$c_1,g$                  \dotfill    & $\cdots$ & $\hatcurLBig{44}$ & $\cdots$ & $\cdots$ \\
~~~$c_2,g$                  \dotfill    & $\cdots$ & $\hatcurLBiig{44}$ & $\cdots$ & $\cdots$ \\
~~~$c_1,r$                  \dotfill    & $\hatcurLBireccen{43}$ & $\hatcurLBir{44}$ & $\hatcurLBir{45}$ & $\hatcurLBir{46}$ \\
~~~$c_2,r$                  \dotfill    & $\hatcurLBiireccen{43}$ & $\hatcurLBiir{44}$ & $\hatcurLBiir{45}$ & $\hatcurLBiir{46}$ \\
~~~$c_1,i$                  \dotfill    & $\hatcurLBiieccen{43}$ & $\hatcurLBii{44}$ & $\hatcurLBii{45}$ & $\hatcurLBii{46}$ \\
~~~$c_2,i$                  \dotfill    & $\hatcurLBiiieccen{43}$ & $\hatcurLBiii{44}$ & $\hatcurLBiii{45}$ & $\hatcurLBiii{46}$ \\

\sidehead{RV parameters}
~~~$K$ (\ms)              \dotfill    & $\hatcurRVKeccen{43}$\phn\phn & $\hatcurRVK{44}$\phn\phn & $\hatcurRVK{45}$\phn\phn & $\hatcurRVK{46}$\phn\phn \\
%
%
~~~$e$ \tablenotemark{f}               \dotfill    & $\hatcurRVecceneccen{43}$ & $\hatcurRVeccentwosiglimeccen{44}$ & $\hatcurRVeccentwosiglimeccen{45}$ & $\hatcurRVeccentwosiglimeccen{46}$ \\
~~~$\omega$ (deg) \dotfill    & $\hatcurRVomegaeccen{43}$ & $\cdots$ & $\cdots$ \\
~~~$\sqrt{e}\cos\omega$               \dotfill    & $\hatcurRVrkeccen{43}$ & $\cdots$ & $\cdots$ & $\cdots$ \\
~~~$\sqrt{e}\sin\omega$               \dotfill    & $\hatcurRVrheccen{43}$ & $\cdots$ & $\cdots$ & $\cdots$ \\
~~~$e\cos\omega$               \dotfill    & $\hatcurRVkeccen{43}$ & $\cdots$ & $\cdots$ & $\cdots$ \\
~~~$e\sin\omega$               \dotfill    & $\hatcurRVheccen{43}$ & $\cdots$ & $\cdots$ & $\cdots$ \\
~~~RV jitter FEROS (\ms) \tablenotemark{g}       \dotfill    & \hatcurRVjitterAeccen{43} & \hatcurRVjitter{44} & \hatcurRVjitterA{45} & \hatcurRVjitterA{46} \\
~~~RV jitter HARPS (\ms)        \dotfill    & \hatcurRVjitterBeccen{43} & $\cdots$ & $\hatcurRVjitterB{45}$ & $\cdots$ \\
~~~RV jitter PFS (\ms)        \dotfill    & $\cdots$ & $\cdots$ & $\cdots$ & \hatcurRVjitterB{46} \\

\sidehead{Planetary parameters}
~~~$\mpl$ ($\mjup$)       \dotfill    & $\hatcurPPmlongeccen{43}$ & $\hatcurPPmlong{44}$ & $\hatcurPPmlong{45}$ & $\hatcurPPmlong{46}$ \\
~~~$\rpl$ ($\rjup$)       \dotfill    & $\hatcurPPrlongeccen{43}$ & $\hatcurPPrlong{44}$ & $\hatcurPPrlong{45}$ & $\hatcurPPrlong{46}$ \\
~~~$C(\mpl,\rpl)$
    \tablenotemark{h}     \dotfill    & $\hatcurPPmrcorreccen{43}$ & $\hatcurPPmrcorr{44}$ & $\hatcurPPmrcorr{45}$ & $\hatcurPPmrcorr{46}$ \\
~~~$\rhopl$ (\gcmc)       \dotfill    & $\hatcurPPrhoeccen{43}$ & $\hatcurPPrho{44}$ & $\hatcurPPrho{45}$ & $\hatcurPPrho{46}$ \\
~~~$\log g_p$ (cgs)       \dotfill    & $\hatcurPPloggeccen{43}$ & $\hatcurPPlogg{44}$ & $\hatcurPPlogg{45}$ & $\hatcurPPlogg{46}$ \\
~~~$a$ (AU)               \dotfill    & $\hatcurPPareleccen{43}$ & $\hatcurPParel{44}$ & $\hatcurPParel{45}$ & $\hatcurPParel{46}$ \\
~~~$T_{\rm eq}$ (K)        \dotfill   & $\hatcurPPteffeccen{43}$ & $\hatcurPPteff{44}$ & $\hatcurPPteff{45}$ & $\hatcurPPteff{46}$ \\
~~~$\Theta$ \tablenotemark{i} \dotfill & $\hatcurPPthetaeccen{43}$ & $\hatcurPPtheta{44}$ & $\hatcurPPtheta{45}$ & $\hatcurPPtheta{46}$ \\
%
~~~$\log_{10}\langle F \rangle$ (cgs) \tablenotemark{j}
                          \dotfill    & $\hatcurPPfluxavglogeccen{43}$ & $\hatcurPPfluxavglog{44}$ & $\hatcurPPfluxavglog{45}$ & $\hatcurPPfluxavglog{46}$ \\ [-1.5ex]
\enddata
\tablenotetext{a}{
    Times are in Barycentric Julian Date calculated directly from UTC {\em without} correction for leap seconds.
    \ensuremath{T_c}: Reference epoch of
    mid transit that minimizes the correlation with the orbital
    period.
    \ensuremath{T_{12}}: total transit duration, time
    between first to last contact;
    \ensuremath{T_{12}=T_{34}}: ingress/egress time, time between first
    and second, or third and fourth contact.
}
\tablecomments{
For \hatcur{43} we adopt a model in which the eccentricity is allowed to vary. For the other three systems we adopt a model in which the orbit is assumed to be circular. See the discussion in Section~\ref{sec:globmod}.
}
\tablenotetext{b}{
   Reciprocal of the half duration of the transit used as a jump parameter in our MCMC analysis in place of $\arstar$. It is related to $\arstar$ by the expression $\zrstar = \arstar(2\pi(1+e\sin\omega))/(P\sqrt{1-b^2}\sqrt{1-e^2})$ \citep{bakos:2010:hat11}.
}
\tablenotetext{d}{
    Scaling factor applied to the model transit that is fit to the HATSouth light curves. This factor accounts for dilution of the transit due to blending from neighboring stars and over-filtering of the light curve.  These factors are varied in the fit. For \hatcur{43} and \hatcur{44} we fix these values to one because the analysis is performed on light curves after applying signal-reconstruction TFA to correct for over-filtering. For \hatcur{46} we list separately the dilution factors adopted for the G755.2 and G754.3 HATSouth light curves.
}
\tablenotetext{e}{
    Values for a quadratic law, adopted from the tabulations by
    \cite{claret:2004} according to the spectroscopic (ZASPE) parameters
    listed in \reftabl{stellar}.
}
\tablenotetext{f}{
    For fixed circular orbit models we list
    the 95\% confidence upper limit on the eccentricity determined
    when $\sqrt{e}\cos\omega$ and $\sqrt{e}\sin\omega$ are allowed to
    vary in the fit.
}
\tablenotetext{g}{
    Term added in quadrature to the formal RV uncertainties for each
    instrument. This is treated as a free parameter in the fitting
    routine. In cases where the jitter is consistent with zero, we
    list its 95\% confidence upper limit.
}
\tablenotetext{h}{
    Correlation coefficient between the planetary mass \mpl\ and radius
    \rpl\ estimated from the posterior parameter distribution.
}
\tablenotetext{i}{
    The Safronov number is given by $\Theta = \frac{1}{2}(V_{\rm
    esc}/V_{\rm orb})^2 = (a/\rpl)(\mpl / \mstar )$
    \citep[see][]{hansen:2007}.
}
\tablenotetext{j}{
    Incoming flux per unit surface area, averaged over the orbit.
}
\ifthenelse{\boolean{emulateapj}}{
    \end{deluxetable*}
}{
    \end{deluxetable}
}



\section{Discussion}
\label{sec:discussion}

We have presented the discovery of four new short period transiting systems from the HATSouth
 network. The systems were identified as planetary candidates using HATSouth photometric light curves
and then confirmed as planetary mass objects by measuring precise radial velocities for the host stars. The
precision of the transit parameters was also improved by using additional follow-up light curves obtained
with 1m-class telescopes.
We found that the four planets have orbital periods shorter than 5 days and masses in the Neptune to Jupiter mass range,
but all of them show radii similar to that of Jupiter.

These four new systems add to the valuable population of extrasolar planets transiting stars with precisely determined masses and radii. In the top panel of Figure~\ref{fig:rad} we show the planet radius as a function of the planet mass for the
population of transiting planets with uncertainties in \mpl\ and \rpl\ at the level of 35\%\footnotetext[1]{query to exoplanets.eu
for systems having reported values of \rstar, \teff, [Fe/H], and $a$}, and we have included our four new systems. The lower panel
of Figure~\ref{fig:rad} uses the same population of planets but in this case we plot the T$_{eq}$--\mpl\ diagram. While both diagrams
show that the physical properties of HATS-43b to HATS-46b are consistent with what is expected based on the distribution of
known transiting planets, we can point out a few interesting properties.

\ifthenelse{\boolean{emulateapj}}{
    \begin{figure*}[!ht]
}{
    \begin{figure}[!ht]
}
\plotone{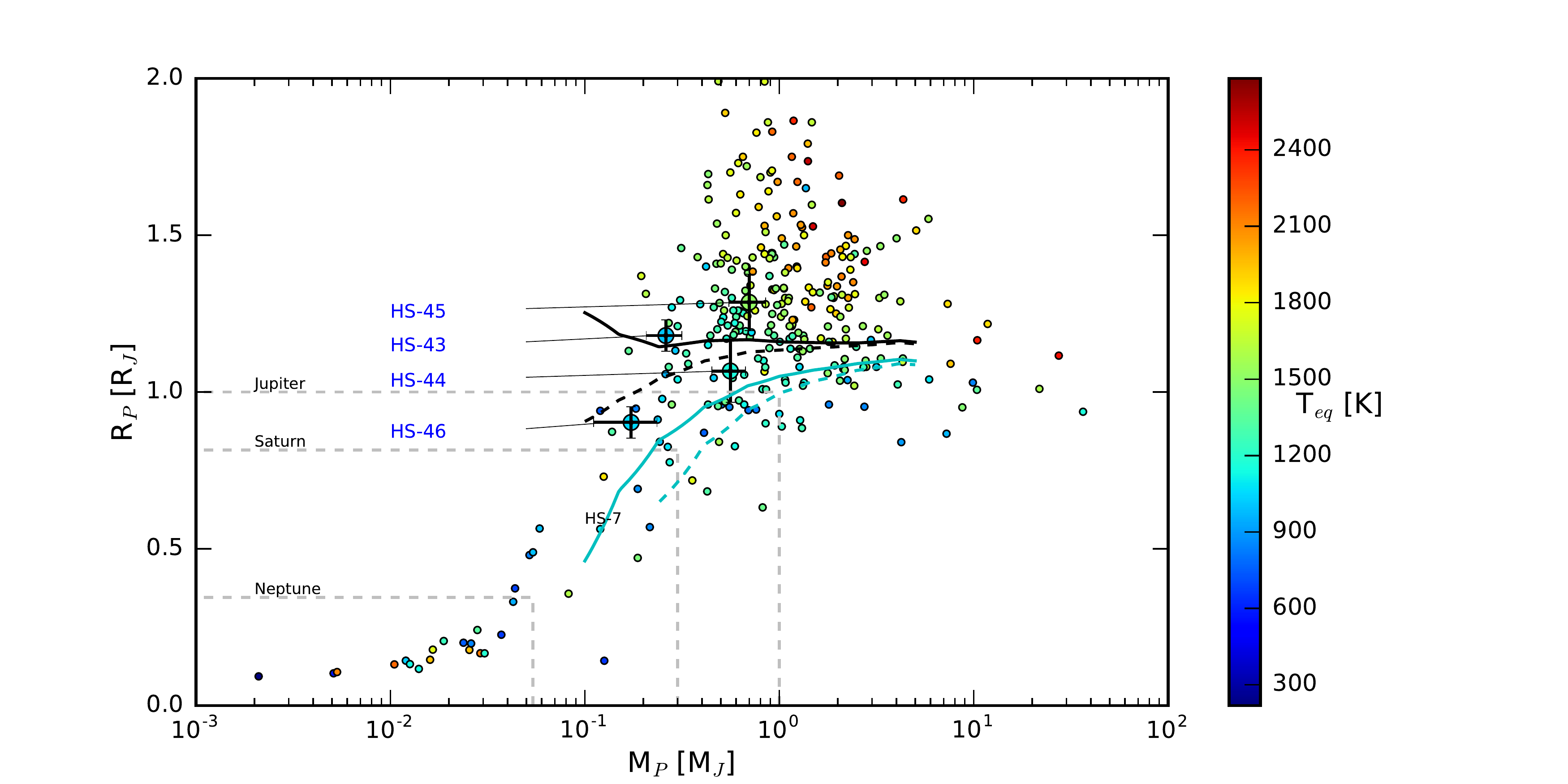}
\plotone{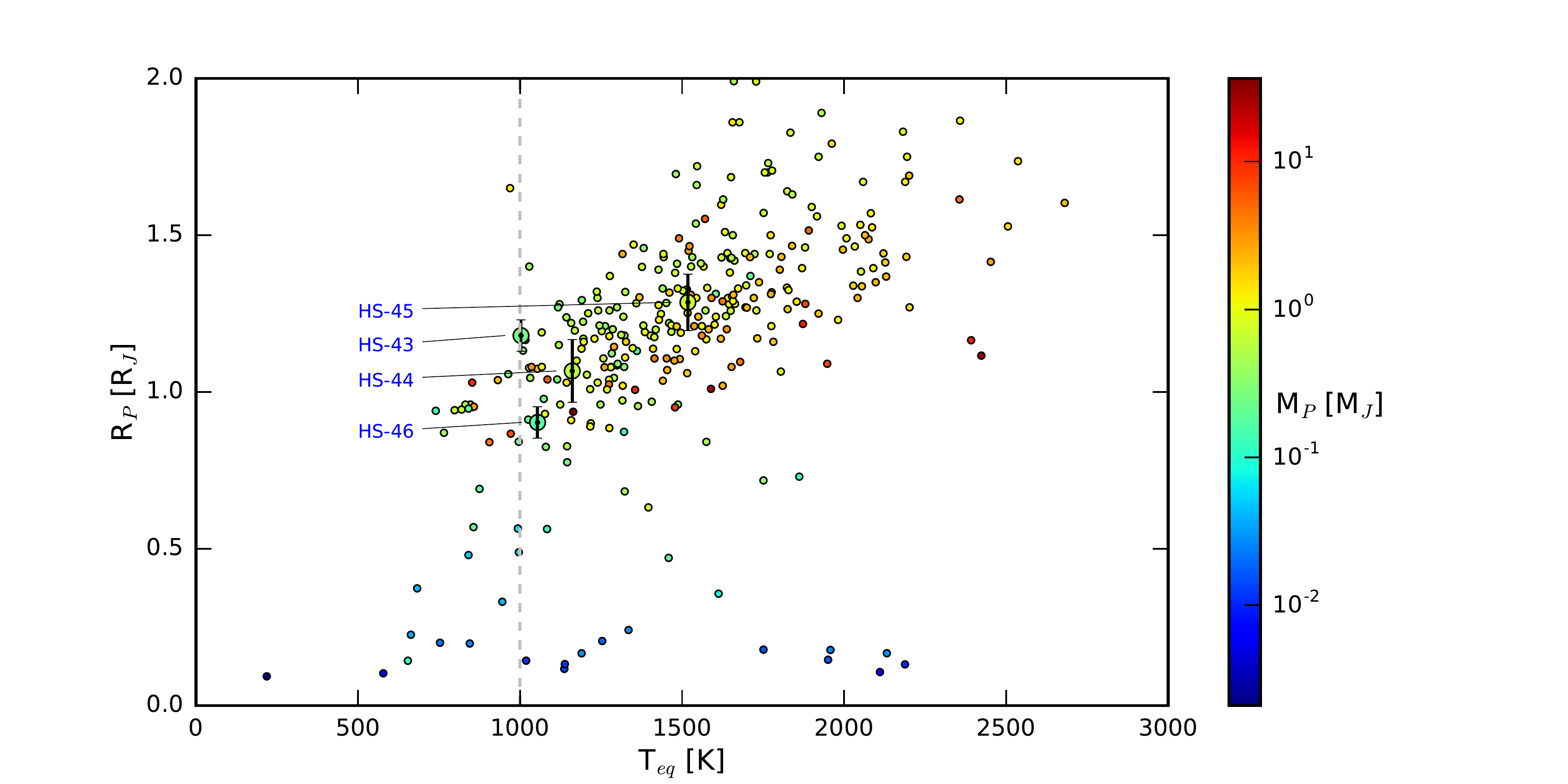}
\caption{
   Top: Planetary mass--radius diagram for the population of well characterised planets with masses and radii measured at the 35\% level.
   The planetary equilibrium temperature is colour coded. The big circles with error bars correspond to \hatcurb{43},  \hatcurb{44},  \hatcurb{45}, and  \hatcurb{46}.
   The plotted lines correspond to the \citet{fortney:2007} models for irradiated planets at 0.045 AU from the host star.
   The black lines represent planets with an age of 1 Gyr, while light blue lines represent planet with an age of 4.5 Gyr. From top to bottom
   the models contain core masses of 0, 10, 25 and 50 earth masses. Bottom: Planet radius as a function of the equilibrium temperature for
   the same systems considered in the upper panel. The dashed grey line corresponds to the temperature limit below which inflation
   mechanisms of hot Jupiters are not expected to play a major role. While the radii for \hatcurb{44}, \hatcurb{45}, and \hatcurb{46} clearly follow
   the empirical trend of increasing radius with the insolation level, \hatcurb{43} has a slightly larger radius that the one predicted from this empirical correlation.
 }

\label{fig:rad}

\ifthenelse{\boolean{emulateapj}}{
    \end{figure*}
}{
    \end{figure}
}

\subsection{\hatcurb{43}}
With a mass of \mpl = $\hatcurPPmlongeccen{43}$ \mjup\ and an equilibrium temperature of $T_{\rm eq}$ =  $\hatcurPPteffeccen{43}$ K that lies close
to the 1000 K limit proposed by \citet{kovacs:2010} below which planet radius is not expected to be strongly affected by stellar insolation, this planet has a radius of  \rpl = $\hatcurPPrlongeccen{43}$ \rjup,
which is particularly large if compared with other systems with similar properties. We can identify four other systems having masses and irradiation levels
consistent with the ones of \hatcurb{43}, namely 
HAT-P-19b \citep{hartman:2011:hat1819},
WASP-29b \citep{hellier:2010},
WASP-69b \citep{anderson:2014},
and HATS-5b \citep{zhou:2014:hats5}.
The properties of these systems are summarised in Table~\ref{sim43}.
\hatcurb{43} has the largest radius of this subsample. Even though the metallicity of \hatcur{43} is the lowest one,
which could hint to the absence of a central solid core and consequently a larger radius, the \citet{fortney:2007} models of planetary
structure predict a radius that is more than $2\sigma$ below the adopted value for \hatcurb{43}. On the other hand, \hatcurb{43}
stands out as the only system of the subsample having an eccentricity greater than 0.1. Given that the physical properties of
the stellar hosts of Table~\ref{sim43} are similar, this enhanced eccentricity directly translates in a greater tidal heating rate \citep{jackson:2008}
that could be the driving source responsible for the large radius of \hatcurb{43}.

The low density of HATS-43b makes this system an interesting target for atmospheric studies. Specifically, its expected transmission
spectroscopy signal of $\delta_{trans} = 2350$ ppm, is among the highest values form the full population of discovered transiting systems.
WASP-39b \citep{faedi:2011}, which  has a similar transmission spectroscopy signal ($\delta_{trans} = 2500$ ppm) and similar physical properties to \hatcurb{43},
has been the target of numerous atmospheric studies \citep{kammer:2015, fischer:2016, nikolov:2016} that find that this planet has a cloud free atmosphere with presence of Rayleigh scattering slope and Na and K absorption lines. \hatcurb{43} is a well suited comparison target to study the atmospheres of Saturn mass planets with even lower temperatures.

\ifthenelse{\boolean{emulateapj}}{
    \begin{deluxetable*}{ccccccc}
}{
    \begin{deluxetable}{ccccccc}
}
\tabletypesize{\scriptsize}
\tablecaption{Discovered transiting planets having reported \mpl\ and $T_{eq}$ values consistent with \hatcurb{43}\label{sim43}}

\tablehead{
    \multicolumn{1}{c}{Name}                   &
    \multicolumn{1}{c}{\mpl [\mjup]}          &
    \multicolumn{1}{c}{$T_{eq}$ [K]}         &
    \multicolumn{1}{c}{\rpl [\rjup]}              &
    \multicolumn{1}{c}{[Fe/H] [dex]}          &
    \multicolumn{1}{c}{e}                           &
    \multicolumn{1}{c}{H [$10^{19}$ W]\tablenotemark{a} }   \\
}
\startdata
\noalign{\vskip -3pt}
& & & & & & \\
~~~WASP-29b    & $ 0.244 \pm 0.020$ & $  980 \pm 40 $ & $  0.792^{+0.056}_{-0.035} $ & $ +0.11 \pm 0.14 $ & $0.03^{+0.05}_{-0.03}$ & $1.18$ \\
~~~WASP-69b    & $ 0.260 \pm 0.017$ & $  963 \pm 18 $ & $  1.057 \pm 0.047 $              & $ +0.15 \pm 0.08 $ & $<0.1\ at\ 2\sigma$       & $<16$             \\
~~~HAT-P-19b    & $ 0.292 \pm 0.020$ & $  1010 \pm 42 $ & $  1.132 \pm 0.072 $            & $ +0.23 \pm 0.08$  & $0.067 \pm 0.042$        & $14$     \\
~~~HATS-5b       & $ 0.237 \pm 0.012$ & $  1025 \pm 17 $ & $  0.912 \pm 0.025 $            & $ +0.19 \pm 0.08$  & $0.019 \pm 0.019$        & $0.2$ \\
& & & & & & \\
~~~HATS-43b      & $0.261 \pm 0.054$ & $  1003 \pm 27 $ & $  1.180 \pm 0.050 $            & $ +0.050 \pm 0.041 $  & $0.173 \pm 0.089$        & $52$ \\

\enddata

\tablenotetext{a}{
  Tidal heating rate \citep{jackson:2008}.
}

\ifthenelse{\boolean{emulateapj}}{
    \end{deluxetable*}
}{
    \end{deluxetable}
}

\subsection{\hatcurb{44} \& \hatcurb{45}}
\hatcurb{44} with a mass of  \mpl = $\hatcurPPmlong{44}$ \mjup\ and  a radius of \rpl = $\hatcurPPrlong{44}$ \rjup, and 
\hatcurb{45} with a mass of  \mpl = $\hatcurPPmlong{45}$ \mjup\ and  a radius of \rpl = $\hatcurPPrlong{45}$ \rjup\ are,
two sub-Jupiter mass planets that lie in relatively densely populated regions of the parameter space of transiting systems (see Figure~\ref{fig:rad}).
The most similar system to \hatcurb{44} in terms of planet mass and irradiation level, is WASP-34 \citep[\mpl=0.59$\pm$0.01 \mjup, $T_{eq}$=1160 K][]{smalley:2011},
which has a slightly larger radius of \rpl=1.22$\pm$0.10 \rjup\ which is in agreement with the proposed anti correlation between the planet radius and the
metallicity of the host start. WASP-34b has a moderately metal poor stellar host star ([Fe/H]=-0.02$\pm$0.10) if compared to \hatcur{44} ([Fe/H]=+0.32$\pm$0.07).
In the case of \hatcurb{45}, the most similar system is HAT-P-9b \citep[\mpl=0.78$\pm$0.09 \mjup, $T_{eq}$=1530 $\pm$ 40 K, ][]{shporer:2009}, which presents
a significantly inflated radius of \rpl=1.4 $\pm$ 0.06 \rjup, slightly larger but consistent with the one of \hatcurb{45}.

While the radius of \hatcurb{44} can be predicted by using the \citet{fortney:2007} models invoking a core-less structure,
the radius of \hatcurb{45} is larger than predicted, which can be expected due to the moderately high irradiation from its
F-type host star, where some of the proposed inflation mechanisms of hot Jupiters might be in play.

Even though both planets have expected transmission signals significantly smaller than \hatcurb{43}
($\delta_{trans} = 860$ ppm and $\delta_{trans} = 660$ ppm, for \hatcurb{44} and \hatcurb{45}, respectively),
there have been previous atmospheric studies of transiting planets with similar values of transmission signal
\citep[e.g. ][]{parviainen:2016, vonessen:2017}. Additionally, the $\vsini=$ 9.90$\pm$0.40 \kms of \hatcur{45}
makes of this system and interesting target for the determination of the obliquity through the measurement of the
Rossiter-McLaughlin effect. The expected semi-amplitude of the radial velocity anomaly for an aligned orbit is of K$_{RM}$ = 30 \ms.

\subsection{\hatcurb{46}}
Having a mass of \mpl = $\hatcurPPmlong{46}$ \mjup, \hatcurb{46} lies in the sparsely populated region of the parameter space of transiting systems
in the Neptune -- Saturn mass range, which corresponds to the transition zone between ice giants and gas giants. According to the core accretion theory
of giant planet formation \citep{pollack:1996}, planetesimals agglomerate to form rocky embryos that when reaching a  threshold mass of $\sim$10 M$_{\oplus}$, generate a
run away accretion of the surrounding gas of the protoplanetary disk that forms thick H/He dominated envelope (90\% in mass). One of the theoretical challenges of this
model is to understand how ice giants (like Uranus and Neptune) can avoid the accretion of the massive gaseous envelope. For this reason, the discovery of
transiting planets in the ice--gas transition range is important for determining which properties of the systems can play a major role in setting their structure and composition, which
can be then linked to different formation models. The large radius of \hatcurb{46} suggest that this planet is probably a low mass gas giant planet, and not a high mass ice giant.
By using the \citet{fortney:2007} models of planetary structure we find that  \hatcurb{46} should have a core mass of M$_c$=12$\pm$8 M$_{\oplus}$ to explain
its mass and radius, implying a $\sim80$\% H/He dominated composition. Among the population of discovered transiting systems orbiting main sequence stars 
with precise mass estimations from RVs, we can identify
Kepler-89d   \citep[0.16 \mjup, 0.98 \rjup, ][]{weiss:2013},
HATS-8b      \citep[0.14 \mjup, 0.87 \rjup, ][]{bayliss:2015}
HAT-P-48b,  \citep[0.17 \mjup, 1.30 \rjup, ][]{bakos:2016},
WASP-139b \citep[0.12 \mjup, 0.80 \rjup, ][]{hellier:2016}, and
WASP-107b \citep[0.12 \mjup, 0.94 \rjup, ][]{anderson:2017}
as other similar low mass gas giants, while
Kepler-101b \citep[0.16 \mjup, 0.51 \rjup, ][]{bonomo:2014}, 
HATS-7b     \citep[0.12 \mjup, 0.56 \rjup, ][]{bakos:2015},
K2-98b        \citep[0.10 \mjup, 0.38 \rjup, ][]{barragan:2016}, and
K2-27b        \citep[0.10 \mjup, 0.40 \rjup, ][]{petigura:2017}
are compatible with being high mass ice giants. These two groups of planets can be associated to different locations and/or times of formation. The first group of planets
could have been formed relatively close to the host star where the high temperature prevents the formation of icy planetesimals that pollute the planet composition.
On the other hand, the second group of planets could have formed farther away or  early in the disk lifetime where rocky planetesimals are still  present in profusion.
This suggested simple classification of these ten planets is based on the amount of heavy elements inferred from classical models of planetary structure, but there are several
additional factors that are not taken into account that can contribute to modify the planetary radius and mislead the determination of the planet metallicity, e.g. evaporation
\citep{owen:2013},
tidal heating \citep{jackson:2008}, or
collisions with other planets \citep{liu:2015}.
On the other hand, studies of atmospheric composition can be used to directly discriminate if these planets
have H/He dominated envelopes or if there is a significant presence of heavier elements, as was recently shown by \citet{wakeford:2017},
where a significantly metal depleted composition was estimated for the Neptune mass planet HAT-P-26b.
In this context, \hatcurb{46} has a prominent expected transmission signal of $\sim$1500 ppm, which should make of this system a valuable
target for atmospheric studies.



\acknowledgements 

Development of the HATSouth
project was funded by NSF MRI grant NSF/AST-0723074, operations have
been supported by NASA grants NNX09AB29G, NNX12AH91H, and NNX17AB61G, and follow-up
observations receive partial support from grant NSF/AST-1108686.
J.H.\ acknowledges support from NASA grant NNX14AE87G.
A.J.\ acknowledges support from FONDECYT project 1171208, BASAL CATA
PFB-06, and project IC120009 ``Millennium Institute of Astrophysics
(MAS)'' of the Millenium Science Initiative, Chilean Ministry of
Economy. N.E.\ is supported by BASAL CATA
PFB-06. R.B.\ and N.E.\ acknowledge support from project
IC120009 ``Millenium Institute of Astrophysics (MAS)'' of the
Millennium Science Initiative, Chilean Ministry of Economy.
V.S.\ acknowledges support form BASAL CATA PFB-06.  
This work is based on observations made with ESO Telescopes at the La
Silla Observatory.
This paper also uses observations obtained with facilities of the Las
Cumbres Observatory Global Telescope.
We acknowledge the use of the AAVSO Photometric All-Sky Survey (APASS),
funded by the Robert Martin Ayers Sciences Fund, and the SIMBAD
database, operated at CDS, Strasbourg, France.
Operations at the MPG~2.2\,m Telescope are jointly performed by the
Max Planck Gesellschaft and the European Southern Observatory.  The
imaging system GROND has been built by the high-energy group of MPE in
collaboration with the LSW Tautenburg and ESO\@.  
We thank the MPG 2.2m telescope support team for their technical assistance during observations.


\clearpage
\bibliographystyle{apj}
\bibliography{hatsbib}

\clearpage
\LongTables

%
%
\tabletypesize{\scriptsize}
\ifthenelse{\boolean{emulateapj}}{
    \begin{deluxetable*}{lrrrrrl}
}{
    \begin{deluxetable}{lrrrrrl}
}
\tablewidth{0pc}
\tablecaption{
    Relative radial velocities and bisector spans for \hatcur{43}--\hatcur{46}.
    \label{tab:rvs}
}
\tablehead{
    \colhead{BJD} &
    \colhead{RV\tablenotemark{a}} &
    \colhead{\ensuremath{\sigma_{\rm RV}}\tablenotemark{b}} &
    \colhead{BS} &
    \colhead{\ensuremath{\sigma_{\rm BS}}} &
    \colhead{Phase} &
    \colhead{Instrument}\\
    \colhead{\hbox{(2,450,000$+$)}} &
    \colhead{(\ms)} &
    \colhead{(\ms)} &
    \colhead{(\ms)} &
    \colhead{(\ms)} &
    \colhead{} &
    \colhead{}
}
\startdata
\multicolumn{7}{c}{\bf HATS-43} \\
\hline\\
$ 7325.68004 $ & \nodata      & \nodata      & $ -115.0 $ & $   17.0 $ & $   0.273 $ & FEROS \\
$ 7403.66458 $ & $    16.66 $ & $    12.00 $ & $   11.0 $ & $   16.0 $ & $   0.042 $ & FEROS \\
$ 7406.78213 $ & $    36.66 $ & $    30.00 $ & $ -176.0 $ & $   37.0 $ & $   0.752 $ & FEROS \\
$ 7498.50193 $ & $   -19.51 $ & $    16.00 $ & $  -21.0 $ & $   27.0 $ & $   0.651 $ & HARPS \\
$ 7637.79837 $ & $   -13.01 $ & $    27.60 $ & $   -7.0 $ & $   36.0 $ & $   0.389 $ & HARPS \\
$ 7638.81141 $ & $    29.19 $ & $    17.50 $ & $   47.0 $ & $   23.0 $ & $   0.620 $ & HARPS \\
$ 7639.85295 $ & $    47.39 $ & $    15.50 $ & $   -1.0 $ & $   20.0 $ & $   0.858 $ & HARPS \\
$ 7640.83377 $ & $     0.79 $ & $    13.80 $ & $   18.0 $ & $   18.0 $ & $   0.081 $ & HARPS \\
$ 7707.75474 $ & $   -47.81 $ & $    20.10 $ & $   36.0 $ & $   26.0 $ & $   0.329 $ & HARPS \\
$ 7708.70649 $ & $    10.49 $ & $    10.20 $ & $  -19.0 $ & $   13.0 $ & $   0.546 $ & HARPS \\
$ 7714.74483 $ & $     2.16 $ & $    15.20 $ & $  -57.0 $ & $   21.0 $ & $   0.922 $ & FEROS \\
$ 7733.71407 $ & $   -20.44 $ & $    10.70 $ & $  -32.0 $ & $   15.0 $ & $   0.244 $ & FEROS \\
$ 7735.81057 $ & $    36.66 $ & $     9.90 $ & $  -10.0 $ & $   15.0 $ & $   0.722 $ & FEROS \\
$ 7736.55523 $ & $    37.86 $ & $    11.80 $ & $   11.0 $ & $   16.0 $ & $   0.891 $ & FEROS \\
$ 7737.83294 $ & $    -4.84 $ & $    11.30 $ & $  -17.0 $ & $   16.0 $ & $   0.182 $ & FEROS \\
$ 7738.62888 $ & $   -46.74 $ & $    11.00 $ & $   46.0 $ & $   16.0 $ & $   0.364 $ & FEROS \\
$ 7739.67457 $ & $   -11.04 $ & $    12.10 $ & $  -68.0 $ & $   17.0 $ & $   0.602 $ & FEROS \\
$ 7740.55295 $ & $    46.66 $ & $    11.40 $ & $  -64.0 $ & $   16.0 $ & $   0.802 $ & FEROS \\
$ 7742.71078 $ & $   -83.74 $ & $    10.70 $ & $  -15.0 $ & $   15.0 $ & $   0.294 $ & FEROS \\
\cutinhead{\bf HATS-44}
$ 7325.65952 $ & $    74.10 $ & $    17.00 $ & $  -12.0 $ & $   23.0 $ & $   0.790 $ & FEROS \\
$ 7328.66737 $ & $   105.10 $ & $    25.00 $ & $    7.0 $ & $   32.0 $ & $   0.886 $ & FEROS \\
$ 7329.69814 $ & $   -70.90 $ & $    24.00 $ & $   -8.0 $ & $   30.0 $ & $   0.262 $ & FEROS \\
$ 7403.68598 $ & $     6.10 $ & $    15.00 $ & $   25.0 $ & $   19.0 $ & $   0.226 $ & FEROS \\
$ 7404.73356 $ & $   100.10 $ & $    14.00 $ & $ -102.0 $ & $   18.0 $ & $   0.608 $ & FEROS \\
$ 7406.58676 $ & $  -143.90 $ & $    16.00 $ & $  -65.0 $ & $   21.0 $ & $   0.284 $ & FEROS \\
$ 7410.60096 $ & $    70.10 $ & $    18.00 $ & $ -199.0 $ & $   23.0 $ & $   0.747 $ & FEROS \\
$ 7714.71960 $ & $    78.70 $ & $    20.90 $ & $  -74.0 $ & $   28.0 $ & $   0.581 $ & FEROS \\
$ 7735.55952 $ & $   -41.90 $ & $    14.90 $ & $  -46.0 $ & $   19.0 $ & $   0.176 $ & FEROS \\
$ 7736.61332 $ & $    -9.20 $ & $    14.00 $ & $ -128.0 $ & $   19.0 $ & $   0.560 $ & FEROS \\
$ 7738.73011 $ & $   -82.20 $ & $    14.10 $ & $  -15.0 $ & $   19.0 $ & $   0.331 $ & FEROS \\
$ 7739.77508 $ & $    88.70 $ & $    15.40 $ & $  -25.0 $ & $   21.0 $ & $   0.712 $ & FEROS \\
$ 7740.78263 $ & $   -57.60 $ & $    20.00 $ & $  227.0 $ & $   26.0 $ & $   0.079 $ & FEROS \\
$ 7741.75344 $ & $  -147.70 $ & $    15.30 $ & $  -10.0 $ & $   21.0 $ & $   0.433 $ & FEROS \\
$ 7742.77786 $ & $    97.90 $ & $    14.80 $ & $  -92.0 $ & $   20.0 $ & $   0.807 $ & FEROS \\
\cutinhead{\bf HATS-45}
$ 7030.85122 $ & $    17.19 $ & $    33.00 $ & $  -20.0 $ & $   14.0 $ & $   0.557 $ & FEROS \\
$ 7031.79435 $ & $   -38.81 $ & $    27.00 $ & $    4.0 $ & $   13.0 $ & $   0.783 $ & FEROS \\
$ 7033.76761 $ & $  -103.81 $ & $    26.00 $ & $   33.0 $ & $   12.0 $ & $   0.254 $ & FEROS \\
$ 7036.75613 $ & $   -19.81 $ & $    27.00 $ & $   67.0 $ & $   13.0 $ & $   0.967 $ & FEROS \\
$ 7049.71493 $ & $   -77.81 $ & $    28.00 $ & $   66.0 $ & $   14.0 $ & $   0.062 $ & FEROS \\
$ 7050.79282 $ & $  -107.81 $ & $    27.00 $ & $   23.0 $ & $   13.0 $ & $   0.319 $ & FEROS \\
$ 7053.79937 $ & $    46.19 $ & $    31.00 $ & $  -20.0 $ & $   14.0 $ & $   0.037 $ & FEROS \\
$ 7054.69765 $ & $  -115.81 $ & $    60.00 $ & \nodata      & \nodata      & $   0.252 $ & FEROS \\
$ 7057.63438 $ & $    -2.81 $ & $    24.00 $ & \nodata      & \nodata      & $   0.953 $ & FEROS \\
$ 7067.58377 $ & $   -71.01 $ & $    35.00 $ & $ -160.0 $ & $   26.0 $ & $   0.329 $ & HARPS \\
$ 7068.58640 $ & $    49.99 $ & $    38.00 $ & $   66.0 $ & $   29.0 $ & $   0.569 $ & HARPS \\
$ 7069.61845 $ & $   104.99 $ & $    36.00 $ & $   59.0 $ & $   26.0 $ & $   0.815 $ & HARPS \\
$ 7070.57795 $ & $   -32.01 $ & $    46.00 $ & $  127.0 $ & $   33.0 $ & $   0.044 $ & HARPS \\
$ 7071.59384 $ & $   -81.01 $ & $    24.00 $ & $   82.0 $ & $   17.0 $ & $   0.287 $ & HARPS \\
$ 7072.57569 $ & $    22.99 $ & $    37.00 $ & $   99.0 $ & $   26.0 $ & $   0.521 $ & HARPS \\
$ 7403.76988 $ & $    26.19 $ & $    27.00 $ & $   55.0 $ & $   13.0 $ & $   0.610 $ & FEROS \\
$ 7404.79556 $ & $   -12.81 $ & $    35.00 $ & $   49.0 $ & $   16.0 $ & $   0.855 $ & FEROS \\
$ 7447.70623 $ & $    52.19 $ & $    31.00 $ & $   80.0 $ & $   15.0 $ & $   0.102 $ & FEROS \\
\cutinhead{\bf HATS-46}
$ 7182.81976 $ & $   -31.91 $ & $    14.00 $ & $  -77.0 $ & $   19.0 $ & $   0.121 $ & FEROS \\
$ 7187.88874 $ & $   -42.91 $ & $    14.00 $ & $  -20.0 $ & $   19.0 $ & $   0.189 $ & FEROS \\
$ 7189.86037 $ & $    23.09 $ & $    12.00 $ & $    6.0 $ & $   17.0 $ & $   0.605 $ & FEROS \\
$ 7192.88690 $ & $    12.09 $ & $    14.00 $ & $   29.0 $ & $   19.0 $ & $   0.243 $ & FEROS \\
$ 7198.91703 $ & $    10.13 $ & $     9.89 $ & \nodata      & \nodata      & $   0.515 $ & PFS \\
$ 7198.92817 $ & $    23.52 $ & $     9.70 $ & \nodata      & \nodata      & $   0.517 $ & PFS \\
$ 7199.92164 $ & $    57.98 $ & $     9.29 $ & \nodata      & \nodata      & $   0.727 $ & PFS \\
$ 7199.93228 $ & $    31.21 $ & $     9.31 $ & \nodata      & \nodata      & $   0.729 $ & PFS \\
$ 7200.92982 $ & $    28.89 $ & $     8.26 $ & \nodata      & \nodata      & $   0.939 $ & PFS \\
$ 7200.94141 $ & $     6.75 $ & $    12.87 $ & \nodata      & \nodata      & $   0.942 $ & PFS \\
$ 7202.90860 $ & $     0.29 $ & $     7.76 $ & \nodata      & \nodata      & $   0.357 $ & PFS \\
$ 7202.91989 $ & $   -25.30 $ & $    10.02 $ & \nodata      & \nodata      & $   0.359 $ & PFS \\
$ 7203.90169 $ & $    -4.99 $ & $     9.79 $ & \nodata      & \nodata      & $   0.566 $ & PFS \\
$ 7203.91219 $ & $   -37.67 $ & $    10.21 $ & \nodata      & \nodata      & $   0.568 $ & PFS \\
$ 7204.94090 $ & $    -2.99 $ & $     7.94 $ & \nodata      & \nodata      & $   0.785 $ & PFS \\
$ 7210.89718 $ & $    13.09 $ & $    13.00 $ & $   36.0 $ & $   17.0 $ & $   0.041 $ & FEROS \\
$ 7211.91925 $ & $    -4.91 $ & $    12.00 $ & $  -30.0 $ & $   16.0 $ & $   0.257 $ & FEROS \\
$ 7212.89415 $ & $   -29.91 $ & $    11.00 $ & $  -30.0 $ & $   15.0 $ & $   0.462 $ & FEROS \\
$ 7220.79108 $ & $   -83.91 $ & $    14.00 $ & $  -47.0 $ & $   19.0 $ & $   0.127 $ & FEROS \\
$ 7228.71214 $ & $    64.09 $ & $    12.00 $ & $   16.0 $ & $   16.0 $ & $   0.798 $ & FEROS \\
$ 7230.86819 $ & $    -7.91 $ & $    10.00 $ & $    0.0 $ & $   13.0 $ & $   0.252 $ & FEROS \\
$ 7327.60315 $ & $    55.09 $ & $    13.00 $ & $  -31.0 $ & $   18.0 $ & $   0.650 $ & FEROS \\
$ 7570.88719 $ & $   -32.91 $ & $    14.00 $ & $   38.0 $ & $   19.0 $ & $   0.950 $ & FEROS \\
$ 7576.86278 $ & $    19.09 $ & $    13.00 $ & $   -3.0 $ & $   18.0 $ & $   0.210 $ & FEROS \\
$ 7585.86394 $ & $    54.49 $ & $    10.90 $ & $   22.0 $ & $   15.0 $ & $   0.108 $ & FEROS \\
$ 7591.85299 $ & $    30.49 $ & $    14.10 $ & $   -9.0 $ & $   16.0 $ & $   0.371 $ & FEROS \\
$ 7593.76470 $ & $     0.89 $ & $    11.10 $ & $  -27.0 $ & $   15.0 $ & $   0.774 $ & FEROS \\
$ 7612.69088 $ & $    27.29 $ & $    12.20 $ & $  -29.0 $ & $   17.0 $ & $   0.765 $ & FEROS \\
$ 7614.64079 $ & $   -45.31 $ & $    12.80 $ & $  -17.0 $ & $   17.0 $ & $   0.176 $ & FEROS \\
$ 7647.80684 $ & $   -26.71 $ & $    12.10 $ & $   -3.0 $ & $   16.0 $ & $   0.170 $ & FEROS \\
$ 7727.68958 $ & $    34.69 $ & $    12.50 $ & $  -95.0 $ & $   17.0 $ & $   0.014 $ & FEROS \\
\enddata
\tablenotetext{a}{
    The zero-point of these velocities is arbitrary. An overall offset
    $\gamma_{\rm rel}$ fitted independently to the velocities from
    each instrument has been subtracted.
}
\tablenotetext{b}{
    Internal errors excluding the component of astrophysical jitter
    considered in \refsecl{globmod}.
}
\ifthenelse{\boolean{rvtablelong}}{
    \tablecomments{
    }
}{
    \tablecomments{
    }
} 
\ifthenelse{\boolean{emulateapj}}{
    \end{deluxetable*}
}{
    \end{deluxetable}
}

\end{document}